\documentclass[12pt,preprint]{aastex}

\shorttitle{Recurrent Novae}
\shortauthors{Schaefer}
\begin{document}

\title{Comprehensive Photometric Histories of All Known Galactic Recurrent Novae}
\author{Bradley E. Schaefer}
\affil{Physics and Astronomy, Louisiana State University,
    Baton Rouge, LA, 70803, USA; schaefer@lsu.edu}

\begin{abstract}

I collect virtually all photometry of the ten known galactic recurrent novae (RNe) and their 37 known eruptions.  This consists of my modern measures of nearly all archival plates (providing the only data for half of 37 known eruptions), my own 10,000 CCD magnitudes from 1987 to present (providing virtually all of the magnitudes in quiescence for seven RNe), over 140,000 visual magnitude estimates recorded by amateur astronomers (who discovered half the known eruptions), and the small scattering of magnitudes from all the literature.  From this, I produce various uniform products; (1) BVRIJHK comparison star magnitudes and BV comparison star sequences to cover the entire range of eruption, (2) complete light curves for all eruptions, (3) best fit B and V light curve templates, (4) orbital periods for all-but-one RN, (5) exhaustive searches for all missed eruptions, (6) measured discovery efficiencies since 1890, (7) true recurrence time scales, (8) predicted next eruption dates, (9) variations on time scales of minutes, hours, days, months, years, decades, and century, (10) uniform distances and extinctions to all RNe, (11) BV colors at peak and UBVRIJHK colors at minimum all with extinction corrections, and (12) the spectral energy distributions over UBVRIJHK.  Highlights of this work include the discoveries of one new RN, six previously-undiscovered eruptions, and the discovery of the orbital periods for half the RNe.  The goal of this work is to provide uniform demographics for answering questions like the `What is the death rate of RNe in our galaxy?' and `Are the white dwarfs gaining or losing mass over each eruption cycle?'.  An important use of this work is for the question of whether RNe can be the progenitors of Type Ia supernovae.

\end{abstract}

\tableofcontents

\section{Introduction}

	Recurrent novae (RNe) are ordinary novae systems for which the recurrence time scale happens to be less than roughly one century.  That is, RNe are binary stars where matter accretes from a donor star onto the surface of a white dwarf, where the accumulated material will eventually start a thermonuclear explosion that makes the nova eruption.  The ordinary nova systems for which only one eruption has been detected are labeled as `classical novae' (CNe), even though some of them might really be RNe for which all-but-one eruption from the last century has been missed.  The nova systems have a continuum of recurrence time scales, ranging from roughly a decade to perhaps 100,000 years or longer, with the division at roughly one century being purely based on observational expedience.
	
		Why do some nova systems have a recurrence time scale ($\tau_{rec}$) of less than a century?  Here, theory provides a confident answer that two conditions are required to get short inter-eruption times.  The first requirement is that the white dwarf mass must be near the Chandrasekhar mass.  The reason is that the high white dwarf mass makes for a high surface gravity which allows even a relatively small amount of matter to reach the kindling temperature for thermonuclear runaway.  The limit is that the mass of the white dwarf ($M_{wd}$) must be greater than something like 1.2 $M_{\odot}$.  The second requirement is that the mass accretion rate ({$\dot{M}$) must be high.  The reason is that a high $\dot{M}$ is the only way to get enough material onto the white dwarf to reach the runaway condition during the short inter-eruption time interval.  Indeed, the RNe have an accretion rate higher than most cataclysmic variables, with values close to $10^{-7}$ $M_{\odot}$ yr$^{-1}$ (Hachisu \& Kato 2000; 2001; Webbink et al. 1987; Patterson et al. 1998; Knigge, King, \& Patterson 2000; Hachisu et al. 2000; Hachisu, Kato, \& Schaefer 2003).
		
	So the RNe have white dwarfs near the Chandrasekhar mass and matter is being piled on at a very high rate.  With this situation, it is possible that the white dwarf will soon be driven to the Chandrasekhar mass and then collapse as a Type Ia supernova.  As such, RNe are a likely answer to the perennial Type Ia progenitor question.  Often in the literature, discussion does not go past this simple argument.  But there is a major loophole in that the mass ejected during each nova event has not been accounted for.  If the white dwarf ejects more mass during an eruption than it has accreted during the previous inter-eruption interval, then the white dwarf will not be gaining mass and will not collapse as a Type Ia explosion.  So the question really comes down to evaluating the recurrence time scale ($\tau_{rec}$), the mass accretion rate ($\dot{M}$), the mass ejected by the nova event ($M_{ejecta}$), and then whether $M_{ejecta}>\tau_{rec}\dot{M}$.  It turns out that the ejecta mass has a very large uncertainty (over two orders of magnitude) from both observational and theoretical work.  Also, my recent work (see below) has demonstrated that  $\tau_{rec}$ generally has errors of a factor of three and $\dot{M}$ has large uncertainties.  Until we can answer the question of whether $M_{ejecta}>\tau_{rec}\dot{M}$, RNe can only remain a good idea as a solution for the progenitor problem.
	
	Another potential problem for the RNe answer is whether there are enough RNe in our galaxy and Local Group to produce the Type Ia events.  That is, is the RN death rate equal to the Type Ia supernova rate?  Published answers to date appear to say `no' both from a theoretical population synthesis (Branch et al. 1995) and an observational basis (della Valle \& Livio 1996).  (Unfortunately, the answer appears to be `no' for all candidate classes of progenitors, see Branch et al. 1995)  But the theoretical basis for the `no' to RNe is suspect as population synthesis analysis has very large uncertainties when dealing with rare evolutionary paths, especially after common envelope binary evolution has occurred.  This distrust of population synthesis for evaluating rates in such cases is further emphasized by studies of the fraction of 3-8 M$_{\odot}$ stars that must undergo Type Ia explosions is orders of magnitude higher than theory would give (Maoz 2008).  And the observational `no' is also suspect because the realities of discovery statistics were ignored and these can easily increase the RN death rate by a factor of 100 or more.  The way to answer the RNe death rate question can only be to get better demographics for RNe (including their numbers, light curves, accretion rates, and recurrence time scales) as well as to determine their real discovery statistics.
	
	The Type Ia progenitor question is an old one, with major reviews from each of the last four decades appearing in Livio (2000), Branch et al. (1995), Trimble (1984), and Whelan \& Iben (1973).  Even today, extensive debates rage as to which of many systems account for most of the Type Ia events.  The primary dividing line is whether the progenitors have one or two white dwarfs in the binary.  The double degenerate idea has two white dwarfs in close orbit for which angular momentum loss by gravitational radiation grinds the orbit down until the two white dwarfs collide and exceed the Chandrasekhar mass resulting in a supernova explosion.  No double white dwarf system with adequate combined mass and a short enough orbital period has ever been identified, despite massive efforts, so we cannot point to any star in the sky as being a progenitor of this class (Robinson \& Shafter 1987; Napiwotzki et al. 2001).  For example, even the star KPD 1930+275 with a white dwarf in a binary of total mass 1.47$\pm$0.01 M$_{\odot}$ (Maxted et al. 2000, but see Geier et al. 2007) will lose enough mass so as to avoid exceeding the Chandrasekhar mass (Ergma et al. 2001).  The single degenerate model has a mass donor star feeding matter onto a white dwarf which eventually exceeds the Chandrasekhar mass also resulting in a supernova explosion.  For this model, we can point to several populous classes of stars which are candidate progenitors including the RNe, symbiotic stars, and super-soft sources.  These classes have substantial overlap within their populations.  With these many possibilities all of which presumably work, the question comes down to which is the dominant progenitor.  And with the possibility of two roughly equal populations of Type Ia supernovae (Sullivan et al. 2006), there might even be two dominant progenitor classes.
	
	The progenitor question has recently come to the forefront as being of high importance.  The reason is that Type Ia supernovae have become a premier tool for cosmology, for example with the Hubble diagram providing the first and strongest evidence for the existence of Dark Energy.  But supernova cosmology is based on the idea that the luminosity-decline relation has no evolution with redshift.  If $z\sim1$ supernovae systematically have lower metallicity (appropriate for their long lookback time) then their envelopes will have systematically lower opacity with an inevitable change in the luminosity-distance relation and a systematic distortion of the derived distances.  Such systematic effects will lead to a change in the shape of the Hubble diagram, with this error perhaps being interpreted as evidence for Dark Energy.  Dom\'inguez et al. (2001) have demonstrated that the range of evolutionary effects (from systematic changes in redshift of both the metallicity and the main sequence mass of the progenitor) can lead to typical errors of $\sim$0.2 mag out to $z=1$.  The supernova researchers only argue against evolution with their evidence that the high-z events have similar spectra to low-z events (Hook et al. 2005), although this is not proof of the lack of evolution at less than the ~0.2 mag level.  The question of evolution is critical as the amount of evolution at $z\sim1$ is comparable to the differences between cosmologies with and without Dark Energy.  Without knowing the identity of the progenitor, evolution calculations are not possible and the effect can significantly change the shape of the Hubble diagram.  So in principle, the progenitor problem is critical for the entire supernova cosmology enterprise.  And it will only become more important, as future experiments (like the SNAP satellite) are seeking to measure the Hubble diagram with an accuracy of ~0.01 mag, a level at which evolution effects can dominate and perhaps make the goal unreachable.  So a solution to the progenitor problem has wide importance.
	
	The solution can come only with a broad investigation of the various candidate progenitors.  A part of this will be the study of RNe, with particular attention to questions relating to whether $M_{ejecta}>\tau_{rec}~\dot{M}$ and the value for the RN death rate.  These studies are essentially demographic investigations that must examine many fine details of all the known RNe systems.
	
	The only comprehensive review or summary of RNe is the influential and superb paper by Webbink et al. (1987).  But a lot has been learned since that time.  Unfortunately, all of the many subsequent observations are widely scattered.  For later reviews, we have a conference proceeding on one specific eruption (Evans et al. 2008), two brief overviews in conference proceedings (Anupama 2002; 2008), and one short summary (Sekiguchi 1995).  A deeper problem is that all of the photometry from before 1970 or so has systematically wrong magnitudes due to universal errors in the magnitude scale, and thus the majority of RNe events have only poorly measured properties.
	
	Over many years, I have been collecting a large database of observations of all the known galactic RNe, both in quiescence and in eruption.  For example, I have measured comparison star magnitudes and sequences so that old and new magnitudes can be correctly placed onto a uniform modern magnitude scale.  I have also remeasured all the archival photographic plates of all the pre-1970 eruptions, so that the light curves are directly comparable.  I have made exhaustive searches of the archival material for previously undiscovered eruptions.  I have extensive long-term light curves for the RNe at quiescence.  I have 47 and 80 eclipse times on both sides of the latest eruptions of U Sco and CI Aql respectively.  The easy-to-point-to results are the discovery of one new RN system, the identification of six previously unknown eruptions, the discovery of five of the orbital periods, the confident measure of the orbital period change across two eruptions, and the prediction of the dates of the next eruptions of U Sco and T Pyx.
	
	Now is the time for a comprehensive review of RNe photometry because of (a) the importance of the progenitor problem, (b) the lack of recent comprehensive reviews, (c) the scattered nature of the data in the literature, (d) the now-corrected calibration problems with the old magnitudes, and (e) the new data.  This paper seeks to fill the need by providing a comprehensive set of new data, remeasured old data, and data from the literature.  The goal is to provide a uniform analysis on the various properties of all known galactic RNe and the demographics of the group.
	
\section{The Known Galactic Recurrent Novae}

	What are the known RNe in our own galaxy?  This is an important question because the best census is needed for valid demographic studies.  The question of which stars are RNe has sometimes been uncertain.  One source of confusion is whether a particular eruption is a nova event (involving thermonuclear runaway on the surface of the white dwarf) or a dwarf nova event (involving a brightening of the accretion disk).  Dwarf novae with rare high-amplitude eruptions can be easily confused with RNe, as can CNe that also display dwarf nova eruptions.
	
	There is a long history of claims for stars being recurrent novae.  One of the first discovered variable stars is Mira, and it was called a nova when its first peak was recognized and then a second eruption was spotted.  The modern idea of `recurrent nova' had to await the separation of nova events as based on spectroscopy.  At various times, several dwarf novae (including WZ Sge, VY Aqr, RZ Leo, and V1195 Oph) have been identified as RNe (Webbink et al. 1987).  The two eruptions of V616 Mon (A0620-00) pointed to an unusual RN (Webbink et al. 1987), but we now know this system actually has a black hole as the compact star.  V529 Ori has a ``remarkably checkered history" and it is now unclear whether the reported events are of the same star or even were nova eruptions (Webbink et al. 1987).
	
	V1017 Sgr is an interesting case, with eruptions in 1901, 1919, 1973, and 1991, with a history and details in Webbink et al. (1987).  The eruption in 1919 has a peak of seventh magnitude (or somewhat brighter) and a light curve shape that is typical for novae events.  But the other three eruptions all peak at fainter than tenth magnitude and have light curve shapes typical of dwarf novae events.  The small brightenings are confirmed as being dwarf nova events by their spectroscopic development (Webbink et al. 1987; Vidal \& Rogers 1974).  The hybrid nature of V1017 Sgr has the strong precedent from a few other hybrid systems.  Most famously GK Per had a prototypical nova eruption in 1901 and has now had 14 dwarf novae events starting in 1966.  So V1017 Sgr is not a RN.
	
	As old novae (labeled as being CN) are having their second nova event discovered, the numbers of known RNe are steadily increasing.  Thus, V394 CrA, V745 Sco, V3890 Sgr, CI Aql, and IM Nor have been recognized as RNe with second eruptions in 1987, 1989, 1990, 2000, and  2002.  I expect that this discovery rate will speed up as more old novae are eligible to get their second eruption.  These newly discovered systems usually involve stars where the first eruption is poorly observed.  The known RN population is now dominated by systems with only moderate coverage and little recognition by researchers.
	
	The currently known RN are T Pyx, IM Nor, CI Aql, V2487 Oph, U Sco, V394 CrA, T CrB, RS Oph, V745 Sco, and V3890 Sgr.  I have ordered these systems by their orbital period (as finalized later in this paper), even though the period of V2487 Oph is only approximately known.  With one of the RN systems being proven only a few months ago (V2487 Oph), I realize that this list will soon have more additions.
	
	Table 1 provides a list of these RNe, along with their primary properties.  The second and third columns give the peak and quiescent brightness in the V-band ($V_{max}$ and $V_{min}$) as taken from Section 5.  The fourth column gives the time it takes for the V-band light curve to fade by three magnitudes from the peak brightness ($t_3$) as taken from Section 6.  The next column gives the orbital period for the binary system as taken from Section 10.  The final column gives the years of all known eruptions as taken from Section 3.
	
	Several properties seen in Table 1 demonstrate a stark contrast between RNe and CNe.  The RNe median amplitude is 8.6 mag (and none larger than 11.2 mag), in strong contrast with the CNe with most having amplitudes of $>$12 mag.   The RNe are shorter in eruption duration (median $t_3$ of 11 days) than the CNe (median $t_3$ of 44 days).  Almost all the classical novae have orbital periods ($P_{orbital}$) of less than 0.3 days, whereas 80\% of the RNe have longer orbital periods (with four of them having values of near one year).  On average, the recurrent and classical nova populations have substantially different properties even if there is much overlap, with the RNe being generally low-amplitude, fast-declining eruptions in systems with long-orbital periods.
	
	\subsection{Subdivision of the Recurrent Novae}
	
	The RNe present us with a wide diversity of properties.  It is tempting to seek divisions and categories that make sense for organizing this diversity.  The divisions might be constructed for two purposes: first to identify the range of properties, and second to seek bimodality (or trimododality) which will indicate separate physical processes or histories.  In the past, workers have tried to make divisions based on the orbital period (or equivalently, the luminosity class of the companion), the eruption duration (or equivalently, the $t_3$ value), and the light curve shape (specifically, the presence of a plateau in the tail of the eruption).  
	
	Schaefer (1990) and Schaefer \& Ringwald (1995) pointed to the fact that the orbital periods come in three separate groups: very short periods with low-mass main sequence star companions (like almost all other cataclysmic variables), periods comparable to one day (with subgiant companions), and very long periods comparable to one year (with red giant companions).  Such a division might be relevant for pointing to separate evolutionary paths leading to RNe, but no discussion or models have been presented to explain the three separate paths that lead to a three-fold division.  A three-fold division might have consequences relating to the interaction of the nova with the circumstellar environment, for example where a red giant companion might have a wind that can produce hard x-rays when the nova shell plows through it.  But even for this, the RNe with red giants have widely varying wind densities, while there is no distinction between the two shorter period classes because both have no sensible winds.  The size of the companion star will determine how much hydrogen appears in the spectrum of Type Ia supernovae, should these systems end with such an explosion.  Finally, a three-fold division will produce differing spectra in quiescence, as the different luminosity classes for the companion star will be revealed by their spectra.  Despite the tempting nature of this three-fold division, I do not know how to evaluate the significance of a tri-modal period distribution (see Table 1 and Figure 1), and with the small number of known RNe, I suspect that the significance is low.  In principle, the size of the companion star is irrelevant to the eruption physics (which likely depends only on the white dwarf mass, the composition of the accreted material, and perhaps on the accretion rate).  Most importantly, the three-fold division does not correlate with any of the eruption properties.  That is, the eruption light curves, the peak absolute magnitudes, the spectral developments, the elemental abundances, the amplitudes, the accretion rates, the speed classes, the recurrence time scales, and the galactic positions all have no relation or correlation with the size of the companion star. In all, the three-fold division is relevant only for peripheral issues concerning RNe.
	
	Most RNe are very fast in decline, so it might be tempting to split off the relatively slow decline events (T Pyx, IM Nor, and CI Aql) into their own class (Duerbeck et al. 2002).  Kato et al. (2002) suggested a variant on this by separating out just IM Nor and CI Aql into a subdivision based on similarities in the light curve and spectra.  Some division based on $t_3$ would be relevant for the eruption physics, but I see no evidence for anything other than a continuous distribution of RNe durations.  Figure 1 displays a plot of $P_{orbital}$ versus $t_3$ for the ten RNe, and I see no significant groupings or bimodality.  
	
	Hachisu et al. (2000a; 2002; 2009) point to a plateau in the tail of light curves as indicating a RN, with the plateau caused by an extended phase of hydrogen burning.  No demographic study of nova plateaus has been made, although I will use my templates in Section 6 to quantify the situation for plateaus in RNe.  The result will be that 60-90\% of the RNe are seen to have plateaus.  Most critically, the RNe with plateaus (T Pyx, IM Nor, CI Aql, V2487 Oph, U Sco, and RS Oph) creates a group that has no correlation with any other categorization ideas.  
	
	Kenyon (1986) has separated out T CrB and RS Oph as being in a separate class of symbiotic nova.  Symbiotic stars are defined by him to be those with absorption features of a late type giant star plus hydrogen or helium emission lines from some hot region.  T CrB, RS Oph, V745 Sco, and V3890 Sgr certainly meet this definition.  Nevertheless, Miko{\l}ajewska (2009) points out that these RNe are at the extreme edge of the range of properties of symbiotic stars, with short orbital periods, high mass white dwarfs, and very small mass ratios.  Symbiotic novae are defined by Kenyon to be a {\it very slow} eruption involving thermonuclear burning of hydrogen on a symbiotic star.  While the RNe might have nova eruptions on (technically) symbiotic stars, they are not symbiotic novae.  Iben (2003) identifies the critical distinction between classical novae and symbiotic novae as the donor star fills its Roche lobe for classical novae and does not fill its Roche lobe for symbiotic novae.  With this as the critical distinction, RS Oph and T CrB are not symbiotic novae (Schaefer 2009).  But the exact definition is not important, because the eruptions of T CrB, RS Oph, V745 Sco, and V3890 Sgr are tremendously different from those of the symbiotic novae.  Symbiotic novae have amplitudes from 2-6 mag and $t_3$ values from years to decades.  Certainly, all symbiotic novae are fundamentally different from all RNe.  That is, the fast eruptions of T CrB and RS Oph have no useful resemblance to the symbiotic novae, so any definition that lumps the two together has no utility for understanding either class.
	
	The RNe can plausibly be divided up into groups based on amplitude (perhaps separating out IM Nor, V394 CrA, and U Sco), or recurrence time scale (separating out  U Sco, RS Oph, T Pyx, and likely V2487 Oph).  An extreme example of the chaos and instability in RN subdivisions is in Rosenbush (2002) wherein {\it four} categories were defined.  The trouble with all these ideas is that the divisions are neither bimodal nor physically motivated, so it looks to me like all the groups are simply the arbitrary selection of some dividing line as applied to some favored property.  Perhaps more importantly, these various divisions do not break up the RNe into consistent groups, which is to say that any one selected dividing line does not have any predictive power for other properties.
	
	\subsection{A New Division}
	
	I propose a new division based on the mechanism by which the RN has achieved its high accretion rate.  The short-period systems (T Pyx and IM Nor) cannot have steady mass transfer at the high required rate, so their accretion is being driven by a declining supersoft x-ray source which heats the companion star (Knigge et al. 2000; Schaefer et al. 2009).  The other RNe all have evolved stars, for which the high accretion rate is being driven by the evolutionary expansion of the donor star.  (Most classical novae do not have the high accretion caused by an evolving secondary star, and those few that do, for example GK Per, do not have large enough white dwarf mass to become recurrent.)  This physically motivated distinction has the advantage of dividing the RNe into two groups based on the underlying cause for the fast recurrence time scale.  
	
	A further advantage is that this division identifies a stark distinction in the evolutionary paths of the systems.  The short period systems with accretion driven by a supersoft x-ray source will have cyclic evolution going from slow accretion, to a normal nova eruption that starts a supersoft source, to a high accretion rate system with recurrent nova eruptions, to a hibernation state, and then starting the million year cycle over again (Schaefer et al. 2009).  The long period systems with accretion driven by the secondary star evolution will be steady as the donor star continues its expansion, with the RN state being long living.
	
	An important advantage of this new division is that it divides the RNe into two groups based on whether the systems will ultimately collapse as a Type Ia supernova.  The short period systems (driven by the supersoft source) will not become supernovae, the reason being that the RN phase only lasts a few centuries out of a million years while the dynamically dominant event (the regular nova eruption that starts the supersoft source) will eject more mass than is accreted (Schaefer et al. 2009).  The long period systems (driven by expansion of the secondary) will become supernovae as the steady accretion will pour more mass onto the white dwarf than is ejected by eruptions (Hachisu \& Kato 2001; Hachisu et al. 2000a; 2002;  Hachisu, Kato, \& Schaefer 2003).  Thus, people interested in Type Ia progenitors should only look at the systems with evolved companions, while the rate calculations for the death of RNe should not include the short period systems.
	
	In summary, this new division of the RNe into short-period and long-period systems has come up with a division that produces the same groups as based on {\it four} widely-different and important physical bases.  The short-period RNe (T Pyx and IM Nor) have short orbital periods (less than a third of a day), have their accretion driven by heating from the white dwarf, have long evolutionary cycles (between RN eruptions and hibernation), and will {\it not} become Type Ia supernovae.  The long-period RNe (the other 8 RNe) have long orbital periods (longer than a third of a day), have their accretion driven by expansion of the companion star, have a steady evolutionary phase as the companion expands, and apparently will become Type Ia supernova.
	
\section{All Known Eruptions}
	
	What are the known eruptions from the galactic RNe?  This list is important for establishing the recurrence time scale, the inter-eruption intervals, and the demographics for discovery statistics.  Many of the eruptions are well known and need no particular discussion.  However, eight eruptions (including one possible eruption) are little known or unpublished, and these will be discussed in detail here.  Table 2 has a detailed listing of all eruptions and full information on the discovery.
		
\subsection{Specific Eruptions}
		
	The 1941 eruption of CI Aql was discovered with a directed search through all the archival plates at Harvard College Observatory (Schaefer 2001b).  The importance of this discovery is that CI Aql is one of the systems for which I have made a good measure of its period change across the 2000 eruption (hence yielding a value for $M_{ejecta}$) so that a confident value for $\tau_{rec}$ can be used to test whether the white dwarf is gaining mass.  Indeed, the recurrence time scale has changed from its original 83 years (1917 to 2000) to 24 and 59/N years, where N might be 1, 2, or 3 to represent any missed eruptions from 1941 to 2000.  The detailed light curve for this eruption is presented for the first time in Section 5.

	The 1900 eruption of V2487 Oph was discovered as a result of intentionally testing a prediction that the system (known as a CN from its 1998 eruption) is actually a RN (Pagnotta, Schaefer, \& Xiao 2008; Pagnotta et al. 2009).  That the star might be a recurrent nova was recognized by Hachisu et al (2002) on the basis of a plateau in the light curve of the 1998 eruption as well as by a model estimate that the system has a high mass white dwarf.  We independently identified V2487 Oph as being a strong candidate RN based on its very high expansion velocity and the presence of high excitation lines in its outburst spectra.  In 2004, we had searched many of the Harvard plates for prior eruptions, but had not been exhaustive in the search.  In the summer of 2008, we again looked through all remaining Harvard plates (and all Sonneberg archival plates) and found one eruption.  This eruption was evidenced by a single Harvard plate (AM505) from June 1900.  As always, when an event is detected on only one plate, we have to be cautious about the eruption.  In this case, the reality of the eruption was obvious immediately.  One reason was that the image was several magnitudes above the plate limit, so the significance of the image is high.  But the immediately obvious clincher was that all star images on the plate had a characteristic dumbbell shape (a result of a double exposure of equal duration with a small offset between exposures) and this exact same image shape was seen in the V2487 Oph image.  This proves that the light recorded by the image came through the telescope and completely rules out any possible plate defect situation as well as most of the other possible alternatives.  The lack of trailing (of the basic dumbbell shape) also eliminated the possibility of an asteroid, while this is further ruled out by a comprehensive search of all asteroids by Brian Marsden.  The position of the transient image was measured to be within a few arc-seconds of the position of V2487 Oph, and this makes the possibility of some other nova extremely unlikely.  In all, V2487 Oph is known to have had an eruption in 1900 with very high confidence.  Other plates in 1900 were also examined, but none were recorded at a time to sufficient depth to have any hope of catching the eruption.  That is, only one of the plates could possibly have recorded the 1900 outburst.  This is actually the usual case for Harvard plates of that era for very fast novae.  (V2487 Oph has a very fast decline, with it fading by 3 magnitudes below peak within 8 days.)  That is, with ordinary gaps in the plate schedule, an eruption most of the time would be completely missed and most of the remainder would be found on only one plate.  This provides a strong lesson to us that most of the early eruptions of fast novae would be completely lost.
	
	The 1917 eruption of U Sco was also discovered with an exhaustive search through the Harvard plates (Schaefer 2004a).  The pre-1940 plates at Harvard had already been examined by Thomas (1940), but she had missed the one plate (AC18624) that shows this eruption, likely because the plate had U Sco only in the extreme corner.  This circumstance allowed for the reality of the image to be confidently made because the normal point spread function for stars at that position has a large and characteristic shape shared by the U Sco image as well as all surrounding stars.  The image of U Sco was bright and roughly three magnitudes above the plate limit, so there is no chance of random grain enhancements being mistaken for a real object in the sky.  Also, the position of the image matches that of U Sco to within a few arc-seconds.  With the perfect image shape, brightness, and position, the identity of this transient object with U Sco is certain.  The magnitude of U Sco in this image corresponds to U Sco being within one day of peak.  All plates around the time of peak were examined, with none others showing U Sco, however, this is not surprising as the poor time coverage and poor limiting magnitude of the closest plates all come nowhere near to being able to show this eruption (see the light curve in Section 5).  Out of the entire month around the peak, a U Sco eruption would have been missed on 20 days of that month.  So it is sheer luck that this U Sco eruption was discovered at all, and the most likely case was for it to have been recorded on only one plate.  (The identical situation arises for the 1906 eruption of U Sco [Thomas 1940 and see Section 5], the 1890 eruption of T Pyx, and for the 1900 eruption of V2487 Oph [Pagnotta, Schaefer, \& Xiao 2008; Pagnotta et al. 2009], all of which are only known by a single plate.)  A strong lesson from this eruption is that it is easy for a U Sco eruption to slip through even the best old coverage in the world.  The light curve information for this eruption (little though there is) is presented in this paper for the first time.
	
	The 1945 eruption of U Sco was also discovered as part of an exhaustive examination of all Harvard plates showing U Sco (Schaefer 2001a).  However in this case, 37 plates show U Sco in outburst.  Nevertheless, the peak was missed as the earliest plates in the light curve only show U Sco at tenth magnitude (roughly 10X fainter than peak).  The light curve is published here for the first time.  This eruption is after the plate search of Thomas (1940), as otherwise she would have easily found it.  The first four eruptions of U Sco in the last century (in 1906, 1917, 1936, and 1945) are known only retrospectively and only from Harvard plates.  Indeed, the Harvard plates provide the {\it only} record for 11 of the 37 RN eruptions, plus the primary record for 8 more eruption, and thus this one archival collection provides roughly half of all known RN eruptions.  The number of plates that record each U Sco eruptions is 1, 1, 12, and 37 respectively, which demonstrates the increase in frequency of Harvard plates.  This rise in frequency explains why the expected ~1896 eruption was missed and it readily suggests that many other RNe could easily have missed eruptions in the early twentieth century.  
	
	The 1969 eruption of U Sco was only recorded late in its tail by two observers and the data have never been published.  However, one of the observers is Albert Jones, who is widely recognized as one of the greatest visual observers of all time.  He is the discoverer of many nova eruptions (including two {\it other} RN events) as well as a co-discoverer of SN1987A.  Janet Mattei has identified him as being the best observer in the world and as having ``photometric eyes", while Daniel Green has identified Jones as the best observer in the world of cometary magnitudes.  It would be hard to question even one positive detection of U Sco by Albert Jones, but he has ten positive observations all of which define a fading tail.  And the second observer is the legendary Frank Bateson, who was the longtime leader of the RASNZ variable star section.  (Bateson told me about this 1969 event during one of his visits to America.)  There is no chance that these two skilled observers would be mistaken about the light curve.  Given that U Sco is always around 17-18 mag with no excursions brighter than that (other than eruptions), the 1969 event can only have been a nova eruption missed by the rest of the world.  This is not surprising as the U Sco peak was in early February and the position was just coming out from behind the Sun deep in the dawn sky.  This illustrates that the solar gap (with U Sco being only 3.6$\degr$ from the Sun every 28 November) will force the missing of some U Sco eruptions.  Again, we have the lesson that the discovery of U Sco eruptions is hit-or-miss and depends on random happenstance such that a significant fraction of the events will be missed entirely.
	
	The 1842 eruption of T CrB has been published by Sir John F. W. Herschel (1866), although this might simply be a report of a nearby constant star.  John Herschel (son of William Herschel) is one of the greatest visual observers of all time, so his report must be taken seriously.  He writes that he had been mapping the stars in 1842 by means of measuring the positions of all stars inside triangular regions which cover the entire sky.  Upon the 1866 discovery of T CrB, he had examined his notes from 9 June 1842 and found that he had recorded a 6.7 magnitude star at the exact same position!  McLaughlin (1939) claims that Herschel's position is one degree off and hence the observation was not of T CrB but rather of BD+25$\degr$3020 (a normal G-type main sequence star).  However, Herschel was confident that his new star was at the same position as the 1866  event and that no mistake was made.  The case depends on the interpretation of the identification of the new star in Herschel's chart.  If Herschel indicated his new star with an asterisk then the position is that of T CrB, while if the editor had added the asterisk to Herschel's chart and the 1842 star is indicated by the 6' notation next to it then the indicated position is indeed far from T CrB and consistent with BD+25$\degr$3020.  The case is not clear from the old article, so Herschel's original charts or the {\it Monthly Notices} records should be examined if they still exist.  A reason to doubt McLaughlin's hypothesis is that BD+25$\degr$3020 is very faint (at a constant $V=7.06$), so that it is unlikely (but not impossible) that Herschel could detect it.  In all, I do not know whether to accept the 1842 eruption as real.  Nonetheless, the confidence in this event is sufficiently low that I should not include it in the rest of this paper.
	
	The 1907 eruption of RS Oph is only known from its post-eruption dip visible in the Harvard data (Schaefer 2004b).  After {\it all} eruptions, RS Oph fades below its normal quiescent magnitude by over one magnitude for a duration of about half a year.  Such fading is {\it only} seen after a known eruption, other than on one occasion in 1907.  In 1907, when RS Oph came out from behind the Sun, it was faint, getting down to B=13.1 mag and fainter.  (Before the solar gap, RS Oph was averaging B=11.8 mag.)  RS Oph remained in this faint state for at least 148 days until the start of the next seasonal gap, after which it was back to normal with an average magnitude around 11.8 mag. The obvious idea is that this is a post-eruption dip and that the outburst itself was lost due to conjunction with the Sun.  This is a reasonable and inevitable conclusion, but it is not `proven'.  While it is always possible that RS Oph faded for half a year in an unprecedented dip unrelated to a nova event, this possibility is contrived and has zero positive evidence.  As such, I take the evidence to point with good confidence to the fact that RS Oph had a nova eruption in the winter of 1907.
	
	The 1945 eruption of RS Oph is only known from AAVSO data, where the star was seen fading from 9.6 mag down into a typical post-eruption dip (Oppenheimer \& Mattei 1993).  Again, the post-eruption dip was just as the star was coming out from the solar gap in the dawn sky.  However, in contrast with the 1907 eruption, the 1945 eruption also shows a definite fading tail during which RS Oph was certainly brighter than at any other time outside of a nova outburst.  The combination of the fading tail and the classical post-eruption dip makes the case for a 1945 nova so strong as to be certain.
		
\subsection{Overview of Eruptions}

	In all, we have 37 eruptions from the ten known galactic RNe.  Half of the RNe have only two known eruptions, although they might have many more in the last century that were missed.  U Sco has a total of nine known outbursts while RS Oph has a total of eight nova events.
	
	The 37 eruptions are listed in Table 2.  The first two columns are the star and the year of the maximum.  The third column gives the Julian Date of the day of peak brightness (as taken from Section 6).  The fourth column gives the time between eruptions in years ($\Delta$T).  The next three columns give the name of the first discoverer (with co-discoverers listed as footnotes), the source or means of that discovery, and a reference.  Columns eight and nine give the Julian Date of the discovery and the time difference (in days) between the peak and the discovery.  The last column indicates whether the eruption is recorded on the Harvard plates, and if so then whether the eruption is recorded exclusively on the Harvard plates.
	
	The shortest known inter-eruption interval is 7.88 years (for U Sco between 1979 and 1987) while the longest is 97.98 years (for V2487 Oph between 1900 and 1998).  A dozen eruptions were discovered with the Harvard plate stacks, while another three were discovered with other plate archives, all long after the eruption had come and gone.  The Harvard plates provides the only known data for 11 eruptions and the bulk of the data for 15 eruptions, such that without the Harvard plates we would not even have realized that the RN phenomenon existed until the first example would have been discovered in 1946 (T CrB).  Only four events were discovered by professional astronomers near peak.  All of the other eruptions (exactly half at 19 events) were discovered by amateur astronomers or with amateur astronomers' data.  (In this count, I have chosen to take the three PROBLICOM discoveries of William Liller as being that of an amateur, despite his having been a professional astronomer of high repute who worked intensively with the Harvard plate collection, because he has retired and no one was paying for his time or the equipment that he was operating from the patio of his house in Vina del Mar; see Liller 1992 and Liller \& Mayer 1990.)  For the 23 non-archival discoveries (i.e, those with a discovery delay of less than 1000 days), 18 were made within four days of the peak brightness, while 15 were made within one day of maximum light.  Eight of the RN events have more than one independent discoverer, while three of the events have three or more discoverers.
	
\section{Comparison Stars}

	The photometric histories of the RNe are always determined by comparing the target system with nearby comparison stars of known magnitudes.  In preparing for this paper, I have had three distinct applications which require comparison stars.  First, I need accurate magnitudes in many bands for just a few relatively faint stars near to the RN to serve as comparison and check stars for differential photometry with CCD images.  Second, I need magnitudes of many stars spread over the entire magnitude range of the outburst to serve as sequences for the eruption light curve on archival photographic plates.  Third, I need  to know the real magnitudes on a modern magnitude scale for various stars used as comparison stars by olden astronomers who have reported eruption light curve magnitudes by direct comparison with these stars.
	
	The magnitude scale was systematized by Pogson (incidentally, the discoverer of the first known RN outburst) in the nineteenth century.  By roughly a century ago, the standard magnitude scale was the `North Polar Sequence' defined with the Harvard plates.  The sequence was extended to faint magnitudes by various methods such as comparing image diameters on double exposures where the ratio of exposure times were known precisely.  We now know that reciprocity failure and other non-linearities in photographic emulsions make for substantial systematic errors that accumulate as fainter and fainter stars are calibrated in this fashion.  Magnitudes for target stars were determined by comparing star image sizes on photographs of fields double exposed on the pole region plus the target field.  By the 1930's, the North Polar Sequence had been transferred to many of the Harvard-Groningen Selected Areas for more convenient placement on the sky.  The trouble with these transfer sequences was that substantial and then-unknown errors were always present.  In general, all magnitude sequences from before the 1950's were systematically distorted (from the modern scale) with the errors getting larger as the stars get fainter.  Typical errors for a tenth magnitude star are from 0.5-1.5 magnitudes.  Sandage (2001) has made a modern study of magnitudes in the primary Selected Areas and finds systematic distortions of 0.6 mag at B=8, 0.8 mag at B=10, 1.0 mag at B=12, and 1.3 mag at B=14.  These systematic errors in the primary standard stars were then transfered to all magnitudes in the literature.  I have made detailed comparison of many old sequences versus modern magnitudes, and I find that there is much scatter on top of this systematic problem.  For comparison stars for use with Pluto, the errors were 0.7 mag in 1934 for a photographic sequence by Walter Baade and 0.01 in 1954 for a photoelectric sequence by Robert Hardie (Schaefer, Buie, \& Smith 2008).  For comparison stars for use with SN1937C, the errors ranged from 0.3 to 0.9 mag in the B-band and 0.7 in the V-band (Schaefer 1994).  For comparison stars for use with SN1960F, the errors ranged from 0.3 to 0.9 mag in the B-band and 0.7 in the V-band (Schaefer 1996a).  For comparison stars for use with SN1974G, the errors ranged from 0.3 to 0.9 mag in the B-band and 0.7 in the V-band (Schaefer 1998).  For comparison stars for use with SN1981B, the errors ranged from 0.3 to 0.9 mag in the B-band and 0.7 in the V-band (Schaefer 1995).  The point of this recitation is to emphasize that any magnitude before $\sim$1950 certainly has big errors while any magnitude before ~1970 or so likely has big errors.  Given this point, a comprehensive photometric history of the old eruptions must somehow overcome this big problem.  This can be done either by remeasuring the magnitudes on the original plate material or by rescaling the old reported magnitudes based on the corrected magnitudes for the comparison stars.
	
	The creation of modern accurate magnitude sequences had to await the widespread usage of linear detectors; the photoelectric photometer and later the CCD.  Early photoelectric photometers could not go faint enough to be relevant, so it was only in the 1950's that comparison stars could be calibrated deep enough to be useful.  Even then, it took many years for useful and accurate standard stars to be determined (Landolt 1973; 1983; 1992) and for RNe sequences to be measured (e.g., Kilkenny et al. 1993; Henden \& Munari 2006).  Many deep sequences had to await the availability of CCD cameras and the increase of telescope availability.  By the 1970's or so, magnitude sequences and RNe magnitudes are all on the modern magnitude scale.  For purposes of this paper, any magnitude in the literature after 1970 is taken to be correct.
	
	The archival photographic plates (when used with modern comparison stars) are exactly on the modern B-magnitude scale.  That is, the color terms are effectively zero.  Another worry arises from the realization that the RNe have emission lines while the comparison stars do not, so some systematic variations could conceivably arise.  But any such effects are miniscule for two reasons.  First, any offset will be constant and thus not effect the shape of the light curve.  For all applications of the photographic plates, we only care about the light curve shape, so an offset is irrelevant to the conclusions.  This includes the comparison between photographic, photoelectric, and CCD magnitudes, with the reason being that these all have essentially constant central wavelengths and equivalent widths of the spectral sensitivity (by design).  Second, the RNe have typically only one percent of their blue flux in emission lines (cf. Williams 1983).  So even in the false case where slight changes in a passband allow the line light to be recorded or not, the variations will only be of order 0.01 mag.  For realistic passbands with nearly constant width and roughly Gaussian shapes, the errors will always be much smaller than 0.01 mag.  Thus, we can have good confidence that the old archival plate RN magnitudes have no measurable systematic bias when compared to ordinary field stars or when compared with photoelectric or CCD magnitudes.
	  
	  For my three applications (see beginning of this section), I have created two sets of comparison stars for each RN.  The first set is three faint stars near to the RN, which I have measured in the bands BVRIJHK.  These stars are designated as `Comp' (typically a few magnitudes brighter than the normal quiescent magnitude for the RN), `Check' (typically close to the quiescent RN in both position and magnitude), and `Check2'.  These stars are all chosen to be well isolated from neighboring stars, and all have been found to be constant over the many observations that I have made.  The observations to calibrate these comparison stars have been made many times over on many telescopes (the 0.9m, 1.0m, and 1.3m telescopes at Cerro Tololo and the 0.8m and 2.1m at McDonald).  These observations have been made with the usual full program of observations of standard stars (Landolt 1992) at high and low airmass on photometric nights.  The typical measurement uncertainty in my measures is likely 0.01-0.02 mag.  A number of these stars have been measured by other workers, including, for example, the T Pyx stars by Arlo Landolt (2005 private communication).  I have JHK calibration for the stars, but the magnitudes that I will adopt are those taken from the 2MASS survey (Cutri et al. 2003\footnote{http://irsa.ipac.caltech.edu/applications/Gator/}).  The magnitudes for these stars are presented in Table 3.  The specific positions of these stars will be listed in Table 4.
	  
	  My second set of comparison stars consists of sequences in B and V for stars spanning the range from the quiescent to peak magnitude.  The reason for restricting myself to B and V for this application is that essentially all of the old literature, archival records, and photographic plates are in one of those two magnitude systems.  Also, the availability of the B-V color then allows for seeking color terms in the conversion to modern magnitudes.  The stars that I have selected are single, isolated, constant, and evenly spaced in magnitude as best as possible.  I have also selected as many stars as possible that have been used by previous workers for their sequences.  With this last point, I then have a modern and a literature value for the same sequence, and thus a reasonable conversion formula can be determined.  Faint stars in the sequences are always near to the RN, and I thus have measured them as part of the calibration program described in the previous paragraph.  Bright stars in the sequence are necessarily far from the RN, so I could independently calibrate only a few of them.  For stars brighter than tenth magnitude or so, well-measured photometry is already available from the literature, which I have accessed and averaged from the SIMBAD database\footnote{http://simbad.u-strasbg.fr/simbad/}.  For many of the RNe, the AAVSO has already constructed well-measured sequences\footnote{http://www.aavso.org/observing/charts/vsp/}, often going quite deep.  A variety of other sources have been used to fill in my sequences, including the Tycho2 catalogue\footnote{http://www.rssd.esa.int/hipparcos} (Hog et al. 2000), the GSC2 catalog\footnote{http://www-gsss.stsci.edu/Catalogs/GSC/GSC2/GSC2.htm} (Lasker et al. 2008), and  the ASAS3 catalogue\footnote{http://archive.princeton.edu/~asas/} (Pojmanski 2001).
	  
	  My magnitude sequences are presented in Table 4.  The first and second columns are a designation for the field and reference name for the star.  My star names are just sequential letters, with a few small gaps because some stars in the series were later identified as being possibly variable or otherwise unsuitable.  The third column lists the positions (in J2000 coordinates) accurate to roughly one arc-second as useful for the unique identification of the star.  The fourth column lists the formal catalog name for the star (if any).  The fifth column gives the star names from the AAVSO name, which is always equal to the V-band magnitude expressed to the nearest 0.1 mag with the decimal dropped, so for example, a V=12.34 mag star would be designated `123'.  The next two columns give the B and V magnitudes of the stars.  The uncertainties in these magnitudes are likely to be 0.01-0.02 mag, although occasionally larger uncertainties are possible.  In some cases, where the magnitude is known to be uncertain by roughly 0.1 mag or larger, the value is only quoted to the nearest tenth.  The eighth column cites the source of the quoted magnitudes.  The notation K1993 refers to Kilkenny et al. (1993) and H\&M 2006 refers to Henden \& Munari (2006).  The last column gives the designations as used earlier in the literature, with a single or double letter indicating the reference followed by a colon and the designation in that reference.  For T CrB, we have A (Ashbrook 1946), W (Wright 1946), B (Bertaud 1947), and Pe (Petit 1946).  For RS Oph, we have S (Shapley 1933), Pr (Prager 1940), and F (Fleming 1907).  For V745 Sco, we have Pl (Plaut 1958).  For V3890 Sgr, we have M (Miller 1991).  Most of the lines in the comparison sequences table are only presented in the on-line version only.
	  
	  Some of the extant magnitudes for old eruptions are now only available as specific reports in the literature.  Most of these explicitly identify their comparison stars and report their adopted magnitudes for these stars.  In general, these old sequences have substantial errors and these are propagated into the RN magnitudes.  Fortunately, it is not too late to correct the old magnitudes.  The general solution is to plot the old adopted magnitudes versus the modern magnitudes for these stars, and to fit some smooth curve so as to establish the relation between the old and modern magnitudes.  With this relation, I then correct the reported RN magnitudes onto a modern magnitude scale.  To this end, I have chosen many of the comparison stars in my sequences as being those relevant for correcting old sequences.
	  
	  For the U Sco sequence of Pogson (Thomas 1940), we have five stars with old and new magnitudes.  The RMS deviation of Pogson's sequences is 0.39 mag.  Nevertheless, the best fit linear relation is that the Pogson magnitudes are close to the modern magnitudes.
	  
	  Various series of magnitudes for the eruptions of T CrB have been published that need correction to the modern magnitude scale.  For the 1866 eruption, we are often not told the comparison stars (nor the adopted magnitudes for the stars used), so I can only take the reported magnitudes at face value.  Distortions in the light curve are likely to be minimal for times when the nova is brighter than seventh magnitude or so as the old magnitude standards were reasonable for bright stars.  (Indeed, the 1866 light curve will be seen to closely match the light curve shape from the 1946 eruption down to around eighth magnitude.)  Vertical compression in the light curve tail is possible, but this will be confused with the effects of the variations in the red giant companion.  The comparison stars of Payne-Gaposhkin \& Wright (1946) are also used by Wright (1946), with a correction of $B=0.8 \times B_{Wright}+1.9$ for $B_{Wright}<11$ and $B=2 \times B_{Wright}-11.3$ for $B_{Wright}>11$.  The magnitudes for the comparison stars of Pettit (1946) have a constant offset from the modern magnitudes, such that $V=V_{Pettit}+0.12$.  The comparison stars of Shapley (1933) were also used by Weber (1961) and are well represented by a constant offset with $B=B_{Shapley}+0.36$.  The comparison stars of Ashbrook (1946) have magnitudes that deviate from those in Table 4 with an RMS scatter of 0.12 mag and no systematic shifts, so I will adopt his visual estimates with no corrections.  Similarly, the comparison stars from Bertaud (1947), except for $\epsilon$ CrB, have an RMS error of 0.11 mag with no evidence for systematic errors.  The AAVSO magnitudes are also correct.  The calibration of the comparison stars in Gordon \& Kron (1979) is modern.
	  
	  RS Oph has a variety of magnitude sequences published for the early eruption with the Harvard plates (Fleming 1907; Prager 1940; Shapley 1933).  These B-band sequences have substantial differences from each other and from the modern magnitudes.  For Fleming, the correction is $B=B_{Fleming}+1.5$ for $B_{Fleming}<10$ and $B=1.6 \times B_{Fleming}-4.5$ for $B_{Fleming}>10$.  For Prager, the correction is $B=B_{Prager}+0.07$ for $B_{Prager}<10$ and $B=1.4 \times B_{Prager}-3.93$ for $B_{Prager}>10$.  For Shapley, the correction is $B=1.13 \times B_{Shapley}-1.0$ for $B_{Shapley}<13$ and $B=0.8 \times B_{Shapley}+3.3$ for $B_{Shapley}>13$.
	  
	  Five of the dozen magnitudes for the 1949 eruption of V745 Sco are from Plaut (1958) with Leiden plates which I have not examined.  Plaut explicitly states his comparison stars (although his stars h, I, and k cannot be identified with his published chart) and his adopted magnitudes, thus his RN magnitudes can be converted to a modern B-band magnitude.  Plaut's magnitudes ($B_{Plaut}$) have a best-fit linear relation with my final magnitudes (see Table 4) such that $B=1.4 \times B_{Plaut}-5.0$.
	  
	  Most of the magnitudes for the 1962 eruption of V3890 Sgr are from the plates at the Maria Mitchell Observatory.  These magnitudes were originally estimated by H. Dinerstein, but Miller (1991) has gone back and remeasured the magnitudes.  I have not examined the Maria Mitchell plates, but we can place the magnitudes from Miller onto a modern magnitude scale by comparison of the sequences.  From her comparison stars, I find that $B=1.02 \times B_{Miller}-0.52$ for $B_{Miller}>14$ and $B=1.23 \times B_{Miller}-3.46$ for $B_{Miller}<14$.  This correspondence will be applied to the Miller magnitudes for V3890 Sgr.  Wenzel (1990) only quotes the magnitude for one of his comparison stars, but this is in the center of the range, so I adopt $B=B_{Wenzel}+0.24$.

\section{Eruption Light Curves}

	The light curves for the 37 eruptions have a wide variety of coverage.  Four of the eruptions have only one plate to form their light curve, while many of the older eruptions have at most a few dozen plates recording the event.  The more recent eruptions have extensive coverage, with the best one being the 2006 outburst of RS Oph.  
	
	Here, I collect all the existing data for most of the eruptions.  (However, for a few of the eruptions with large amounts of data, my tabulations may not be complete, as some data sources may have been missed.  Fortunately, in these cases, the magnitudes collected will provide a full light curve anyway.)  I will present all these magnitudes explicitly in tables in this paper, because most of the magnitudes have never been published, while the published light curves are widely scattered in the literature and have been extensively revised onto a modern magnitude scale as part of this paper.
	  
	  A large source of data is from the visual observations in the databases of the American Association of Variable Star Observers (AAVSO), Royal Astronomical Society of New Zealand (RASNZ), Variable Star Observers League of Japan (VSOLJ), and the Variable Star Network (VSNET).  These observations are almost entirely in the visual band.  (Almost all AAVSO work is performed with the color vision, that is with the cones in the retina of the eye for which the effective wavelength equals that of the Johnson V-band, as the target is more than 0.8 mag brighter than the limiting magnitude, see Schaefer 1996b.)  The modern AAVSO magnitude sequences are all accurate, so the only question is with the older AAVSO data.  For estimates where the star is brighter than roughly ninth magnitude, the deviations in the comparison star magnitudes from their modern values are relatively small.  For cases where it matters, I have gotten the sequences used at the time and have made the corrections if needed.  The typical uncertainty in these visual magnitudes is 0.15, although these will change with circumstances and observer, from roughly 0.1 to 0.3 mag.
	
	The AAVSO database contains $\sim83\%$ of all magnitudes of RNe in eruption.  I will not include the AAVSO magnitudes in my ten tables in the printed version of this paper, because the tabulation would take many pages of journal space and because the AAVSO provides a reliable long-term archive that is freely available to everyone.  Nevertheless, the complete list of AAVSO eruption magnitudes will be presented in Table 5, with most of the 12,115 magnitudes only visible in the on-line version of this paper.  In this table, the first column identifies the RN, the second column gives the Julian Date of the observation, the third column gives the observing band, the fourth column gives the observed magnitude with its error bar (taken to be 0.15 mag in general), and the last column cites the source.
	
	For many of the eruptions (also during quiescence) the primary data source is the archival photographic plates at Harvard College Observatory.  This collection has roughly 500,000 plates from 1890 to 1953, with a smaller collection covering around the 1980's.  Each plate is a thin glass rectangle (usually 8.5x10 inches in size) with a photographic emulsion on one side.  Almost all plates have blue sensitive emulsion.  The Harvard plates provided the original definition of the B-band, and later scales for photoelectric and CCD magnitudes kept the same spectral sensitivity.  So if the comparison stars that set the magnitude scale are in modern B magnitudes, then the resultant RN magnitude will also be on the modern magnitude scale.  (The old magnitudes taken from the plates have been called 'photographic' magnitudes, with the difference to the modern B magnitudes being only that the old comparison star magnitudes have offsets compared to modern measures, see Section 4.)  Thus, with my modern comparison star sequences, the Harvard plates provide a wonderful `time machine' to get RNe light curves as far back as 1890.
	
	The plates are taken in series, with one series for each telescope used.  The plates from each series are numbered sequentially, with a capital letter series identifier followed by the sequential number.  For example, one of the deeper plate series (taken with the 24-inch Bruce Doublet telescope) is designated with the letter `A', such that plate A1 was taken in 1893 and the last A plate (A27504) was taken in 1950.  The better A plates have a limiting magnitude of roughly 18, with of order 30 useable plates that show the field of any one RNe.  The primary deep series (with limits of 16-18 mag) are the A, MC, and MF, while there are many tens-of thousands of plates with less deep limits (with limits of 13-16 mag) in the B and I series.  Harvard also has several series (chiefly RH, RB, AM, AC, and the Damons) that are patrol plates, where large areas of the sky are regularly photographed to moderate limits (typically 12-15 mag).  The entire sky is covered roughly equally, with typically 1000-3000 plates covering any particular RN.  Further details are available in Collazzi et al. (2009) or at the Harvard archive web site\footnote{http://tdc-www.harvard.edu/plates/}.
	  
	  The one-sigma uncertainty in the magnitude depends substantially on the plate material, with various field effects increasing the uncertainty towards the edge of the plate.  For a well exposed field with a good sequence of comparison stars, measurements with iris diaphragm photometers can give a median accuracy of 0.08 mag, and can get as good as 0.04 mag (Schaefer, Buie, \& Smith 2008).  However, most of the plates were measured by direct visual comparison of image sizes by means of a view through a loupe (a handheld magnifier).  I have a very large amount of experience in this, and I also have performed a variety of experiments over the years for the accuracy of these comparisons.  The median accuracy is 0.15 mag, with typical accuracy ranging from 0.1 to 0.2 mag.  It is difficult to evaluate the uncertainty for any one individual plate, so I have used a default one-sigma error of 0.15 mag unless my notes point to some other value as quoted in the tables.  While the measurement uncertainty for plates is roughly a factor of ten worse than usually obtainable with photoelectric or CCD techniques, this uncertainty is comparable or smaller than the usual random flickering.  Thus, for many purposes, the moderate accuracy of photographic magnitudes is perfectly fine and just as good as more 'modern' measures.
	  
	  The eruption light curves are collected together into one table for each RN (Tables 6-15).  The format for each line will come in one of two forms, depending on whether the B-band and V-band magnitudes were made effectively simultaneously.  For most observations, the columns will list the Julian date of the observation, the band, the observed magnitude (along with its uncertainty), and the source of the data.  If B and V magnitudes are from the same time, then the second and third columns will instead report the B-band and V-band magnitudes in that order.  (This is simply to make the tables more compact.)  Each eruption will have a header specifying the year of the peak.  The source of the data might be the Harvard plate number (e.g., B5289, AM1172), the organization reporting the data, or the literature reference.  The observations are listed in time order.  The complete eruption light curves consist of those magnitudes from each of the tables for individual RNe (Tables 6-15) plus the AAVSO magnitudes for many individual eruptions collected into Table 5 (most of which is only visible in the on-line table).  For the next ten paragraphs in this section, I will describe the data sources specific to each of the ten galactic RNe.

	T Pyx has five known eruptions	, with photometry for the first four known almost entirely from the Harvard plates.  I have remeasured the brightness of all the eruption plates at Harvard with the comparison stars from Table 4.  The 1966 eruption has three large data sources, including the AAVSO,  Landolt (1970), and Eggen et al. (1967).  These data essentially stopped at day 80 after the peak.  However, within the last year, the original data logbooks of Albert Jones (see Section 3.1) have been transcribed and added to the AAVSO data collection.  This `new' data has a wonderful coverage from day 80-300, and suddenly reveals a sharp drop-off in the light curve followed by a plateau, never previously suspected.  All data sets show a day-to-day variation of 0.5-1.0 mag. These variations are all much larger than any uncertainty (with the best illustration of this being the time resolved light curve of Landolt 1970), and are certainly intrinsic to T Pyx.  T Pyx is the only RN to display such flares/oscillations.  The light curves for the eruptions are presented in Tables 5-6 and Figures 2-5.
	
	IM Nor has two eruptions, with the first only known from eight Harvard plates.  A light curve for the 1920 outburst was never published, although Elliot \& Liller (1972) remeasured the magnitudes and plotted a low resolution curve with no numbers.  Neither J. Elliot nor W. Liller could locate their original magnitudes, and I could not find the original notebook with Ida Woods' measures at Harvard.  So the measures of the 1920 eruption light curve are entirely from my own viewing of the original plates along with the comparison star sequence in Table 4.  One of the plates has a V-band spectral sensitivity.  For the 2002 eruption, Liller's first two magnitudes are in a red bandpass and they represent the initial rise to peak.  Almost all the remaining light curve is in the V-band, taken from data reported either in the {\it IAU Circulars} or by the AAVSO.  The exception is the average magnitudes on five nights late in the tail as measured with CCD photometry by Woudt \& Warner (2003), for which I only quote the nightly average.  Significantly, the long times series by B. Monard (taken from the AAVSO database) well covers the late tail of the eruption, although in a non-standard band-pass (see Table 22).  The light curve for both eruptions are presented in Tables 5, 7, and 23 plus Figures 6-7.  The top panel in Figure 7 shows just V-band observations, while the lower panel adds in the unfiltered-CCD magnitudes of Monard with a 0.6 mag offset to make the them match in the large time intervals of close overlap.  By this combination of V and unfiltered magnitudes, we get a single light curve that covers from before the peak all the way until the nova returns to quiescence.  A potential trouble with this combination is that the differential color might change over time (say, due to increasing relative strength of the H$\alpha$ line).
	
	CI Aql has three known eruptions, with the first in 1917 being covered primarily with the Harvard plates plus a few from Heidelberg.  Williams (2000) reports a modern measure of the Harvard plates from the 1917 eruption, although in this paper I will report on my independent measures based on the comparison sequence in Table 4.  The system had its second discovered eruption in 2000, with a fairly well observed light curve and the realization that yet another little-known nova was actually a RN.  The 2000 eruption has wonderful coverage by the AAVSO observers, giving good measures of the light curve fluctuations in the visual band from just before peak until very late times.  Additional observations are reported in the {\it IAU Circulars} and Matsumoto et al. (2001).  The light curve for the 1941 eruption is entirely from the Harvard plates and is reported here for the first time.  The light curves for the three eruptions are presented in Tables 5 and 8 plus Figures 8-10.
	
	V2487 Oph has two eruptions, with the first in 1900 being known only from a single Harvard plate.  In 1900, the discovery plate shows the nova near its peak magnitude, which only lasts a day or two.  There are a variety of other Harvard plates taken that month, but none of them go deep enough to come anywhere near recording this eruption.  The light curve from 1998 reveals one of the fastest known nova eruptions.  The light curves for both eruptions are listed in Table 9, while the 1998 eruption light curve is plotted in Figure 11.
	
	U Sco has a total of nine known eruptions.  The first observed eruption was in 1863, where only one observer recorded the event.  This was one of the earliest novae that was recorded in modern times.  The light curve observations by Pogson have been collected by Thomas (1940).  Pogson's comparison star magnitudes are consistent with the modern magnitudes, so his magnitudes are taken exactly.  The next four eruptions (in 1906, 1917, 1936, and 1945) are only known from the Harvard plates, and the magnitudes given in Table 10 are entirely from my own measures of the original plates with the modern magnitude sequence from Table 4.  The 1917 and 1945 eruption light curves are published here for the first time.  The 1906 and 1917 eruptions have only one magnitude, while there are many limits from around the same time for which these are all fully consistent with the usual light curve.  I will not give a figure plotting these one-point eruption light curves.  The 1945 eruption is the only one to record the sudden drop in B-band brightness starting 33 days after peak.  Other than the serendipitous archival plates from the 1945 eruption, only four magnitudes have been recorded for U Sco in eruption after the sharp drop at 33 days.  All four magnitudes are in the V-band and they do {\it not} show any evidence for the drop.  It might be possible that the drop did not occur in the 1979 and 1999 events, or it might be that the drop is only prominent in the blue light.  The 1969 eruption is known only from 16 visual observations, now residing in the archives of the {\it Royal Astronomical Society of New Zealand} (RASNZ).  The observations are published here for the first time.  I have examined the magnitude sequence used by Jones and Bateson (courtesy of P. Loader) and find that their magnitudes are close to the modern magnitude scale.  Almost all of the magnitudes from the 1979, 1987, and 1999 eruptions are from the AAVSO.  A variety of sources in the literature have reported a few additional magnitudes, and I am presuming these to be on the modern magnitude scale even though they do not report on their comparison stars.  The magnitudes from Budzinovskaya et al. (1992) are not useable because their detector was a TV system with unknown color sensitivity, and their quoted comparison star magnitudes deviate from Table 4 unsystematically with errors ranging from -1.02 to +0.22 magnitudes.  U Sco has been seen with certainty on its rising branch only three times; once on a Harvard plate in 1936 at B=10.75 mag 0.25 day before maximum, once by Dr. N. W. Taylor (of the RASNZ) in 1987 at V=14.0 mag in the day before maximum, and once by P. Schmeer in 1999 at V=9.5 mag 0.3 day before maximum.  The final rise to maximum in the 1936 event is at a rate of 19 magnitudes per day.  With this rate for the entire rising branch, U Sco will go from minimum to maximum in roughly 6-12 hours.  In the 1999 eruption, we know that U Sco rose from below V=14.3 to peak in less than 0.46 days.  U Sco goes from minimum to peak to one magnitude below peak all in under one day, and this is a daunting constraint for fast response in the upcoming U Sco eruption.  The light curves for all nine eruptions are listed in Table 10, while most of these are plotted in Figures 12-18.
	
	V394 CrA has two poorly observed eruptions.  The 1949 eruption has a few magnitudes appear scattered in the literature, and seven Harvard plates have detections.  The 1987 eruption has a few AAVSO magnitudes, a variety of magnitudes presented in the {\it IAU Circulars}, plus 14 B and V magnitudes from Sekiguchi et al. (1989).  The photographic observations in the late tail of the 1949 eruption have a large scatter, while the visual observations early in the 1987 eruption have a large scatter.  In both cases the scatter might indicate very large oscillations in the light curve, but I do not think so.  The good photoelectric observations of Sekiguchi et al. (1989) do not show such oscillations.  Instead, I take the scatter in 1949 to be the usual problems with faint comparison stars in the 1940's and I take the scatter in 1987 to be the usual problems with observers early in an eruption using makeshift comparison stars before standardized sequences are available.  I have not been able to personally examine the non-Harvard plates for the 1949 event, but my broad experience suggests that such problems are common.  The light curves for the two eruptions are given in Table 11 and Figures 19-20.
	
	T CrB has two eruptions (or maybe three if we accept the 1842 event).  The two eruptions were very well observed, likely due to its brightening up to the second magnitude.  The first was in 1866, where T CrB was the earliest nova event that was well studied.  The quiescent nova was identified as a tenth magnitude variable source, so various groups kept track of the variations.  In 1920, the great Leslie Peltier started regularly monitoring the brightness.  (Peltier and Albert Jones are justifiably considered as the greatest visual observers of the last century.)  Based on a generic suspicion that old nova might erupt again, he started to regularly monitor  several old novae with bright quiescent levels, including T CrB, RS Oph, and GK Per.  At the time he started, no RN had been established and the concept was not known.  As told in a famous passage from his inspirational book {\it Starlight Nights} (Peltier 1965), he watched T CrB for decades, only to miss the one night when it finally erupted, so that another person made the anticipated discovery in 1946.  Peltier says ``We had been friends for many years; on thousands of nights I had watched over it as it slept" and then after the eruption that ``There is no warmth between us anymore".  An important product of Peltier's work is a good long-term light curve before the eruption, which displays a unique pre-eruption dip in the light curve (Peltier 1945).  The 1946 eruption was heavily observed by amateurs and professionals, in both B-band and V-band.  The post-eruption light curve displays a unique brightening that lasted a hundred days, starting after the light curve had already returned to minimum and been there for fifty days.  A check of the old light curve from 1866 showed people that a similar post-eruption event occurred previously and hence was not some disconnected happenstance.  I have heard a variety of theoretical models (e.g. Webbink et al. 1987, Selvelli et al. 1992; Hachisu \& Kato 2001), but I have also heard convincing counter-arguments to these various ideas, so at this time I do not think that we know the cause of the re-brightening.  The light curves are given in Table 12 and Figures 21-23.  As the time of interest for the light curves is so long and the AAVSO database records so many magnitudes, I have presented the AAVSO magnitudes away from the main peak as averages over 0.01 year time bins.
	
	RS Oph has eight recorded eruptions in the last century.  The first two (in 1898 and 1907) are only recorded on Harvard plates.  The remaining six eruptions have almost all their light curve information coming from the AAVSO database, with only occasional other magnitudes from the professional literature (mainly scattered in the {\it IAU Circulars}).  The AAVSO has roughly 8000 observations covering the times of the eruptions, and to save space in the tables and to reduce the usual observational scatter, I have constructed 3-day averages for time intervals where the light curve is not changing rapidly and where there are typically $>$4 values per interval.  The unique property of the RS Oph light curve (not seen in any other RN or classical nova) is the post-eruption dip.  That is, from day 100 to 500, or so, RS Oph is significantly fainter by over one magnitude than its normal quiescence level.  This dip appears after all eruptions.  The exact level of the dip varies slightly from eruption-to-eruption, and this is easily attributed to the known long-period variations of the red giant companion star.  This same effect makes for small variations in the late tail of the decline from the nova event.  Similarly, the pre-eruption level changes somewhat with the normal oscillations of the companion.  The light curves are given in Tables 5 and 13 plus Figures 24-31.
	
	V745 Sco has two poorly-observed eruptions.  For the 1937 eruption light curve, seven magnitudes come from my measurements of the Harvard plates.  I have not examined the five archival plates at Leiden (Plaut 1958), but I have corrected the magnitudes reported by Plaut based on the his magnitudes for comparison stars (see Section 4).  The 1989 eruption was not well observed, with few visual observations (and no CCD measures), likely due to the faintness and speed of the decline.  The visual magnitudes in the 1989 decline show scatter at about the half-magnitude level, likely due to the usual problems with comparison stars and color terms.  The light curve is presented in Table 14 and Figures 32 and 33.

	V3890 Sgr has two eruptions.  For the 1962 eruption, we only have B-band magnitudes from archival plates at the Maria Mitchell Observatory (Dinerstein 1973; Miller 1991) and the Sonneberg Observatory (Wenzel 1990).  The 1990 eruption has a V-band light curve that is well defined by visual estimates.  Note the occasional large scatter, which I take to be the usual systematic problems.  Buckley et al. (1990) report on UBV photometry, with this providing our only color information, and also provides the way to compare the two eruptions.  The light curve for both eruptions is given in Table 15 and Figures 34 and 35.

\section{Light Curve Shapes}

	The light curves based on individual data points show the usual amount of scatter and gaps.  A smoothed light curve can better represent the real variations of the eruption.  Such a template light curve is also better for display as well as deriving light curve properties (like $t_3$).  In this section, I will describe the templates for the ten known galactic RNe.
	  
\subsection{Construction of Templates}
	
	The templates are simply smoothed light curves with the gaps filled in.  A fundamental problem (at least in principle) concerns the best time scale for the smoothing, as too short a time scale will result in random measurement noise being taken as a real variation in the nova, while too long a time scale could smooth out real changes in slope of the decline.  In practice, this is not a problem, as the data frequency is high enough to well-sample the relatively slow variations in the decline.  That is, with one exception, the underlying variations are smooth on time scales that are a fairly small fraction of the time since the peak.  The one exception is the rapid oscillations in the light curve of T Pyx, which we see from Landolt (1970) to be $>0.8$ mag on time scales of under one day.  In this case where we only have brief snips from the oscillations, all we can do is to create a template that tracks through the middle of these oscillations.
	
	Another fundamental problem is whether a separate template should be created for each eruption.  As discussed in the next subsection, the eruption light appears to follow the same template for all eruptions, and this justifies the creation of a single template applicable to all eruptions.  This means that gaps in one eruption can be filled in with observations from other eruptions.   Gaps in one band can also be filled in with information from another band, provided that some reasonable color can be adopted.

	  For my templates, I created piecewise-linear light curves that tracked through the middle of the data.  I used Occam's Razor to try to have the templates follow the observations with a minimal number of points.  The templates were constructed to minimize the deviations for all the eruptions of each individual RN.  I also made small adjustments (within the photometric uncertainties) so as to keep a fairly smooth B-V color curve.
	
	The templates are expressed as a light curve with B and V magnitudes as a function of time from the peak for many specific times.  For times between the given times, a simple linear interpolation should be used to give the magnitude at that time.  The templates are presented for all ten RNe in Table 16.  The first column gives the time in days from the peak, while the last two columns give the B-band and V-band magnitudes at the specified time.  The templates all start with the average quiescent magnitude for a time nominally 999 days before the peak.  Generally, there are no data late in the tail of the eruption, so the templates stop at the times of the last available data in these cases.
	  
\subsection{Are the Templates Constant?}
	
	For a given RN, does its light curve change from eruption-to-eruption?  That is, does the template change with each eruption?  This can be tested by looking to see if there are any significant deviations from my average templates.  For this, we can examine the eruption light curves in each figure to see where the observations deviate from the template.  A substantial problem with this task is that we have to worry about the usual photometric errors.  Another problem relates to the variability of the additive brightness of the companion star, as this component changes greatly for T CrB and RS Oph and dominates late in the tail of the eruption light curve.  Finally, we have to worry about the flickering, eclipses, and daily random fluctuations displayed by various RNe.  Here is a nova-by-nova evaluation of whether the eruption light curves are constant:
	
	For T Pyx, we have the substantial problem that its light curve has large amplitude oscillations on the time scale of a day (Landolt 1970).  This will create an ambiguity in that deviations from the template could arise either from the oscillations, from eruption-to-eruption changes, or from ordinary photometric errors.  Given the known oscillations, I see no significant deviations from the average template for any of the first four eruptions, which are all based on Harvard data.  For the 1966 eruption, the pre-maximum magnitudes from AAVSO observers are an average of one magnitude below the template.  These early magnitudes are certainly incorrect as seen by comparison with the photoelectric magnitudes of Eggen et al. (1967), as is likely a simple problem with the comparison stars available in the first days of the eruption.  In all, T Pyx appears to have an unvarying eruption light curve, although with fast oscillations superposed.  
		
	For IM Nor, we have the problem that the 1920 eruption is known only from B-band magnitudes (except for one plate) while the 2002 eruption is known only with V-band magnitudes.  So all I can do is construct the two templates and see whether the color curve is reasonable.  The templates have a B-V color of 0.7 mag throughout, and this is consistent with a reasonable color and extinction.  In all, as far as we can tell, the two eruptions had the same light curve.
		
	For CI Aql, we have the problem that the first two outbursts have no V-band magnitudes while the third eruption has little B-band data.  In addition, the light curves for the first two eruptions have no overlap in time after peak.  Thus, we can only make poor tests for constancy of the light curve shapes.  In all, it looks like the light curves for all three eruptions are identical to within the normal measurement problems.  The one possible exception to this concerns the two brightest magnitudes from the 1917 eruption.  These magnitudes are B=8.7, while the 2000 eruption has B=9.5 at the same phase in the eruption.  We cannot blame a single plate flaw for this difference as we have two plates in 1917.  And the 2000 B-band template is unlikely to be so far off as this would require an extreme B-V color that changes fast.  Nor can we simply move the date of maximum in 1917 as this is well-constrained by the faint first point on the rising branch.  With CI Aql not showing fast oscillations, the two 1917 plates make a plausible case for a substantial difference in light curve shape between the 1917 and 2000 eruptions.  Nevertheless, I am not convinced by such a conclusion.  The problem is that the plate quality for these plates is low, the stars are near the edge, and strong field effects (in particular, the changing image size and shape with distance from the plate center) can lead to substantial systematic errors when the comparison stars are relatively far away (as will be the case when the nova is near peak).  In all, I do not judge the evidence to be sufficiently strong to force a conclusion that the light curves change.
		
	For V2487 Oph, we only have one magnitude from the first eruption, so we cannot make any real comparison between light curves. 
		
	For U Sco, we have many eruptions, but the overlaps for particular times and bands is such that only several of the events can be well compared.  The constructed template provides a reasonable representation for all the light curves.  We have the usual photometric systematic problems, for example with the V-band magnitudes of Kiyota (1999) being systematically fainter than those of the AAVSO from 8-25 days after the 1999 peak.  Small deviations from the template can be pointed to (e.g., days 29 and 30 in the 1936 eruption and around day 10 in the 1987 eruption).  But I judge these to be too small to be significant given the usual photometric systematic problems, eclipses, and flickering.  In all, I take the U Sco light curve to be constant from eruption-to-eruption.
		
	For V394 CrA, we have a relatively small amount of poor data.  The problems with the scatter (likely due to ordinary problems with comparison star calibration) means that the derived template will have a larger uncertainty than for other RNe.  Within this larger uncertainty, there is no evidence that the light curve shape was different in 1949 from 1987.
		
	For T CrB, I have constructed the templates from the very well observed 1946 outburst.  The 1866 eruption fits the 1946 template closely up until the star gets close to minimum.  Around the time that the RN returns to the quiescent level, we see a series of magnitudes from a single observer that are bright (compared to everyone else) by up to one magnitude from day 21 to 80.  This is certainly some usual error (perhaps a poor sequence of comparison stars or perhaps a wrong star identification) by Bird (1866).  Another error is the report that T CrB was seen at V=5.1 on JD 2431856 (i.e., 4 days before anyone else saw any rise) as recorded in the AAVSO database.  This report would have constituted the real discovery of the 1946 outburst if correct.  But discussions at the time (e.g., Pettit 1946) give the discovery to N. Knight (Knight 1946) even though the Siberian observer A. Kamenchuk saw the rise earlier (Kamenchuk 1946), and so there must have been some known reason to discount the observation.  I expect that the quoted date should have been reported as JD2431865, with the last two digits switched.  During the secondary maximum, the 1866 light curve is consistently a third of a magnitude brighter than at the same time in the 1946 light curve.  Could this be a significant difference between the two eruptions?  Possibly.  A possible resolution might simply be due to errors in the early comparison stars.  Alternatively, the red giant companion star has long term variations with an amplitude of roughly one magnitude, and the two secondary maxima could well simply be at different phases in the red giant cycle, resulting in the total system having some apparent offset in the light curve when the system is near quiescence.  That is, the differences between the light curves are likely just due to a different luminosity of the red giant.  Either explanation would have the entire nova eruption luminosity history plus any accretion disk brightening history being identical.  Likely the luminosity of the red giant would be considered an incidental and confusing light source that can be ignored for purposes of understanding the RN phenomenon and for purposes of understanding the specific eruption physics.  As such, the light curve for the nova event itself (plus any possible accretion disk phenomenon) is apparently constant from eruption-to-eruption.
		
	For RS Oph, the template provides a good fit to all eight light curves.  But the match is not perfect.  When RS Oph is faint, the light curve is often somewhat bright or dim compared to the template.  The deviations from the template vary slowly with respect to the duration of the eruption, and the deviations are the largest during the post-eruption dips.  During the post-eruption dips, the nova eruption light has faded away, and the accretion has stopped with an empty disk (Worters et al. 2007), so the only significant optical light source at the bottom of the post-eruption dip is the red giant.  So variations from the template must come from the red giant.  Such slow variations of the red giant are already known, for example from the modulations on the orbital period (Gromadzki et al. 2008) and from the power density spectrum of the variations in quiescence (see Section 12.3).  These variations have a typical amplitude of half a magnitude, and this is exactly what we see in the deviations from the template.  So we have a consistent picture that the nova light always follows the same template while the red giant varies slowly with moderate amplitude, such that the nova-plus-red-giant brightness will show deviations as observed from the average template.  (After the end of the post-eruption dip, when accretion resumes, the quiescent light will be from both the varying red giant plus the varying accretion disk and hot spot.)  Therefore, the nova part of the light curve is consistent with being constant from eruption-to-eruption.
		
	For V745 Sco, any comparison of the 1937 and 1989 eruption light curves has the problem that the first was only measured in B-band and the second was only measured in V-band.  So all I can do is to construct templates for the two colors and then evaluate whether the resultant color curve is reasonable.  With this, B-V varies from 1.1 near peak to 1.9 later in the tail, and I judge this to be reasonable because other RNe have similar color behavior.  The V-band light curve has substantial problems due to scatter, so there is a large uncertainty in the derived color curve.  As such, the two light curves are apparently consistent with each other.
		
	For V3890 Sgr, we have little direct comparison between the two eruption light curves.  Nevertheless, a simple template can be constructed with a constant B-V color that agrees with the light curves from both events.  As such, it appears that the light curve shape is constant for the two outbursts of V3890 Sgr.
		
		In all, I have found no significant change in the nova light curves from eruption-to-eruption.  The only confident exception to this rather strong statement is that T Pyx has fast oscillations which likely do not repeat exactly from eruption-to-eruption.  Other possible exceptions, as noted above, all arise from obvious or likely errors in the photometry.  Also, note that the non-nova light, in particular the light from the red giant companion star can make small eruption-to-eruption changes in the total system brightness.  My conclusion that the nova light always has identical light curves is a fairly strong statement, involving 33 eruptions.  What this is telling us is that the  eruption light curve depends on system parameters (like the white dwarf mass and the composition) which do not vary from eruption-to-eruption.  Also, given that the accretion rates vary greatly over time, the constancy of the eruption light curves suggests that the eruption depends little on the accretion rate.

	\subsection{Light Curve Properties}

	The templates can be used to estimate the properties of the RNe light curves.  That is, we can get a better value for peaks and durations from the templates than can be gotten by considering individual light curves with all their various gaps and the scatter of individual magnitudes.  With a template, we are averaging over all nearby magnitudes in a single light curve and averaging over all light curves.
	
	From the templates in Table 16, I have evaluated a variety of light curve properties for all ten known galactic RNe.  These are in presented in Table 17.  The first two properties ($B_{peak}$ and $V_{peak}$) are the brightest magnitudes in the light curve for the B- and V-bands, while the third line gives the V-band magnitude 15 days after peak.  Lines 4 and 5 give the B-V color at peak ($B-V_{peak}$) and at a time when the V-band light curve has faded by 2 magnitudes from peak ($B-V_{Vpeak+2}$).  The next two lines give the amplitude (the difference between the peak and quiescent magnitudes) in the B- and V-bands.  The next eight lines give the time in days from the peak until when the V-band light curve has faded by some integral number of magnitudes (as indicated in the subscript).  So for example, $t_2$ and $t_8$ are the times from the peak until when the template has faded by 2 and 8 magnitudes from the peak.  The value of $t_{-3}$ is the rise time from when the nova is 3 magnitudes fainter than peak until the time of the peak.  Some of the values are not given for the cases where the amplitude of the light curve is too small for the `t' value to be defined or for the cases where the light curve is not followed late enough to allow a measure.  The next dozen lines give the time (in days) for which the light curves are brighter than the magnitude indicated in the subscript (e.g., $D_{B=9mag}$).  These eruption durations are based on the apparent magnitude (instead of the magnitude relative to the peak).  The purpose in defining these durations is for use in discovery statistics, where it matters how long the eruption is brighter than some discovery threshold.  The first half of these magnitudes are for the B-band, while the last half are for the V-band.  The last two lines give equivalent durations for the optical light curves ($D_B$ and $D_V$) in the B- and V-bands.  These are calculated with the integral under the light curve.  The optical flux is used (not the magnitude) in this integral, so the output will be proportional to the optical energy.  The flux has the quiescent flux subtracted out, so the output is referring to the energy of the nova eruption.  The flux has been normalized to that of the peak, so the values are distance independent.  The quoted values have units of `days', and they are like an effective duration.  The total optical energy of the nova eruptions are equal to that of an idealized square-shaped light curve with the same peak and the given equivalent duration.  These equivalent durations are useful for calculating the energetics of the outburst.

	\subsection{Comparison of Light Curves}
	
	How do the light curves of the RNe compare?  In Figure 36, I have simply plotted all the templates from Table 16.  We see an apparent wide range of peaks, durations, and quiescence levels.  By construction, the peak times all come in the same position.  The peak magnitudes are governed in part by distances and extinctions (poorly known) that are irrelevant to the eruption physics.  So a better comparison might be to vertically shift all the templates so that they match at the peak magnitude, as shown in Figure 37.  We still see a wide variety of durations and decline rates.  To compare the {\it shapes} of the light curves, we can scale the horizontal axis by $t_3$ so as normalize out the durations, as in Figure 38.  Now, all the light curves are starting to look similar.  Part of this is by construction, as the scaled templates must match at two points.  Nevertheless, the light curves are indeed not greatly different.
	
	Are RN light curves different from classical nova light curves?  This question cannot be answered with confidence, as I know of no comparable study of light curve shapes for classical novae.  (Payne-Gaposhkin 1964 and Duerbeck 1981 give a fairly broad categorization of nova light curve shapes, but there is little quantification or plots for a real comparison.)  Nevertheless, from perusing various papers with light curves, I think that the RN light curves are similar to those of many novae.  A possible difference is that the RNe do not have the diversity of CNe, for example there are no RNe with DQ-Her-like dips nor any RNe with slow symbiotic-like light curves.
	
	A comparison of the B-band and the V-band templates for an individual RN will give the B-V color curve.  The color curves for the ten RNe have been collected in Figure 39.  I am somewhat surprised by the small color evolution for many of the RNe.  I would have expected that RNe would behave like CNe, starting out with a negative B-V before peak, B-V$\sim$0 around peak, and substantial reddening as the nova fades.  (Of course, interstellar extinction will redden this whole idealized color curve.)  Instead, T Pyx, IM Nor,V2487 Oph, U Sco, V394 CrA,  and V3890 Sgr have nearly constant colors.  And the color evolutions for T CrB and RS Oph are apparently the simple transition from a constant blue nova color to a constant red companion color as the nova light fades.  (The color curve for V745 Sco is poorly determined, see Figures 32 and 33.)  With this, it appears that the color of the nova light is roughly constant (or at least does not vary much) throughout the eruptions for 90\% of the RNe.  The one exception is CI Aql, which has a brief reddening interval followed by a return to a blue color followed by a slow reddening.  This is neither the nearly-constant color pattern of the other RNe nor the monotonic-reddening pattern of CNe.

	\subsection{Broken Power Law Light Curves}
	
	Hachisu \& Kato (2006; 2007) present a general model for nova light curves and conclude that they have found a 'universal decline law' based on the ordinary physics of the expansion of the nova shells.  They derive that the flux should fall off proportional to $t^{-1.75}$ when the nova is declining from roughly 2 to 6 magnitudes below peak, then break to a decline proportional to $t^{-3.5}$, followed soon by shifting to a decline proportional to $t^{-3.0}$.  This is a tremendously useful generalization.  The break from the $t^{-1.75}$ behavior occurs at a time which varies somewhat with composition, but mainly with the mass of the white dwarf.  The break time varies from one year or more for white dwarf masses of 0.6 $M_{\odot}$, to $\sim100$ days for 1.0 $M_{\odot}$, to $\sim25$ days for 1.3 $M_{\odot}$.  This result provides a practical means of estimating the mass of the white dwarf from the light curve alone.  The RNe (necessarily with $M_{WD}> 1.2$ $M_{\odot}$) must have a fast break.
	
	This result has three limitations.  First it does not account for the light curve before or near peak.  Second, a variety of effects can be superposed on the universal decline.  Thus, flares and oscillations prominent in the light curve do not have their rapid variability described, and we can only fit power law models through the middle of the light curves.  Also, dust formation is not addressed, and we can only ignore the portions of the light curve dominated by dust absorption.  The ordinary luminosity at quiescence is not part of the universal decline, so this light must be subtracted out to get the eruption light.
	
	The third limitation of the universal decline law is that it is addressing the continuum flux alone.  That is, the emission line flux is not included, and this flux rises rapidly and starts dominating during the transition phase of the nova event.  Hachisu \& Kato point out that the Stromgren `y' band filter is narrow and does not include many of the prominent emission lines, so y-band light curves should follow their universal decline law.  But the broader V-band filter includes emission lines which start becoming prominent before the break and raise the observed V-band light curve above the universal decline law.  They point out that the V-band light curves have a transition from one universal decline law to a parallel light curve with identical slopes and break times yet brighter so as to include the emission flux.  This transition will make the application of the universal light curve into a messy fit for V-band data.  Unfortunately, all we have for the recurrent novae are V-band light curves, so we can only make do.  Nevertheless, I will report here on fits of the RN templates to broken power laws.  Despite the messy transition caused by the emission lines, the V-band light curves should start out with a $t^{-1.75}$ slope, have a break around 25 days, and end up declining as $t^{-3.0}$.
	
	When constructing a power law for the light curve, we must specify some zero for the time.  That is, the flux will be proportional to $(t-t_0)^{-1.75}$, where $t$ is the time of the observation and $t_0$ is the zero time.  Hachisu \& Kato take the zero time to be the instant that the eruption expansion starts.  Operationally, I will take this as the time when the nova brightness starts its rise from quiescence.  The uncertainties in the power law slope will not be significant for the times after the nova has faded by two magnitudes from peak.  With the usual conversion from flux to magnitudes, the power law segments should obey an equation of $V=V_{break} - 2.5 \times \alpha \times \log[(t-t_0)/(t_{break}-t_0)]$, where $t_{break}$ is the break time,  $V_{break}$ is the V-band magnitude at that break time, and $\alpha$ is either -1.75 (for $t<t_{break}$) or -3.0 (for $t>t_{break}$).  Again, given the complexity of the emission line flux rising during the transition period, the details around the time of the break should not closely follow the universal decline law.  Also, the quiescent flux must be subtracted to give the V-band magnitude of the nova light alone.
	
	I have fitted the RN light curve templates to a simple broken power law.  The fit is described by four parameters; the slope for times before the break ($\alpha_{early}$), the $t-t_0$ time of the break ($t_{break}$), the V-band magnitude of the break ($V_{break}$), and the slope for times after the break ($\alpha_{late}$).  These values are presented in Table 17.  With my use of the V-band, we can only acknowledge that the fits are only a messy version of the universal decline law.  For example, with T Pyx,  the break apparently occurs before the peak and the plateau is a significant bump that is blindly fitted over.  A substantial problem is how to handle the plateaus (which are claimed to be unique to RNe), as some choices in how to handle them can improve the comparison between theory and observation.  Nevertheless, the median values are $\langle \alpha_{early} \rangle = -1.45$ and $\langle \alpha_{late} \rangle = -2.9$.  I take these median values (despite their substantial scatter) to be in reasonable agreement with the theoretical universal decline law.
	
	The universal decline law provides a nice and specific theoretical prediction to be tested.  (It also provides a nice organizing principle for understanding the declines and for estimating the white dwarf masses.)  I have plotted the declining portion of the RN templates (after subtracting the quiescent fluxes) on a logarithmic time axis in Figures 40 and 41.  The first of these figures has the light curves without any vertical shifting so that individual RNe can be picked out.  The second of these figures has all the light curves shifted vertically and horizontally so that they all their fitted breaks at the same point.  This figure also has line segments representing $\alpha=-1.75$ and $\alpha=-3.0$ which pass through the origin, as a representation of the universal decline law.  We see that there is the usual amount of scatter, but that the light curves generally are close to $\alpha=-1.75$ for early times and are close to $\alpha=-3.0$ at late times.  That is, the universal decline law of Hachisu \& Kato gives a good description of the behavior of RN light curves. Indeed, it is astonishing that the complexity and variety of light curves presented in Figure 36 can be reduced to a single uniform shape as in Figure 41.  This match with theory is by no means perfect.  For example, the late-time slope for CI Aql is $\alpha_{early}=-1.8$ or else $t_{break}>200$ days, while the apparent break time for IM Nor is at 235 days.  Other than these two cases, the RNe have fast break times (much faster than ordinary classical novae) as appropriate for their higher mass white dwarfs.  (RS Oph has a break time at 77 days and is within the range allowed for RNe masses.)  And four of the RNe have break times apparently $\sim6$ days, which indicate that the white dwarf is very close to the Chandrasekhar mass.  On this basis, I would identify V2487 Oph, U Sco, V745 Sco, and T CrB as the systems most likely to collapse as a Type Ia supernova the soonest.

	\subsection{Sharp Drops}

	In the last year, three of the RNe have been recognized to have sharp drops in their light curves.  Sharp drops in nova light curves (whether CN or RN) have not been anticipated, have no precedent, have no prior suggestion, and have no prior theoretical prediction.  What has happened in the last year is that three sets of archival data have finally been examined.  The first case is that the 1966 T Pyx data from Albert Jones has finally been taken out of his log books and made available through the AAVSO data base.  The second case is that I have taken B. Monard's recent unfiltered CCD data from the AAVSO data base and combined them with the regular AAVSO V-band data (see Figure 7).  The third case is that I have measured the modern magnitudes for all plates that show the 1945 U Sco eruption and the final light curve had the sharp drop apparent only in the preparation of the template for this paper.  It is a sobering thought for our community that there had been no other effective photometry after day 80 for T Pyx and after day 30 for U Sco.
	
	T Pyx has a sudden decline starting on day 85 in both the B-band and V-band.  In 20 days, the brightness declines by about 2.0 magnitudes.  Then the light curve goes nearly flat for roughly 65 days in the V-band and apparently for only 30 days in the B-band.
	
	IM Nor has a sudden decline starting on day 235, fading by two magnitudes in 40 days.  The slope of the light curve is substantially flatter both before and after the sharp drop.  For this conclusion, the primary evidence for the drop is a comparison of the slope soon after day 235 with unfiltered CCD light versus the slope soon before day 235 with visual observations.  A worry is that the unfiltered magnitudes might have a substantial color correction term (to get to V magnitudes) that changes greatly with time, and that this change might produce the sharp drop as an artifact.  In particular, the unfiltered CCD magnitudes would include emission lines (primarily the $H\alpha$ line) that will start rising in prominence relative to the continuum around the start of the transition region.  This worry can be largely refuted by two arguments.  First, after a constant offset, the long light curve from the unfiltered CCD magnitudes matches closely the V-band magnitudes at the same time.  This includes 6 nights of unfiltered magnitudes from the peak up until the sharp break matching the high-time-resolution V-band light curve plus the V-band photometry of Woudt \& Warner (2003) matching the light curve from the unfiltered CCD magnitudes after the sharp drop.  Based on this, it appears that any color correction term must be nearly constant.  Second, the proposed changes in the color correction term go in the opposite direction to that needed to produce the sharp break.  That is, as the $H\alpha$ flux increases relative to the continuum during the transition region, the unfiltered CCD magnitudes will include the growing $H\alpha$ flux and will relatively brighten.  In all, there are good reasons for taking the sharp drop at face value.
	
	U Sco sudden decline starting on day 33 and declining in the B-band by 1.6 mag over the next two days.  This decline is attested to by many good quality plates measured on the modern B-band magnitude scale, so there is no doubt about the existence of the sharp drop.  The V-band light curve after day 33 has only four measures (one in 1979 and three in 1999), so it is poorly defined.  Nevertheless, the V-band light curve apparently does not have the same sharp drop so apparent in the 1945 B-band data.  Possible reconciliations include that U Sco has the sharp drop only in blue colors (from dust formation?), the V-band data is not accurate so as to mask the drop, or U Sco had a sharp drop in 1945 that was not repeated in 1979 or 1999.
	
	Similar sharp drops could possibly occur late in the light curves of V394 CrA, V745 Sco, and V3890 Sgr plus maybe V2487 Oph; although the available light curves are not adequate to see these drops should they exist.  Sharp drops do not occur for CI Aql, T CrB, and RS Oph.  Perhaps it is significant that the sharp drops tend to occur in the systems with the shorter orbital periods.
	
	What is the cause of the sudden drops?  Dust formation is not consistent with the largely achromatic dimming in T Pyx, nor with the prominence of the IM Nor sharp drop in such a red band, nor with the consistency between the Woudt \& Warner (2003) V-band magnitude and the unfiltered CCD magnitudes at the bottom of the drop.  Within the framework of the Hachisu \& Kato model for plateaus (see the next subsection), another possible explanation is that the sharp drops correspond to the end of the supersoft phase (with nuclear burning near the white dwarf) such that the reprocessed light is suddenly lost from the light curve.  Presumably, we will soon have a detailed theoretical model to explain the sharp drops.
	
	\subsection{Plateaus}
	
	Do the RNe light curves have plateaus?  This is important because plateaus are being identified as a diagnostic of the RN status.  Specifically, this has been claimed for CI Aql (Hachisu, Kato, \& Schaefer 2003), V2487 Oph (Hachisu et al. 2002), and U Sco (Hachisu et al. 2000a).  The claim is that the plateau phase is characteristic of the U Sco subclass of RNe.  And recently the plateau idea has been extended to RS Oph (Hachisu et al. 2009).  The idea is that the plateau is caused ``by the combination of a slightly irradiated [companion star] and a fully irradiated flaring-up disk with a radius $\sim$1.4 times the Roche lobe size" (Hachisu et al. 2000a).  The irradiation of the disk by the supersoft emission (from the nuclear burning near the white dwarf after the wind has stopped) leads to a fairly constant optical flux added with the steadily declining light from the shell leads to a flattening of the light curve until the time when the nuclear burning turns off.  The claim is that these `true plateaus' will only occur under the conditions that also produce a recurrent nova.  As such, only RNe can have true plateaus, although not all RNe need to have plateaus.
	
	How can we recognize a plateau in a light curve?  Presumably, the plateau will be a substantial time interval during which the light curve has near zero slope followed by a relatively steep decline.  In this definition, I do not know of any formal limits for my words ``substantial", ``near", and ``relatively", although we should be able to recognize a plateau when we see it.  A real problem is that all nova events start out with a steeply falling brightness that slows down, and this could be taken as a plateau.  A practical problem is that many eruptions do not have sufficiently long coverage to reveal either the plateau or the sudden drop after the plateau.  Unfortunately, I do not know of any test with optical data (say, from a spectrum) to identify a plateau, so we are only left with looking for a stretch with a near zero slope followed by a steep drop.  (If true plateaus are caused by the end of the supersoft phase, then presumably we can check the x-ray spectrum to identify the plateaus.)  We can look through Figures 2-35 and Table 16 to see if we can spot any plateaus.
	
	A substantial problem is that `false plateaus' might also occur in nova light curves.  These `false plateaus' (so-called by Hachisu \& Kato) arise from a different mechanism from true plateaus.  False plateaus are when a broad bandpass (like the V-band) includes emission line flux which increases in brightness (compared to the continuum) such that the the continuum-plus-line flux is relatively constant.  The false plateaus can be in CN or RN.  How can the two types of plateaus be distinguished?  One possibility is to use a relatively narrow bandpass (such as the Stromgren y-band) which has little contamination from emission lines, such that a false plateau will not appear.  Another possibility is to compare the light curve shape around the plateau for different filters (with presumably different contributions from emission lines), such that a false plateau will have substantial differences from band-to-band.
	
	For T Pyx, we have good coverage going to near the quiescence level.  The light curves display an obvious plateau starting around 105 days after the peak.  This plateau comes immediately after a sharp drop in the light curve, and this is different from the other plateaus, as these have the flat portion starting during an apparently normal decline.  Another unique feature of the T Pyx plateau is that the duration is greatly different between B-band ($\lesssim$30 days) and V-band (65 days).  Within the framework of the Hachisu \& Kato model, we have an inconsistency because the supersoft phase should end when the plateau stops (around day 170) and when the sudden drop starts (around day 85).  Another problem is that the V-band and B-band light curve shapes are substantially and significantly different, with different durations and strong color evolution, with this implying that T Pyx has a false plateau.
	
	The 2002 eruption of IM Nor has an interval from day 120-235 with a flatter slope.  This interval has a slope of 0.008 mag/day (as compared to 0.033 mag/day before and 0.048 mag/day after).  This slope is small compared to the other plateau slopes (see below), but this is deceptive since IM Nor has such a long duration that the before and after slopes would also qualify as being ``near zero". A plateau from 120-235 days is consistent with the sharp drop at 235 days being the turn off time for the supersoft phase.  Unfortunately, we do not have any color information to distinguish between true and false plateaus.  In all, I conclude that IM Nor has a plateau, and the sharp drop suggests that this is a true plateau.
	
	CI Aql has {\it rises} over days 14-17 and 36-55, but these are neither near-zero slopes nor substantially long.  Hachisu, Kato, \& Schaefer (2003) point to the plateau as being at $V\sim14$ mag from days 150-300.  Indeed, my templates show the CI Aql light curve to go nearly flat with a slope of 0.003 mag/day for days 80-190.  After this, there is a somewhat steeper slope (0.005 mag/day), even though this slope is still very flat (see Figure 10).  Color information is not adequate to distinguish between the true and false classes.  However, CI Aql does have x-ray observations which show it to be a supersoft source during the plateau phase, so it is a true plateau (Hachisu, Kato, \& Schaefer 2003).  In all, CI Aql does appear to have a true plateau in its light curve.

	V2487 Oph has a definite plateau from days 14-22.  The slope during this plateau is 0.013 mag/day, while the slope before was 0.100 mag/day and the slope after was 0.067 mag/day.  Thus, the plateau is sharply distinct from the general decline and there is a definite steepening of the decline.  As such, I accept this as a plateau.  Unfortunately, we do not have color information or x-ray information from the time of the plateau, so the cause of the plateau is not known.  Hachisu et al. (2002) point to a plateau phase from days 10-30 after peak, and my analysis confirms this result.
	
	U Sco has a definite flattening during days 10-33 after peak.  The slope during the plateau is 0.05 mag/day, which is still fairly steep compared to other plateaus.  Before the plateau the slope is around 0.4 mag/day, while after the plateau the slope is 0.8 mag/day in the B-band.  This shift is large enough and sharp enough while the slope is sufficiently flat that the plateau designation seems more natural than a simple slowing of the decline.  The plateau is identical in shape between B-band and V-band, which argues that flattening is a true plateau.  U Sco has been examined in the X-ray regime only one time (20 days after peak) by {\it BeppoSAX} from 0.2-2.0 keV, where it was found to be a supersoft source  (Kahabka et al. 1999), which further points to the plateau being `true'.   With this, the case for a true plateau for U Sco is confident.
	
	V394 CrA is claimed to have a plateau from 10-30 days (Hachisu \& Kato 2000).  Its light curve has a definite flattening around day 20, going to a slope of 0.026 mag/day in the V-band and 0.018 mag/day in the B-band.  This might be a plateau, but it could also be an ordinary slow-down in the decline.  A decline at the last observed rate would put V394 CrA back to the quiescent magnitude around 320 days after peak.  (This is consistent with V394 CrA being back at minimum by July 1989; Schaefer 1990.)  The coverage of the light curve does not go past 64 days after peak, so we cannot know whether there was any sudden drop at the end of a plateau.  In all, only a weak case can be made for the flattening in the light curve being a plateau, and I judge the existence of the plateau to be possible but not conclusive.
	
	T CrB shows a smooth decline with no evidence of any plateau.  (The secondary maxima are not related to plateaus.)  

	RS Oph has a smooth decline on its way to the post-eruption dip, but there is a definite flattening from days 50-76.  The slope is 0.015 mag/day, while the slopes are 0.052 and 0.041 mag/day before and after.  So there is a distinct flattening followed by a distinct drop-off.  But the slope is definitely not zero, and it is unclear whether 0.015 mag/day is near enough to zero to count as a plateau.  The lack of color changes throughout the plateau and the supersoft phase during the plateau points to this being a `true plateau'.  Hachisu et al. (2009) identify this flattening as a plateau with the same mechanism as for the U-Sco-like RNe.
	
	V745 Sco has a poorly observed break in its decline starting at around 10 days after peak.  The slope changes from 0.35 mag/day to 0.035 mag/day.  This could well be a plateau, but doubts can arise from the sparseness of the post-break observations, the photometric problems with the light curve, and the lack of an observed drop-off after the plateau.  The RN would return to its quiescent level around 170 days after peak if it continued to decline at the rate of 0.035 mag/day, and this is plausible for the lack of any drop-off.  As such, the existence of a plateau is possible but not conclusive.
	
	The V3890 Sgr light curve has no near-flat segment, and the coverage gets close to the quiescent level.  Some might point to the lesser slope from days 4-17 (0.12 mag/day) as being a plateau, followed by a steepening (0.24 mag/day).  We are left with a situation much as for RS Oph, which Hachisu et al. 2009 claims to have a plateau.  The question is how flat does a plateau have to be?  Is the important feature of a plateau the fact that the time interval has a flat light curve or that the time interval has a distinctly flatter slope than the intervals before and after?  Until this ambiguity can be resolved, I will consider a plateau in the V3890 Sgr light curve as `possible'.
	
	So we are left with six RNe (T Pyx, IM Nor CI Aql, V2487 Oph, U Sco, and RS Oph) all showing definite plateaus, three RNe (V394 CrA, V745 Sco, and V3890 Sgr) have possible-but-inconclusive plateaus, and only one RN (T CrB) certainly has no plateau.  Of those with plateaus, apparently all are `true' plateaus other than that for T Pyx.  These conclusions are tabulated in Table 17.  Among the U-Sco-like systems (i.e., those with $P_{orbital}\sim1$ day), all four have or might-have plateaus.  For RNe, $60-90\%$ have plateaus.
	
	Do classical nova light curves display plateaus?  Unfortunately, this question cannot be answered as there is no study of the shapes of CN light curves. Without such a study, we do not know if CN plateaus are common, occasional, or non-existent.  It might be that all systems with small $t_3$ have plateaus or that a substantial fraction of all nova events of all types have plateaus, in which case the RN plateaus would not be diagnostic.  Or it might be that the occasional plateaus might point to RNe hiding out as CNe because of missed eruptions.

\section{The Missed Eruptions}

	A nova eruption can be missed for many reasons.  Perhaps the eruption occurred around the time of its conjunction with the Sun (during the solar gap).  Perhaps the eruption peaked and faded around the time of Full Moon when no one was looking (during the lunar gap).  Perhaps the eruption peaked and faded during a time when no one was watching (during an observational gap).  Perhaps the eruption occurred during a time when someone was watching but the watchers did not go deep enough to catch the event.  Perhaps the eruption occurred at a time when people were looking deep but the eruption never got bright enough to be detected.  Perhaps the eruption occurred with people looking deep enough, but no one recognized the event.  A missed eruption can come from one of these reasons or from some combination of reasons.
	
	We have a variety of reasons for knowing that many RN eruptions in the last century have been missed.  Four of the known RN eruptions are known from only one Harvard plate, and the light curves and distribution of shapes demonstrate that these systems have low likelihoods for detection of the events, with the implication that other events have been missed.  The RNe are fast eruptions that easily fit into the solar and lunar gaps, so again we know that a significant fraction of RN eruptions will be missed.  In the last century, U Sco has eruption intervals of 10.8, 19.3, 8.9, 23.7, 10.4, 7.9, and 11.8 years, so it appears that it has a regular eruption cycle of $10\pm2$ years with missed eruption around 1927 and 1957.  When a concerted effort was made to examine archival data worldwide, I found six missed eruptions while another group found another missed eruption.  Few of the eruptions had independent discoveries (see Table 2), and this implies a fairly low discovery efficiency.
	
	An understanding of the missed eruptions is necessary for measuring the recurrence time scale, and for the questions of the number of RNe in our galaxy as well as for whether the white dwarf is gaining mass.  Indeed, these are the questions that have motivated my extensive time commitment to searching through old archives worldwide.  
	
	Even though many eruptions were missed, we can quantify how many eruptions were missed in a statistical manner.  For example, even though V2487 Oph was only seen to go off in 1900 and 1998, we can use the discovery efficiencies to estimate the number of missed events.  So we must evaluate the discovery efficiencies for each RN.  And this must be done both for undirected nova searches (when the observer is scanning large areas of the sky for any nova that might appear) and for directed searches (when the observer is looking for a nova event from some particular star).  These efficiencies will be estimated in the next two subsections.  The last subsection will use the efficiencies to estimate the likelihood and number of missed eruptions.
	
	\subsection{Undirected Search Efficiency}
	
	The most general nova search is where the observer is looking for new stars anywhere in a large field.  Typically, this will be an amateur scanning the sky with binoculars, the blink comparison of plates, or the automated comparison of CCD images for variables.  All first-discovered RN events will be discovered with undirected searches.  And, since no one is monitoring old novae with any regularity, all second-discovered RN events were discovered as undirected searches, with the exception of the discovery of the 1900 outburst of V2487 Oph.  (Peltier was monitoring T CrB with only one then-known eruption, but the actual discovery was made by an amateur spotting a bright star in the sky.) The discovery efficiency will depend on the peak brightness of the nova, its position, and its duration.
	
	The undirected discovery efficiency as a function of peak magnitude can be evaluated with a calculation by Shafter (2002).  He has constructed a theoretical peak magnitude distribution from a detailed model of a realistic galaxy (including the space distribution and galactic extinction) and a typical peak absolute magnitude.  I have constructed an observed peak magnitude distribution from 182 cataloged events (Downes et al. 1997) while Shafter (2002) has constructed a similar distribution.  The model magnitude distribution is normalized for classical novae that peak brighter than second magnitude (with an implicit assumption that {\it all} bright nova will be discovered).  With this normalization, the model distribution reflects what the observed distribution should be provided all events are discovered.  As such, the ratio of the number of observed events to the model events will reveal the discovery efficiency.  Thus, with Shafter's model, I can calculate the discovery efficiency as a function of magnitude, $f_{disc}(V_{peak})$.  I find values of 1.0, 0.35, 0.22, 0.14, and 0.09 for peak magnitudes of 2, 4, 6, 8, and 10 respectively.  This result might be startling to some people, as it shows that most novae are missed.  Indeed, as most novae have faint peaks, something like 90\% of outbursts are lost.  Also, three-quarters of well-placed naked-eye events are missed.  But such poor efficiencies are not surprising to me, based on detailed examination of gaps and threshold from several nova hunters, AAVSO data, and Harvard plates.
	
	The discovery efficiency as calculated by Shafter does not include the effects of the solar gap.  How long is the solar gap?  More usefully, what is the probability that a randomly timed nova eruption will be lost due to the solar gap?  I will evaluate this by two means.  (a) From detailed examination of dates of observations from AAVSO and Harvard plates, I have quantified the gap duration in days for many RNe for many years.  The median solar gap is 150 days.  For a median duration for the RN being brighter than the discovery threshold of 30 days, there will be 120 days out of the year for which the solar gap would make for a missed eruption.  This gives a discovery efficiency of $(365-120)/365=0.67$ from the solar gap alone.  (b) For randomly timed nova peaks, we would expect them to be evenly distributed throughout the year.  Also, we would expect an even distribution in terms of the number of days from opposition.  To evaluate the distribution of nova peaks with respect to the time from opposition, I have tabulated all first and second-class data in Table 1.1 of Payne-Gaposhkin (1964).  I have created histograms for novae with the days from opposition divided into six equal bins (close to one month in size).  For all 72 novae, the numbers are 20, 15, 22, 9, 3, and 3 for the six bins.  We see a sharp drop off in numbers with most discoveries in the half-year centered on opposition.  I will take the uniform discovery rate to equal the average of the first three bins, 19 in this case.  With this uniform rate, we would expect $6\times19=114$, whereas only 72 were discovered for an efficiency of $72/114=0.63$ (with a one-sigma uncertainty of 0.08).  I find that this efficiency does not depend on the decade of the nova.  I also find that this efficiency has only a small dependency of the peak magnitude, being constant (to within the Poisson statistics) for all novae with peaks fainter than fourth magnitude.  (For bright novae peaking above second magnitude, the efficiency is $0.80$, but the statistics are so poor that this could still be 0.63 with no significant magnitude dependency of the efficiency.)  So, from both means, I find the solar gap to lead to a discovery efficiency of close to 2/3.  This means that the number of missed novae events is about half of those observed.
	
	One implication of this result is that the nova rate in our Milky Way galaxy (as calculated by Shafter 2002) must be increased by 25-50\%.  The reason is that Shafter's normalization of the theoretical magnitude distribution is based on equating the observed and theoretical rates for novae brighter than second magnitude, and this presumes that {\it all} bright novae are discovered.  But the solar gap is large and even very bright novae are missed due to solar proximity.  For a discovery efficiency of 67-80\%, the nova rate must be increased by 25-50\%, and this would increase the nova rate in our Milky Way by 25-50\%.
	
	The discovery efficiency also depends strongly on the duration of the nova event.  A nova with a large $t_3$ will have many chances to be discovered, whereas a fast nova can likely fit into the gaps and be missed.  To a good approximation, the discovery efficiency will be proportional to $t_3$.  The discovery efficiencies from the Shafter calculation and for the solar gaps are both based on ordinary classical novae, for which the median $t_3$ value is 44 days (Shafter 1997).  Thus, we have a multiplicative factor for the discovery efficiency of an individual nova of $(t_3/44)$ with units of days.
	
	The observed nova rate does not vary greatly over time within the last century, as can be seen in plots of the nova discovery rate in Shafter (1997) and Della Valle \& Livio (1996).  From Table 2, I have constructed a similar histogram for the discoveries of RNe, and again find that the rate does not vary significantly with decade.  As such, I will take the undirected search efficiency to not vary with time between 1890 and present.  
	
	We can now put these results together to get a formula for calculating the discovery fraction in an undirected search for any particular nova.  We have $F_{disc}=f_{disc}(V_{peak})\times 0.67 \times (t_3/44)$.  This can now be applied to each of the ten RNe for all of the years 1890 to 2008.  The results of this are presented in Table 18, with one line for each RN.  The median efficiency is 4\% (with a full range of 0.6-19\%).  In other words, the general all-sky searches for novae eruptions will only discover around 1-out-of-25 RN events.  This means that most RNe remain undiscovered.
	
	\subsection{Directed Search Efficiency}
	
	When a system is known to be recurrent, searchers will specifically monitor the star and examine old plates, with these procedures substantially increasing the chance of discovering an old or new eruption.  Fortunately, we can usually get good lists of the dates and limits of all the observations.  For example, the AAVSO, RASNZ, and VSNET databases record the full observation details.  Additionally, the various plate archives also have dates for all plates and I have recorded the limiting magnitudes.  These can then be used to calculate the probability an eruption can slip through the directed search.
	
	In Table 19, I have tabulated the gaps and limits for all of the searches with archival plates that have been made by myself and coworkers.  These constitute all of the archival plate searches.  (Earlier searches have been made, but these are superseded by the searches reported in this paper, as evidenced by the three eruptions discovered despite previous searches of the same archives.)  I have also tabulated the gaps and limits from the AAVSO and RASNZ databases, as well as Liller's PROBLICOM search published for CI Aql.  In Table 19, I have ordered the searches by the RN (given in the first column) in my usual order by orbital period.  The second and third columns give the source of the archival data and the applicable years.  The next column gives the typical limiting magnitude for whether the RN would have been detected had it been in outburst ($m_{lim}$).  These thresholds vary substantially, typically over a range of one-or-more magnitudes, however plates with poor limiting magnitudes are generally not counted even though the plates were examined.  These limits were not recorded for all plates, but instead were recorded for many hundreds of uniformly sampled plates for each target.  The next column gives the approximate number of observations or plates used in the search ($N_{obs}$), with an accuracy of about 10\%.  Columns 6 and 7 give the average length (in days) for the solar gap ($S_g$) and other gaps ($S_o$), as taken from detailed tabulations of observation dates from many years.  The gaps vary substantially in number and size from year-to-year, so the quoted number and duration of the gaps can only represent the average case.  The next-to-last column gives the number of days for which the RN is brighter than $m_{lim}$ ($T_{vis}$).  These values are tabulated from the brightest value of $m_{lim}$ (so as to provide conservative limits) and for the durations used in Table 19.

	We can now calculate the discovery efficiency.  This is taken as the fraction of a year where the event would have been discovered for a randomly placed peak.  Let me give some example to illustrate how this works.  If deep observations are made every day in a year (for a circumpolar target) then all eruptions would have been detected and the fraction detected ($F_{disc}$) will be unity.  Now let us take the case of a RN that is brighter than tenth magnitude for 30 days and a set of observations with a 120 day solar gap plus no other gaps as long as 30 days.  Then, for 365 days as possible peak dates throughout the year, the fraction of peak dates for which an eruption will be discovered is $F_{disc}=[365-(120-30)]/365=0.75$.  For the general case, the solar gap is $G_s$, the other gaps have average length $G_o$ and number $N_g$, and the RN has a time of $T_{vis}$ brighter than the discovery threshold.  In this case, we have $F_{disc}=[365-(G_s-T_{vis})-N_g(G_o-T_{vis})]/365$, where the subtractions are not allowed to go negative.  With this, my calculated $F_{disc}$ values are tabulated for each search in the last column of Table 19.
	
	How should we combine the $F_{disc}$ values for multiple searches in the same year?  Unfortunately, we cannot simply add the $F_{disc}$ values as the observations are largely duplicative rather than complimentary.  That is because all data sources have largely the same gap structure.  Both amateurs and patrol plate series tend to have a much larger solar gap than could be possible if the observers have pushed hard to fill the gaps.  Also, both amateurs and patrol plates avoid times of the Full Moon.  For example, this last summer at the Sonneberg Observatory, I saw that the currently ongoing photographic patrol program did not try to observe for any time when the moon is up.  (The Sonneberg program is currently the only all-sky year-round program that is recording deep images for many uses.)  The implication is that the combined $F_{disc}$ should be approximately equal to the largest value from the searches for that year.
	
	With this, we can construct the values of $F_{disc}$ for every year from 1890 to present for each of the ten RNe.  These are tabulated in Table 20.  The last line gives the discovery efficiency averaged over the years 1890-2008 ($<F_{disc}>$).
	
	The number of missed eruptions can be estimated from $<F_{disc}>$ and the observed number of eruptions ($E_{obs}$).  The actual number of eruptions in the time interval should be $E_{total}\sim E_{obs}/<F_{disc}>$.  The number of missed eruptions is $E_{missed}=(E_{total}-E_{obs})$.  For cases with low $<F_{disc}>$, we can have a substantial $E_{missed}$, with the best example being V2487 Oph ($<F_{disc}>=0.30$ and $E_{obs}=2$) with likely 4-5 missed eruptions.  With extra information, we have to take each RN on a case-by-case basis.

\section{Recurrence Time Scales}

	The recurrence time scale ($\tau_{rec}$) is important for RN demographics and for knowing whether $M_{ejecta}>\tau_{rec}\dot{M}$.  A naive means to evaluate $\tau_{rec}$ would be to simply take the eruptions tabulated in Table 2, and divide some time interval by the number of eruptions.  But we have to be careful in choosing the time interval and we have to account for the missed eruptions.  This must be done on a case-by-case basis.
	
	For T Pyx, the time intervals between eruptions are 11.9, 17.9, 24.6, and 22.1 years, for an average of  19 years.  This led to the common expectation that T Pyx would have its next eruption around 1986 (i.e., 1967+19), so many people were keeping close track trying to make the discovery.  There has been increasing frustration as T Pyx did not erupt.  The time since the last eruption is now 42 years and counting.   For T Pyx, the duration of the eruption is sufficiently long that the eruption would be discovered even if it went off at the start of a seasonal gap.  In such a case, just a few observations per year will discover all eruptions.  As T Pyx has been under close observations every year since 1890, we can be confident that no eruptions have been missed.  We now realize that what has happened is that the T Pyx accretion rate dropped substantially soon after the time of the last eruption, so it will be a long time until the next eruption (Schaefer 2005).  Indeed, a more detailed accounting that includes the recent declines since 2005 plus the associated larger trigger mass implies that T Pyx won't erupt for many centuries (Schaefer et al. 2009).  Also, with the likely continuing decline in accretion, T Pyx will soon be going into hibernation, and thus will not suffer any further RN events for almost a million years (Schaefer et al. 2009).  That is, T Pyx has stopped being a recurrent nova.  So how do we define the 'recurrence time' in such a case?  A further realization is that the recurrence time scale changes from eruption-to-eruption (from 11.9 to $>$42 years in the case of T Pyx).  I think that a reasonable way to characterize the T Pyx recurrence time scale is to take the average inter-eruption interval over the time when T Pyx was a RN (that is, to not include the time after the 1967 eruption when the accretion rate dropped to the point where T Pyx no longer was a RN), and that is 19 years.
	
	For IM Nor, we have two known eruptions (1920 and 2002) for a simplistic recurrence time scale of 81.5 years.  But there could easily have been missed eruptions in the many decades between.  For the interval with a plausible missed event (1930-1991), $<F_{disc}>=0.85$, with most of the chances for a missed eruption being from 1955-1977.  As such, the most likely case is for no missed eruption, although one missed eruption is a real possibility.  With this, the average $\tau_{rec}$ is 82 years, or perhaps 41 years.
	
	For CI Aql, we have inter-eruption intervals of 23.9 and 59.1 years.  This suggests that one-or-two eruptions were missed, resulting in a fairly constant interval.  Experience with other RNe (particularly T Pyx and RS Oph) shows that factor-of-three variations are common, which removes the necessity to invoke missed eruptions to have a near-constant interval.  Additionally, the coverage from 1950-1990 (with the Harvard, Sonneberg, Maria Mitchell, and PROBLICOM photographs) is pretty good (86\%), so a missed eruption is unlikely (but possible).  However, the quiescent magnitude from 1917-1941 is close to that between 1941-2000, so with a relatively constant accretion rate, we must have missed one or two eruptions between 1941 and 2000.  The interval before the 2000 eruption is critical for CI Aql, as I have measured the change of eclipse period across the last eruption (hence giving $M_{ejecta}$) so we need to know $\Delta$T to evaluate whether $M_{ejecta}>\Delta T~\dot{M}$.  With one missed eruption (say, around 1970) we have $\Delta T\approx 30$ years, while with two missed eruptions we have $\Delta T\approx 20$ years.  The most likely average $\tau_{rec}$ is something like $(2009-1890)/5=24$ years.
	
	For V2487 Oph, we have a case where many eruptions certainly were missed.  In this case, we cannot use the simplistic recurrence time scale (98 years), but must instead correct for the missed outbursts.  The $\langle F_{disc} \rangle$ is 0.30, so the total number of eruptions in the interval 1890-2008 should be approximately $2/0.30=6.7$.  With this, the best estimate is that V2487 Oph had 6-7 eruptions (4-5 missed), for average recurrence time scale of roughly $(2009-1890)/6.7=18$ years.

	For U Sco, we have nine eruptions, with five inter-eruption time intervals of near ten years.  Two of the inter-eruption time intervals are close to two-times-ten years, and this suggests that one eruption was missed in these intervals.  This is the exact situation expected since a significant fraction of U Sco eruptions must be lost due to the solar gap.  This would mean missed eruptions around 1927 and 1957.  The inter-eruption interval varies at least from 7.9 to 11.8 years.  This variation has been shown (Schaefer 2005) to be a simple result of secular changes in the mass accretion rate resulting in differing times required to accumulate the trigger mass on the surface of the white dwarf.  The average recurrence time scale is $(1999-1906)/9=10.3$ years.

	For V394 CrA, we have two eruptions so a simplistic recurrence time scale is the time between the two eruptions, 38.3 years (see Table 2).  The $\langle F_{disc} \rangle$ is 0.48 with two eruptions discovered, which suggests that $E_{total}$ is four (for two missed eruptions).  With this, the best estimate average recurrence time scale is $(2008-1890)/4=30$ years.
	
	For T CrB, the nova peak is short, but the eruptions are recognizable for a long time by the post-eruption rebrightening.  Between the Harvard plates and the AAVSO observers, T CrB has been closely monitored since 1890, and there is no possibility of any missed eruption.  From 1866 to 1890, T CrB has been monitored with reasonable frequency, so there is likely no missed eruptions in that interval.  So in all, we can be confident that T CrB  had only two eruptions from 1866 to present.  The recurrence time equals the one observed interval of 80 years.
	
	For RS Oph, the post-eruption dip is recognizable for almost a year after the short peak.  This means that just a few observations per year are sufficient to recognize an event.  Between the Harvard plates and the AAVSO observers, RS Oph has been closely monitored since 1890.  So we can be confident that no eruptions have been missed since 1890.  The inter-eruption intervals vary from 8.6 to 26.6 years.  The average recurrence time is $(2009-1890)/8=14.7$ years.
	
	For V745 Sco, we have two eruptions with an inter-eruption time interval of 52.2 years.  But many eruptions have likely been missed due to the faint peak ($V_{peak}=9.4$) and fast decline ($t_3=9$ days).  With the average $F_{disc}=0.36$ and $E_{obs}=2$, we have $E_{total}=5.6$ (for 3 or 4 missed eruptions).  As such, our best estimate of the average recurrence time scale is $(2009-1890)/5.6=21$ years.
	
	For V3890 Sgr, there are two eruptions with an inter-eruption interval of 27.9 years.  But there are likely a number of missed eruptions.  (For example, with a recurrence time scale of 27.9 years, we would have missed eruptions in 1934 and 1906.)  With $F_{disc}=0.43$ and $E_{obs}=2$, we have $E_{total}=4.7$ (for 2 or 3 missed eruptions).  With this, we have the best estimate for the average recurrence time scale to be $(2009-1890)/4.7=25$ years.
	
	The results from this section are collected in Table 21.  The second column gives the inter-eruption time intervals ($\Delta$T) as expected for missing events.  It is important to realize that the inter-eruption intervals can change significantly.  For well-known cases, the ratio of the longest-to-shortest $\Delta$T is $\geq$3.4 for T Pyx, $\geq$1.5 for U Sco, and 3.1 for RS Oph.  The third column gives the average recurrence time scale.

\section{The Next Eruption}
	
	With the results from the last section, we can make some predictions as to the dates of the next eruptions for each of the known RNe.  This task has some utility in planning upcoming observations and for promoting close monitoring as appropriate.
	
	Schaefer (2005) has made predictions for the next eruptions of T Pyx and U Sco.  The physical basis is that the eruptions should be triggered when some constant amount of mass has accumulated on the surface of the white dwarf, and this constant can be calibrated for each RN by looking at the B-band flux, which is a measure of the accretion rate.  The eruption will be triggered when the accumulated material ($<\dot{M}>\Delta T$) reaches some critical ignition mass ($M_{ign}$).  The accretion rate averaged over the pre-eruption interval ($<\dot{M}>$) will be proportional to a known power of the B-band flux ($F_B^{1.5}$), which is directly observable on many nights in the pre-eruption interval.  We then get the time from the previous eruption to the next eruption to be $\Delta T \propto M_{ign} F_B^{-1.5}$ with $M_{ign}$ being a constant for a particular RN so $\Delta T \propto F_B^{-1.5}$.  The constant of proportionality can be directly determined by looking at previous inter-eruption intervals where $\Delta T$ is known.  This method was confirmed by the good correlation between the inter-eruption interval and the average magnitude during that interval.  That is, when the quiescent star is bright (i.e., with high accretion rates) the next eruption comes fast, while when the RN is faint between eruptions the wait is long.  This provided a simple explanation for why T Pyx has already gone 42 years since its 1967 eruption whereas its previous average $\Delta$T was 19 years, because the B-band brightness dropped by a magnitude around the time of the 1967 eruption.  With the T Pyx accretion largely turned off, we will have to wait a long time for the white dwarf to accumulate enough mass to trigger the next thermonuclear runaway.  The quantitative analysis gives the next eruption date of 2052$\pm$3 (Schaefer 2005), while a more detailed analysis shows that the next eruption is either a millennium or a million years away (Schaefer et al. 2009).  Similarly, for U Sco, the upcoming eruption date is 2009.3$\pm$1.0.  The primary uncertainty in this result is that the accretion rate could easily change after the date of 2005, whereas the predicted dates are assuming that the accretion continues at the average rate for the earlier portions of the current inter-eruption time interval.  My continued and frequent monitoring of the two systems (with the SMARTS 1.3m telescope on Cerro Tololo) shows that U Sco continues from 2005-2009 with the same B-band brightnesses.
	
	Selvelli et al. (2008) have revisited the question of the predicted date for the next eruption of T Pyx.  Their analysis is based on a very detailed reconstruction of all system parameters, with a premier place for the ultraviolet observations with the {\it International Ultraviolet Explorer}.  They too are using the basis that $\Delta T \propto M_{ign} / \dot{M}$.  Their $\dot{M}$ is evaluated from the accretion luminosity, which is dominated by flux in the ultraviolet as they well measure.  Their $M_{ign}$ is taken from theoretical models as a strong function of the white dwarf mass and the accretion rate.  Their conclusion is that the next eruption will be in the year 2025 (for a recurrence time of 58 years), without stating any uncertainty.  Unfortunately, their real error bar on the next eruption date is very large.  Their $M_{ign}$ depends strongly on the uncertain mass of the white dwarf, such that for their stated model results of 1.36-1.38 $M_{\odot}$, their ignition mass changes by a factor of two and hence the $\Delta T$ changes by a factor of two.  Other researchers are not so confident in the claimed accuracy for the determination of the white dwarf mass, so the associated uncertainty in $\Delta T$ will be even larger than a factor of two. Another problem comes from the strong inverse-square dependancy of $\dot{M}$ on the distance.  Even for their optimistically small range of distances ($3500\pm500$ pc), their associated uncertainty in $\dot{M}$ and hence $\Delta T$ is nearly a factor of two.  Again, other researchers evaluate the distance uncertainty to be even larger (see below), so the real uncertainty from this one source is likely substantially larger than their implicit factor of two.  With two independent sources of error providing one-sigma ranges of a factor of two (or more) each, the overall range of uncertainty should be roughly a factor of three (or more).  As such, the one-sigma uncertainty range for the $\Delta T$ estimate of Selvelli et al. (2008) is really $\sim 33-100$ years (or more), so that their predicted date for the next eruption is anytime from 2000-2067.  As such, the Selvelli et al. (2008) estimate is fully consistent with the much more accurate estimate of $2052\pm3$ from Schaefer (2005).  The reason for the much greater accuracy of my estimate is simply that all the unknowns are collected into one constant of proportionality which is evaluated empirically (from many prior eruptions) with no presumptions on models or distance.
	
	van Loon (2009) has pointed to a pattern in the inter-eruption intervals of RS Oph to make a prediction that the next eruption should occur around the year 2015.  The idea is that the $\Delta T$ values apparently follow a cycle of roughly 9-22-22 years, so the next eruption after 2006 should be nine years later.  There is no physical basis for any such cycling and it is hard to conceive how the system could keep track of where it is in a cycle of three, so I take this idea to be essentially numerology.  A convincing refutation is simply to recognize that the 1945 eruption was overlooked, so the real series of inter-eruption times (from Table 2) is 8.6, 26.6, 12.3, 12.6, 9.3, 17.3, and 21.0, with no pattern apparent.
	
	For the eight RNe without the accurate physics-based predictions, we can estimate the next eruption date by simply adding the average recurrence time scale to the latest eruption date.  Such predictions are neither reliable nor accurate.  I have listed predicted dates for the next eruptions in Table 21.  Five of the ten RNe have expected outbursts in the next decade.  This provides an urge for our community to keep these RNe under intense monitoring.
	
	With U Sco, we have a confident case where we know well in advance exactly which star will suffer a nova eruption and approximately when.  This allows for the possibility of planning a detailed observing program to take advantage of the prediction.  Indeed, for this opportunity, coworkers and myself have a set of plans and target-of-opportunity observations in place, with coverage in x-ray/ultraviolet/optical/infrared with photometry and spectroscopy.  Goals are to get  hourly monitoring for an early discovery as well as to resolve the rise, daily BVRIJHK photometry for many months, fast photometry to catch eclipse times in the tail, and the spectral energy distribution from the x-ray to the infrared on many nights.  A wonderful opportunity arises with U Sco's total eclipses with an accurately known ephemeris, so we can get an eclipse mapping of the supersoft source with the XRT detector on {\it Swift}.  A bad case will be if U Sco happens to rise during the solar gap.  With good circumstances, we hope to make the imminent U Sco eruption into the best observed nova of all times.
	
	Other than the inevitable and imminent eruption of U Sco, the next RN eruption could well come from a `classical nova' whose previous eruption was only poorly observed.  In recent years, we have been surprised by second outbursts from V394 CrA (in 1987), V745 Sco (in 1989), V3890 Sgr (in 1990), CI Aql (in 2000), and IM Nor (in 2002).  Soon enough, one of the many forgotten old-novae will pop off and reveal a new RN system.
	
	We can estimate the rate at which new RNe will be discovered.  From earlier, I estimated that $\sim$80 RNe are hiding in the catalogs as ordinary CNe, and these are the systems that will be identified when a second eruption is discovered.  From Table 21, the median recurrence time scale for RNe is 24 years, so there should be of order 80/24=3.3 RNe eruptions every year.  With a typical discovery efficiency of 1-in-25 (for RNe in an undirected nova search), we will only catch a second eruption every $25/3.3=7.5$ years.  This is comparable to the observed rate for new RNe based on undirected searches of once per 4.2 years (5 new RNe since 1987).  Indeed, with the observed rate for discovering second eruptions from old `classical novae' being larger than predicted, it looks like there might be many more than 80 RNe hiding in the CN lists.
	
\section{Photometric Variations on the Orbital Period}

	So far in this paper, I have been primarily giving the photometric history of the RN eruptions.  To be comprehensive, I must also deal fully with the photometric histories during quiescence.  For the next three sections, I will in turn give comprehensive new data sets plus analysis for the photometric behavior during quiescence of all ten RNe, first for their variability associated with the orbital period, then their fast variations, and finally for their long-term secular changes.
	
	Part of a comprehensive history is to present the detailed photometry, partly to display the basis for the results and partly to allow later investigators access to the original data for further investigations.  To this end, in Table 22, I am presenting complete lists of all known magnitudes during quiescence for V2487 Oph, V394 CrA, V745 Sco, and V3890 Sgr.  (The columns are the RN, the heliocentric Julian Date of the middle of the observation, the band for the quoted magnitude, the magnitude with its one-sigma error bar, and the source for the measure.)  This table is long, so only a stub is presented in the print version of this article, while the whole table is available in the on-line version.  The T Pyx quiescence light curve is not included as it has already been exhaustively presented in Schaefer et al. (1992), Patterson et al. (1998), and Schaefer (2005).  IM Nor is not included in Table 22 because the only existing data are from the old sky surveys (see below), my three BVRIJH measures (see Table 25), and Monard's fast time series photometry (see Table 5 and 23).  I have vast amounts of data in quiescence for CI Aql and U Sco, with these being aimed at eclipse timings, and these magnitudes being reserved for a later paper on the change of their orbital periods.  I have not included (in Table 22) the list of the incredibly large number of quiescent magnitudes for T CrB and RS Oph, as these are available in the AAVSO data archive (but see Figures 62 and 63).  The data reported in this paper  and as part of my long-term program now constitute virtually all of the known quiescent magnitudes for IM Nor, CI Aql, V2487 Oph, U Sco, V394 CrA, V745 Sco, and V3890 Sgr, plus the large majority of the published T Pyx quiescent magnitudes.
	
	Most cataclysmic variables show periodic photometric modulations, with the period arising from the orbit.  These can arise from eclipses, ellipsoidal variations, irradiation effects, and asymmetric emission from the hot spot.  As such, the modulations can provide information of a wide variety for the properties of the system.  The single most important property of any cataclysmic variable system is the orbital period.  As such, we can learn a lot from seeking and measuring the orbital modulations.  This is especially true for RNe, as they uniquely have an incredibly wide range of periods (from 0.076 to 519 days; over a factor of 6000).  In this section, I will present the case for each of the galactic RNe.  In summary, periods are now confidently known for nine systems (five discovered as a part of the programs reported in this paper), of which three systems are eclipsing (CI Aql, U Sco, and V394 CrA), one system apparently has shallow eclipses (V3890 Sgr), and another system has likely eclipses (IM Nor).
	
	\subsection{T Pyx}
	
	Early work on T Pyx gave claimed orbital periods of $100\pm5$ minutes (0.069 days, Szkody \& Feinswog 1988) and 3.439 hours (0.1433 days, Barrera \& Vogt 1989).  Schaefer (1990) reports on 372 B-band magnitudes from 1988-1989 and found a highly significant and stable modulation with a period of 2.3783 hours (0.099096 days).  The problem with this was that the Fourier transform had severe daily alias structures, so it was hard to pick out the correct alias and other aliases were possible (specifically including the alias at 0.076 days).  Schaefer et al. (1992) collected 1713 magnitudes from 1966 to 1990, and found that the true period is $0.07616\pm0.00017$ days.  The modulation was found to be variable in time, with no explanation suggested.  Patterson et al. (1998) collected a wonderful set of time series observations (185 hours total on 38 nights) on telescopes in Chile and South Africa from 1996-1997.  Their excellent time coverage allowed them to see the more complicated picture with multiple periodicities.  The dominant periodicity is at  $0.0762264\pm0.0000004$ days with high stability, and they identify this as the orbital period.  Their averaged light curve folded on this period is shown in Figure 42.  Collecting all the data together, they find that the orbital period is increasing with $\dot{P} = 6\times 10^{-10}$ and $P/\dot{P} = 300,000$ years.  (This period change reconciles the two accurate periodicities in Schaefer et al. 1992 and in Patterson et al. 1998.)  Patterson et al. also recognized weaker signals with periods of 0.1098 days and 1.24 days, with these varying in amplitude and of unknown stability.  With these transient periodic signals beating with the orbital modulation, they provide a ready explanation for why previous workers had reported night-to-night changes in the light curve shape.
	
	 The underlying cause of these modulations is not certain.  The 0.076 day periodicity dominates and is very stable, so this seems very likely to be the true orbital period.  Patterson et al. suggests that the 1.24 day periodicity might arise from precession in the accretion disk.  They have no suggestion for the origin of the 0.1098 day period, although they mention the white dwarf rotation period in a related context.  The shape of the light curve (see Figure 42) is a flat top lasting half the orbit and then a dip that lasts the other half of the orbit.  The duration of the dip is much too long for it to be an eclipse.  Patterson et al. (1998) suggest that the dip is caused by heating effects on the companion star for a fairly low inclination ($\sim 10\degr-20\degr$) system.  They point out that the observed change in the orbital period could easily arise from conservative mass transfer in the binary at a rate of $\dot{M}=2\times 10^{-6}~M_{\odot}$ yr$^{-1}$.  However, they caution that cataclysmic variables have a long history of deception by period changes (which often is dominated by some cause other than conservative mass transfer) and they point out that any theoretical understanding of the high accretion rate (in such a short-orbit system) is lacking.  An obvious observational task is to check the T Pyx orbital period to see if it is indeed increasing linearly (with a quadratic O-C curve) as predicted by their ephemeris.
	 
	 With T Pyx having a distinct orbital modulation, it is a RN system for which we can now get a pre-eruption orbital period, and then also get a post-eruption orbital period after the next eruption.  A substantial problem is that the modulations on the orbital period are not distinct when viewing just a few cycles (due to the usual flickering plus other periodicities being superposed), so many cycles are needed to give a good epoch.  Also, even with a well-measured averaged light curve from around one epoch (e.g., Figure 42), the phase of minimum light is poorly defined when compared to the minimum of an eclipsing system, and this will then require long runs of observations to determine the orbital period to adequate accuracy.  Another daunting problem with this task is that T Pyx is expected to next erupt $10^3$ to $10^6$ years from now (see Section 9), and that is a long time to wait.
	
	\subsection{IM Nor}
	
	IM Nor was largely ignored until it erupted for a second time in 2002.  I then realized that a good run of time series photometry was highly desirable so as to get an orbital period, and even had telescope observations scheduled from Cerro Tololo.  But Woudt \& Warner (2003) were the first, and they did a nice job at the Sutherland site of the South African Astronomical Observatory.  In February 2003, late in the tail of the 2002 eruption, they discovered an orbital period of 2.462 hours (0.1026 days) manifested as periodic dips in brightness (see top panel of Figure 43).  At first glance, the dips look like they are eclipses, but Woudt \& Warner point out that the duration of the dips is too long for being eclipses of either the accretion disk or the (possibly bloated) companion. They consider the possibility that the dips are caused by superhumps, but they reject this possibility due to the shape and amplitude of the dips not matching those of superhumps in other systems.  Then, taking CI Aql as an analogy, they conclude that IM Nor has a light curve where the dips are primarily caused by a reflection effect from the heated face of the secondary star, plus partial eclipse of the accretion disk.  With their CI Aql analogy, Woudt \& Warner might expect that IM Nor in quiescence would have only an eclipse of the normal width, as the irradiation of the companion star will be greatly reduced.
	
	The AAVSO database has a wonderful collection of fast time series photometry of IM Nor by Berto Monard of Bronberg Observatory (Pretoria South Africa).  He reports 6256 measures on 17 nights with typical intergration times of 30 seconds, plus 15 measures on scattered nights from JD2452460-2453360 (all given in Table 5).  These measures are taken with an unfiltered CCD camera and the zero point for the magnitudes are taken for a red band.  As such, we cannot place these magnitudes onto any of the standard magnitude systems, but it is nevertheless highly useful for defining the shape of the late tail of the light curve and for tracking the periodic modulations.  The nightly magnitudes (or the nightly average magnitudes) are given in Table 23.  We see the usual slowing-decline, with the nova reaching its quiescent steady state around JD2453050 (761 days after the peak).
	
	The Monard light curves display a significant modulation with the 0.1026 day period (see bottom panel of Figure 43).  When the nova is bright, the modulation has a low amplitude.  The amplitude steadily increases as the system fades.  The amplitudes for each night where Monard has a long time series are presented in Table 23.  To the best of my knowledge, this is the first time that anyone has measured the change of eclipse amplitude throughout the tail of a nova eruption.  Presumably, these amplitudes should be used by theorists and modelers to test their models concerning the optical depth all the way to the center of the nova shell.  Also, it might be possible to translate these optical depths as a function of time into a measure of the total mass ejected by the nova event.
	  
	  The dip in the folded quiescent light curve (Fig. 43) lasts roughly 0.3 in phase, so it is impossible that it is entirely caused by an eclipse.  The light curve shape has not changed greatly from the time of Woudt \& Warner (2003) until quiescence.  With the suggestion of Woudt \& Warner that the light curve shape is caused in part by irradiation, then apparently the degree of irradiation has changed little from the time in the tail of the light curve until a time 129 days after quiescence has been reached.  One possible explanation is that IM Nor has a persistent supersoft x-ray source which heats its companion star so as to drive the high mass accretion rate (like for T Pyx, see Knigge et al. 2000, Schaefer et al. 2009).  That is, IM Nor has a short orbital period (inside the period gap), so ordinary angular loss mechanisms cannot drive the accretion at the high rate required to power the fast recurrence time scale, and the supersoft source hypothesis provides a ready explanation.  And this supersoft source can provide the fairly constant irradiation to account for the light curves in Figure 43.
	
	The Monard light curves can also be used to produce a very accurate set of eclipse times.  I have made folds and fits and have derived a set of heliocentric JDs for the photometric minima for 12 night (see Table 23).  The ephemeris of Woudt \& Warner (2003) has the minima at HJD $2452696.53+N\times0.1026$, while the accuracy of this is not adequate to cover the whole time interval.  With the minima from Monard, I have improved the ephemeris.  The integer for the cycle count ($N$) and the $O-C$ are given in Table 23, with the one-sigma scatter in the minima times equalling 0.002 days.  The better period is $0.10263312\pm0.00000027$ days and the epoch is HJD $2452696.52538\pm0.00055$.  A substantial trouble with taking this accurate period as the precise orbital period is that the center-of-light in the system (which corresponds to the phase of minimum) shifts from the tail of the eruption (where the center-of-light is near the white dwarf) to quiescence (where the center-of-light is near the hot spot).  Another question relates to whether the minima are associated with eclipses or with the irradiated side of the companion star.  Until this question is answered, we cannot make any real corrections to my accurate period.
	
	With IM Nor possibly showing eclipses (or at least some dip tied to the orbital period), this system can be used to measure the orbital period change across the next eruption.  This would require, starting soon, a long series of eclipse timings.  The disappointing likelihood is that the next IM Nor eruption will be a long time coming ($\sim$2084 or perhaps $\sim$2043), and this is a long time to wait.
	
	\subsection{CI Aql}
	
	CI Aql was identified as an eclipsing binary (Mennickent \& Honeycutt 1995) long before its second known eruption in 2000.  The orbital period was 0.62 days, and their folded light curve shows deep partial eclipses superposed on prominent ellipsoidal variations.  With the eruption in 2000, it was obvious that an accurate measure of the post-eruption orbital period could lead to a good measure of the period change caused by the mass loss due to the ejected shell of material.  The pre-eruption period can be well-measured because the Mennickent \& Honeycutt  light curves go back to 1991 and because an accurate post-eruption O-C curve can be extrapolated back to the date of the eruption so as to give a good effective eclipse time at the end of the pre-eruption interval.  With this in mind, in 2001, I started a prolonged program to get eclipse timings of CI Aql.  I used various telescopes at McDonald Observatory and at Cerro Tololo.  I have also found eclipses going back as early as 1926 as taken from the Harvard plates.  I now have a total of 80 eclipse times based on 4500 individual magnitudes, with this program ongoing.  My best orbital period is $0.6183609\pm0.0000005$ days.  The analysis of the full O-C curve is beyond the scope of this paper, and will be reserved for a later paper.
	
	Lederle \& Kimeswenger (2003) used the Mennickent \& Honeycutt data, along with their own eclipse timings, to claim a measure of the orbital period change, which they then ascribed to be caused by the usual conservative mass transfer in a binary.  This conclusion has many grave problems.  First, they ignore the effects of the mass loss during the eruption, so their derived $\dot{M}$ can only be a limit.  Second, both of their eclipse timings are on the tail of the eruption, and this means that the center of light was shifted (to near the white dwarf) from its normal quiescent position (near the hot spot) which introduces a systematic offset in the time of eclipse minimum.  The size of this shift is up to 0.006 days for CI Aql and around 0.015 days for U Sco, with these effects completely swamping the claimed effect.  Third, the claimed period shift is based on eclipse times over small baselines, so they must be poor in accuracy, and indeed the claimed period change is only roughly 2-sigma in significance (even ignoring the systematic problems).  In all, the claimed period change is certainly wrong.
	
	The light curve of CI Aql in the late tail of its 2000 eruption changes its shape as it fades.  The peak V magnitude in the light curve varies as 14.7 mag on JD 2452019 (347 days after peak), 15.3 mag on JD 2452125 (453 days after peak), 15.9 on JD 2452224 (552 days after peak), 15.9 mag on JD 2452426 (754 days after peak), and has fallen to the quiescent level with a maximum of 16.0 mag on JD 2452769 (1097 days after peak).  The secondary minimum is not visible before day 500 or so.  The depth of the primary eclipse is 1.0 mag on day 453, 0.8 mag on day 552, 0.8 mag on day 754, and 0.7 mag in quiescence.  The marked asymmetry between the brightness at phases 0.25 and 0.75 is first apparent in the light curve on day 754.
	
	The light curve of CI Aql in quiescence is shown in Figure 44.  A number of points can be drawn from this figure:  First, the light curve repeats from orbit-to-orbit with fair accuracy, with the exception of around the phase 0.1-0.2.  I would guess that these variations are associated with the usual changes in the hot spot brightness, where sometimes the hot spot is bright and sometimes it is dim.  Second, the system is brighter at phase 0.25 than at phase 0.75 by 0.12 mag.  I know of no precedent for such an asymmetry.  The maximum deviation from a symmetric light curve is around phases 0.1-0.2, and this suggests that the asymmetry is caused by the extra light from the hot spot being preferentially beamed in one direction.  But I know of no theoretical expectation for the phasing of this extra light.  Third, the secondary minimum is apparent with a depth of 0.1 mag or slightly smaller.  Fourth, the usual flickering in the light curve has an amplitude of under 0.1 mag, and the flickering has apparently disappeared during the eclipses.
	
	\subsection{V2487 Oph}
	
	V2487 Oph does not have a known orbital period.  Hachisu et al. (2002) have likened this system to the U Sco `subclass' and hence suggest a period range between 0.3-3.0 days or so.  With this system only discovered as an RN a few months ago, no one has taken a long close look at the system.  However, I have taken a brief look back in May/June 2002 and May 2003.  With the McDonald Observatory 82-inch and 0.8-m telescopes, I took 68 CCD images in BVRI colors on 8 nights.  I have also recently taken photometry on 21 nights during September/October 2008 with the SMARTS 1.3-m telescope in Chile, resulting in 20 BVRI magnitudes plus simultaneously 21 J-band magnitudes.  In the spirit of this paper being a comprehensive photometric history and these sketchy observations being all the photometry in quiescence, I have tabulated all my 109 magnitudes in Table 22.  Unfortunately, no significant periodic modulation was seen.  V2487 Oph is an obvious candidate for a detailed intensive study.
	
	\subsection{U Sco}
	
	U Sco was discovered to have a deep eclipse with an orbital period of 1.23 days (Schaefer 1990).  I immediately realized that this provides a good opportunity to get an accurate pre-eruption orbital period leading to the measure of the mass of the ejected shell.  With the lure of a big science return, I started to regularly take time series photometry so as to measure eclipse times.  The idea was to measure the pre-eruption orbital period, await the next eruption, and then measure the post-eruption orbital period.  It would take perhaps a decade to measure the periods with sufficient accuracy, so I knew back in the late 1980's that I was starting on a very long-term program.  The next eruption occurred in 1999, and I have since been measuring further eclipse times.  I currently have 47 eclipse times based on 2300 individual magnitudes.  The analysis of this large data set is beyond the scope of this paper, and will be reserved for a later paper.
	
	Schaefer \& Ringwald (1995) used the early eclipses to derive an orbital period of $1.2305631\pm0.0000030$ days.  Matsumoto, Kato, \& Hachisu (2003b) have tried to use these data along with an eclipse time during the 2000 eruption to deduce a change in the orbital period.  This procedure gives a completely erroneous result.  One primary reason is that there is a shift in the phase of minimum light associated with the expected change in the position of the center of light from the offset hot spot to a region centered on the white dwarf during the eruption.  My long series of eclipse timings demonstrates that this effect is 0.015 days (with systematic drifting as the nova fades) that I can see in both the 1987 and 1999 eruptions.  That is, the primary measure that goes into their putative period change suffers a systematic error of 22 minutes simply because it is during an eruption, and this completely invalidates their derived period change.  Another reason is that their claimed eclipse time based on the alleged detection of a secondary minimum is spurious because they are merely pointing to an insignificant bump in the light curve with the usual flickering (with many such bumps recognizable).
	
	Various light curves for U Sco are presented in Figures 45-47.  We see a prominent and deep eclipse.  These eclipses appear to be flat-bottomed, and so the eclipse is total.  For the primary eclipse, the average contact times have phases $-0.0857\pm0.0036$, $-0.0132\pm0.0014$, $+0.0121\pm0.0021$, and $+0.0915\pm0.0224$ for the first, second, third, and fourth contacts respectively.  The phases are with respect to the time of minimum light in the eclipse as based on parabolic fits to the faint half of the eclipse.  The fourth contact has a substantial scatter which is larger than the measurement uncertainty.  At quadrature (i.e., phases -0.25 and +0.25), the system brightness varies considerably, but the average brightness is B=18.5, V=17.8, and I=17.3 mag.  The scatter in the magnitudes during eclipse is substantially smaller (presumably because the variable hot spot is not visible), while the average brightness is B=19.9, V=18.9, and I=18.1 mag.  The B-band light curve outside of eclipse is apparently flat with the usual flickering superposed.  The I-band light curve displays a prominent secondary minimum (with typical depth of 0.3 mag) and possible ellipsoidal variations.  The contact times for the secondary minimum (based on only three secondary minima) are at phases of -0.40 and +0.36.  The reason why the secondary minimum is visible in the I-band light curve (but not the B-band light curve) is because the companion star is much redder than the accretion disk, so in red light the loss of light during the secondary eclipse is a substantial portion of the total system light.
	
	\subsection{V394 CrA}
	
	V394 CrA has a photometric period of 0.7577 days with a roughly sinusoidal modulation of amplitude $\sim$0.5 mag (Schaefer 1990).  This periodicity is highly significant.  But the phased light curve shows much scatter.  And my later data (see Schaefer 2009 and Table 22) have years that display a lower amplitude.  What appears to be going on is that the periodicity is prominent when V394 CrA is faint and the periodicity displays a lower amplitude when the star is bright.  This could arise by there being some extra unmodulated light that is variable on a long time scale, such that when the extra light is bright it damps the periodic modulations.  Detailed analysis of the light curves demonstrates a consistency with this idea (that the amplitude decreases appropriately as extra light is added), although the variations are sufficiently large that this is not a fine conclusion.  Even during times when V394 CrA is bright, I can see the basic sinusoidal pattern rising and falling.  So a folded light curve will always show large scatter even though the basic sinusoidal pattern is stable.
	
	Schaefer (2009) presents a detailed analysis of the 500 magnitudes from 1989 to 2008 for V394 CrA.  The photometric modulation is stable, implying that it is related to the orbital period.  However, the folded light curve shows an asymmetry between the odd and even cycles, and this is a clear indication that the orbital period is actually twice the photometric period.  With this, the orbital period is $1.515682\pm0.000008$ days, which is accurate enough for use in seeking its change across the next eruption (presumably about a decade from now).  The ephemeris for the primary minimum is $HJD_{primary}=2453660.81 + N\times 1.515682$.  The folded and averaged light curve (Fig. 48) shows what looks like broad primary and secondary eclipses, although (like IM Nor and T Pyx) the durations might be too long to be true eclipses.  The depths are 0.40 and 0.28 mag for the primary and secondary eclipses (from the average brightness outside minima).  There might be some ellipsoidal modulations with the eclipses superposed.  The out-of-eclipse phases display an asymmetry with the system brighter in the elongation after the primary eclipse than after the secondary eclipse.  This behavior is similar to that seen in CI Aql.
	
	\subsection{T CrB}
	
	T CrB has a red giant companion star (M3 III), so it must have a long orbital period.  Sanford (1949) discovered the period with a radial velocity curve, and this period has been refined by Kraft (1958), Kenyon \& Garcia (1986), and Fekel et al. (2000) to $P_{orb}=227.5687\pm0.0099$ days and an epoch for zero phase (when the red giant is in conjunction in front of the white dwarf) is $2447861.73\pm  4.6$.  With a 40-year light curve, Leibowitz, Ofek, \& Mattei (1997) independently find the orbital period to be $227.532\pm0.170$ days.  This period implies that the red giant companion star is close to that of a canonical isolated M3 III star if it fills its Roche lobe.
	
	The mass of the T CrB hot star has been a critical question in the history of this system.  Kraft (1958) originally determined this mass to be $\ge$2.1 $M_{\odot}$, which is certainly much larger than the Chandrasekhar mass.  Webbink (1976) and Webbink et al. (1987) took this to heart, decided that the hot star has to be a main sequence star, and constructed a model for the nova eruption based on an accretion instability.  This model has always left many researchers uncomfortable, partly because the nova event is so characteristic of thermonuclear events in classical novae and partly because there is no precedent for Webbink's scenario.  And the entire basis for the model (the high mass for the accreting star) is not regarded as confident.  Part of the reason is that the mass is even today based on only seven photographic radial velocities (with only two plates near quadrature) for the hot component taken by Kraft in the 1950's, and Kraft has twice cast doubts on his own measures (Selvelli, Cassatella, \& Gilmozzi 1992).  Part of the reason is that the radial velocity curves of the hot component are notoriously unreliable because the emission lines from the disk do not accurately follow the motion of the white dwarf (cf. Wade 1985; Robinson 1992; Schaefer \& Ringwald 1995).  Part of the reason is that Kraft's mass is only two-sigma from the Chandrasekhar mass.  Part of the reason is that a further study by Kenyon \& Garcia (1986) has lowered the limit to simply $>1.6 M_{\odot}$, and only small changes in the mass ratio will get the limit below the Chandrasekhar mass.  All of these reasons combine to make it reasonable to think that the mass of the hot component could well be that of a white dwarf.  On top of this, Selvelli, Cassatella, \& Gilmozzi (1992) make a strong case that T CrB must have a white dwarf.  In particular, T CrB is emitting a very high luminosity mostly in the ultraviolet, its spectrum contains high-excitation emission lines like He II and N V, these emission lines have profiles with very large rotational broadening, the system shows flickering even down to the 10-second time scale, and the system is a bright x-ray source.  All of these observed properties are essentially impossible within the main sequence accretor model of Webbink, yet are normal for a white dwarf with an accretion disk.  They make many further points showing how the 1946 eruption of T CrB is inconsistent with predictions of the Webbink model and is consistent with the usual thermonuclear runaway model for RNe.  In view of all this, one of the authors of the Webbink et al. (1987) paper has already admitted that T CrB has a white dwarf (Livio 1992a).  So, despite not being able to point to where Kraft made an error of 8 km/s in his radial velocity on one plate, our community is now confident that T CrB does have a white dwarf and its eruptions are powered by thermonuclear runaway. 
	
	The orbital period is strongly manifest in the light curve as ellipsoidal variations on the red giant (Zamanov et al. 2004; Leibowitz, Okek, \& Mattei 1997; Belczy\'nski \& Miko{\l}ajewska 1998).  The folded light curve is shown in Figure 49, taken directly from Figure 1 of Zamanov et al. (2004).  The folded light curve in Leibowitz, Ofek, \& Mattei is identical in shape (sinusoidal) and average (V=10.1 mag), although it has a smaller full amplitude (0.14 mag instead of 0.33-0.39 mag).    The orbital inclination has an upper limit of  close to 70$\degr$ based on the lack of eclipses.  Belczy\'nski \& Miko{\l}ajewska (1998) perform a complete analysis of the ellipsoidal variations and the radial velocity curve, with their conclusions being that the mass of the white dwarf is $1.2\pm0.2$ M$_{\odot}$, the mass of the red giant is $0.7\pm0.2$ M$_{\odot}$, the mass ratio is $0.6\pm0.2$, the orbital inclination is $60\degr \pm 5 \degr$, and the red giant is rotating synchronously.
	
	Leibowitz, Ofek, and Mattei (1997) use a 40-year light curve to claim the existence of ``an oscillation, possibly periodic with a period of 9840 d, with an amplitude of 0.09 mag."  Unfortunately, their 40 years of data only contains less than 1.5 cycles.  As such, no claim for such a periodicity should have been made.  They undoubtedly have seen ups-and-downs in the light curve at the 0.1 mag level with a time scale of decades, but this is a greatly different case than a ``periodicity".
	
	The spectral type of the red giant companion star has been reported from M0 to M5.  In Table 24, I have collected these reported spectral types.  A substantial worry for any such compilation is that the various spectral types are based on different diagnostics, with these inconsistencies perhaps producing substantial scatter.  For T CrB, with its radial velocity curve, we can reliably phase the measures.  In principle, there must be a reflection effect at some level, wherein the hemisphere facing the white dwarf is heated and displays an earlier spectral type than does the outward facing hemisphere.  If this effect is detectable in T CrB, we might see the reported spectral types correlated with phase in Table 24.  The zero phase for the ephemeris in Table 24 is for when the red giant is in conjunction in front of the white dwarf, so the spectral type will be the coldest around zero phase and hottest around 0.5 phase.  However, I do not see any such phase-type correlation, so any reflection effect must produce a heating roughly smaller than the equivalent of 2 subclasses.  This fairly weak conclusion is in agreement with the observational evidence that the reflection effect does not appear in the optical light curve and the theoretical evidence that the luminosity of the white dwarf is too small to produce a measurable reflection effect (Belczy\'nski \& Miko{\l}ajewska 1998).
	
	\subsection{RS Oph}
	
	RS Oph has a long orbital period.  This is known because the spectral features of an M giant are seen in its spectrum.  Garcia (1986) measured 8 points on a radial velocity curve over 479 days and suggested that RS Oph has a period of $230\pm10$ days, but I see no significant periodicity in his noisy data (even with hindsight).   Dobrzycka \& Kenyon (1994) measured 47 radial velocities for the absorption features and they got a fine sinusoidal curve with a period of $460\pm5$ days.  Fekel et al. (2000) added 15 more radial velocities and substantially extended the time interval with data, and thus improved the orbital period measure to $455.72\pm0.83$ days.  Brandi et al. (2009) have added spectra from 1998 to 2008 and fine-tune the orbital period to 453.6$\pm$0.4 days.  The JD epoch when the red giant is in conjunction in front of the white dwarf (zero phase) is $2445043.54\pm5$.  Significantly, Brandi et al. (2009) also measure the radial velocity curve for the wings of the H$\alpha$ line, and derive a mass ratio (for the red giant and the white dwarf) to be $0.59\pm0.05$.
	
	The orbital period has only weak manifestation in the light curve.  Oppenheimer \& Mattei (1993) were only able to find a weak signal with a period of 471 days only in the 1985-1993 time interval.  Their Fourier analysis was of the extensive collection of AAVSO magnitudes from 1933-1993.  No signal at 230 days was ever seen.  The reason for this small modulation on the orbital period is likely because the large non-orbital variations (see Section 12.2) mask the orbital effects and because RS Oph has a moderately-low inclination ($\leq 35 \degr$, Dobrzycka \& Kenyon 1994; $49\degr - 52\degr$, Brandi et al. 2009).  Gromadzki et al. (2008) have performed essentially the same analysis, except they have also taken out the long term trends.  They find a significant periodicity at $452.9\pm2$ days, and no significant signal at the half-orbital period.  They find roughly sinusoidal photometric modulation in the V-band with amplitude of 0.4-0.6 mag from 1935-1998, but an amplitude of less than 0.2 mag from 1998-2003.  The lack of any signal at the half-orbital period again rules out any significant ellipsoidal effect.  The variations at the orbital period might be from reflection effects or from asymmetries in the emission of light from the disk.  The timing of the photometric peaks has the minima occurring when the red giant is in front, and this is consistent with a reflection effect.  They point out that the minima are deeper when the star is brighter, with the implication that the companion star is responsible for the brightness fluctuations in quiescence.
	
	The RS Oph system might have a reflection effect that is manifest in the spectral classification of the red giant companion star.  In particular, the star should appear colder when the red giant is in conjunction in front of the white dwarf (near zero phase) and should appear hotter when the inward hemisphere is visible (near 0.5 phase).  In Table 24, I have collected the many reported measured spectral types for RS Oph, as well as the V-band magnitude for that date (as taken from the AAVSO database).  The spectral type has no correlation at all with the orbital phase.  This is easy to understand because of the low inclination so that the viewing geometry does not change greatly with orbital phase.  The spectral type has been plotted versus the V-band magnitude in Figure 50.  (For the vertical axis of the plot, the spectral types are translated to numbers, M0 to 0, M1 to 1, and so on, with the modern convention that eliminates the K6-K9 classes for giants, so K5 is set to -1, and K4 set to -2.)  Here we see a strong correlation where the spectral type changes from K5 to M4 roughly linearly as the V-band magnitude changes from 10.7 to 12.2 mag. This effect will be discussed in Section 12.2 as evidence for giant convection cells on the red giant dominating the long-term variations.
	
	The referee has expressed distrust of the three RS Oph spectral measures by Walker (1979), Bohigas et al. (1989), and Sherrington \& Jameson (1983) as based on narrow procedural grounds.  This points to a generic problem with collecting measures from diverse sources (as in Table 24). The referee further suggests that only the spectral types from Anupama \& Miko{\l}ajewska (1999) and Dobrzycka et al. (1996) form a homogenous data set of sufficient size and quality to be used for the purposes of seeking correlations.  With these rejections and selections, all remaining measures show RS Oph to have only small variability between K5 and M0.  In this case, the correlation of spectral type with magnitude goes away, while the spectral types still show no correlation with orbital phase.
	
	In some references, the spectral type of RS Oph is stated to vary from G5 to M4.  The apparent basis for the G5 extreme is the measure reported in Adams, Humason, \& Joy (1927).  However, this classification has been later denied by one of the original observers after a re-examination of the original spectral plates (Humason 1938).  As such, we should now correctly state that the spectral type of RS Oph only varies from K4 to M4, or maybe just K5 to M0.

	\subsection{V745 Sco}
	
	V745 Sco must have a very long orbital period because its spectrum shows the companion star to be a M4 III red giant (Harrison, Johnson, \& Spyromilio 1993).  Schaefer (2009) reports on a series of 516 R-band magnitudes and 98 J-band magnitudes from 2004 June to 2008 August plus 7 BVI magnitudes, for a total of 621 measures (see Table 22).  With this, a highly significant photometric periodicity of 255$\pm$10 days was found.  The systematic variations in the light curve between odd and even cycles demonstrates that the orbital period is twice the photometric period.  The folded and binned light curve is displayed in Figure 51.
	
	For 510 day ellipsoidal oscillations, we are left with an ambiguity as to what phase corresponds to the red giant being in conjunction in front of the white dwarf.  Presumably, this zero phase will correspond to one of the two minima in brightness, as the conjunction will be when the small cross section of the red giant is pointing towards Earth.  But which of the minima has the red giant in front?  This can be distinguished by looking at the depths of the minima as well as the measured spectral types for the red giant as a function of orbital phase.  The conjunction with the red giant in front of the white dwarf (what I am taking as zero phase) should have the heated hemisphere away from Earth so we will see the cool hemisphere and a relatively late spectral type and the deepest minimum.  At a phase of 0.5, the heated hemisphere will be pointed towards the Earth and the minimum will not be deep due to the reflection effects, and the spectral type will be relatively early.  Taking the deepest minimum to be zero phase, I get an ephemeris for the minima to be JD $2453800+N\times510$.  In Table 24, I have collected all the measured spectral types and calculated the orbital phases.  Unfortunately, all observations were fortuitously taken near elongation (phase ~0.25 and ~0.75), and indeed the spectral types change little, being M6$\pm$2.  As such, the spectral types are not helpful for confirming the zero of the orbital phase.
	
	\subsection{V3890 Sgr}
	
	V3890 Sgr has a M5III red giant companion star discovered with infrared spectroscopy (Harrison, Johnson, \& Spyromilio 1993).  Schaefer (2009) reports on 374 optical magnitudes over 138 nights from June 2004 to September 2008 and 105 J-band magnitudes with the SMARTS telescopes in Chile, 465 magnitudes effectively in the I-band on 340 nights with the ROTSE robotic telescope in Australia and Texas, 41 B-band magnitudes and 13 deep limits from 1899 to 1939 with the Harvard plates, 207 B-band magnitudes from 1956-1991 measured from archival plates in the collection of the Maria Mitchell Observatory on Nantucket (Robinson, Clayton, \& Schaefer 2006), and 68 visual magnitudes (not counting limits) from 1995 to 2004 reported by amateur astronomers from around the world that appear in the AAVSO database.  This collection of 374+105+465+54+207+68=1273 magnitudes has been used to search for photometric periodicities by means of a discrete Fourier transform.  Two photometric periodicities were found.
	
	The first periodicity, at $103.8\pm0.4$ days, was independently identified in the SMARTS J-band and ROTSE I-band light curves.  In both data sets, the folded light curve shows a simple sine wave with little scatter (Fig. 52).  The full amplitude is 0.14 mag in the J-band, and close to 0.26 mag in the I-band, R-band data, and V-band.  The RMS scatter around the best fit 103.8 day sine waves are 0.04, 0.15, 0.23, 0.44, and 0.57 mag for the J, I, R, V, and B bands respectively.  In the B-band, the large scatter is dominated by the usual flickering (which is much larger than in redder bands) and not by photometric measurement errors (which are substantially smaller than the observed scatter), so that the presumed amplitude of 0.26 mag is lost in the noise.  The spectral energy distribution shows that the light from the red giant companion star dominates by a factor of a thousand in the infrared, so the 103.8 day periodicity must arise from the red giant.  Schaefer (2009) attributed this periodicity to pulsations in the red giant, because pulsations of comparable amplitude and period are common in red giants (Fraser, Hawley, \& Cook 2008).  
	
	The second photometric period is $259.85\pm 0.15$ days, with strong signals in the SMARTS, AAVSO, and Harvard data sets (Schaefer 2009).  This periodicity is stable from 1899-2008, and this demonstrates that this is tied to the orbital period.  The J-band and I-band data do not show any significant signal at a period near 259.85 days.  I take this to mean that this period is associated with the bluer light and hence with the accretion disk.  This is further supported by the systematic rise in amplitude from the red towards the blue.  Unfortunately, the rise in amplitude of this signal (towards the blue) is competing with the rise in the flickering towards the blue, so that the signal is most significant around the V-band and R-band.  In the V-band, the folded light curve appears as roughly a sine wave with an amplitude of half a magnitude and an RMS scatter of 0.3 mag.  Again, the odd and even numbered minima and maxima are systematically different form each other (see Fig. 53), so the orbital period must be twice the dominant photometric period, so $P_{orb}=519.7\pm0.3$ days (Schaefer 2009).  The SMARTS light curve, folded on a 519.7 day period and then bined in phase is shown in Figure 53.  We see a pronounced difference between the depths of the primary and secondary minima. With this, the ephemeris for the primary minimum is JD $2454730+N\times519.7$.
	
	Schaefer (2009) suggests that the variations seen in Figure 53 arise from ellipsoidal variations (to explain the primary photometric periodicity at half the orbital period) with a shallow eclipse (to explain the substantially deeper primary minimum.  Several complications (or problems) cloud this scenario.  First, the AAVSO data show the SMARTS primary minimum to be less deep than the secondary minimum.  That is, in the V-band, the depths of the two minima are reversed, and this would require some additional effect on top of the eclipse.  Second, ellipsoidal modulations should be easily detectable in the I and J bands and they should dominate in the near-infrared, but they are not visible.  Third, the flickering shown in Fig. 57 is at an orbital phase of 0.03, where the hot source might be eclipsed.  Any final conclusion must await more data and some clear theoretical understanding.
	
	In principle, the irradiation effects on the red giant might be apparent with a systematic change of its spectral class with orbital phase.  Indeed, reports of the spectral class vary from K5.5 to M8 (see Table 24).  However,  both measures involving a K5.5 classification are at orbital phases 0.97 and 0.62, while the latest spectral classification (M8) has a phase near to elongation.  As such, there is no correlation between the spectral class and the orbital phase.  After finding no phase/spectrum correlation for any of the four RNe with red giant companions, we realize that the principle is too weak to have any utility. 
	
\section{Short-term Variability in Quiescence}
	
	In quiescence, what is the photometric behavior of the RNe on short-time scales?  Here, I am taking the term `short-time scale' to mean faster than the orbital period.  For comparison, all cataclysmic variables show flickering and some show coherent periodicities and some show quasi-periodicities.  The flickering is generally on the time scale of minutes and is always very blue in color, with the likely cause being due to blobs in the accretion stream creating momentary brightenings of the hot spot or possibly due to fluctuations in the inner accretion disk.  The coherent periodicities might be associated with the rotation of a white dwarf with a high magnetic field.  Quasi-periodicities might be associated with superhumps and eccentric accretion disks.  So what is the answer for the ten galactic RNe?  In short, they all display the usual flickering and nothing else.  Here is a star-by-star account:
	
	T Pyx suffers fast and irregular flares with typical amplitude of 0.1 mag.  Two typical examples are displayed in Figure 54.  The typical duration of the flares is $\sim10$ minutes.  The flares are more prominent in blue light.  The flare amplitude is roughly equal to the orbital modulation, so the flaring makes for a large scatter in folded light curves and it often hides the general shape of the periodic signal.  For example, compare the average light curve in Figure 42 with the light curve in Figure 54 (with both covering a similar duration).  Patterson et al. (1998) find no periodicity (or quasi-periodicity) that is faster then the orbit.  Thus, T Pyx has normal flickering only.
	
	The only time series studies of IM Nor is those of Woudt \& Warner (2003) with 14.23 hours of high-speed V-band photometry and of Monard (in this paper) with 135 hours of high speed unfiltered CCD photometry.  We see frequent jittering at the 0.03 mag level and occasional flares up to the 0.1 mag level.  Perhaps the jittering goes away during the dips in the light curve.  There is no apparent periodicity or quasi-periodicity in the variations, although the data stream is too short to make too fine a measure of this.  In all, it looks like IM Nor has the usual flickering as seen in all cataclysmic variables.
	
	I have many long time series in the V-band on CI Aql, especially with coverage of the eclipses.  Outside of eclipses, CI Aql shows variability on a wide range of time scales faster than the orbital period.  This is readily seen in Figure 44, where the deviations from the average curve represent the fast variability.  We see flares on time scales of ten minutes with amplitudes of order 0.05 mag, and we see hour-long brightenings at the 0.15 mag level.  In all, I judge the flickering to be less than that seen for T Pyx and IM Nor.  The flickering is apparently absent during the eclipses.
	
	Little is known about the short-term variability of V2487 Oph.  However, my brief time series in 2002 and 2003 displayed significant short term variations.  In 2002, V2487 Oph was varying between V-band magnitudes of 17.2-17.4 mag over five nights, but also had a flare of at least 0.6 mag amplitude (getting to V=16.73) with a rise time of twenty minutes.  In three nights in 2003 (with 0.5, 0.8, and 3.8 hours respectively), V2487 Oph varied from 17.13-17.53 mag (V-band) with monotonic rises and falls on each night.  This is typical of flickering in cataclysmic variables.
	
	I have many long time series of U Sco, with most of these covering the eclipses.  Nevertheless, there is substantial coverage with reasonable time resolution outside of eclipses.  Some of these time series can be seen in Figures 45 and 47, where the deviations from the average light curve will show the fast variations.  In these figures, we can see strings of nearly connected points that were all taken in rapid succession.  For example, in the upper left of Figure 47 from phase -0.45 to -0.3, we see a time series which displays the usual flickering with time scales of minutes to hours and amplitudes up to 0.1 mag.  The flickering can largely disappear at times, as is apparent in the upper time series of Figure 47  for phases 0.05 to 0.1 and 0.35 to 0.45.  The out-of-eclipse level varies substantially from orbit-to-orbit, and we can see up-to half a magnitude changes in both the B-band and the I-band.  The brightness at mid-eclipse varies by half a magnitude in the B-band but by less than 0.2 mag in the I-band.
	
	I do not have good long times series photometry of V394 CrA.  Part of the reason is that the RN is faint (down to B=20 mag) so that the Cerro Tololo 0.9-m telescope is not big enough for useful short exposures and part of the reason is that I had chosen to sample the light curve on many nights instead of dedicating a few nights.  All of my photometry is tabulated in Table 22, and I know of no other photometry.  My best time series are from nights in my first run in 1989, and the best of these is displayed in Figure 55.  We see brightness changes of 0.07 mag on times scales of a dozen minutes and a flare with duration of two hours and an amplitude of half a magnitude.  At other times the light curve appears smooth, for example on JD 2447716 we see no significant variations superposed on a linear decline of 1.0 mag/day lasting for nearly seven hours.  This all looks like the normal flickering on a cataclysmic variable.
	
	T CrB shows fast flickering on the time scale of $\sim5$ minutes from peak-to-peak with typical amplitudes of 0.1-0.2 mag in the V-band (Walker 1954; Walker 1977; Oskanian 1983; Dobrzycka, Kenyon, \& Milone 1996; Zamanov et al. 2004).  This flickering is more prominent in blue light, with amplitudes typically half a magnitude in the ultraviolet and near-zero in the V-band and R-band.  The flickering displays the usual distribution of peak amplitudes with frequent low-amplitude events and less-common high-amplitude events.  Zamanov et al. (2004) has constructed a power density spectrum which shows the power to vary as the frequency to the -1.46 power (typical of cataclysmic variables, Yonehara et al. 1997; Bruch 1992) from seconds to hours.  No periodicity is seen.  The flickering is visible throughout the entire orbit.  On occasion, the flickering goes away entirely, even in ultraviolet light.  The amplitude of the flickering is linearly proportional to the average brightness (after the red giant contribution is subtracted out).  This fast flickering is typical of many cataclysmic variables.
	
	RS Oph shows fast flickering with a typical peak-to-peak time scale of $\sim 8$ minutes with a typical amplitude of 0.1 mag (Walker 1954; Walker 1977; Dobrzycka, Kenyon, \& Milone 1996; Simon, Hudec, \& Hroch 2004).  The amplitude of the flickering increases with the average intensity level.  Dobrzycka, Kenyon, \& Milone claim that RS Oph has its flickering riding on top of an 82 minute sinusoidal modulation, but this claimed periodicity is certainly not significant given their short time series and this claim is not supported by other data sets.  The flickering does not depend on the orbital phase, and this is not surprising given the likely low inclination of the orbit.  The flickering light is blue in color.  The distribution of flickering amplitudes shows many low-amplitude events and few high-amplitude events.  This flickering is not present during the post-eruption dips, presumably because the accretion disk has emptied out during the nova event so neither the accretion disk nor the hot spot is present at such times (Worters et al. 2007).  
	
	V745 Sco shows fast flickering (see Figure 56).  The flares appears to be continuous, with durations of half-an-hour to one hour, and with amplitudes of 0.1-0.2 mag.  For the one day shown (JD2453185, see left panel), the flickering is superposed on a general decline at the rate of 0.6 mag/day.  The average magnitude level changes substantially from night-to-night (see right panel), with daily changes of order 0.1 mag.  The total range for the mean magnitude of V745 Sco in this observing run is from 15.5 to 14.9 mag.  These variations appear to be correlated over a time scale of 5 days.  In all, V745 Sco displays typical flickering as seen in all cataclysmic variables.
	
	V3890 Sgr shows fast flickering, as shown in Figure 57.  We see continuous flickering with typical durations of half-an-hour and typical amplitudes of 0.03 mag.  The night-to-night variations show a rising trend of 0.1 mag/day and a superposed changes of $\sim0.1$ mag.  The flickering appears to have a smaller amplitude than is characteristic of other RNe and other cataclysmic variables.  This might be due to generally low amplitudes in the R-band, or it might be due to the effectively constant light from the red giant companion star diluting the flare light.
		
\section{Long-term Variability in Quiescence}
	
	In quiescence, what is the photometric behavior of the RNe on long -time scales?  Here, I am taking the term `long-time scale' to mean slower than the orbital period.  Long-term variability is important for a variety of reasons, including (a) a test for whether the total mass of material accreted between eruptions is constant, (b) a search for pre-eruption rises as claimed for roughly half of the novae, (c) a test for whether novae slowly decline for decades after the eruption, and (d) a measure of accretion rate variability on long-time scales.  These questions play into larger issues relating to the hibernation model.  The big advantage of RN for these questions is that the systems can be monitored intensively {\it before} eruptions and they generally have been intensively followed for long times {\it after} eruptions.
	
	\subsection{Individual Systems}
	
	T Pyx has an excellent well-sampled quiescent light curve from 1892 to present.  This is possible because the system is bright enough at minimum to be seen on many Harvard plates and because of intense amateur and professional attention ever since its 1944 eruption.  I have exhaustively searched the Harvard plates and measured magnitudes during quiescence on a modern magnitude scale (i.e., from Tables 3 and 4).  I have also collected and corrected a large number of B-band magnitudes from the literature.  This quiescent light curve for T Pyx is displayed in Figure 58.  (This figure is updated from that presented in Schaefer 2005.)  We see a long term trend for T Pyx to dim, from 13.8 mag in 1892 to 15.7 mag in the last year.  This trend is highly significant and is large (nearly a factor of six in luminosity).  The trend has no particular or repeated connection to the times of the eruptions.  For example, we don't see pre-eruption rises or consistent declines after eruptions.  Superposed on the declining trend is the usual flickering which leads to the observed scatter.  The scatter from 1925-1953 is definitely much larger than the scatter after 1977, and this effect is too large to be observational error in the Harvard magnitudes.  We see that when T Pyx is bright we have short inter-eruption intervals and that when T Pyx is faint we see long inter-eruption intervals.  When the system has a high accretion rate, the disk will be bright and the critical mass will collect on the surface of the white dwarf quickly.  When the system has a low accretion rate, the disk will be dim and it will take a long time to collect the critical mass.  A detailed analysis for T Pyx and U Sco (Schaefer 2005) shows that the average B-band flux (actually, a power of the flux so as to be proportional to the accretion rate) times the duration of the inter-eruption interval (to be proportional to the total mass accreted onto the surface of the white dwarf) is a constant from interval-to-interval.  Thus, we see that the drop in brightness (and hence the slowing of the accretion) after 1970 means that T Pyx will take a very long time to accrete a critical mass onto the surface of the white dwarf.  If the system continues at the same rate then we have to wait centuries before the next eruption (Schaefer et al. 2009).  If the accretion rate keeps falling off as it has over the last century, then T Pyx will be moving into a state of hibernation and it will be almost a million years until its next eruption (Schaefer et al. 2009).  In either case, T Pyx has stopped being a RN.

	IM Nor has no published photometry during quiescence.  (Woudt \& Warner's 2003 light curve and period come from the late tail of the 2002 eruption when the V-band magnitude was at 16.5 mag around 409 days after the peak.)  The magnitudes during quiescence that I can pull out are my own BVRIJH sets on three days (see Table 25), two sky survey magnitudes (see below), and Monard's unfiltered magnitudes (see Figure 43 and Tables 5 and 23).  From my photometry, I find R-band magnitudes of 17.80, 17.75, and 17.71 on Julian Dates of 2453590.58 (2005 Aug 8), 2453607.52 (2005 Aug 25), and 2453804.81 (2006 March 10) respectively.  From the AAO-SES sky survey plate on JD 2448827.9 I find R=17.4, while from the SERC-J sky survey plate on JD 2443960.3 I find V=18.5 mag and hence R=17.9 for the usual V-R color, both with an estimated error bar of $\pm 0.3$ mag.  While the sampling is bad, the brightness does not have any significant variations on the one-year time scale or before/after eruption, with the small differences seen being comparable to the orbital modulation and measurement uncertainty.  Monard's magnitudes (on a non-standard red magnitude system) show a variation between 17.0-17.7 with an RMS scatter of 0.24 mags (0.12 mag if one outlier datum is ignored) for 8 nights during quiescence.  In all, the only information shows a nearly constant star with variations of $\sim0.2$ mag or less.
	
	CI Aql has a variety of difficulties in producing a long-term light curve that is consistent.  One problem is that the magnitudes from eclipses will only add noise, so I have systematically deleted the observations within 0.1 phase of eclipse minima.  (The two exceptions in the figure will be the old eclipses from the Harvard plates, included to illustrate the old times.)  To avoid biasing the long-term average towards those nights on which I happen to have many magnitudes, I have taken nightly averages.  We also have to worry about catching tails of eruptions and confusing them for variability during quiescence.  Another problem is that all the early magnitudes are in the B-band and most of the later magnitudes are in the V-band, so I have converted all the B-band magnitudes to V-band with an average $B-V=1.03$ mag.  A final problem is that the magnitudes from Robocam (Mennickent \& Honeycut 1995) might need a small additive offset applied to all the magnitudes.  With these complications, I have constructed a long-term light curve from the Harvard plates, Schmidt sky surveys, the Robocam series, and my own CCD magnitudes from several telescopes at each of McDonald Observatory and Cerro Tololo Inter-American Observatory (see Figure 59).  While ignoring the old eclipses and the two eruptions, we see that CI Aql is nearly constant at V=16.1 mag, with the usual flickering causing scatter about this average.  Given the gap from 1996-2000 and the uncertainty in the normalization of the Robocam data, this light curve is not useful for determining whether the 2000 eruption had a pre-eruption rise on a time scale of less than five years.  Unfortunately, the poor coverage does not allow a real check of whether the accretion rate (as measured by the system brightness) varies inversely with the inter-eruption interval.  Nevertheless, we have an apparent inconsistency between a constant quiescent brightness and the varying intervals between observed eruptions (23.6 years from 1917-1941 and 59.3 years from 1941-2000).  In particular, with no missed eruptions from 1941-2000, we would expect that the average B-band magnitude in this interval to be 1.0 mag fainter than the earlier time interval, and Figure 59 shows that this is not so (despite gaps in the coverage).  This discrepancy could be resolved if there was an eruption around 1970 (or perhaps two eruptions around 1960 and 1980).  All throughout this interval the estimated probability for detecting an eruption is $83\%$ (see Table 20), so eruptions could well have been missed.  Maybe the best resolution for the inconsistency would have one or two  missed eruptions, with this possibility having important consequences for testing $M_{ejecta}>\tau_{rec}\dot{M}$ for CI Aql.
	
	V2487 Oph does not have much useful long-term information on its quiescent magnitude.  All I know is that the average V-band magnitude was 17.26 in 2002, 17.38 in 2003, and 17.43 in 2008.  This is consistent with normal variations due to flickering on top of no secular changes.
	
	U Sco has a fairly good long-term light curve in quiescence.  Most of this is provided by my regular monitoring program started in 1988.  Before that time, I have made use of magnitudes from deep Schmidt survey plates.  A complete tabulation of my data is presented in Figure 60.  This is basically an updated version of a figure in Schaefer (2005).  That study demonstrated that the material accreted onto the white dwarf is a constant over each inter-eruption interval, and also made the prediction that the next eruption will be in the year $2009.3\pm1.0$.  The updated light curve shows that the average brightness from 2005-2008 is close to the same as from 1999-2004, and this means that the predicted date is still good.  We see that U Sco varies substantially, from roughly 18 to 19 mag.  But this variation has no apparent correlation with the time of eruption.  That is, we see no pre-eruption rise and no systematic decline after eruption.
	
	V394 CrA has many B-band magnitudes from my observing runs at Cerro Tololo (with the 0.9-m and 1.0-m telescopes) in 1989, 1994, 1995, 1996, 2004, and 2005, plus many B-band and R-band magnitudes from many nights throughout the observing seasons of 2004-2008 on the Cerro Tololo 1.3-m telescope.  In all, I have 298 and 160 magnitudes in B and R respectively.  These can provide a good measure of the long term activity of V394 CrA, and they have been plotted in Figure 61.  (For better comparison, the R-band magnitudes have been converted to B-band magnitude using B-R=1.2 mag.)  We see a large amplitude long-term variability.  V394 CrA goes from 18.3 to 20.5 mag, almost a factor of ten difference in luminosity.  This wide variation occurs on time scales of hours (see Figure 55), on time scales of tens of days (see Figure 48), and on time scales of years (see Figure 61).
	
	T CrB has extensive photometry of its quiescent state since 1890.  I have my measures of the brightness of T CrB from 1890 to 1953 on the Harvard plates, Leslie Peltier kept frequent track of the visual magnitude from 1920 to 1946, and   many AAVSO observers have covered it from the 1946 eruption to present.  The AAVSO record is particularly impressive with over 80,000 observations (averaging 1300 magnitudes per year) with essentially no seasonal gap.  Leibowitz, Ofek, \& Mattei (1997) have analyzed the AAVSO record for 40 years, from 1956 to 1995, with results already given in Section 10.9.  Here, I have extracted the AAVSO magnitudes from 1948 (just after all the rebrightenings associated with the 1946 eruption are over) until 2004.  This light curve is displayed in Figure 62, with the AAVSO magnitudes binned into 0.01 year intervals.  We see an overall decline from 9.9 in 1948 to 10.25 in 2004 (averaging 0.0062 mag per year), which is small but highly significant.  But the structure in Figure 62 is too complicated to characterize it as simply a decline at some rate.  Rather, a better description would be that there was a linear decline of roughly 0.014 mag per year from the end of the eruption until 1970, then a flat topped rebrightening by 0.2 mag lasting from 1970 to 1992, followed by a flat light curve from 1992 to 2004.  We clearly see that T CrB is varying up and down on decadal time scales by $\sim0.2$ mag.  With such variations, we cannot come to any real conclusions about whether the apparent decline after the eruption is some sort of a systematic effect (perhaps associated with a cooling white dwarf).  In the face of such up-and-down variations, we can only seek systematic effects after eruptions by a statistical comparison of many events.
	
	RS Oph has an excellent visual light curve from 1933 to present from the AAVSO.  I have constructed a light curve from 47,000 individual magnitudes, binned into 0.01 year intervals, from 1934 to 2004, with the eruptions and the post-eruption dips clipped out (see Figure 63).  We see a chaotic mess with variability on time scales of months to years to decades where the character of that variability is constantly changing.  We see flares from one-month to two-years in duration, trends lasting for a decade, and decades of relative calm.  The total amplitude of variation is over two magnitudes, from roughly 9.9-12.4 mag.  In no case is there any apparent variation associated with an eruption.  (Recall, the post-eruption dips have been clipped out of this light curve.)  An imprudent observer might point to the rises in 1967 and 2005 as `pre-eruption rises', but a look at the entire light curve shows that this is just normal variation, and a selection effect where the flat light curves before the 1933 and 1945 eruptions as well as the decline before the 1958 eruption are ignored.  I have made a list of 30 significant local maxima in the quiescent light curve and the time between peaks varies from 0.7 to 9.5 years with no preferred value.  Oppenheimer \& Mattei (1993) made a period search for each inter-eruption interval and found nothing significant.  They also took a Fourier transform of the entire data set and noted a periodicity of 2016 days.  I have repeated this analysis on my longer data stream and can reproduce this result with a period closer to 2030 days.  However, a look at Figure 63 immediately shows that any such period is poor, being created by a rough alignment of a handful of peaks.  In such a case, the significance of a peak in a Fourier transform is not the relevant question, but rather we should be asking at what probability we would get a false periodicity if random flare times are scattered about.  (To realize that a Fourier transform can yield an apparently significant peak with no real periodicity, consider a star with occasional long and bright flares at random times, where a Fourier transform of some time interval with just two random flares will always give a strong nominally-significant Fourier peak for a `period' equal to the separation time of the two flares.  When more peaks are added, more possibilities for periods arise.  After long experience with this question in relation to low-mass x-ray binary systems, a general conclusion is that more than nine peaks have to be seen in a shot-noise situation before a periodicity can be believed.  This rule can be overturned with {\it independent} information available, for example if a fairly specific light curve shape is expected and observed.)  I have already created a statistical method for exactly this question in the context of alleged periodicities in Gamma-Ray Burst light curves (Schaefer \& Desai 1988), with this method being exactly relevant for the question at hand.  I find that the existence of some apparent-but-false periodicity is a likely result of random flare times for the number of flares as seen in Figure 63.  That is, when our human pattern-recognition computer is allowed to freely move around the zero phase and period to best match the peaks which we can define by {\it a posteriori} criteria, then we can usually find some periodicity, especially if we are allowed to skip predicted peaks and are allowed to have sloppy matches between observed and predicted peak times.  In Figure 63, we see missing peaks at 1954.2 and 1987.6 and sloppy matches at 1948.7 and 1959.8 and at other times.  (And the eruptions occur at all phases of this false periodicity.)  The model for looking at flare times is very flexible (that is the problem), but let me present one typical example and its calculated probability of chance periodicity.  Let us take a peak to be when any small time interval has two or more points brighter than 10.4 mag in Figure 63 (so as to include all the peaks that match the 2030 day claim), which gives us a total of 12 peaks from 1939.6-1992.9.  Of these dozen peaks, eight of them match the best-fit 2030 day periodicity to within 1.5 years, even though two predicted times (not in a row) have missed eruptions.  With this, the value of $P(12,8,0.03,1)$ is found from interpolation in Table 5 of Schaefer \& Desai (1988) to be near 50\%.  That is, it is even-odds that a periodicity as significant as observed will be created by the random placement of flares in the RS Oph light curve.  This is exactly as expected for the `random case', and it is certainly too poor to consider the 2030 day `periodicity' as being significant.  In all, we can be confident that RS Oph in quiescence does not have any significant periodicity.
	
	V745 Sco has a long-term light curve that consists only of my CCD magnitudes taken at Cerro Tololo from 2004-2008, almost all in the R-band (see Figure 64).  What we see is a star that varies by up to 2.5 magnitudes on all time scales, from minutes (see Figure 56 left panel) to days (Figure 56 right panel) to months and years (see Figure 64).  The long-term variations show prominent peaks of over one magnitude amplitude with durations of 40-200 days.  And there are even longer term variations, as V745 Sco generally faded from 2004 to 2005-7 only to have a fast rise (0.006 mag/day) from 2007 to 2008.  With variations like this, I see no way to connect such with any effects of the 1989 eruption.
	
	V3890 Sgr has many magnitudes from 1899 to present, but I have various problems in constructing a consistent long-term light curve for quiescence.  One problem is that the various sources are made with different bands, so long-term trends might be confused with color terms.  Nevertheless, the colors of V3890 Sgr do not change greatly, so I have converted the B-band magnitudes to the V-band with $B-V=0.9$ mag and I have converted the R-band magnitudes to the V-band with $V-R=1.25$.  But for this, I cannot convert the ROTSE magnitudes, as ROTSE uses an unfiltered CCD so that I do not know how to reliably make the conversion.  Another big problem is that three of the data sets (Harvard, Maria Mitchell, and AAVSO) have limiting magnitudes such that V3890 Sgr is frequently invisible.  In this case, we are only seeing the brighter parts of the light curve and get an incorrect idea of the average magnitude.  For V3890 Sgr varying substantially, our derived average magnitude will always be just a bit brighter than the typical limit, and will not be simply related to the true mean brightness level.  I do not know how to solve this problem except to point it out to the reader to be wary on this point.  With this warning and corrections to V-band magnitude, I have constructed a long-term light curve for V3890 Sgr in quiescence from 1899 to present (see Figure 65).  A problem with this composite light curve is that all four data sets are largely disjoint in time, so we cannot get a good idea of any offsets in the magnitudes.  The only data set without the limiting magnitude problem is the SMARTS data (after JD 2453180) and this displays a relatively tight magnitude range of mostly 15.2-15.8 (after correction to the V-band).  This is in marked contrast with the AAVSO data (JD 2449788-2453068) with a range from 14.1-15.8 even with truncation around fifteenth magnitude, is in marked contrast with the Maria Mitchell plates (JD 2435695-2449573) with a range of around 15.2-16.7 even with truncation around 16.5 mag, and is also in marked contrast with the Harvard plates (JD 2414846-2429465) with a range of 16.1-18.4 even with a truncation of 17.5 mag and deeper.  Why do the SMARTS data have a completely different and smaller range than all the other data sets?  One possibility is that V3890 Sgr has been `calming down' in recent years, but this seems unlikely as the observed ranges are too closely related to the data sources.  Another possibility (my preferred possibility) is that the flickering and variability are relatively small in amplitude in red light and have a relatively large amplitude in blue light.  With this, the Harvard blue data will have the largest amplitude of variation as the deep plates can pick up V3890 Sgr when it is faint.  My long-term light curve can also be used to address questions of the relation of the quiescent brightness to the two eruptions (marked by vertical lines in Figure 65).  The saving circumstance is that both eruptions have one data set (from the Maria Mitchell plates) both before and after.  One question is whether the light curve shows any anticipatory rises before the eruptions?  Indeed, looking at Figure 65, an incautious observer might suggest a pre-eruption dip before the 1962 event and a pre-eruption rise before the 1990 event.  But I judge the existence of a mechanism that alternates rise and dips to be elusive and I judge both possibilities to be consistent with the ordinary scatter in the light curve.  The differences in the data sets and the possibility of systematic changes in the limiting magnitudes over time makes problematic any search for secular drifts during quiescence.
	
	\subsection{RN are Highly Variable on All Time Scales in Quiescence}
	
	A default view of novae in quiescence is that they are largely constant (excepting eclipses and orbital variations as well as the usual flickering), and this extends to the long time scales.   But a strong result from this section is that seven-out-of-nine RNe are varying by more than a magnitude on time scales longer than their orbital period.  That is, on time scales of years to decades to a century, these novae are varying apparently-chaotically with amplitudes from 1-2.5 magnitudes. This is huge, with the luminosity changing irregularly by factors of from 2.5-10 year-by-year and decade-by-decade.  Two of the RNe (IM Nor and CI Aql) apparently have only small long-term variations while one RN (V2487 Oph) has little information on long-term variability.  Now combine this with the fact that {\it all} RNe show fast flickering (see Section 11) and {\it all} RNe show large amplitude variations on the orbital time scales (see Section 10).  What we have is a strong and somewhat surprising conclusion that RNe display large amplitude changes on time scales of minutes, hours, days, weeks, months, years, and decades.
	
	This strong and surprising result needs explanation.  As far as I know, no theorist has ever tried explaining most of the variations reported in this paper.  A few of the variations have ready explanations (like the ellipsoidal effects and the eclipses), while other can have reasonable old models tried (like having the flickering come from variable accretion onto the hot spot).  But how can we explain the large amplitude decadal variations?  Why do five of the RNe have an asymmetric eclipse light curve, with it being brightest during the elongation immediately {\it after} the primary minimum?  Why does the folded light curve for T Pyx (and perhaps IM Nor) show a dip that is much too long to be an eclipse?
	
	Another question is whether classical novae have similar variability on all time scales as do RNe?  I expect that most CN researchers would be surprised to find that many normal nova are behaving as chaotically as the RNe.  If CNe are different from RNe (perhaps with less variability), then we need to understand how the difference arises.  To answer these questions, we need a comprehensive study of CNe to complement the study of RNe reported in this paper.  A quality study along these lines is Duerbeck (1992), with an artful mixture of long-term visual observations, photometry from the literature, plus his own CCD observations, yielding long-term light curves with large numbers of magnitudes for a handful of the best-known novae.  I think that this impressive work should be extended to many more CN.
	
	\subsection{Power Density Spectra}
	
	A power density spectrum (PDS) is a depiction of the Fourier transform of the light curve with both the Fourier power and frequency depicted on logarithmic axes.  For this to produce useful results, the time series should be uniformly sampled and cover a wide range of time.  Generally, the PDS can be represented by power laws, possibly with breaks, superposed periodicities, and quasi-periodicities.  Cataclysmic variables (including novae) generally show power laws with indices typically from -1.5 to -2.0 (Bruch 1992; Yonehara et al. 1997), and often there are breaks at low frequencies where the PDS rolls over to a near-zero slope.  Roughly, the PDSs show the usual $1/f$ noise that breaks to white noise at low frequency.  For cataclysmic variables, the PDS is only ever measured for frequencies corresponding to times shorter than a long night.  The PDS has only been published for one RN, and that for T CrB for frequencies corresponding to less than an hour with the index averaging -1.46 (Zamanov et al. 2004; Dobrotka et al. 2009).
	
	The AAVSO has a wonderful collection of data for T CrB and RS Oph that is good for making PDS to extremely low frequencies.  I have created uniformly sampled light curves with time bins of 0.01 year duration with each bin usually having a dozen averaged magnitudes.   For T CrB, we have essentially complete coverage (i.e., with no solar gaps) from 1947 to present, thus allowing frequencies corresponding to periods of up to decades to be measured.  For RS Oph, we have coverage from 1934 to present, however there are yearly solar gaps averaging a quarter of a year in duration plus five eruption gaps (including the post-eruption dips) lasting about one year. The gaps have been filled in by taking adjacent data streams, and for RS Oph this will create unknown distortions (likely small and non-systematic) for frequencies corresponding to periods shorter than a fraction of a year.  The Fourier transforms for the uniformly sampled light curves were then calculated and averaged into equal logarithmic bins in frequency.  The resultant PDSs are plotted for T CrB and RS Oph in Figures 66 and 67.
	
	The T CrB PDS has a power law index of -0.9 for frequencies above $10^{-4}$ cycles per day, i.e., 7 nanoHertz.  (The one high point close to the log-frequency of -2.05 corresponds to the ellipsoidal modulation at half the orbital period.)  Below frequencies of $\sim 10^{-3.5}$ cycles per day, the random scatter will inevitably get large due to the few cycles involved in the light curve for these frequencies.  That is, the particular realization of the light curve (and hence the PDS) for the time interval 1947-2008 will have ordinary deviations from any long-term average PDS.  An obvious idea is that the power law PDS in Figure 66 connects to the power law PDS at much higher frequencies as reported by Zamanov et al. (2004) and Dobrotka et al. (2009).  But this idea can only be tested by comparing the absolute normalization of the powers, and the idea is suspect because the amplitude of fast flickering is comparable to the amplitude of the long-term trends. 
	
	The RS Oph PDS has a power law index of -1.8 for frequencies above $10^{-3.3}$ cycles per day.  The low frequency slope is roughly flat below this frequency (corresponding to white noise), with this whole PDS looking characteristic of other cataclysmic variables.  However, the break to white noise is likely not of high significance due to the particular realization of any underlying PDS resulting in expected variations at the lowest frequencies.
	
	  For both T CrB and RS Oph, the normal quiescent light appears to have comparable contributions from both the red giant and the accretion disk.  With this, either the red giant or the accretion component can be the seat of the long-term variations.  One or both of two physical mechanism might causes the long-term variations and the power law PDS spectra for T CrB and RS Oph:  
	
	The first mechanism is accretion instability, whether due to instabilities in the accretion disk (Yonehara et al. 1997; Dobrotka et al. 2009) or due to the variations in the rate of material falling of the companion star.  We already know that RS Oph and T CrB have variable accretion rates (both because of the observed flickering and because of the precedent for variability on all time scales for short period cataclysmic variables), so the first mechanism must be operating.  The accretion instability mechanism is known to produce power law PDS spectra (Yonehara et al. 1997; Dobrotka et al. 2009).  Anupama \& Miko{\l}ajewska (1999) have found that T CrB and RS Oph both have a scattered (but significant) correlation between the Balmer line flux and the V-band magnitude, so that some fraction of the long-term brightness changes must be associated with the accretion.  The correlation between the spectral type and brightness in RS Oph (see Fig. 50) could arise because the varying luminosity of the accretion component can veil the absorption bands of the red giant.  In addition, some fraction of the optical spectrum appears as a cF-type shell absorption spectrum (associated with ongoing accretion), and this fraction of the light cannot have variations associated with the red giant (Gromadzki et al. 2008).  One specific type of accretion instability is the dwarf nova eruption, and we know that novae with long orbital periods have dwarf nova events.  The flares in the light curve of RS Oph (Fig. 63) have some similarities, yet their shape is not that of dwarf novae events (most are too pointed at the top), the individual flares are not similar to each other (they vary greatly in duration, shape, and amplitude), and the flares are not even moderately evenly spaced in time.  So the flares in RS Oph are not dwarf nova events.  For a model of PDS power law indices arising in accretion disks, the variability can only occur on dynamical time scales within the disk, which are all shorter than the rotation period (Yonehara et al. 1997).  With T CrB and RS Oph both having variability on the time scale of a decade and longer, this implies that the long term variability is not caused by accretion instability.
	
	The second mechanism (to explain the power law PDS) is random variations arising from convection cells bringing hotter material to the surface, with a power law distribution of cell sizes and the larger cells varying over longer time scales (Schwarzschild 1975).  We know that the second mechanism is likely to be operating because a wide variety of red giants and supergiants (both isolated and in interacting binaries, both symbiotic and non-symbiotic, both Miras and non-pulsating stars) have ubiquitous power law PDS to very low frequencies (Kiss et al. 2006; Stello et al. 2008; Templeton \& Karovska 2009) which is identified with the convection cells on the surface (Antia et al. 1984; Hayes 1984; Gray 2001; 2008; Bedding 2003; Freytag et al. 2002).  The PDS of other red giants is strikingly similar to the PDS of RS Oph and T CrB, so it is natural to expect the the RNe red giants will by themselves produce the long-term variability.  These variations are thought to arise from ordinary convection in the red giant, where cells of different size bring hot and bright material from the interior to the surface.  Note that the radius of the red giant does not change, so the accretion rate will not fluctuate by a large amount from this cause.  The faster lower-amplitude fluctuations will be caused by the small convection cells which cover only a small fraction of the surface and which have a relatively fast turnover time.  The longest time scale variations will be caused by the largest convection cells which can provide a large fraction of the star's surface with hot material.  The possible break to white noise at the lowest frequencies in the PDS for RS Oph could perhaps be caused by some maximum size to the convection cells.  A prediction of this model is that the surface temperature of the red giant should be correlated with brightness.  For example, when a large convection cell brings a large area of hot material to the surface, the hot temperature of that large area will make for a hotter spectral classification based on spectral lines as well as a higher luminosity (by Boltzmann's law).  This prediction is exactly fulfilled by the spectral class versus V-band magnitude correlation already identified (see Figure 50, Table 24, and Section 10.8).  Indeed, a detailed calculation can reproduce the slope observed in Figure 50.  A problem with the second mechanism is that it cannot account for the correlation between brightness and the Balmer line flux (Anupama \& Miko{\l}ajewska 1999).
	
	With the arguments presented in the last two paragraphs, I conclude that both mechanisms are operating.  A way of estimating the relative contribution is by looking at Figures 5 and 6 of Anupama \& Miko{\l}ajewska (1999), where the Balmer line fluxes are plotted versus the V-band magnitudes.  For both T CrB and RS Oph, the confident existence of a correlation reveals the long-term variations associated with the accretion, while the large scatter about these fit lines reveals the long-term variations  associated with some separate mechanism (presumably convection cells on the red giant).  The variance in these plots associated with the correlation is similar to the variance about the correlation, so that means that the two mechanisms have comparable effects.  That is, apparently, both mechanisms have comparable effects (at least in the V-band) and neither dominates.
	
	The other RNe do not have well-sampled long-term light curves, so a PDS cannot be constructed.  Nevertheless, the long-term light curves for T Pyx (Fig. 58), U Sco (Fig. 60), V394 CrA (Fig. 61), V745 Sco (Fig. 64), and V3890 Sgr (Fig. 65) all display variations on the longest observed time scales from many years up to a century.  With this, it appears that a common characteristic of RNe is to have significant power in their PDS down to very low frequencies.  It is reasonable that the very-low-frequency variations for the four systems with red giant companions (T CrB, RS Oph, V745 Sco, and V3890 Sgr) are caused by normal fluctuations in the red giant.  However, this explanation does not work for the systems that have high-amplitude very-low-frequency variations that do not have red giant companions (T Pyx, U Sco, and V394 CrA).  Presumably, for these three systems, there must be some secular changes in the system that leads to long-term variations in the accretion rate.  For T Pyx, with the white dwarf being a long-lasting supersoft source of high luminosity (Patterson et al. 1998), the decline might be related to the supersoft emission.  U Sco and V394 CrA are not supersoft sources (other than late in their eruption tails as the wind fails and before the nuclear burning stops), so they do not have any known explanation for their long-term variations.
	
	\subsection{The Secular Decline of T Pyx}
	
	T Pyx has been systematically declining in brightness from 1890 to 2009, fading from 13.8 to 15.7 mag in the B-band.  This is highly significant.  This secular decline is also completely unprecedented for RNe, CNe, or any cataclysmic variable.  This mysterious decline has a variety of implications.
	
	Schaefer et al. (2009) reports on a single image from the year 2007 of the T Pyx shell as taken by the {\it Hubble Space Telescope}.  When compared with similar images from 1994 and 1995, the individual knots are seen to be expanding homologously with velocity $\sim600$ km s$^{-1}$.  The lack of deviations from this expansion demonstrates that none of the knots has experienced significant deceleration.  With this, the nova eruption that ejected the visible knots was in the year $1866\pm5$.  The mass in the `1866' ejecta is $\sim10^{-4.5}$ M$_{\odot}$.  Given the mass and ejection velocity, the `1866' eruption can only be a regular nova eruption (i.e., not a RN eruption) where the prior accretion rate was around $4 \times 10^{-11}$ $M_{\odot}$ yr$^{-1}$ (as appropriate for an ordinary cataclysmic variable below the period gap driven only by angular momentum losses from gravitational radiation) for a time interval of roughly $\sim$750,000 years.  The transition from this low accretion state to the high observed accretion state for the many decades after 1890 ($\dot{M}\gtrsim 10^{-7}$ $M_{\odot}$ yr$^{-1}$) was caused by the ignition of a luminous supersoft source on the white dwarf which heats the atmosphere of the companion star so as to drive the high accretion (as suggested by Knigge et al. 2000).
	
	Schaefer et al. (2009) take the secular decline in T Pyx to demonstrate that the driven accretion is not adequate to keep the supersoft source self-sustained.  By 2009, the accretion rate has fallen to only 3\% of its value in 1890.  This provides a natural explanation for why T Pyx was not seen by {\it XMM-Newton} to be a supersoft source in 2006 (Selvelli et al. 2008).  The secular decline also explains why T Pyx has not had its long-expected eruption in the late 1980's, why T Pyx will not have another eruption for a very long time, and that T Pyx has stopped being a recurrent nova.  With the continuing secular decline, the accretion in T Pyx will virtually stop within decades from now, and the system will enter a state of hibernation, calculated to have a duration of order 2,600,000 years.
	
	Schaefer et al. (2009) then puts together a full picture of the evolution of the T Pyx system.  It starts out with the system being an ordinary cataclysmic variable (CV) with low accretion rate appropriate for a system below the period gap, until an ordinary nova eruption ejects a massive shell and ignites a supersoft source which drives a high accretion rate so as to power fast RN eruptions, with the secular decline in the accretion stopping the high accretion after one or two centuries, only to have the system fall into a state of hibernation until the relentless losses from gravitational radiation forces the system back into contact and back to the state of an ordinary CV again.  Thus, T Pyx will have cyclic evolution, evolving between ordinary CV state, the RN state, and the hibernation state, with a time scale of order 3,300,000 years.  Changes in the mass of the white dwarf are dominated by the regular nova eruptions (like the `1866' event) for which more mass is ejected than accreted (Yaron et al. 2005), with the RN episodes too short to make any difference.  Thus, T Pyx will {\it not} become a Type Ia supernova.
	
	\subsection{Pre-eruption Rises and Dips}
	
	Robinson (1975) has made the only study of pre-eruption magnitudes for CN, with his data source being only old literature papers.  One of his main results is to identify a previously-unrecognized phenomenon as being common, the pre-eruption rise.  He finds that 5-out-of-11 CN show a rise in their light curve starting from 1-15 years before the eruption and brightening by 0.25-1.5 mag.  This statement and conclusion is echoed in various nova reviews (e.g., Warner 2002) even until recently as part of our general and critical knowledge on CN.  The claimed effect is astounding, as theorists have no plausible idea as to how the accretion rate (governed by the companion star) `knows' in advance to anticipate that the eruption trigger (governed by the depth of material on the surface of the white dwarf) is getting near critical.  Given this disconnect with theory and the claimed high frequency (roughly half of CNe), I am surprised that no theoretical or observational follow-up has been made on Robinson's work.
	
	To help solve the observational question, during the summer of 2008, I led a group from Louisiana State University to visit the Harvard and Sonneberg archival plate collections to examine the original data for all the claimed pre-eruption rises.  The idea was that we would use our modern magnitude sequences and examine all available plates.  Our results (Collazzi et al. 2009) are that all of the claimed pre-eruption rises are seen to not be real or significant rises, with a problem being simple errors in the early literature.  The one exception is that one of Robinson's pre-eruption rise cases has been confirmed.  V533 Her displays a very well sampled rise starting roughly 1.5 years before the fast eruption with a steady rise by up to about 1.5 mags brightening.  A very long and well-sampled light curve both before and after eruption shows the pre-eruption rise to be completely unique and far outside the normal behavior of the system.  Many other classical novae were also examined for pre-eruption rises, and we found and confirmed one other case, for V1500 Cyg, which has already been well documented (Kukarkin \& Kholopov 1975; Alksne \& Platais 1975; Samus 1975; Duerbeck 1987; Rosino \& Tempesti 1977; Wade 1987).  This pre-eruption rise went from B=21.5 to B=13.5 in the month before the eruption.  So, again, theorists have the task of explaining how well-observed pre-eruption rises can occur.
	
	RNe light curves should also be examined for pre-eruption rises.  This class of nova has the big advantage that we can know in advance which star will explode so pre-eruption magnitudes are readily available.  (This is in contrast to the case for CN, where the only way to get pre-eruption magnitudes is from archival photographs.)  RNe share the identical system configurations, identical physics in the accretion stream and disk, and identical physics of the nova eruption, so any physical mechanism that might cause pre-eruption rises for CNe should also cause identical pre-eruption rises on RNe.  In this case, we should examine the long-term light curves of RNe in quiescence.
	
	Collazzi et al. (2009) have examined the quiescent light curves presented in this paper for pre-eruption rises before 11 eruptions of 5 RNe.  In all 11 eruptions, no pre-eruption rise was found.  Including 16 CNe and RNe, pre-eruption rises have been found for only two systems (V533 Her and V1500 Cyg).
	
	While not a pre-eruption {\it rise}, the unique pre-eruption behavior of T CrB before its 1946 eruption must be highlighted.  Figure 23 shows that T CrB had a prominent pre-eruption {\it dip} starting a year before the fast rise of the eruption itself.  T CrB had been systematically monitored by the AAVSO (primarily by Leslie Peltier) since 1919, with scattered observations going back to the 1860's.  The dip is unique in T CrB's history now of 145 years, and the observers back in 1945 realized that something special was going on (Peltier 1945).  The dip below the normal quiescent magnitude started close to one year before the eruption, with the fairly sharp decline going from roughly 1-2 mag fainter than T CrB's previous normal state.  Such dips (by 1-2 mag below the steady level) have not occurred before or after the 1946 eruption (c.f. Figure 62).  The faintest ever magnitude occurred 29 days before the eruption, when T CrB was close to two magnitudes faint.  This time coincidence (29 days for an event that is unique out of 145 years) is what makes me confident that there must be some causal connection between the dip and the eruption.  This dip cannot be associated with ellipsoidal variations because the colors vary strongly in ways contrary to the idea, the duration of the minimum is much longer than possible for ellipsoidal effects, such effects are not visible in the 1919-1945 AAVSO light curve, and the amplitude is much deeper than ever observed.  The dip cannot be due to an ejection of gas which forms a shrouding dust shell, because the dip amplitude is comparable in B-band and V-band, while the color evolution is much different from that of an R CrB star.  The brightness in the minimum of the dip is comparable to the brightness of the red giant alone, which would suggest that accretion has somehow turned off, but how does the companion star `know' to decrease the rate of material being pushed over the Roche lobe one year before the time when the pressure at the base of the hydrogen layer on the white dwarf reaches the trigger point.  The orbital phases (based on the ephemeris of Fekel et al. 2000) of the dip covers 1.3 orbital periods, with the dip starting at phase 0.28, the dip being deepest in the V-band at phase 0.54 (with the white dwarf being in front of the red giant), the dip ending in the V-band at phase 0.16 (after the red giant has passed around in front of the white dwarf), and the dip being deepest in the B-band at phase 0.56.  With no reasonable explanations, I think that the pre-eruption dip sets a challenge to theorists.

	\subsection{Pre-eruption Dwarf Nova Events}
	
	A small number of novae certainly show dwarf nova outbursts starting long after the nova eruption has faded away.  The classic case is GK Per (Nova Per 1901), which has now had 14 small eruptions starting in 1966.  Another case is V446 Her (Nova Her 1960), which started showing flares of 5-11 day durations and 1.5 mag amplitudes starting around 1990 (Honeycutt, Robertson, \& Turner 1995).  Discussions of these hybrid systems (e.g., Livio 1992b; Vogt 1989) include V3890 Sgr as a hybrid system.  The basis for this is a few points on a light curve presented by D. Hoffleit in an addendum to Dinerstein (1973) that show an apparent flare in 1939 with a duration of less than a hundred days and with an amplitude of around 1.5 mag.  These data are from the Harvard plates and I have independently examined these plates and many more.  I find the same `flare' that Hoffleit describes, even though my magnitudes have a systematic offset from hers due to the usual differences in comparison star magnitudes.  Hoffleit merely looked at a relatively small number of plates and noticed one `flare'.  Later workers looked at her light curve and declared this event to be a dwarf nova outburst.  But, when I look over a large number of plates, I find many other `flare' events in the Harvard and Maria Mitchell data (see Figure 65).  With V3890 Sgr always varying chaotically, apparent flares (which could be confused with a dwarf nova eruption if viewed in isolation) are continually happening.  This is not the behavior of a dwarf nova, rather it is just the normal variations on all time scales.  As such, V3890 Sgr must be taken off the lists of novae with dwarf novae events.
	
	V3890 Sgr was considered important as being one of three putative examples of nova systems that display dwarf nova outbursts {\it before} the nova eruption.  With V3890 Sgr now eliminated from this list, we have to consider the other systems.  V446 Her certainly shows dwarf nova events long after its 1960 eruption (Honeycutt, Robertson, \& Turner 1995), but the only evidence for outbursts before the nova event is a small section out of a long light curve which shows a rise and a fall (Fig. 4 of Robinson 1975; Fig. 1a of Stienon 1971).  By itself, this `flare' has little conviction as it does not rise significantly above the average magnitude.  Robinson himself emphatically denies that this `flare' is a dwarf nova event, but later readers have not read closely and so the case is now presented in the literature as being a pre-eruption dwarf nova event (Livio 1992b; Vogt 1989).  And it gets worse when the whole light curve (Fig. 3 of Robinson 1975) is seen to display similar ups and downs frequently and on all time scales, with the putative `dwarf nova event' being just one of many examples all happening frequently.  So even these data alone merely show that V446 Her does not have pre-eruption dwarf nova events, but rather is just varying in the usual manner.  In addition, Collazzi et al. (2009) have examined all the original plates at Harvard and Sonneberg and have measured a much longer light curve in quiescence, both before and after the nova event.  We find that V446 Her is varying up and down on all time scales.  So again, we see that a normal variation when viewed in isolation was mistaken to be a dwarf nova event.  As such, V446 Her should be taken off the lists of novae with {\it pre-eruption} dwarf novae events.
	
	Livio (1992b) mentions a third case of pre-eruption dwarf nova events, for the weird and unique very-slow nova PU Vul.  (PU Vul is a symbiotic nova, so it is unclear whether any example from this event has application to classical novae.)  The basis for the claimed dwarf nova eruptions is a light curve from the Harvard plates from 1890 to 1979, with this showing definite variability from 16.5 to 15.0 mag, including one brightening by 1.5 mag that lasted roughly 100 days (Liller \& Liller 1979).  But such an event is much too long in duration and much too small in amplitude to be a dwarf nova eruption.  In addition, the frequency of magnitudes has the star well above its minimum roughly half the time, and this is not consistent with the behavior of dwarf novae.  Instead, the pre-eruption light curve is consistent with the usual variability displayed by many novae both before and after the eruption (Collazzi et al. 2009).  As such, there is good evidence to discount the possibility that PU Vul displayed pre-eruption dwarf nova eruptions.
	
	So the three examples of novae with pre-eruption dwarf nova events are wrong.  But there is another unheralded case; V1017 Sgr.  As mentioned in Secion 2, it had a classical nova event in 1919 plus dwarf nova events in 1901, 1973, and 1991 (Webbink et al. 1987).  The 1901 event is a good example of a pre-eruption dwarf nova.  Likely, it is significant that V1017 Sgr has an unusually long orbital period of 5.7 days (Sekiguchi 1992).  (This period is based on relatively few points in a radial velocity curve and has never been confirmed.  Indeed, I have made a sustained effort to detect photometric modulations, but these have returned no significant variations  with the claimed orbital period.)  The long orbital period might be a condition to get dwarf nova events in a system with such a hot central source, with the precedent including GK Per at 1.99 days and Q Cyg at 0.42 days. 
	
	\subsection{Post-eruption Declines?}
	
	Long after the nova eruption is over (perhaps 1-10 years later), does the quiescent nova slowly fade in brightness?  The standard expectation by nova researchers is that the system will remain essentially constant in brightness.  (This is excepting the known effects of orbital modulation, dwarf nova eruptions, and flickering, all on too short a time scale to be important for whether the nova systems fade away over the decades or centuries.)  This expectation arises in part because the quiescent brightness should be driven by the mass transfer rate which largely should be unaware of the depth of accreted material on the white dwarf.  Also, Robinson (1975) has established that the pre-eruption magnitude generally is close to the post-eruption magnitude, which argues against any fading.  Finally, occasional non-systematic magnitude measures have not been realized to show any significant fading in a few systems, so no famous case is known to highlight the question.
	
	The expectation of a constant quiescent level was first questioned about two decades ago.  The basis was the hibernation model for the evolution of nova, with long-term systematic changes in the accretion rate.  The picture is that the mass loss from a nova event will cause a slight separation in the binary, and the high $\dot{M}$ is maintained by the hot white dwarf irradiating the companion star, puffing up its atmosphere, and driving a non-steady-state high accretion.  As the white dwarf cools down from the eruption, the irradiation declines, the accretion rate falls, and the systems fades.  Kovetz, Prialnik, \& Shara (1988) made a specific prediction within the hibernation model that a typical slow decline rate in the first century after outburst should be of order 0.012 mag per year (for bolometric magnitudes).  We now have a specific prediction to test.
	
	The first search for a systematic decline rate was a paper by Vogt (1990).  But the methodology used a one time brightness measure (the magnitude at quiescence minus the eruption peak magnitude) decades after the eruption for many novae, where the measure is a tangle of many effects leading to a huge scatter.  The second search (Duerbeck 1992) was a well-planned use of quality long-term data from the literature and contemporary CCD magnitudes for 15 of the best observed novae up to 72 years after the outburst.  Well-sampled light curves were only presented for six novae.  Duerbeck found that the old novae all showed variability on under one-year time scales with amplitudes ranging from 0.2-2 mags.  (This is similar to the case for RNe.)  For long-term variations, his set of CN showed all types of variations (i.e., rising, declining, flat, falling-then-rising, and with flares) with amplitudes ranging from near-zero to 1 mag.  (This is similar to the case for the RNe, although the RNe systems usually have a higher amplitude of long-term variations.)  Despite having the light curves going in all directions, the overall average was for a small decline at a rate of $0.010\pm0.003$ mag per year.  I interpret this as a reasonable agreement with the prediction of fading from the hibernation model, but that other mechanisms are also present making for a substantial variation around this average decline rate.
	
	In principle, the hibernation model should apply to the RNe, and so we can test to see if they have a systematic decline after eruption.  The short recurrence times will make it difficult to distinguish a secular fading in quiescence from a possibly lingering tail of the eruption.  The short recurrence times will also often prevent any long stretch of time for which a slope can be better measured.  T Pyx appears to be in a century-long decline at an average rate of 0.016 mag per year (Figure 58) that makes it hard to attribute to any one nova outburst.  Such a rate must have started from 1850-1890 and it cannot keep up for long (at least with T Pyx remaining a RN), so we realize that the average decline rate can at best be episodic.  CI Aql has a poorly-sampled but flat light curve with zero post-eruption declines (Figure 59).  U Sco has what might be a post-eruption decline from 1988-1995 (at a rate of around 0.04 mag per year), although the next inter-eruption interval has a flat average light curve with zero decline (Figure 60).  V394 CrA has good coverage from 1989 to 2008 after its 1987 eruption, and there are substantial variations superposed on an overall rise of 0.05 mag per year or so (Figure 61).  T CrB started out by declining at a rate of 0.014 mag per year for the first 22 years, then had a flare lasting 22 years with an amplitude of 0.2 mag, and then the light curve has been flat for the last 16 years (Figure 62).  RS Oph has a complex light curve (Fig. 63) with chaotic rises and falls on all time scales.  The time intervals from 1948-1959, 1969-1982, and 1986-2002 might be claimed to display post-eruption declines (plus many added short-term flares).  But these declines are small (about a third of a magnitude) compared to the variations in the light curve, the whole post-eruption time intervals 1969-1985 and 1986-2006 are actually more-or-less flat (with flares), while the other post-eruption intervals do not show declines (1935-1945 is flat and 1959-1967 is chaotic with a rising trend).  So the apparent declines in RS Oph are neither significant nor systematic.  V745 Sco has only observations from 2004-2008, long after its 1989 outburst, yet we still see rises and falls of up to 2.5 magnitudes (Fig. 64), so there is certainly no simple decline.  V3890 Sgr has a problematic light curve (Figure 65), but there is no large decline before or after the 1962 and 1990 eruptions.  We do not have any useful information on the post-eruption decline for IM Nor or V2487 Oph.
	
	Just as with Duerbeck's CN study, my RN study finds a very mixed set of behaviors for the post-eruption decline.  Some of the RNe have no decline (CI Aql, U Sco after 1999, and V3890 Sgr), others have an apparent decline (U Sco after 1987, and RS Oph after the 1945 eruption), one has a systematic decline that is not affected by eruptions and must be transitory (T Pyx), one has a systematic rise (V394 CrA), and some have chaotic rising and falling (T CrB, RS Oph after the 1958 eruption, and V745 Sco).  If I blindly average the decline rates, I get 0.002 mag per year.  While there might be a fading due to hibernation in RNe, there must be larger forces changing the decline rate in a chaotic manner.

\section{Colors in Quiescence}
	
	The colors of the RNe are diagnostic of the accretion disk, the companion star, and the extinction.  Some colors for most of the RNe have already been published in the literature.  However I have colors for UBVRIJHK for all the RNe on many dates each.  In this section, I will report on my new colors and add in the colors from the literature.  These have been collected in Table 25.  The B-I versus I-K colors have been plotted for a single epoch in Figure 68.
	
	A complication is that the colors are changing on all time scales.  We have already seen that (a) the blue flickering light (and its color) changes on time scales of minutes and longer, (b) the colors change on the orbital time scale due to eclipses, irradiation of the secondary, and ellipsoidal effects, and (c) the brightness and presumably the color changes on the time scale of years and decades for unknown reasons.  This complexity means that the color is ever changing, so we can hardly quote one value.  This is the reason that I have included all the individual colors and their dates.  A reasonable but risky simplification is to take some average of the individual measures and use this value for model comparisons.
	
	A further complication is that the colors are affected by extinction, whereas we need the intrinsic colors for comparison with models and for using the spectral energy distribution to derive the accretion luminosity.  For this, we must have some independent measure of the extinction.
	
\section{Extinction and Distance}
	
	To convert the observed photometry into luminosity and energy (and hence to answer many physics questions about the system), we need both the distance and extinction.  Distances are always hard-to-get for cataclysmic variables.  The problem is that there are no standard candles, no redshifts, and no associated stars/clusters to use, while measures involving the intervening interstellar material are always too crude.  A further problem is that researchers often select one or two of the various distance estimates while ignoring other valid measures, with biased selections possibly distorting the best estimate.  For this paper with its emphasis on collecting observational results, a trap to be avoided is to use distance estimates based on theoretical models, as otherwise we would have later models being tested against earlier models instead of against the observations.  For this paper, I will provide new extinctions and distances to all the RNe based on intrinsic colors in the light curve and on the maximum magnitude-rate of decline (MMRD) relation based on my new templates.  I will also provide new distances based on the flux of the companion star alone and the assumption that the star fills its Roche lobe, with these providing the best distances for the five stars with such measures.  I will also provide an unbiased review of the literature for observational measures of both distance and extinction.
	
	The first subsection will systematically present the evidence for the extinction.  The second subsection will systematically use three relations based on the light curve to get the absolute magnitude of the nova at peak and hence the distances for all ten RNe.  The third subsection will isolate the flux from the companion star (either by total eclipse or by going to the infrared) so as to get a distance measure to five RNe.  The fourth subsection will go through the RNe one at a time, reviewing the literature and deriving a consensus distance estimate.  The fifth subsection will use the distance to RS Oph to address the question (critical to system models) of whether the red giant star fills its Roche lobe.
	
	\subsection{Extinction}
	
	Extinction can be measured in various ways, including the depths of interstellar absorption lines, the Balmer decrement, and a comparison of an observed color to some presumed intrinsic color.  These methods are notoriously unreliable with large scatter and questionable assumptions.  So we have to realize that high accuracy will not be possible.
	
	One method for estimating extinction is to use the observed color of the erupting nova around the time of the peak.  van den Bergh \& Younger (1987) demonstrated that the intrinsic $B-V$ color of nova events is $+0.23\pm0.16$ at peak and $-0.02\pm0.12$ when the V-band has faded by two magnitudes from peak.  With this, the excess color ($E_{B-V}$) can be derived from the difference between the intrinsic and observed colors.  To this end, I have used my light curve templates to evaluate the required colors (see Table 17).  These derived extinctions (expressed as $E_{B-V}$ in magnitudes) are collected into Table 26.  I have also collected the various published extinctions for inclusion in the table.  The bottom line of Table 26 is a concordance extinction, with the error bars roughly covering the spread of estimates.  I will use these final extinction estimates in the calculations for distances and spectral energy distributions below.

	\subsection{Distances from MMRD Relations}
	
	The `maximum magnitude versus rate of decline' (MMRD) relations connect the light curves to the peak absolute magnitudes of nova events.  van den Bergh \& Younger (1987) show that nova eruptions have a nearly constant absolute magnitude at 15 days after peak with $M_V(15)=-5.23\pm 0.39$.  The distance modulus (i.e., $\mu=V_{15days}-M_V(15)$) can then be calculated.  A substantial worry for many of the RNe is that these results are not for fast nova events, whereas the few fast novae might be exceptions, suggesting that there might be some dependancy on the time it takes for the light curve to decline by two or three magnitudes from its peak ($t_2$ or $t_3$).  The MMRD relations that explicitly use the rate of decline come in various versions.   I will use two from Downes \& Duerbeck (2000) that make use of $t_2$ and $t_3$ and separate out the fast events (Duerbeck's `A' class).  For the fast events they give $M_{Vpeak}=-11.26+1.58\log (t_3)$ and $M_{Vpeak}=-10.79+1.53\log (t_2)$, while for the slow events (T Pyx, IM Nor, and CI Aql) they give $M_{Vpeak}=-8.13+0.57\log (t_3)$ and $M_{Vpeak}=-8.71+1.03\log (t_2)$.  The scatter for any one nova about these relations is $\sim0.5$ mag.  With the $t_2$ and $t_3$ values from Table 17, we can calculate $M_{Vpeak}$ and then the distance modulus ($\mu=V_{peak}-M_{Vpeak}$) for each of the three relations.  These three $\mu$ values are averaged together and then converted to distances with the usual equation $\mu=5\log (d_{pc}) - 5 + 3.1E_{B-V}$ where $d_{pc}$ is the distance in parsecs.  These three values are not independent, so the average is an attempt to smooth out the variations from any one relation.  The  light curve inputs, distance moduli for all three relations, the averaged distance moduli, and the MMRD distances are presented in Table 27 for all ten RNe.  The formal error in $\log d_{pc}$ is near 0.2, but the systematic uncertainties are likely much larger.
	
	From Table 27, we see that three of the RNe have calculated distances of $>24$ kiloparsecs.  These are not reasonable distances, as it places the systems outside the far edge of our galaxy. This is alerting us to the likelihood of large systematic errors, at least in these cases.
	
	\subsection{Distances from Companion Stars}
	
	The best of the distance methods is to use the properties of the companion star.  In particular, with the known size of the companion star and a measured surface temperature, we can derive a blackbody distance to the companion.  But this only works when the light from the companion can be isolated.  This can be done either by looking in the infrared (where the red giant dominates the flux for T CrB, RS Oph, V745 Sco, and V3890 Sgr) or during {\it total} eclipse (for U Sco).  The distances can then be derived by three methods, which are all using the same essential physics, yet which differ in the input measures.  I have no reason to prefer any one method or any one set of inputs.  So I will take the distance to be the average of all three values.  The scatter amongst these three values will be a reasonable measure of the measurement uncertainty

	If we can get the brightness, temperature, and size of the companion, then we can use the blackbody surface flux and the inverse-square law of light to derive a distance.  For this, the only way to get the size of the companion is to take the size of the Roche lobe as based on the orbital period (with the value having only weak dependency on the stellar masses in the system).  For RNe, the Roche lobe filling case is confidently known as the means to get matter accreting onto the white dwarf, yet nevertheless the Roche lobe filling assumption has been questioned for RS Oph (see Section 14.5).  The temperature can either be gotten from looking for the peak of the spectral energy distribution or by a classical spectral classification of absorption lines on the companion.  To get the brightness of the companion star, we must have some way of isolating its flux from the rest of the system.  This can be done by measuring the system brightness during a total eclipse (for U Sco) or by looking in the infrared where the red giant dominates (for T CrB, RS Oph, V745 Sco, and V3890 Sgr).  If there happens to be additional light in the system not realized in this calculation, then the real distance should be {\it farther} than the derived distance.  The necessary  conditions are available for only these five RNe.  For these systems, the distances determined with the companion star are much more accurate and reliable than all other distance measures.
	
	The first method for getting distances to the companion stars (see Schaefer 2009 for details) uses the spectral type to get the effective surface temperature of the companion star.  Problems with this include the fill-in of lines with disk light, the uncertainties in the spectral class measures, imperfect calibration of the temperatures for a given spectral class, and the non-simultaneity of the spectral class measurement with the magnitude measurement.
	
	The second method (see Schaefer 2009 for details) uses the frequency of maximum spectral flux and Wien's Law to get the effective surface temperature of the companion star.  A problem with this is that the companion's light is not a perfect blackbody.
	
	The third method uses the Barnes-Evans relation (Barnes \& Evans 1976).  The physics of this is the same as for calculating the blackbody distances, but the calibration is purely empirical and varies with spectral type and luminosity class.  An advantage of this method is that it is closely independent of reddening.  A useful version of this relation is $F_K = 4.22 - 0.1m_K - 0.5 \log(\phi)$, where $m_K$ is the observed K-band magnitude (as taken from Table 25) and $\phi$ is the angular diameter of the star in milli-arc-seconds.  The value of $F_K$ is $3.84\pm0.2$ for normal red giants in the range of spectral types covered by the RNe (Cahn 1980).  With the Roche lobe radius (determined with the orbital period), we can convert the angular diameter of the companion star to a distance.
	
	Livio, Truran, \& Webbink (1986) used the first method to get a distance to RS Oph of 3180-3290 pc, while Barry et al. (2006), Brandi et al. (2009), and Schaefer (2009) reproduce the same distance.  Belczy\'nski \& Miko{\l}ajewska (1998) also used the first method for T CrB to get a distance of $960\pm150$ pc.  Schaefer (2009) has presented detailed calculations for the distances to the four RNe with red giant companions by both the first and second methods.  The distance to the companion stars by all three methods are given in Table 27.  The RMS scatter in the values produced by the three methods is indicative of the uncertainties associated with the blackbody distances.  For the four RNe with red giant companions, the fractional one-sigma uncertainties are 21\%, 27\%, 13\%, and 31\%, with an average of 23\% error.  For these four RNe, I will take the best estimate of the distances to be the average from all three methods with a 23\% uncertainty.
	
	The distance to U Sco can be confidently measured because we can determine the temperature and flux from the companion star {\it alone} by looking during the time of its total eclipse.  The equations for the first two methods are presented in Schaefer (2009).  The input, intermediate results, and final distance are presented in Table 28 in the same format as Table 1 of Schaefer (2009).  The mass of the white dwarfs in RN systems must be close to 1.35 M$_{\odot}$.  Based on radial velocity information and models for the companion star, $M_{comp}$ is close to 1.0 M$_{\odot}$ (Johnston \& Kulkarni 1992; Schaefer \& Ringwald 1995).  The uncertainty on the companion mass is likely around 30\%, and this translates into an uncertainty on $R_{Roche}$ of only 10\%.  Thus, we know the size of the Roche lobe to all needed accuracy.  For U Sco, the companion was determined to be G3-6 III-IV (Webbink et al. 1987), K2 IV (Anupama \& Dewangan 2000), F8$\pm$2 (Johnston \& Kulkarni 1992), and G0$\pm$5 (Hanes 1985), so I will take the median spectral type to be G5 IV.  The effective temperatures for the average spectral type (G5 IV) is taken from Drilling \& Landolt (2000).  The range of variation (or uncertainty) in spectral type corresponds to an uncertainty in $T_{comp}$ of 300 degrees or less.  The V-band magnitude for the companion star alone in the U Sco system is taken from the depth of the total eclipse (see Figure 46).  The bolometric magnitude for U Sco is $m_{bol}=m-A+BC$ with BC being the usual bolometric correction of -0.3 for a G5 IV star (Drilling \& Landolt 2000).  For U Sco, the observed colors at minimum are B-V=1.02 and V-I=0.79, and these correspond to intrinsic colors of 0.8 and 0.5 mag respectively.  The B-I color suggests that the companion has an effective temperature of around 6000 K (Johnson 1966).  For this temperature, $\nu_{max}$ is close to the center of the I-band, so I will use my observation that $I=18.1$ mag during the total eclipse (see Figure 45) and convert this to get an extinction-corrected $f_{\nu}(\nu_{max})$ for use in Table 28.  For the two similar methods (with independent input), I get two distances (included in Table 27) which average to 12,000 pc.  A comparison of the distances from the first two methods for all five RNe gives an uncertainty of 17\%.  So the distance to U Sco is $12000\pm2000$ pc, which places U Sco at the far side of the Milky Way's bulge.
	
	One implication from these distances is to have confident evidence that the MMRD distance for U Sco and T CrB have large errors, by factors of 3.1 and 4.0 respectively.  These errors are large, corresponding to 2.5 and 3.0 mag in the absolute magnitude.  With this, these two very-fast RNe become by-far the biggest outliers in the MMRD relations (Downes \& Duerbeck 2000).  It is unclear to me whether the problem lies with the incredible speed of the decline or some quirk related to the short recurrence time scales.  Both possibilities are poor because the other three systems are also fast RNe yet the MMRD works acceptably for them. (Their error ratios are 0.5, 1.9, and 1.3 for RS Oph, V745 Sco, and V3890 Sgr.)  I will adopt a factor-of-two error for MMRD distances, and I will be happy if I can find alternative evidence if the MMRD distance appears unreasonable.
	
	\subsection{Distances to Individual Systems}
	
	For T Pyx, with its prominent shell, we could hope to get a distance with an expansion parallax.  But this hope is stymied because the shell is expanding at a relatively slow velocity (with no deceleration) from an ordinary nova event (presumably around 1866) for which the date is not independently known and for which the expansion velocity is not known (Schaefer et al. 2009).  Webbink et al. (1987), Patterson et al. (1998), and Selvelli et al.  (2008) evaluate and summarize T Pyx distance measures.  Webbink et al. (1987) could only place limits of $>1050$ pc (from the calcium lines) and $<4500$ pc (from the rate-of-decline relation).  Patterson et al. (1998) consider several measures (Ca II lines, the speed of decline relation, the 2175A feature, and the Eddington limit) without providing details, then concluding that the distance is 2500-4500 pc.  Selvelli et al. (2008) give the same basic input and relations, along with full details, and conclude that T Pyx has a distance of $3500\pm350$ pc. The MMRD gives a distance of 3200 pc.  Given the poor observational constraints and the lack of consensus, I will adopt the middle ground offered by Patterson et al. (1998), with T Pyx having a distance between 2500-4500 pc, or $3000\pm1000$.
	
	For IM Nor, we do not have any good constraints on distance.  Duerbeck et al. (2002) used the strengths of the interstellar Ca II, Na I, K I, and diffuse interstellar lines to conclude that $E_{B-V}\ge0.8$ and $d\ge2500$ pc.   Orio et al. (2005) used the equivalent width of the Na I line to find $E_{B-V}$ from 0.5-1.1 mag and then the MMRD relation to get distances between 1500-6400 pc.  The Orio et al. distance is based on a $t_2$ value that is a factor of two smaller than given in Table 17, and so their distance is superseded by my MMRD distance of 3400 pc (Table 27).  So, with nothing better in sight, I will adopt the distance as from Table 27, $d=3400^{+3400}_{-1700}$ pc.
	
	For CI Aql, we again only have moderate quality extinction information and MMRD to get the distance.  (Theoretical models of CI Aql (Hachisu \& Kato 2003; Hachisu, Kato, \& Schaefer 2003; Lederle \& Kimeswenger 2003) uniformly give values close to $E_{B-V}=1.0$ mag and $d=1500$ pc.  But for this paper, I will only consider model-independent results.)   The only observationally-based distance is my MMRD results (Table 27) with $d=5000^{+5000}_{-2500}$ pc.
	
	For V2487 Oph, we again have only the MMRD relation (Table 27) with $d=32,400$ pc.  For a galactic latitude of $+7.8\degr$, the nova is 3.2 kpc above the disk and is on the far outer edge of the galaxy.  Such an extreme position is worrisome that the distance is greatly over-estimated.  Indeed, I would place an upper limit on the distance of perhaps 25 kpc for any plausible association with our Milky Way.  I can think of two reasonable suggestions to avoid the implausibly large distance suggested from the MMRD relation.  One resolution is that the peak was missed, so that a brighter $V_{peak}$ would give a closer distance.  But to get a distance equal to that to the galactic center, we would have to have $V_{peak}=4.5$ mag, and this is not plausible.  For the latest upper limit on the pre-eruption light curve, where the nova was not seen 8 days before the discovery, the extrapolated peak must have been fainter than 6.8 mag.  For an extinction of 0.5 mag and a peak at 6.8 mag, the shortest possible distance is 14,100 pc.  The second resolution is to note that U Sco is apparently a similar system whose MMRD distance is 3.1 times farther than a reliable distance based on the secondary (Section 14.3), so we could try to apply the same correction factor to V2487 Oph.  With this, the MMRD distance of 32,400 pc gets reduced to 10,000 pc.  In all, the range of plausible distances is from 10,000 to 25,000 pc, with the lower part of this range preferred simply due to the higher star density.  I will express this large uncertainty as $d=12,000^{+13,000}_{-2,000}$ pc.

	For U Sco, we have many papers that present observationally-based distances.  Conclusions include 22-95 kpc (Webbink 1978), $3.5\pm1.5$ kpc (Hanes 1985), 8-29 kpc (Schaefer 1990), 17 kpc (Duerbeck \& Seitter 1980), 5-25 kpc (Webbink et al. 1987), 14 kpc (Warner 1995), and 60 kpc (Warner 1987).  (Model-based estimates tend to be with shorter distances, including 4.1-6.1 kpc from Hachisu et al. 2000a; 6-8 kpc from Hachisu et al. 2000b; and 3.3-8.6 kpc from Kato 1990.)  My MMRD distance (Table 27) is 37,700 pc.  The many large distances in the literature are just versions of this latest MMRD value.  Webbink et al. (1987) puts an upper limit on the distance of $\sim 25$ kpc ``if it is plausibly to be a member of our galaxy".  In this confusion, we are lucky that the system has total eclipses, as this allows us to derive a reliable and fairly accurate distance, and this is $12,000\pm2000$ pc.
	
	For V394 CrA, we have little information on distance.  Duerbeck (1988) gave a distance limit based on the presumption that the peak absolute magnitude is $\leq -7$ (i.e., a fast nova will be super-Eddington) to get a distance $\geq5000$ pc.  My  MMRD distance from Table 27 is 24,400 pc.  For a galactic latitude of $-7.7\degr$, the nova would be 3 kiloparsecs above the plane and on the far edge of the Milky Way.  We can place an upper limit on the distance as $\sim25,000$ pc for the system to be associated with our Milky Way.  The far out MMRD position can be moderated by an increase in the extinction, but the colors are already too blue.  An alternative way to moderate the disturbingly-large MMRD distance is to realize that the similar U Sco system has a MMRD distance that is 3.1 times too large (see above), so perhaps a better MMRD distance for V394 CrA (as calibrated with U Sco) would be 7,800 pc.  The observational situation is that we only have limits of 7,800-25,000 pc, while the MMRD distance has substantial calibration questions so that any distance within the allowed range is possible.  Within the preferred range, the shorter distance scale is much more likely as the star density gets rather thin for the higher distances.  In all, I will take the distance to be $10,000^{+15,000}_{-3,000}$ pc.  Hachisu \& Kato (2000) use their theoretical model to conclude that the distance is something like 4200-6100 pc.  But again, it is important to realize that such a result is model dependent and is not the observation-based distance sought in this paper.  A substantial problem is that the Hachisu \& Kato model requires the extinction during eruption to be near zero yet also to be $E_{B-V}=1.10$ mag in quiescence.  They resolve this large difference in extinction by one sentence ``we may suggest that the intrinsic absorber of V394 CrA is blown off during the outburst".  Apparently, the idea is that the eruption radiation destroys nearby dust (leading to low extinction during outburst) and the ejected material later forms new dust of its own (leading to high extinction during quiescence).  But this idea won't work because the observations at quiescence are taken at such a late time that the dust will have dispersed to invisibility.  And my detailed calculations of dust formation shows no detectable dust formation.  And the theoretical model has the critical inconsistency that the extinction during the eruption (or anytime) cannot be smaller than $E_{B-V}\sim 0.25$ mag due to the intervening ISM in the first kiloparsec.  In all, I do not see a consistent theoretical scenario.  So we are in an unsatisfactory situation where there is no acceptable theoretical model and the observational situation only has broad limits.  I will adopt the unsatisfactory observational results that the distance is $10,000^{+15,000}_{-3,000}$ pc.
	
	For T CrB, the presence of the red giant companion star  can be used to get a distance.  Harrison, Johnson, \& Spyromilio (1993) used the infrared colors and the K-band magnitude to estimate $E_{B-V}=0.10\pm0.02$ mag and a distance of 1020 pc.  Patterson (1984) gives a distance to T CrB of 1200 pc based on measures of the secondary star.  Bailey (1981) used an empirical calibration of the K-band surface brightness as a function of V-K color to derive a distance of 1180 pc.  Krautter et al. (1981) used the spectral type of the red giant to get an absolute magnitude and a distance of 1300 pc.  Belczy\'nski \& Miko{\l}ajewska (1998) derived a distance based on the red giant (assuming it fills its Roche lobe) to be $960\pm150$ pc, which is similar to my value by the same method in Table 27.  I adopt the distance from the average of the three methods in Table 27 (and the mean 23\% uncertainty) of $900\pm200$ pc.
	
	For RS Oph, distance estimates in the previous literature has been summarized by Barry et al. (2009), while a critical review is given by Schaefer (2009).  Historically, RS Oph distance has been estimated to be over a rather wide range ($\lesssim$ 540 to 3180-3290 pc), however the most cited value has been 1600 pc from Hjellming (1986) as supported by a limit of $\lesssim$2000 by Cassatella et al. (1985).  However, Schaefer (2009) has corrected errors and added the conservative effects of systematic uncertainties to find that the literature distances are 1050-4200 pc (for the MMRD relations), no limit (Monnier et al. 2006), 1200-4900 pc (Rupern, Mioduszewski, \& Sokoloski 2008), $>$1000 pc (Hjellming et al. 1986), no limit (Cassatella et al. 1985), and `3200?' pc (Livio, Truran, \& Webbink 1986).  These independent distance evidences have not decided whether the RS Oph is sufficiently far away so as to require that the red giant fill its Roche lobe (with a distance of 4200$\pm$900 pc or $\gtrsim$ 3200 pc) or not fill its Roche lobe (with distances $<$3200 pc).  Schaefer (2009) demonstrates that the wind accretion model for RS Oph cannot account for the observed rate of mass accumulation on the white dwarf by a factor of 100,000$\times$.  Thus, the only way to get a large enough accretion rate is for the red giant to fill its Roche lobe.  With this, the best distance measures are based on the companion star (Table 27), for which I average the three methods and adopt a 23\% error.  The best estimate distance is 4200$\pm$900 pc.
		
	For V745 Sco, the presence of the red giant companion star can be used to get the distance.  My MMRD analysis gives a distance of 14,100 pc.  Such a large distance is relatively unlikely as the RN would then be well outside the galactic bulge (with this being by far the highest density of stars along the line of sight).  Harrison, Johnson, \& Spyromilio (1993) used the assumption that a M4 III star would have a K-band absolute magnitude of -5.5 to derive a distance of 4600 pc, although they point out that such a canonical luminosity could well be far off for any particular star (there is much vertical scatter amongst red giants in the H-R diagram), especially one that is in an interacting binary.  With my orbital period for V745 Sco, we can now get a reliable size for the red giant and derive the distance to the companion star as the average of the three methods in Table 27.  With this, we have a distance of $7800\pm1800$ pc, and V745 Sco lies in the middle of the galactic bulge.
	
	For V3890 Sgr, we have many measures of the extinction (cf. Table 26), but few measures of the distance.  Harrison, Johnson, \& Spyromilio (1993) estimate $d\approx5200$ pc as based on the assumption that the interacting red giant star has the same absolute magnitude as a canonical isolated red giant, with this assumption possibly being far wrong as we know that the red giants have a lot of vertical scatter in the H-R diagram.  My MMRD analysis (Table 27) gives a distance of 7,600 pc.  Fortunately, again the discovery of the orbital period allows for a reliable size of the red giant and a distance to the companion star (Table 28).  With this, the distance to V3890 Sgr is $7000\pm1600$ pc and it resides in the bulge of our Milky Way galaxy.
	
	All the distance and extinction estimates from this section have been collected into the top of Table 29.
	
	\subsection{Is the RS Oph Red Giant Filling Its Roche Lobe?}
	
	The general presumption is that the companion stars in all nova systems are filling their Roche lobes.  This strong result arises mainly from the need to get matter falling into the accretion disk at a high rate.  This imperative is particularly strong for RNe, where the accretion rate onto the white dwarf must be something like $10^{-7}$ M$_{\odot}$ yr$^{-1}$.  However, this imperative might not be required in the case where the companion star is a red giant.  Red giants have substantial stellar winds, some fraction of which will accrete onto the white dwarf, so the Roche lobe can possibly be underfilled and yet the system can still have significant accretion.  
	
	Of the four RNe with red giant companions, a case for an underfilled Roche lobe has only been made for RS Oph.  In this case, the accretion onto the white dwarf would have to be driven by a stellar wind perhaps mediated through a ring around the companion star.  Garcia (1986) interpreted profile changes in the Fe II line as being like those seen in a Be star, which would imply that the red giant cannot fill its Roche lobe.  But this model hardly seems unique and it is based only on an analogy, so the underfilling of the Roche lobe was just a suggestion.  The primary reason given by Garcia for RS Oph not filling its Roche lobe comes from a simple comparison of the calculated Roche lobe radius (for presumed masses) with the canonical radius of an M0 giant.  But this argument was critically based on the reckless presumption that the star in a weird binary just happens to have the same radius as some isolated star with the same temperature.  Garcia quoted the Roche lobe radius as 74 $R_{\odot}$ while his canonical M0 giant has a radius of 50 $R_{\odot}$ and hence concluded that the Roche lobe is not filled.  But I see no reason why the real companion star can't be just a little bit higher up in an H-R diagram and totally filling its Roche lobe.  That is, red giants of the same temperature (all along a vertical line in an H-R diagram) come in a wide range of sizes (spreading up and down along that vertical line), so there is no real constraint on the size.  As such, the original argument against Roche lobe filling is far too weak for serious conclusions.
	
	A more sophisticated version of the same argument comes from detailed modeling of the fluxes and temperatures of the companion star (e.g., Hachisu \& Kato 2001).  The derived size of the companion star critically depends on the adopted distance to the star.  Hachisu \& Kato adopted a distance of 600 pc and concluded that the red giant only fills 25\% of the Roche lobe (by radius), while Dobrzycka et al. (1996) adopted a distance of 1500 pc and concluded that the red giant only fills 60\% of its Roche lobe.  If the distance to RS Oph is truly as close as these old assumptions, then everyone would agree that the Roche lobe is underfilled.  So the primary question really comes down to the distance.  If RS Oph is $\gtrsim$3000 pc away, then it is filling its Roche lobe, while if RS Oph is nearer then it will not be filling its Roche lobe.
	
	Historically, the RS Oph distance question went horribly awry.  The first (apparently) reasonable distance estimates to RS Oph were by Casatella et al. (1985) and Hjellming et al. (1986), and so these are always the papers that later works cited.  No one in the later literature spotted the simple and killing error that completely invalidated these distance estimates.  (In particular, for a galactic latitude of 19.5$\degr$, the line of sight leaves the gas and dust in the disk within a kiloparsec of Earth, so there would be no absorption from the Carina Arm in any case and the naive linear relation between distance and absorption is far wrong.)  The Hjellming et al. paper and its claimed distance of 1,600 pc has been cited 63 times, always as a primary reference for the distance to RS Oph.  And then, no accurate or reliable method for distance determination came along (see previous section) as a substitute for the invalid 1,600 pc distance estimate.  So our community has been left with the original (invalid) distance of 1,600 pc plus a number of inaccurate and unreliable measures that convinced no one.  By virtue of repetition and nothing better, our community has taken the distance of 1,600 pc as the default answer.  At the conference on cataclysmic variables in Tucson in March 2009, I systematically asked all four speaker who talked about RS Oph as to what they thought the distance was, and they all answered a hesitant ``1,600 pc".  This shows the legacy of the old error.  And the error is reinforced by the circular argument of modelers adopting a short distance scale, finding that the red giant does not fill its Roche lobe, developing models with wind accretion, and then later researchers look back on these models as justification for the short distance scale.  So, historically, the current default distance of 1,600 pc (and the conclusion that the red giant does not fill its Roche lobe) has arisen largely from a bandwagon effect (cf. Schaefer 2008) that was started as a simple error in an old paper.  
	
	The stellar wind accretion hypothesis (required if the red giant underfills its Roche lobe) can be tested.  We know from various independent methods what the mass loss rate of the red giant is and what fraction will accrete onto the white dwarf between eruptions.  We also know from multiple independent methods how much accreted material is required to trigger the eruption.  We can then compare the two, with the wind accretion hypothesis being plausible only if the mass accreted onto the white dwarf is comparable or larger than required for the eruption.  Schaefer (2009) has performed this analysis, finding that the wind from the red giant in the RS Oph system has a mass loss rate of $\sim 3.7\times10^{-8}$ M$_{\odot}$ yr$^{-1}$, with $\sim 5.4\times10^{-10}$ M$_{\odot}$ falling onto the white dwarf in the average inter-eruption interval of 14.7 years, while roughly $5.9\times10^{-5}$ M$_{\odot}$ is required as the trigger mass.  The wind accretion model fails by a factor of $100,000\times$.  Thus, the only way to feed the white dwarf fast enough is with Roche lobe overflow.  So I confidently conclude that the red giant fills its Roche lobe.  
		
\section{Galactic Distribution}
	
	What is the galactic distribution for the RNe?  We will not be able to make a high precision measure of the distribution, because we only have ten stars and because of the usual interstellar medium masking.  To quantify the positions of the known RNe, I have collected the distances (from the previous section) and the galactic coordinates ($\ell$, $b$, and the angle from the galactic center $\theta_{GC}$) into Table 29.  From this, I calculate the height above the galactic plane for the best estimate distances, $Z_{best}$.  I have also tabulated the heights above the galactic plane for the minimum acceptable distances, $Z_{min}$.  A top view and a side view of our Milky Way, along with the projected positions of the RNe (and the uncertainty associated with the distances) are displayed in Figure 69.
	
	The RMS scatter in the $Z_{best}$ values is 1600 pc, while the RMS scatter in the $Z_{min}$ values is 1300 pc.  This is very large compared to the observed value of 190 pc made from a large demographic study of cataclysmic variables, while the intrinsic value (after correction for interstellar absorption) is more like 150 pc (Patterson 1984).  Patterson's sample is composed mainly of relatively nearby systems, so this would be representative of the scale height of the disk population.  As such, we see that the RNe are not a simple disk population.  Shafter (2002) and others have pointed out that the classical novae are a combination of a disk and a bulge population.  With the large scale height for three RNe and the concentration on the sky towards the galactic center for eight RNe, we see that the RNe must have a large contribution from a bulge population.  But RNe cannot be entirely a bulge population, because no more than 50\% can be within 3 kpc of the galactic center and because the remaining RNe are similar to a disk population with its RMS scatter (400 pc) much closer to that of general CVs.  Another measure is that we see 4-out-of-10 RNe $>30\degr$ from the galactic center, while close to half of all CNe are $>30\degr$ from the galactic center, which implies that the bulge-to-disk ratio is comparable for RNe and CNe.  While estimates are necessarily crude due to the low numbers, it appears that RNe are roughly evenly divided between thick disk and bulge populations, in a situation similar to that amongst CNe.
		
\section{Absolute Magnitudes}
	
	In the preceding sections, I have been reporting on the best values for the peak magnitudes, the quiescent magnitudes, the distances, and the extinctions.  With this, we can now calculate the absolute magnitudes for the RNe.  We can use the standard relation that the absolute magnitude is $M=m-5\log(d)+5-A$, where $m$ is the observed apparent magnitude, $d$ is the distance in parsecs, and $A$ is the absorption from the interstellar medium in magnitudes.  The absorption is a function of the observed color excess ($E_{B-V}=A_B-A_V$), with $A_V=3.1~E_{B-V}$ for the V-band, and with $A_B=4.1~E_{B-V}$ for the B-band.  The intrinsic color ($(B-V)_{0}$) will equal the observed color corrected for the color excess, as $(B-V)_0=(B-V)-E_{B-V}$.  As an intermediate step, I have tabulated the observed distance moduli ($\mu=m-M=5\log(d)-5+A$) for both the B-band and the V-band (see Table 29).
	
	In Table 29, I collect the various magnitudes for the ten RNe.  These include the peak B-band magnitude ($B_{peak}$), the peak V-band magnitude ($V_{peak}$), the average B-band magnitude in quiescence ($\langle B_q \rangle$), the average V-band magnitude in quiescence ($\langle V_q \rangle$), and the range of $V_q$.  I have converted all these apparent magnitudes to absolute magnitude with the observed distance modulus, and these are also placed into Table 29.  I have also added the intrinsic colors at peak and at quiescence.

	The peak absolute magnitudes in the V-band have an average value of -8.0 mag, which is close to the value for CNe (Shafter 2002).  The intrinsic colors at peak have an average value of $(B-V)_0=-0.1$ mag, with this being close to the value for CNe.  Thus, I am seeing no difference between RNe and CNe at the peak.  This is surprising to me, as I would expect that the greatly different mass and velocity of the ejected envelope (between RNe and CNe) would have made for systematically different peak conditions.
	
	The average quiescent absolute magnitudes vary substantially, with $-4.1\leq M_V \leq 3.2$ mag.  As expected, the most luminous quiescent systems are those with red giant companions.  The canonical $M_V$ for a M0-M5 red giant is around -0.3 mag.  Both V3890 Sgr and RS Oph are substantially brighter than this, suggestive that there might be a substantial contribution from the accretion disk, and this is confirmed by the relatively blue $B-V_0$ values at quiescence.  Both V745 Sco and T CrB are less luminous than the canonical red giant, but this is not a cause for concern because we have no reason to think that the companions are canonical.  (Again, red giants show a lot of vertical scatter in the H-R diagram.)  The intrinsic colors for these two stars suggests that the contributions from the accretion disks are relatively small.  For the six remaining systems that do not have a red giant companion, the range of $M_V$ in quiescence is 0.0 to 3.2 mag with a median of 2.2 mag.  For comparison with the CNe with reliable distance measures (Shafter 1997), the range of $M_V$ in quiescence is 1.1 to 7.0 mag with a median of 4.6 mag.  Thus, in quiescence, RNe are substantially and significantly more luminous than CNe, by almost an order of magnitude.  This is easy to understand, because theory requires that RNe have a very high accretion rate, and many of the RNe have red giant companion stars, with both of these effects making for brighter systems.

\section{Spectral Energy Distributions in Quiescence}
	
	What is the spectral energy distribution (SED) for the galactic RNe in quiescence?  That is, how does the spectral flux (say, in units of jansky) change across the spectrum.  For most of the RNe, I have measured magnitudes in the UBVRIJHK bands, and with the extinction estimates, I can calculate the SED.  For the short period systems like T Pyx and IM Nor we will see the accretion disk contribution alone, while for the long period systems we will see the accretion disk plus the red giant companion star.  By getting the flux across a wide range, we can separate out the disk and companion components.  An integral under the accretion disk component will give the accretion flux, which can then be converted to the accretion luminosity with the distance estimates given in Section 14.  From the accretion luminosity and the mass of the white dwarf, we can derive the accretion rate ($\dot{M}$).  The RN accretion rates are important for measuring whether $M_{ejecta}>\tau_{rec}~\dot{M}$ and for determining the lifetime (and hence the death rate) as $\sim 0.2 M_{\odot} / \dot{M}$.  With both of these questions, the determination of the RN accretion rate is critical for the most important questions.
	
	This path to $\dot{M}$ has two substantial uncertainties.  The first big uncertainty is that much of the accretion luminosity comes out in the far ultraviolet and the extreme ultraviolet.  The SEDs in this paper will only cover from around 0.36-2.2 microns of wavelength, which does not cover where most of the energy comes out.  Nevertheless, the observed SED will rise and fall with the accretion rate, so model fits can still produce reasonable values.  Also, in some cases, far ultraviolet observations might be available to substantially extend the SED so as to greatly increase the accuracy of model fits.  In particular, I have GALEX observations of three RNe in quiescence, for which I have simultaneous BVRIJHK data from the SMARTS 1.3-m telescope.  The derivation of the needed optical SEDs (for joining with the GALEX SEDs) requires good comparison stars, simultaneous photometry in quiescence, and the best possible distances and extinctions, and has been a primary motivation for much of the work reported in this paper.  The second big uncertainty is that the derived accretion rate depends on the square of the distance and the distance is often poorly determined.  It is for this reason that I have done so much preparation leading up to the results in Section 14.  
	
	In this section, I will only derive the SED, with the further task of deriving the $\dot{M}$ reserved for a later paper.  The first part of this task is simply to collect magnitudes and colors for the ten RNe.  These should all come from nearly simultaneous observations.  For this, I have selected SMARTS observations with UBVRI or BVRIJHK observations of nine of the RNe (V2487 Oph was not known as a RN when these series of observations were made) that were all taken within a 15-minute time interval.  (The GALEX observations were also taken at the same times.)  The U-V colors were taken at other times, yet were nevertheless taken to be representative of the U-V color at the quoted times.  This is a reasonable equality as we can see from Table 25 that the colors do not change greatly.  For some stars, I also had to take the V-H and V-K colors from some other date, and again this is reasonable because the colors do not change greatly.  My collected magnitudes and colors are given in Table 30.  The magnitudes do change on the usual flickering time scales, so even my 15-minute interval of observation will have some error due to non-simultaneity, and I will estimate that the one-sigma uncertainty is typically 0.05 mag for all the RNe.  With this, the errors associated with variability will usually dominate over measurement errors.  For the few cases where the measurement errors are significant, I will correctly propagate the uncertainties into the next table.
	
	The second part of the task (constructing the SEDs) is to correct for the interstellar extinction so as to get the extinction-corrected magnitudes.  For this, we have to use the estimates of the $E_{B-V}$ values from Table 26.  From these, we can calculate the extinction-corrected magnitudes by subtracting the extinction for each band, $A(\lambda)$ in magnitudes, with $A(\lambda)=[A(\lambda)/A(V)]\times 3.1 E_{B-V}$.  Here I have taken the canonical numerical constant of 3.1 applicable to the ordinary extinction in our Milky Way.  The quantity $A(\lambda)/A(V)$ depends on the bandpass, and these are tabulated in Table 31 along with other standard properties of each band (Mathis 2000).  With this, we can convert all the colors in Table 30 into extinction-corrected magnitudes (denoted with the subscript `0'), along with their one-sigma error bars, and I have tabulated these in Table 32.  The error bars are being dominated by the uncertainties in the extinction.  So the errors for individual points are correlated. 
	
	The third part of the task is to convert the extinction-corrected magnitudes into physical units.  I have chosen to convert them into flux densities ($f_{\nu}$) with units of jansky (Jy), with this flux unit equal to $10^{-26}$ watt per square meter per hertz or $10^{-23}$ erg s$^{-1}$ cm$^{-2}$ Hz$^{-1}$.  For this conversion, I have tabulated the flux in jansky for a zero-magnitude star in each bandpass (see Table 31, as taken from Bessel 1979; Campins, Reike, \& Lebovsky 1985).  A simple scaling from the zero-magnitude flux to the magnitude reported in Table 32 will give the flux density, with these values reported in Table 33.  Both T CrB and RS Oph have been detected with the $\it IRAS$ satellite, with T CrB having a 12 $\micron$ flux of 0.70 Jy ($6\%$ error bars) and a 25 $\micron$ flux of 0.3 Jy ($14\%$ error bars), and with RS Oph having a 12 $\micron$ flux of 0.43 Jy ($10\%$ error bars) (Schaefer 1986).  These flux densities as a function of frequency constitute the SED for each RN.  I have plotted these SEDs in Figure 70. 
	
	The fourth part of this task is to correct for the distances, that is, to convert from fluxes to luminosities.  For this, we have $L_{\nu}=4\pi d^2 f_{\nu}$, where the distances are taken from Table 29.  I have taken $\log(L_{\nu})$ in units of watt per hertz and presented these in Figure 71.  I have not propagated the uncertainty in the distances, which can be substantial for V394 CrA and V2487 Oph.  With this figure, we can finally see the RNe relative to each other with the distances taken out.
	
	The SEDs show the expected features of the red giant companions and the accretion disks.  The four systems with red giant companions are all bright in the near infrared and have peaks $\sim$14.3 in $\log (\nu)$.  With the equivalent of Wein's displacement law for the peak of the flux distribution corresponding to temperatures of around $3,400\degr$, as appropriate for M giant stars.  RS Oph has another peak in its SED at around $\log (\nu)=14.7$ (in the V-band), likely due to continuum from the cF-type source (associated with the accretion stream) and nebular emission lines (Anupama \& Miko{\l}ajewska 1999; Brandi et al. 2009).  The CI Aql, T Pyx, IM Nor, and U Sco systems all display a power law rise with the classic signature of an accretion disk ($f_{\nu} \propto \nu^{1/3}$).  I see no evidence of the companion star.  However, the presence of a moderate secondary eclipse on U Sco in the I-band and the ellipsoidal effects on CI Aql demonstrates that there are indeed contributions to the flux from the secondary stars.  All of the RNe have their SEDs falling from the B-band to the U-band, and this cannot be due to measurement errors because everyone sees these same colors and it cannot be due to under-correcting the extinction because the B-band fluxes would rise correspondingly and the size of the extinction error bars are too small to lead to a rise towards the ultraviolet.  I think that the systematic fall towards the U-band is real, and I do not understand it.
	
	All RNe in quiescence vary on all time scales (minutes-hours-days-weeks-months-years-decades) usually with amplitudes of over one magnitude (see Sections 10-12 and Figures 42-67).  So how can one SED and its derived $\dot{M}$ represent the accretion rate over a whole eruption cycle.  One adequate answer is that the single derived $\dot{M}$ will provide a typical value that is approximately correct, and this is a lot better than the alternatives.  But we can do better.  With the colors not changing much (outside of eclipses and eruptions), the SED can be scaled by the V-band magnitude.  Or rather, the SED scaling should go as the V-band flux ($\propto 10^{-0.4V}$).  With many magnitude estimates throughout an inter-eruption interval, we can take each V-band flux, scale the reported SED to each V-band flux, calculate the $\dot{M}$ values for each scaled SED, average all the $\dot{M}$ values to get the average accretion rate over the inter-eruption time interval, and then multiply by the time of that interval to get the mass accreted between eruptions.  This plan is straight forward, practical, and follows the complexity of the RNe variability.

\section{Open Questions for Observation and Theory}
	
	A variety of questions have been raised in this work, and I will itemize these questions here.  I will not separate out the various higher level questions.  Such questions as ``What are the average $\dot{M}$ values?",  ``Are the white dwarfs gaining mass?", and ``What is the RN death rate in our Local Group?" will be addressed in other papers.  Instead, I will select questions with some relevance for photometric issues, be it concerning observational facts or theoretical interpretation:
	
\subsection{Observational Questions}
	
	(O1) What is going on with the sudden fading around day 33 in the tail of the U Sco light curve?  Is the fading temporary, only to have the nova rebrighten and return to its prior fading decline (as would arise from the formation and dispersal of dust in the ejecta shell), or is the fading monotonic (as would arise from a turn off in the nuclear burning)?  And how does this fading behavior depend on color?  Does the x-ray brightness turn off at that time?  All of the extant data for prior events have already been examined, so the only answer will come from detailed photometry of the upcoming eruption.
	
	(O2) What is the orbital period for V2487 Oph?  The single most important datum for any cataclysmic variable is its orbital period, and this is especially true for RNe due to their very wide range of $P_{orb}$.  V2487 Oph likely has a period of near one day, but this is guessed only on a perceived similarity with the RNe like U Sco.
	
	(O3) What is the frequency of plateaus in the light curves of classical novae?  If they are common, then maybe their presence in RN light curves does not mean much.  If plateaus are in only a fraction of classical novae, then this might provide an indication of the fraction of classical novae that are really RNe.  Also, plateaus amongst the classical novae light curves could point to good RN candidates, for example, for use in archival plate searches for the earlier eruptions.
	
	(O4) Another observational imperative is to create a long series of accurate minima times for both T Pyx and IM Nor.  This task has three science returns:  First, with an accurate orbital period, we can await the next eruption and then measure the post-eruption orbital period so as to measure the change in its orbital period.  This is an important task that wil require a large and sustained observing program.  Unfortunately, both RNe are expected to have their next eruptions only many decades or longer from now (see Table 21).  Second, we can measure whether the T Pyx period is increasing linearly.  That is, does the O-C curve rise quadratically as predicted if the period change were caused by conservative mass transfer?  If so, then this will force models to take the high derived $\dot{M}$ at face value, despite theoretical difficulties in understanding.  Third, we can measure the period for IM Nor using only minimum times during quiescence, and this will provide a good ephemeris to see the deviations in eclipse times during the tail of the 2002 eruption.  For U Sco and CI Aql, I see large deviations in eclipse times made during the eruption with these being caused by the shift in the center of light in the fading nova.  With the excellent data of Monard for the 2002 eruption, we can trace the deviations throughout the eruption, for direct comparison with theoretical models.
	
	(O5) Someone really has to make a modern radial velocity curve for the white dwarf in T CrB.  This is to replace the 1950's radial velocity curve of Kraft (1958) for which the amplitude is based on only two plates.  The study of Kenyon \& Garcia (1986) had high spectral resolution and good time coverage, but did not cover any emission lines.  As the velocities are small, high spectral resolution is required, the spectral range should be carefully selected to include emission lines, and the observations should be well-spaced over a several year time span.
	
	(O6) I am predicting that five of the RNe (V394 CrA, V2487 Oph, U Sco, V3890 Sgr, and V745 Sco) will go off in the next decade.  With the expected uncertainties in such predictions, any one of these could go off any night now.  This creates an observational imperative to keep regular daily monitoring of all five so as to promptly catch their next eruptions.
	
	(O7) How do the long-term variations in the quiescent light curve compare between CNe and RNe?  That is, do the CNe have long-term variations of high amplitude as shown by most of the RNe?  Do the CNe show any systematic rise or fall after nova events?
	
	(O8) How do the very-low-frequency PDSs connect with the higher-frequency PDSs?  That is, for at least RS Oph and T CrB, we see $1/f$ noise for time scales longer than a week and shorter than a few hours, but we don't know how they connect.  In principle, with correct normalization of the PDSs, we can see whether the high and low frequency behaviors are all part of one continuous power law.  Alternatively, we could construct a PDS from a time series with nearly uniform coverage from time scales of hours to months, with the possibility of breaks between.  Within the ideas mentioned in this paper, the short time scale variations should only extend to something like the orbital period, while the convection cells on the red giant will provide little power on short time scales.
	
	(O9) Is IM Nor a supersoft source and does it have a shell from an earlier ordinary nova event?  Like T Pyx, IM Nor has a very high accretion rate for which ordinary mechanisms cannot account, hence making it likely that the system is also driven by a supersoft source after an ordinary nova eruption.  Deep observations with {\it Chandra} or {\it XMM-Newton} might be able to detect a residual supersoft source, while the {\t Hubble Space Telescope} might be required to find a shell.  Problems for this task are that any IM Nor supersoft source is turning off and the source is near the galactic plane (latitude $+2.5\degr$) with moderate extinction.
	
\subsection{Theoretical Questions}
	
	(T1) Why did T CrB suffer a distinct, significant, and unique fading in the year {\it before} its 1946 eruption?  And why would this fading behavior be different in the B and V bands?  The fading is by around one magnitude below the usual level of the system, with this going to two magnitudes below the usual level in the B-band at a time 29 days before the eruption.  My first thought is that the accretion turned off (for unknown reason) hence making the system lose the light from the accretion disk, but maybe the depth of the drop will require the red giant companion to be dimmed somehow.  And what is the physical connection between this fading and the subsequent nova eruption?  That is, how can the turning off of accretion {\it anticipate} or {\it trigger} the nova event?
	
	(T2) Why did T CrB have a secondary maximum?  This event carries a large amount of energy.  The maximum started after the system had returned to quiescence for fifty days, so why the delay?  Are there any other novae with secondary maxima?
	
	(T3) Why do the quiescent light curves of half the RNe (CI Aql, U Sco in the red, V394 CrA, V745 Sco, and V3890 Sgr) have an asymmetry where the maximum just after the primary minimum is higher than the maximum preceding the primary minimum?  I do not know of any precedent for this amongst the classical novae, while other types of cataclysmic variables that have an asymmetry have the brightest phase just {\it before} the primary eclipse\footnote{http://cbastro.org/cataclysmics/atlas/}.  I suggest that the extra light arises from the hot spot, and if so, then we need an explanation for why it appears brightest around phase 0.15.
	
	(T4) Why do most of the RNe display large amplitude variations on long time scales (years to decades)?
	
	(T5) How does the eclipse amplitude and phase change throughout the eruption.  This will depend in detail on the optical depths of the envelope as a function of time.  We already have amplitudes and phases for three RNe (IM Nor, CI Aql, and U Sco), and we can anticipate wonderful measures from the upcoming eruption of U Sco.  In principle, the observations can be used to test theoretical models by providing an observed optical depth all the way to the center of the nova shell throughout the entire eruption.  Also, I hope that this optical depth information can be combined with spectral data so as to derive a value for the total mass ejected during the eruption.
	
\section{Conclusions}
	
	This paper is addressing a fairly restricted topic (observational issues arising from photometry of one small class of stars with only ten known members).  Nevertheless, these data provide a fundamental basis for front-line questions of broad importance (hibernation and the Type Ia progenitors).  To address these big questions, I repeatedly realized that I needed comprehensive and correct photometric measures of the few known RNe.  But I found that the published data were widely scattered and not-modern, while the majority of the data have not been published (not even counting the huge AAVSO data sets).  So this paper represents my attempt to collect all RNe photometry and put it together in a consistent and modern way.  I will be using the results from this paper as the fundamental observational basis for a variety of future papers, while I expect that a wide variety of researchers will also be able to pull out the best numbers on RNe from this paper.
	
	Here I will summarize the new points.  Many of the new results are also tabulated in Table 34.  Additionally, I will itemize the new points for each of the ten RNe, and then summarize the broader conclusions:
	
	(T Pyx) T Pyx is unique for having a flagrant plateau in its eruption light curve that starts after a precipitous  drop (2.0 mag in 20 days).  T Pyx is unique in having a secular decline from 1890 to present across many eruptions.  

	(IM Nor) The light curve shows a sharp drop after a plateau.  For the first time, the depth of eclipse in a nova system has been measured throughout the entire eruption all the way to quiescence.  The amplitude increased from 0.01 mag at a time 110 days after peak (3 magnitudes below the maximum), to 0.25 mag in the late tail, to 0.4 mag in quiescence.
	
	(CI Aql)  The 1941 eruption of CI Aql was discovered with an exhaustive search through the Harvard plate collection.  This changes the recurrence time scale from 83 years to 24 and 59/N years (for N=1, 2, or 3).  With the quiescent magnitude changing relatively little over the last century, the best idea is that the accretion rate and recurrence time scale has been roughly constant, suggesting N=2 or N=3, with this being consistent with a discovery efficiency of 86\% over the time between 1941 and 2000.  The light curve of CI Aql has a mysterious asymmetry in that it is brighter in the time interval just after the primary eclipse than in the time interval just after the secondary eclipse.  I have measured 80 eclipse times for CI Aql, with the best period being $0.61836090\pm0.0000005$ days.  Previous claims to having measured a period change for CI Aql are certainly wrong because they were using eclipse times from during the eruption and these times are systematically shifted earlier by 0.006 days, with this producing the entire claimed effect.
	
	(V2487 Oph)  As part of our work, we predicted that the CN was actually a RN, and we tested this prediction at Harvard and Sonneberg.  The prediction proved to be correct when we discovered the eruption in 1900 (Pagnotta, Schaefer, \& Xiao 2008; Pagnotta et al. 2009), and this provides confidence that our indicators (high excitation lines and high expansion velocities during eruption) are valid pointers to RNe.  Only $30\%$ of the eruptions are detected, so the real recurrence time scale cannot be the naive 1998-1900=98 years, but instead must be more like 18 years.  To keep the distance to V2487 Oph reasonable (i.e., in our galaxy and preferably not far from the bulge), the peak magnitude might have to be several magnitudes brighter than observed, perhaps as bright as seventh mag.
		
	(U Sco)  I discovered two previously-unrecognized eruptions on the Harvard plates (in 1917 and 1945), while another eruption in 1969 was found in the archives of the RASNZ.  With nine known eruptions, we see that U Sco erupts every $10\pm2$ years with missed eruptions around 1927 and 1957.  U Sco rises from quiescence to peak in roughly half a day.  With $t_3=2.6$ days, U Sco is the fastest known nova of any type.  In the new 1945 light curve from the Harvard plates, I find a sudden and steep drop in the B-band magnitudes starting 33 days after the peak and dropping by over two magnitudes within the next several days.  Other than this 1945 eruption, only 4 magnitudes have been recorded after day 33 (these on days 44-57 for the 1979 and 1999 eruptions) and these V-band data do {\it not} show evidence for the sudden drop in blue brightness on day 33.  U Sco is predicted to have its next eruption in the spring of 2009 (within a year), and this provides a wonderful opportunity to prepare an observing campaign from x-ray to infrared.  U Sco has a classic eclipsing light curve, with a total eclipse at the primary minimum and ellipsoidal effects visible in red light.  I have 47 eclipse times for U Sco from 1989 to present.  Prior claims to have found a period change are certainly wrong as a key role was given to eclipse times during the eruption for which systematic shifts of up to 0.015 days earlier occur.  The light curve for the uneclipsed system is relatively flat in the B-band, but it displays ellipsoidal modulations plus a secondary minimum (with amplitude 0.3 mag) in the I-band.  The light curve during eclipse has relatively little flickering and apparently has a flat bottom with duration of $0.0253\pm0.0025$ in phase.  Using the visual magnitude of the companion star alone (during the total eclipse), the companion star temperature (from the observed spectral type), and the companion star radius (from the Roche lobe diameter), I derive a distance of $12,000\pm2000$ pc.
	
	(V394 CrA) V394 CrA has its average brightness level changing up and down by roughly one magnitude.  Prominent sinusoidal oscillations appear when the system is faint, and they get lost in the flickering when the system is bright.  That is, the slowly changing brightness (presumably from variable accretion) is superposed on top of unchanging periodic modulations (presumably from ellipsoidal, reflection, and eclipse effects).  The accurate orbital period is $1.515682\pm0.000008$ days.  The folded light curve shows a primary minima that appears to be an eclipse, while a shallower secondary minimum is also visible.  Outside of eclipse, the system is systematically brighter soon after the primary minimum.
	
	(T CrB) An old question is whether the mass accreting star in the system is a white dwarf, but by now many very strong arguments  and proofs have been presented that T CrB has a white dwarf and the nova is caused by the usual thermonuclear runaway on its surface.  In the year before the eruption, the brightness dipped with large color swings by 1-2 magnitude to unprecedentedly faint levels (with its all-time faintest level being 29 days before the eruption).  This is too close a coincidence to be by chance, so there is likely some causal connection by which the fading is physically connected with the eruption.  During quiescence, the claimed periodicity of 9840 days (Leibowitz, Ofek, and Mattei 1997) is certainly wrong.  In quiescence, the power density spectrum is rising as a power law from seconds to hours and from days to decades.  The observed power law index (-0.8) is somewhat shallower than that of variations on a red giant (with index -1.37 to -1.54) or variations on cataclysmic variables (with indices typically -1.5 to -2.0).
	
	(RS Oph) Two additional eruptions were identified in 1907 (by myself) and in 1945 (Oppenheimer \& Mattei 1993) on the basis of seeing the characteristic post-eruption dip as the star came out from behind the Sun.  These post-eruption dips last for 100-500 days after the peak, with the exact levels likely determined by the brightness of the red giant companion at the time.  During quiescence, the system brightness behaves chaotically, with fast and slow flares, with secular rises and declines, and with decades of relative constancy.  The power density spectrum shows a $1/f$ noise power law rise (with index -1.8) down to extremely low frequencies corresponding to many years.  This PDS shape is closely similar to those of many other red giants, which suggests that the long-term variations are arising in part from the red giant companion.  But a scattered correlation between V-band magnitude and the Balmer line flux shows that at least part of the variations are associated with the accretion.  So apparently, the long-term variability and the power law PDS are caused by both mechanisms, convection cells on the red giant and accretion instabilities, with comparable contributions.  The suggested periodicity of 2016 days (Oppenheimer \& Mattei 1993) is certainly wrong.  Prior claims that the red giant gets as `hot' as G5 (Adams, Humason, \& Joy 1927) are mistakenly based on a retracted measurement (Humason 1938), so the entire observed range is from K5 to M4 (or maybe just K5 to M0).  Over this range, the spectral type varies linearly with the V-band brightness from 10.7 to 12.2 mag.
	
	(V745 Sco)  V745 Sco has sinusoidal oscillations with an amplitude of 0.4 mag and period of 255 days.  These variations have alternating high and low maxima plus deep and shallow minima, so the real orbital period is $510\pm 20$ days.
	
	(V3890 Sgr)  V3890 Sgr has been claimed to have pre-eruption dwarf nova outbursts, but this claim has been proven to be wrong.  (In addition, the putative pre-eruption dwarf nova events for V446 Her and PU Vul are shown to not be dwarf nova events.)  I have discovered two photometric periodicities; 103.8 days likely from pulsations in the red giant star, and $519.7\pm0.3$ days which equals the orbital period.  The system was apparently fainter on average (by roughly one magnitude) in the first half of the last century as compared to the last half-century.
	
	(1)  I propose that the RNe can be usefully divided into those with short orbital periods (T Pyx and IM Nor) and long orbital periods (the other RNe).  This division separates the RNe by the cause of the high mass accretion rate (a continuing supersoft source heating the companion versus evolutionary expansion of the companion star) and by whether the system will become a Type Ia supernova (no and yes, respectively).
	
	(2) The RNe have distinctly different average properties from the CNe, even though their distributions have substantial overlap.  The RNe have a median $t_3$ of 11 days (with a full range of 2.6-80 days, including the all-time fastest nova U Sco), while the CNe have a median $t_3$ of 44 days (with a full range of 3.2-900 days).  The RNe have a median amplitude of 8.5 mag (with a full range of 6.2-11.2), while the CNe have a median amplitude of around 12.5 mag (with a full range from around 7-16 mag).  The RNe have a median orbital period of 1.4 days (with a full range of 0.076-519.7 days), while the CNe have a median period of 0.17 days  (with a full range of 0.059-5.8 days).  The RNe have a median absolute magnitude in quiescence of 0.8 mag (with a full range of 3.2 to -4.1 mag), while the CNe have a median $M_V$ in quiescence of 4.6 mag (with a full range of around 7.0 to 1.1 mag).  However, both RNe and CNe have an essentially identical absolute magnitude at peak ($-8.0$ mag) and color at peak ($\sim 0.0$ mag).  Thus, RNe are generally faster, brighter in quiescence, and with longer orbital periods when compared to the CNe.
	
	(3) All the eruption light curves from a single RN are consistent with a single invariant template.  That is, the light curves are always the same from eruption-to-eruption.  This tells us that the eruption light curve depends on system parameters (like the white dwarf mass and the composition) that do not vary from eruption-to-eruption (like $\dot{M}$).
	
	(4) The RNe light curves are similar in shape to those of many classical novae, except that RNe lack the diversity of CN light curves (e.g., with no DQ-Her-like dust dips and no long and chaotic symbiotic nova episodes).  The eruption light curve shapes scaled by $t_3$ for the ten RNe are approximately the same.  That is, other than an overall time scale, the declines from peak are similar for all the systems.  There are definite differences between RNe (such as the oscillations for T Pyx, the sharp drop for U Sco, and the secondary maximum of T CrB), but largely the light curves lie on top of each other in a log-log plot with the magnitude scaled to some common epoch.  In particular, the early declines have power law indices of around -1.75 while the late declines have indices of around -3.0, in reasonable agreement with the theoretical prediction of a 'universal decline law' by Hachisu \& Kato.  The break times for all the RNe (except for IM Nor and RS Oph) is very fast (as short as 6 days) which points to the RNe having white dwarf masses close to the Chandrasekkhar mass.
	
	(5) The presence of plateaus in RN light curves is mixed, with six showing definite plateaus (T Pyx, IM Nor, CI Aql, V2487 Oph, U Sco, and RS Oph), one definitely showing no plateau (T CrB), and the remaining three having possible-but-inconclusive plateaus.
	
	(6) The discovery efficiency for RN events is horrifyingly low.  Undirected searches have an efficiency of from roughly 0.6\% to 19\% with a median of 4\%.  This implies that most galactic RNe with peaks brighter than tenth magnitude never even have one event discovered, that a small fraction have only one event discovered (and the system is cataloged as a classical nova), and a much smaller fraction have two-or-more eruptions discovered (and labeled as RNe).  The ratio of RNe-mislabeled-as-CNe to RNe is roughly a factor of 6-10, which implies that 60-100 `CNe' already in the nova lists have really had multiple eruptions within the last century.  
	
	(7) Directed searches have average efficiencies (from 1890 to 2008) ranging from 30\% to 100\% with a median of 60\%.  This implies that the majority of RNe likely have additional nova eruptions missed over the last century and that their real recurrence time scale is substantially shorter than the naive calculation based only on the discovered eruption dates.
	
	(8) Five RNe (V394 CrA, V2487 Oph, U Sco, V3890 Sgr, and V745 Sco) have predicted dates of the next eruptions within the next decade.  Other than for the case of U Sco (which should reliably go off in the next year or so), the uncertainty on these predictions is large, so the other four could go off any night now or not for two more decades.
	
	(9) Five RNe (CI Aql, U Sco, V394 CrA, V745 Sco, and V3890 Sgr) have an identical pattern in their folded light curve that the maximum following the primary (deeper) minimum is significantly higher than the maximum following the secondary minimum.
	
	(10) All RNe display the usual flickering on time scales from minutes-to-hours and with amplitudes up to half a magnitude.

	(11) All RNe display surprisingly large amplitude variations on time scales of months-years-decades-century.  This behavior has many types of morphology.  Seven-out-of-nine RNe (with long-term light curves) have variations of greater than one magnitude on time scales of from 10-100 years.  Four of these RNe (those with red giant companions) have a power density spectrum for these long time scales that might arise from the normal variations on red giants.  T Pyx has a large-amplitude secular decline over the 119 years likely caused by a feedback mechanism where the supersoft luminosity of the white dwarf is fading.  And I have no idea as to why U Sco and V394 CrA have variations on the time scale of decades.
	
	(12) For fourteen eruptions with good pre-eruption light curves, zero showed a pre-eruption rise while one even showed a significant pre-eruption dip.
	
	(13) The RNe do not show any consistent secular trends after the eruptions.  Instead, some rise, some fall, some rise and fall, one stays constant, and several vary chaotically.  This demonstrates that some other mechanism or mechanisms (in addition to the possible hibernation slow turn-off) are operating and are dominating the behavior in quiescence.  With the existence of these apparently chaotic competing effects (which dominate over any trends from hibernation), it will be difficult to test the hibernation model by seeking slow post-eruption declines.
	
	(14) The RNe are apparently divided with about half as a thick disk population (with a scale height of 380 pc) and half as a bulge population (within 3 kpc of the galactic center).  This is the same as for CNe.
	
	(15) The spectral energy distributions of the RNe can be divided into two groups.  Four RNe have a prominent peak in the IR caused by their red giant companion star.  Six of the RNe have the classic $f_{\nu} \propto \nu ^{1/3}$ as expected for accretion disks.
	
	(16) The RNe in quiescence have a wide range of luminosities with $-4.1\leq M_V \leq 3.2$ mag.  The four brightest are those with red giant companions.  Of the remainder, all are dominated by the accretion disk (as demonstrated by the $\nu ^{1/3}$ spectral energy distribution), and these have a median absolute magnitude of 2.2 mag (and a full range of 0.0 to 3.2 mag).  This is greatly brighter than CNe (with median absolute magnitude of 4.6 mag), likely because RNe must have a high accretion rate.

\section{Acknowledgments}

	In any such paper as this, I have had help and used observations beyond those with citations in the text.  The largest source of data is that from the dedicated people with the AAVSO.  The many observers are quality astronomers producing a wonderfully useful data set.  Indeed, the AAVSO data are so voluminous and useful that it now provides about 98\% of all RNe light curve measures.  Without the heartfelt work of the amateurs, the whole topic of recurrent novae would be barren, both for the lack of known eruptions (only 11 known eruptions from 3 RNe would be now known without amateur work) and for the lack of data on those eruptions.  I thank the many hundreds of amateur astronomers over the last century who have sacrificed their sleep with no immediate reward.  The AAVSO organization provided many items of great use for this paper.  In particular, I have made frequent and extensive use of the various finder charts, and the comparison star sequences have taken many magnitudes from AAVSO sequences, most measured by the current AAVSO Director Arne Henden.  The second source of data and discoveries is the Harvard plate collection.  Harvard plates provide the only data on 11 eruptions and the primary data on 4 more.  Without Harvard data, the discovery of the RN phenomenon would have been delayed until 1946 and the fourth member of the class would have been found in 1987.  The historical plate archives are still producing front-line science, yet they are under a variety of threats, and this is not acceptable.  I would like to thank the many observers and curators who contributed to the Harvard collection over the last 120 years.  Without the amateur nova searchers and the Harvard plate collection, we would even today not know about {\it any} RN.  For my own observations, I have used a vast amount of `small' telescope time over the last 21 years, and this requires a wide variety of support.  So I would like to thank the observers, allocation committees, night assistants, and managers of the SMARTS consortium, Cerro Tololo Inter-American Observatory, McDonald Observatory, and the ROTSE collaboration.  In particular, Suzanne Tourtellotte, Charles Bailyn, Michelle Buxton, and Rebeccah Winnick provided valuable long-term close support.  A large article like this one needs to have knowledgeable readers to help improve the text, and so I thank Arlo Landolt, Ron Webbink, Joe Patterson, and the anonymous referee for spending longer-than-normal time at checking the manuscript.  The National Science Foundation and the National Aeronautics and Space Administration provided funds under grants AST-0708079 and 05-GALEX05-23.

\clearpage

\begin{deluxetable}{llllll}
\tabletypesize{\scriptsize}
\tablecaption{RN Summary
\label{tbl1}}
\tablewidth{0pt}
\tablehead{
\colhead{RN}   &
\colhead{$V_{peak}$ (mag)}   &
\colhead{$V_{min}$ (mag)}   &
\colhead{$t_3$ (days)}  &
\colhead{$P_{orb}$ (days)}   &
\colhead{Eruption years}
}
\startdata

T Pyx	&	6.4	&	15.5	&	62	&	0.076	&	1890, 1902, 1920, 1944, 1967	\\
IM Nor	&	8.5	&	18.3	&	80	&	0.102	&	1920, 2002	\\
CI Aql	&	9.0	&	16.7	&	32	&	0.62	&	1917, 1941, 2000	\\
V2487 Oph	&	9.5	&	17.3	&	8	&	$\sim1$	&	1900, 1998	\\
U Sco	&	7.5	&	17.6	&	2.6	&	1.23	&	1863, 1906, 1917, 1936, 1945, 1969, 1979, 1987, 1999	\\
V394 CrA	&	7.2	&	18.4	&	5.2	&	1.52	&	1949, 1987	\\
T CrB	&	2.5	&	9.8	&	6	&	228	&	1866, 1946	\\
RS Oph	&	4.8	&	11	&	14	&	457	&	1898, 1907, 1933, 1945, 1958, 1967, 1985, 2006	\\
V745 Sco	&	9.4	&	18.6	&	9	&	510	&	1937, 1989	\\
V3890 Sgr	&	8.1	&	15.5	&	14	&	519.7	&	1962, 1990	\\

\enddata
    
\end{deluxetable}

\clearpage

\begin{deluxetable}{llllllllll}
\tabletypesize{\scriptsize}
\tablecaption{RN Eruptions
\label{tbl2}}
\tablewidth{0pt}
\tablehead{
\colhead{RN}   &
\colhead{Year}  &
\colhead{JD$_{max}$}   &
\colhead{$\Delta$T}   &
\colhead{Source}   &
\colhead{Discoverer}   &
\colhead{Ref.\tablenotemark{i}}   &
\colhead{JD$_{disc}$}   &
\colhead{JD$_{disc}$-JD$_{max}$}   &
\colhead{Harvard?}
}
\startdata

T Pyx	&	1890	&	2411547	&	\ldots	&	Harvard plates	&	H. Leavitt	&	1	&	2419900	&	8353	&	Yes (only)	\\
T Pyx	&	1902	&	2415880	&	11.9	&	Harvard plates	&	H. Leavitt	&	1	&	2419900	&	4020	&	Yes (only)	\\
T Pyx	&	1920	&	2422425	&	17.9	&	Harvard plates	&	I. Woods	&	2	&	2422422	&	-3	&	Yes	\\
T Pyx	&	1944	&	2431418	&	24.6	&	Mount Wilson	&	A. Joy	&	3	&	2431549	&	131	&	Yes	\\
T Pyx	&	1967	&	2439506	&	22.1	&	Amateur	&	A. Jones	&	4	&	2439467	&	-39	&	No	\\
IM Nor	&	1920	&	2422529	&	\ldots	&	Harvard Plates	&	I. Woods	&	5	&	2422658	&	129	&	Yes (only)	\\
IM Nor	&	2002	&	2452289	&	81.5	&	PROBLICOM	&	W. Liller	&	6	&	2452284	&	-5	&	No	\\
CI Aql	&	1917	&	2421376	&	\ldots	&	Heidelberg plates	&	K. Reinmuth	&	7	&	2424356	&	2980	&	Yes	\\
\enddata

\tablenotetext{a}{Independently discovered by M. Yamamoto.}
\tablenotetext{b}{Independently discovered by J. Birmingham and J. Schmidt.}
\tablenotetext{c}{Independently discovered by N. Knight and A. Deutsch.}
\tablenotetext{d}{Independently discovered by L. Peltier.}
\tablenotetext{e}{Independently discovered by C. Fernald.}
\tablenotetext{f}{Independently discovered by W. Lowder and G. E. D. Alcock.}
\tablenotetext{g}{Independently discovered by several people.}
\tablenotetext{h}{Independently discovered by W. Liller.}
\tablenotetext{i}{References: 1.	Pickering (1913).
2.	Bailey (1920a).
3.	Joy (1945).
4.	Jones (1966).
5.	Bailey (1920b).
6.	Liller (2002).
7.	Reinmuth (1925).
8.	Schaefer (2001a).
9.	Takamizawa (2000).
10.	Pagnotta, Schaefer, \& Xiao (2008); Pagnotta et al. (2009).
11.	Takamizawa (1998).
12.	Pogson (1865).
13.	Thomas (1940).
14.	Schaefer (2004a).
15.	Schaefer (2001b).
16.	This paper.
17.	Narumi (1979).
18.	Overbeek (1987).
19.	Schmeer (1999).
20.	Erro (1949a).
21.	Liller (1987).
22.	Walter (1867).
23.	Kamenchuk (1946).
24.	Pickering (1905).
25.	Schaefer (2004b).
26.	Loretta (1933).
27.	Oppenheimer \& Mattei (1993).
28.	Fernald (1958).
29.	Beyer (1967).
30.	Morrison (1985).
31.	Narumi (2006).
32.	Plaut (1958).
33.	Liller (1989).
34.	Dinerstein (1973).
35.	Jones (1990).}

\end{deluxetable}

\clearpage

\begin{deluxetable}{lllllllll}
\tabletypesize{\scriptsize}
\tablecaption{Comparison Star Magnitudes 
\label{tbl3}}
\tablewidth{0pt}
\tablehead{
\colhead{RN}  &
\colhead{Star}  &
\colhead{B (mag)} &
\colhead{V (mag)}  &
\colhead{R (mag)} &
\colhead{I (mag)}  &
\colhead{J (mag)}  &
\colhead{H (mag)}  &
\colhead{K (mag)} 
}
\startdata

T Pyx	&	Comp	&	14.91	&	14.32	&	13.94	&	13.57	&	13.03	&	12.86	&	12.81	\\
T Pyx	&	Check	&	16.56	&	15.81	&	15.33	&	14.95	&	14.31	&	13.94	&	13.89	\\
T Pyx	&	Check2	&	14.68	&	13.72	&	13.20	&	12.73	&	12.09	&	11.67	&	11.58	\\
IM Nor	&	Comp	&	16.61	&	15.70	&	15.13	&	14.60	&	13.83	&	13.43	&	13.26	\\
IM Nor	&	Check	&	20.15	&	18.30	&	17.19	&	16.18	&	14.80	&	14.18	&	13.97	\\
IM Nor	&	Check2	&	20.34	&	18.90	&	18.06	&	17.24	&	15.83	&	15.40	&	15.39	\\
CI Aql	&	Comp	&	16.70	&	15.32	&	14.43	&	13.54	&	12.09	&	11.63	&	11.33	\\
CI Aql	&	Check	&	16.70	&	15.47	&	14.69	&	13.94	&	12.80	&	12.39	&	12.19	\\
\enddata

\end{deluxetable}

\clearpage

\begin{deluxetable}{lllllllll}
\tabletypesize{\scriptsize}
\tablecaption{Magnitude Sequences for all Ten RNe
\label{tbl4}}
\tablewidth{0pt}
\tablehead{
\colhead{Field}   &
\colhead{Star}   &
\colhead{J2000 position}   &
\colhead{Catalog ID}   &
\colhead{AAVSO ID}  &
\colhead{B (mag)}   &
\colhead{V (mag)}   &
\colhead{Source}  &
\colhead{Other Designations}
}
\startdata

T Pyx	&	Comp	&	09:04:40.1  -32:22:03	&	\ldots	&	143	&	14.91	&	14.32	&	A. Landolt	&	\ldots	\\
T Pyx	&	Check	&	09:04:41.5  -32:23:00	&	\ldots	&	158	&	16.56	&	15.81	&	A. Landolt	&	\ldots	\\
T Pyx	&	Check2	&	09:04:40.6  -32:20:50	&	\ldots	&	137	&	14.68	&	13.72	&	A. Landolt	&	\ldots	\\
T Pyx	&	A	&	09:04:40.4  -32:21:17	&	\ldots	&	\ldots	&	16.37	&	15.60	&	This paper	&	\ldots	\\
T Pyx	&	B	&	09:04:45.1  -32:22:16	&	\ldots	&	154	&	16.21	&	15.43	&	AAVSO	&	\ldots	\\
T Pyx	&	C	&	09:04:49.1  -32:23:53	&	\ldots	&	152	&	16.06	&	15.17	&	AAVSO	&	\ldots	\\
T Pyx	&	D	&	09:04:43.0  -32:20:28	&	\ldots	&	151	&	15.72	&	15.06	&	AAVSO	&	\ldots	\\
T Pyx	&	E	&	09:04:45.3  -32:21:06	&	\ldots	&	146	&	15.21	&	14.57	&	AAVSO	&	\ldots	\\
\enddata  
\end{deluxetable}

\clearpage
\begin{deluxetable}{lllll}
\tabletypesize{\scriptsize}
\tablecaption{AAVSO Magnitudes for 18 Eruptions
\label{tbl5}}
\tablewidth{0pt}
\tablehead{
\colhead{RN}   &
\colhead{JD}   &
\colhead{Band}   &
\colhead{Magnitude}  &
\colhead{Source}
}
\startdata

T Pyx	&	2422529.9	&		V			&		11.60	$\pm$	0.15	&	AAVSO	\\
T Pyx	&	2439467.14	&		V			&		12.9	$\pm$	0.15	&	AAVSO	\\
T Pyx	&	2439467.15	&		V			&		12.7	$\pm$	0.15	&	AAVSO	\\
T Pyx	&	2439468.94	&		V			&		9.2	$\pm$	0.15	&	AAVSO	\\
T Pyx	&	2439468.95	&		V			&		9.3	$\pm$	0.15	&	AAVSO	\\
T Pyx	&	2439468.96	&		V			&		9.2	$\pm$	0.15	&	AAVSO	\\
T Pyx	&	2439468.97	&		V			&		9.2	$\pm$	0.15	&	AAVSO	\\
T Pyx	&	2439468.98	&		V			&		9.2	$\pm$	0.15	&	AAVSO	\\
T Pyx	&	2439468.99	&		V			&		9.2	$\pm$	0.15	&	AAVSO	\\
T Pyx	&	2439469.01	&		V			&		9.2	$\pm$	0.15	&	AAVSO	\\

\enddata
\end{deluxetable}

\clearpage
\begin{deluxetable}{llll}
\tabletypesize{\scriptsize}
\tablecaption{Eruption Light Curve for T Pyx
\label{tbl6}}
\tablewidth{0pt}
\tablehead{
\colhead{Julian Date}   &
\colhead{Band\tablenotemark{a}}   &
\colhead{Magnitude}   &
\colhead{Source}
}
\startdata

\underline{1890 eruption}	&					&					&		\\
2411493.6	&		B			&	$>$	13.5			&	HCO 	\\
2411516.6	&		B			&		8.1	$\pm$	0.15	&	B5289	\\
\underline{1902 eruption}	&					&					&		\\
2415812.644	&		B			&	$>$	10.7			&	AC2381	\\
2415821.589	&		B			&	$>$	9.8			&	AM1160	\\
2415828.6	&		B			&	$>$	10.0			&	AM1162	\\
2415842.577	&		B			&		9.03	$\pm$	0.15	&	AM1172	\\
\enddata    
\tablenotetext{a}{The filter or band used for the reported magnitude is given.  If a number is given, then that number is a B-band magnitude, while the value in the magnitude column is the V-band magnitude.  This is just a device to report nearly simultaneous B and V magnitudes in a compact manner.}

\end{deluxetable}

\clearpage
\begin{deluxetable}{llll}
\tabletypesize{\scriptsize}
\tablecaption{Eruption Light Curve for IM Nor
\label{tbl7}}
\tablewidth{0pt}
\tablehead{
\colhead{Julian Date}   &
\colhead{Band}   &
\colhead{Magnitude}   &
\colhead{Source}
}
\startdata

\underline{1920 eruption}	&					&					&		\\
2422455.704	&		B			&	$>$	12.71			&	AM15167	\\
2422456.000	&		B			&	$>$	16.61			&	AM15171	\\
2422458.714	&		B			&	$>$	12.71			&	AM15182	\\
2422488.633	&		B			&	$>$	11.45			&	AM15253	\\
2422492.637	&		B			&	$>$	11.45			&	AM15257	\\
2422514.666	&		B			&		9.97	$\pm$	0.15	&	AM15290	\\
2422516.000	&		B			&		10.03	$\pm$	0.15	&	AM15293	\\
\enddata
\end{deluxetable}

\clearpage
\begin{deluxetable}{llll}
\tabletypesize{\scriptsize}
\tablecaption{Eruption Light Curve for CI Aql
\label{tbl8}}
\tablewidth{0pt}
\tablehead{
\colhead{Julian Date}   &
\colhead{Band}   &
\colhead{Magnitude}   &
\colhead{Source}
}
\startdata

\underline{1917 eruption}	&					&					&		\\
2421312.887	&		B			&	$>$	11.31			&	AC18711	\\
2421322.880	&		B			&	$>$	11.31			&	AC18753	\\
2421343.825	&		B			&	$>$	11.80			&	AC18802	\\
2421344.840	&		B			&	$>$	11.31			&	AC18808	\\
2421364.749	&		B			&	$>$	9.72			&	AC18837	\\
2421366.805	&		B			&		10.91	$\pm$	0.15	&	AI16006	\\
2421368.768	&		B			&		8.74	$\pm$	0.15	&	AM13168	\\
\enddata    
\end{deluxetable}

\clearpage
\begin{deluxetable}{llll}
\tabletypesize{\scriptsize}
\tablecaption{Eruption Light Curve for V2487 Oph
\label{tbl9}}
\tablewidth{0pt}
\tablehead{
\colhead{Julian Date}   &
\colhead{Band}   &
\colhead{Magnitude}   &
\colhead{Source}
}
\startdata

\underline{1900 eruption}	&					&					&		\\
2415191	&		B			&		10.3	$\pm$	0.15	&	AM505	\\
\underline{1998 eruption}	&					&					&		\\
2450953	&		V			&	$>$	15.0			&	K. Takamizawa IAUC 6941	\\
2450971.7	&		V			&	$>$	11.5			&	Liller \& Jones IBVS 4771	\\
2450980.061	&		V			&		9.5	$\pm$	0.50	&	K. Takamizawa IAUC 6941	\\
2450980.6	&		V			&		9.5	$\pm$	0.15	&	Liller \& Jones IBVS 4771	\\
2450981.031	&		V			&		10.0	$\pm$	0.50	&	R. \& Y. Kushida IAUC 6941	\\
\enddata    
\end{deluxetable}

\clearpage
\begin{deluxetable}{llll}
\tabletypesize{\scriptsize}
\tablecaption{Eruption Light Curve for U Sco
\label{tbl10}}
\tablewidth{0pt}
\tablehead{
\colhead{Julian Date}   &
\colhead{Band\tablenotemark{a}}   &
\colhead{Magnitude}   &
\colhead{Source}
}
\startdata

\underline{1863 eruption}	&					&					&		\\
2401646	&		V			&		9.1	$\pm$	0.2	&	Thomas 1940	\\
2401647	&		V			&		9.6	$\pm$	0.2	&	Thomas 1940	\\
2401648	&		V			&		10.1	$\pm$	0.2	&	Thomas 1940	\\
2401650	&		V			&		10.5	$\pm$	0.2	&	Thomas 1940	\\
2401651	&		V			&		11.0	$\pm$	0.2	&	Thomas 1940	\\
2401652	&		V			&		11.5	$\pm$	0.2	&	Thomas 1940	\\
2401653	&		V			&		11.9	$\pm$	0.2	&	Thomas 1940	\\
\enddata    
\tablenotetext{a}{The filter or band used for the reported magnitude is given.  If a number is given, then that number is a B-band magnitude, while the value in the magnitude column is the V-band magnitude.  This is just a device to report nearly simultaneous B and V magnitudes in a compact manner.}

\end{deluxetable}

\clearpage
\begin{deluxetable}{llll}
\tabletypesize{\scriptsize}
\tablecaption{Eruption Light Curve for V394 CrA
\label{tbl11}}
\tablewidth{0pt}
\tablehead{
\colhead{Julian Date}   &
\colhead{Band\tablenotemark{a}}   &
\colhead{Magnitude}   &
\colhead{Source}
}
\startdata

\underline{1949 eruption}	&					&					&		\\
2432998	&		B			&	$>$	12.50			&	Erro 1949a	\\
2432999	&		B			&		7.50	$\pm$	0.15	&	Erro 1949b	\\
2433000	&		B			&		8.22	$\pm$	0.15	&	AM27357	\\
2433001	&		B			&		9.39	$\pm$	0.15	&	B74552	\\
2433005	&		B			&		9.50	$\pm$	0.15	&	Erro 1949b	\\
2433013	&		B			&		12.00	$\pm$	0.15	&	RB16210	\\
2433014	&		B			&		11.50	$\pm$	0.15	&	Johnson IAUC 1211	\\
\enddata
\tablenotetext{a}{The filter or band used for the reported magnitude is given.  If a number is given, then that number is a B-band magnitude, while the value in the magnitude column is the V-band magnitude.  This is just a device to report nearly simultaneous B and V magnitudes in a compact manner.}

\end{deluxetable}

\clearpage
\begin{deluxetable}{llll}
\tabletypesize{\scriptsize}
\tablecaption{Eruption Light Curve for T CrB
\label{tbl12}}
\tablewidth{0pt}
\tablehead{
\colhead{Julian Date}   &
\colhead{Band}   &
\colhead{Magnitude}   &
\colhead{Source}
}
\startdata

\underline{1866 eruption}	&					&					&		\\
2402734.3	&		V			&	$>$	6.0			&	Schmidt 1866	\\
2402734.5	&		V			&		2.5	$\pm$	0.5	&	Lynn 1866	\\
2402735.3	&		V			&		2.5	$\pm$	0.3	&	Lynn 1866	\\
2402736.6	&		V			&		3.4	$\pm$	0.3	&	Lynn 1866	\\
2402736.6	&		V			&		4.0	$\pm$	0.3	&	Davis 1866	\\
2402737.4	&		V			&		3.7	$\pm$	0.3	&	Baxendall 1866	\\
2402738.4	&		V			&		4.2	$\pm$	0.3	&	Baxendall 1866	\\
\enddata
    
\end{deluxetable}

\clearpage

\begin{deluxetable}{llll}
\tabletypesize{\scriptsize}
\tablecaption{Eruption Light Curve for RS Oph
\label{tbl13}}
\tablewidth{0pt}
\tablehead{
\colhead{Julian Date}   &
\colhead{Band\tablenotemark{a}}   &
\colhead{Magnitude}   &
\colhead{Source}
}
\startdata

\underline{1898 eruption}	&					&					&		\\
2414437.6	&		B			&	$>$	11.5			&	B21513	\\
2414441.6	&		B			&		12.92	$\pm$	0.15	&	B21549	\\
2414457.732	&		B			&		4.86	$\pm$	0.15	&	AC8	\\
2414467.757	&		B			&		7.97	$\pm$	0.15	&	AC11	\\
2414471.6	&		B			&		8.66	$\pm$	0.15	&	I20924	\\
2414485.6	&		B			&		9.5	$\pm$	0.15	&	Shapley 1933	\\
2414486.6	&		B			&		9.6	$\pm$	0.15	&	Shapley 1933	\\
\enddata
\tablenotetext{a}{The filter or band used for the reported magnitude is given.  If a number is given, then that number is a B-band magnitude, while the value in the magnitude column is the V-band magnitude.  This is just a device to report nearly simultaneous B and V magnitudes in a compact manner.}

\end{deluxetable}

\clearpage
\begin{deluxetable}{llll}
\tabletypesize{\scriptsize}
\tablecaption{Eruption Light Curve for V745 Sco
\label{tbl14}}
\tablewidth{0pt}
\tablehead{
\colhead{Julian Date}   &
\colhead{Band}   &
\colhead{Magnitude}   &
\colhead{Source}
}
\startdata

\underline{1937 eruption}	&					&					&		\\
2428645	&		B			&	$>$	15.1			&	B61889	\\
2428662	&		B			&		15.89	$\pm$	0.15	&	Plaut 1958	\\
2428664	&		B			&		10.41	$\pm$	0.15	&	B61925	\\
2428664	&		B			&		10.75	$\pm$	0.15	&	Plaut 1958	\\
2428668	&		B			&		12.28	$\pm$	0.15	&	RB7347	\\
2428668	&		B			&		12.40	$\pm$	0.15	&	B61938	\\
2428670	&		B			&		13.11	$\pm$	0.15	&	B61952	\\
\enddata
    
\end{deluxetable}

\clearpage
\begin{deluxetable}{llll}
\tabletypesize{\scriptsize}
\tablecaption{Eruption Light Curve for V3890 Sgr
\label{tbl15}}
\tablewidth{0pt}
\tablehead{
\colhead{Julian Date}   &
\colhead{Band}   &
\colhead{Magnitude}   &
\colhead{Source}
}
\startdata

\underline{1962 eruption}	&					&					&		\\
2437795	&		B			&	$>$	15.2			&	Dinerstein 1973	\\
2437818.52	&		B			&		10.74	$\pm$	0.15	&	Wenzel 1990	\\
2437818.54	&		B			&		10.84	$\pm$	0.15	&	Wenzel 1990	\\
2437818.8	&		B			&		10.60	$\pm$	0.15	&	Miller 1991	\\
2437819.8	&		B			&		10.78	$\pm$	0.15	&	Miller 1991	\\
2437820.52	&		B			&		11.14	$\pm$	0.15	&	Wenzel 1990	\\
2437821.49	&		B			&		11.04	$\pm$	0.15	&	Wenzel 1990	\\
\enddata
\end{deluxetable}

\clearpage
\begin{deluxetable}{lll}
\tabletypesize{\scriptsize}
\tablecaption{RN Eruption Light Curve Templates
\label{tbl16}}
\tablewidth{0pt}
\tablehead{
\colhead{$T-T_{peak}$ (days)}   &
\colhead{$B$ (mag)}   &
\colhead{$V$ (mag)}   
}
\startdata

\underline{T Pyx}	&		&		\\
-999	&	15.6	&	15.5	\\
-40	&	15.6	&	15.5	\\
-38	&	10.1	&	9.9	\\
-35	&	8.4	&	8.2	\\
-30	&	8.1	&	7.8	\\
-10	&	7.4	&	7.1	\\
0	&	6.7	&	6.4	\\
\enddata
\end{deluxetable}

\clearpage
\begin{deluxetable}{lllllllllll}
\tabletypesize{\scriptsize}
\tablecaption{RN Light Curve Properties
\label{tbl17}}
\tablewidth{0pt}
\tablehead{
\colhead{Property}   &
\colhead{T Pyx}   &
\colhead{IM Nor}   &
\colhead{CI Aql}   &
\colhead{V2487 Oph}   &
\colhead{U Sco}   &
\colhead{V394 CrA}   &
\colhead{T CrB}   &
\colhead{RS Oph}   &
\colhead{V745 Sco}   &
\colhead{V3890 Sgr}   
}
\startdata

$B_{peak}$	&	6.7	&	9.2	&	9.4	&	10.1	&	7.8	&	7.0	&	2.5	&	5.3	&	10.5	&	8.9	\\
$V_{peak}$	&	6.4	&	8.5	&	9.0	&	9.5	&	7.5	&	7.2	&	2.5	&	4.8	&	9.4	&	8.1	\\
$V_{15days}$	&	7.7	&	9.2	&	10.6	&	13.7	&	14.0	&	12.7	&	8.4	&	7.9	&	13.2	&	11.2	\\
$B-V_{peak}$	&	0.3	&	0.7	&	0.4	&	0.6	&	0.3	&	-0.2	&	0.0	&	0.5	&	1.1	&	0.8	\\
$B-V_{Vpeak+2}$	&	-0.2	&	0.7	&	0.8	&	0.6	&	0.1	&	-0.2	&	0.4	&	0.9	&	1.7	&	0.8	\\
$Amp_B$	&	8.9	&	10.2	&	8.0	&	8.2	&	10.3	&	12.0	&	7.9	&	6.9	&	9.5	&	7.4	\\
$Amp_V$	&	9.1	&	9.8	&	7.7	&	8.2	&	10.1	&	11.2	&	7.3	&	6.2	&	9.2	&	7.4	\\
$t_{-3}$	&	37	&	-17.6	&	10.0	&	0.4	&	0.4	&	0.4	&	0.4	&	0.6	&	1.0	&	1.0	\\
\enddata
 \tablenotetext{a}{The integral under the light curve is made for only the primary nova peak, and the subtracted quiescent flux is that prior to the pre-eruption dip.}
 \tablenotetext{b}{The integral under the light curve includes times only up until the return to the pre-eruption quiescent level, and the subtracted quiescent flux is for the pre-eruption level.}
\end{deluxetable}

\clearpage
\begin{deluxetable}{lllll}
\tabletypesize{\scriptsize}
\tablecaption{Undirected Search Discovery Efficiency
\label{tbl18}}
\tablewidth{0pt}
\tablehead{
\colhead{RN}   &
\colhead{$V_{peak}$ (mag)}   &
\colhead{$f_{disc}$}   &
\colhead{$t_3$ (days)}  &
\colhead{$F_{disc}$}
}
\startdata

T Pyx	&	6.4	&	0.20	&	62	&	0.19	\\
IM Nor	&	8.5	&	0.12	&	80	&	0.15	\\
CI Aql	&	9.0	&	0.10	&	32	&	0.05	\\
V2487 Oph	&	9.5	&	0.10	&	8	&	0.012	\\
U Sco	&	7.5	&	0.16	&	2.6	&	0.006	\\
V394 CrA	&	7.2	&	0.17	&	5.2	&	0.013	\\
T CrB	&	2.5	&	1.00	&	6	&	0.09	\\
RS Oph	&	4.8	&	0.30	&	14	&	0.06	\\
V745 Sco	&	9.4	&	0.10	&	9	&	0.014	\\
V3890 Sgr	&	8.1	&	0.14	&	14	&	0.03	\\

\enddata
\end{deluxetable}

\clearpage
\begin{deluxetable}{lllllllll}
\tabletypesize{\scriptsize}
\tablecaption{Directed Search Discovery Efficiency
\label{tbl19}}
\tablewidth{0pt}
\tablehead{
\colhead{RN}   &
\colhead{Source}   &
\colhead{Years}   &
\colhead{$m_{lim}$ (mag)}  &
\colhead{$N_{obs}$}   &
\colhead{$S_g$ (days)}   &
\colhead{$S_o$ (days)}   &
\colhead{$T_{vis}$ (days)}   &
\colhead{$F_{disc}$}
}
\startdata

T Pyx	&	Harvard plates	&	1890-1954	&	 11-16	&	1000	&	100	&	7X25	&	156	&	1.00	\\
T Pyx	&	AAVSO	&	1929-1942	&	12.5-13.3	&	300	&	165	&	7X20	&	218	&	1.00	\\
T Pyx	&	AAVSO	&	1955-1987	&	13.3	&	2300	&	107	&	6X25	&	270	&	1.00	\\
T Pyx	&	Harvard plates	&	1970-1972, 1978-1989	&	13.5-14.5	&	50	&	150	&	7X20	&	268	&	1.00	\\
T Pyx	&	AAVSO	&	1988-present	&	13.3-15.0	&	15200	&	0	&	0	&	270	&	1.00	\\
IM Nor	&	Harvard plates	&	1894-1954	&	 12-15	&	1000	&	100	&	7X25	&	92	&	0.98	\\
IM Nor	&	Harvard plates	&	1970-1971, 1978-1989	&	14	&	100	&	150	&	7X20	&	203	&	1.00	\\
CI Aql	&	Harvard plates	&	1895-1951	&	14-16	&	1000	&	150	&	5X20	&	91	&	0.84	\\
\enddata
\end{deluxetable}

\clearpage
\begin{deluxetable}{lllllllllll}
\tabletypesize{\scriptsize}
\tablecaption{Yearly Discovery Efficiency
\label{tbl20}}
\tablewidth{0pt}
\tablehead{
\colhead{Year}   &
\colhead{T Pyx}   &
\colhead{IM Nor}   &
\colhead{CI Aql}   &
\colhead{V2487 Oph}   &
\colhead{U Sco}   &
\colhead{V394 CrA}   &
\colhead{T CrB}   &
\colhead{RS Oph}   &
\colhead{V745 Sco}   &
\colhead{V3890 Sgr}
}
\startdata

1890	&	1	&	0.15	&	0.05	&	0.012	&	0.43	&	0.013	&	0.09	&	1	&	0.014	&	0.03	\\
1891	&	1	&	0.15	&	0.05	&	0.33	&	0.43	&	0.013	&	1	&	1	&	0.014	&	0.03	\\
1892	&	1	&	0.15	&	0.05	&	0.33	&	0.43	&	0.013	&	1	&	1	&	0.014	&	0.03	\\
1893	&	1	&	0.15	&	0.05	&	0.33	&	0.43	&	0.013	&	1	&	1	&	0.014	&	0.03	\\
1894	&	1	&	0.98	&	0.05	&	0.33	&	0.43	&	0.013	&	1	&	1	&	0.014	&	0.03	\\
1895	&	1	&	0.98	&	0.84	&	0.33	&	0.43	&	0.013	&	1	&	1	&	0.014	&	0.03	\\
1896	&	1	&	0.98	&	0.84	&	0.33	&	0.43	&	0.013	&	1	&	1	&	0.014	&	0.03	\\
1897	&	1	&	0.98	&	0.84	&	0.33	&	0.43	&	0.013	&	1	&	1	&	0.014	&	0.03	\\
\enddata
\end{deluxetable}

\clearpage
\begin{deluxetable}{llll}
\tabletypesize{\scriptsize}
\tablecaption{Recurrence Time Scales
\label{tbl21}}
\tablewidth{0pt}
\tablehead{
\colhead{RN}   &
\colhead{$\Delta T$ (years)}   &
\colhead{$\tau_{rec}$ (years)}   &
\colhead{Next Eruption}
}
\startdata

T Pyx	&	11.9, 17.9, 24.6, 22.1	&	19	&	$10^3$ to $10^6$ years from now	\\
IM Nor	&	81.5 (or maybe 41)	&	82 (or maybe 41)	&	2084 (2043?)	\\
CI Aql	&	23.9, 59.1/N (N=2 or 3)	&	24	&	2024	\\
V2487 Oph	&	$\sim$118/6.7	&	18	&	2016	\\
U Sco	&	43.0/4, 10.8, 19.3/2, 8.9, 23.7/2, 10.4, 7.9, 11.8	&	10.3	&	2009.3$\pm$1.0	\\
V394 CrA	&	38.3/2	&	30	&	2017	\\
T CrB	&	79.7	&	80	&	2026	\\
RS Oph	&	8.6, 26.6, 12.3, 12.6, 9.3, 17.3, 21.0	&	14.7	&	2021	\\
V745 Sco	&	52.2/2	&	21	&	2010	\\
V3890 Sgr	&	27.9	&	25	&	2015	\\

\enddata 
\end{deluxetable}

\clearpage
\begin{deluxetable}{lllll}
\tabletypesize{\scriptsize}
\tablecaption{Quiescent Light Curves for V2487 Oph, V394 CrA, V745 Sco, and V3890 Sgr
\label{tbl22}}
\tablewidth{0pt}
\tablehead{
\colhead{RN}   &
\colhead{HJD}   &
\colhead{Band}   &
\colhead{Magnitude}   &
\colhead{Source}   
}
\startdata

V2487 Oph	&	2452425.8385	&	V	&	17.37	$\pm$	0.05	&	McDonald 0.8-m	\\
V2487 Oph	&	2452425.8394	&	V	&	17.30	$\pm$	0.03	&	McDonald 0.8-m	\\
V2487 Oph	&	2452425.8404	&	V	&	17.28	$\pm$	0.03	&	McDonald 0.8-m	\\
V2487 Oph	&	2452425.8432	&	V	&	17.34	$\pm$	0.03	&	McDonald 0.8-m	\\
V2487 Oph	&	2452425.8476	&	V	&	17.33	$\pm$	0.03	&	McDonald 0.8-m	\\
V2487 Oph	&	2452425.8500	&	B	&	18.00	$\pm$	0.08	&	McDonald 0.8-m	\\
V2487 Oph	&	2452425.8508	&	B	&	18.20	$\pm$	0.06	&	McDonald 0.8-m	\\
V2487 Oph	&	2452425.8518	&	B	&	18.07	$\pm$	0.04	&	McDonald 0.8-m	\\
V2487 Oph	&	2452425.8545	&	B	&	18.13	$\pm$	0.03	&	McDonald 0.8-m	\\
V2487 Oph	&	2452426.7482	&	B	&	18.30	$\pm$	0.02	&	McDonald 0.8-m	\\
V2487 Oph	&	2452426.7563	&	B	&	18.27	$\pm$	0.02	&	McDonald 0.8-m	\\
V2487 Oph	&	2452426.7641	&	B	&	18.28	$\pm$	0.02	&	McDonald 0.8-m	\\
V2487 Oph	&	2452426.7720	&	V	&	17.34	$\pm$	0.02	&	McDonald 0.8-m	\\
V2487 Oph	&	2452426.7794	&	V	&	17.32	$\pm$	0.02	&	McDonald 0.8-m	\\
V2487 Oph	&	2452426.7864	&	V	&	17.31	$\pm$	0.02	&	McDonald 0.8-m	\\

\enddata
\end{deluxetable}

\clearpage
\begin{deluxetable}{llllll}
\tabletypesize{\scriptsize}
\tablecaption{IM Nor eruption tail and minimum times
\label{tb23}}
\tablewidth{0pt}
\tablehead{
\colhead{JD}   &
\colhead{Magnitude}   &
\colhead{Amplitude}   &
\colhead{HJD minimum}  &
\colhead{$N$\tablenotemark{a}}   &
\colhead{$O-C$\tablenotemark{b} (days)}
}
\startdata

2452398.6	&	11.58	&	0.01	&	\ldots	&	\ldots	&	\ldots	\\
2452460.4	&	12.82	&	0.03	&	2452460.368	&	-2301	&	0.001	\\
2452537.3	&	14.31	&	$<$0.05	&	\ldots	&	\ldots	&	\ldots	\\
2452561.25	&	14.9	&	\ldots	&	\ldots	&	\ldots	&	\ldots	\\
2452561.3	&	14.75	&	\ldots	&	\ldots	&	\ldots	&	\ldots	\\
2452569.24	&	15.0	&	\ldots	&	\ldots	&	\ldots	&	\ldots	\\
2452655.5	&	15.6	&	\ldots	&	\ldots	&	\ldots	&	\ldots	\\
2452674.517	&	15.8	&	\ldots	&	\ldots	&	\ldots	&	\ldots	\\
\enddata
\tablenotetext{a}{The ephemeris for the heliocentric Julian Date of the photometric minimum is $T_{model}=2452696.52538+N\times 0.10263312$.}
\tablenotetext{b}{Observed-minus-calculated minima times, that is $T_{observed}-T_{model}$ where $T_{observed}$ is the HJD of the minimum as given in the fourth column.}
\end{deluxetable}

\clearpage
\begin{deluxetable}{llllll}
\tabletypesize{\scriptsize}
\tablecaption{Spectral Type Measures for Red Giant Companions
\label{tbl24}}
\tablewidth{0pt}
\tablehead{
\colhead{RN}   &
\colhead{Julian Date}   &
\colhead{Phase\tablenotemark{a}}   &
\colhead{Spectral Type}   &
\colhead{Source\tablenotemark{b}}   &
\colhead{V (mag)}      
}
\startdata

T CrB	&	2447377	&	0.87	&	M5	&	 1. 	&	\ldots			\\
T CrB	&	2447507	&	0.44	&	M4.5	&	 1. 	&	\ldots			\\
T CrB	&	2447664	&	0.13	&	M4.5	&	 1. 	&	\ldots			\\
T CrB	&	2448797	&	0.11	&	M0-M2 III	&	 2. 	&	\ldots			\\
T CrB	&	2450180	&	0.19	&	M4 III	&	 3. 	&	\ldots			\\
T CrB	&	2450559	&	0.85	&	M3/4 III	&	 3. 	&	\ldots			\\
T CrB	&	2450587	&	0.98	&	M4 III	&	 3. 	&	\ldots			\\
T CrB	&	2450893	&	0.32	&	M3 III	&	 3. 	&	\ldots			\\
\enddata
\tablenotetext{a}{The phases are for ephemerides with the zero phase at the time of conjnction when the red giant star is in front of the white dwarf.  The ephemerides are $2447861.73+N\times227.5687$ for T CrB, $2450040.17+N\times455.72$ for RS Oph, $2453800+N\times510$ for V745 Sco, and $2454730+N\times519.7$ for V3890 Sgr.}
\tablenotetext{b}{  1. 	Murset \& Schmid 1999.
 2. 	Harrison, Johnson, \& Spyromilio 1993.
 3. 	Anupama \& Miko{\l}ajewska 1999.
 4. 	Adams, Humason, \& Joy 1927.
 5. 	Humason 1938.
 6. 	Swings \& Struve 1941.
 7. 	Barbon et al. (1968).
 8. 	Sanduleak \& Stephenson 1973.
 9. 	Walker 1979.
 10. 	Rosino et al. 1982.
 11. 	Sherrington \& Jameson 1983.
 12. 	Dobrzycka et al. 1996.
 13. 	Evans et al. 1988.
 14. 	Bohigas et al. 1989.
 15. 	Duerbeck, Schwarz, \& Augensteijn 1989.
 16. 	Sekiguchi et al. 1990.
 17. 	Williams et al. 1991.}
 
\end{deluxetable}

\clearpage

\begin{deluxetable}{llllllllllll}
\tabletypesize{\scriptsize}
\tablecaption{Recurrent Nova Colors
\label{tbl25}}
\tablewidth{0pt}
\tablehead{
\colhead{RN}   &
\colhead{Julian Date}   &
\colhead{V}   &
\colhead{U-B}   &
\colhead{B-V}   &
\colhead{V-R}   &
\colhead{V-I}   &
\colhead{B-I}   &
\colhead{V-J}   &
\colhead{J-H}   &
\colhead{J-K}   &
\colhead{Source\tablenotemark{d}}     
}
\startdata

T Pyx	&	\ldots	&	14.93	&	-1.04	&	0.12	&	\ldots	&	\ldots	&	\ldots	&	\ldots	&	\ldots	&	\ldots	&	1	\\
T Pyx	&	\ldots	&	15.53	&	-0.95	&	0.08	&	-0.17	&	\ldots	&	\ldots	&	\ldots	&	\ldots	&	\ldots	&	2	\\
T Pyx	&	\ldots	&	\ldots	&	\ldots	&	\ldots	&	\ldots	&	\ldots	&	\ldots	&	\ldots	&	0.42	&	0.56	&	3	\\
T Pyx	&	\ldots	&	\ldots	&	\ldots	&	\ldots	&	\ldots	&	\ldots	&	\ldots	&	\ldots	&	0.19	&	0.47	&	4	\\
T Pyx	&	2447141.06	&	15.34	&	\ldots	&	\ldots	&	\ldots	&	\ldots	&	\ldots	&	1.00	&	0.28	&	0.56	&	5	\\
T Pyx	&	2447141.15	&	\ldots	&	\ldots	&	\ldots	&	\ldots	&	\ldots	&	\ldots	&	\ldots	&	\ldots	&	0.66	&	5	\\
T Pyx	&	2448524.21	&	\ldots	&	\ldots	&	\ldots	&	\ldots	&	\ldots	&	\ldots	&	\ldots	&	-0.02	&	0.43	&	6	\\
T Pyx	&	2453187.48	&	15.53	&	\ldots	&	0.10	&	0.20	&	0.42	&	0.52	&	\ldots	&	\ldots	&	\ldots	&	This paper	\\
\enddata
 
\tablenotetext{a}{At mid-eclipse}
\tablenotetext{b}{ During eclipse}
\tablenotetext{c}{In eruption}
\tablenotetext{d}{References: 1. Mumford(1971)
2. Landolt (1977)
3. Weight et al. (1994)
4. http://irsa.ipac.caltech.edu/Missions/2mass.html
5. Szkody \& Feinswog (1988)
6. Harrison, Johnson, \& Spyromilio (1993)
7. Szkody (1994)
8. Feast \& Glass (1974)
9. Walker (1977)
10. Walker (1957)
11. Szkody (1977)
}
 
\end{deluxetable}

\clearpage
\begin{deluxetable}{lllllllllll}
\tabletypesize{\scriptsize}
\rotate
\tablecaption{Extinction ($E_{B-V}$ in mag)
\label{tbl26}}
\tablewidth{0pt}
\tablehead{
\colhead{}   &
\colhead{T Pyx}   &
\colhead{IM Nor}   &
\colhead{CI Aql}   &
\colhead{V2487 Oph}   &
\colhead{U Sco}   &
\colhead{V394 CrA}   &
\colhead{T CrB}   &
\colhead{RS Oph}   &
\colhead{V745 Sco}   &
\colhead{V3890 Sgr} 
}
\startdata

For $B-V=+0.23$ at peak:	&	0.1	&	0.5	&	0.2	&	0.4	&	0.1	&	-0.4	&	-0.2	&	0.3	&	0.9	&	0.6	\\
For $B-V=-0.02$ at $V_{peak}+2$:	&	-0.1	&	0.7	&	0.8	&	0.6	&	0.1	&	-0.2	&	0.4	&	0.9	&	1.7	&	0.8	\\
Other methods:	&	0.25$\pm$0.02\tablenotemark{a}	&	$\geq$0.8\tablenotemark{b}	&	0.85$\pm$0.3\tablenotemark{d}	&	0.38$\pm$0.08\tablenotemark{e}	&	$\approx$0.2\tablenotemark{f}	&	$\sim$0.25\tablenotemark{h}	&	0.15\tablenotemark{i}	&	0.73\tablenotemark{k}	&	1.0\tablenotemark{l}	&	1.15$\pm$0.20\tablenotemark{n}	\\
	&		&	0.5-1.1\tablenotemark{c}	&		&	0.62\tablenotemark{e}	&	$\approx$0.35\tablenotemark{f}	&		&	0.10$\pm$0.02\tablenotemark{j}	&		&	1.0\tablenotemark{m}	&	0.90\tablenotemark{n}	\\
	&		&		&		&		&	0.24\tablenotemark{g}	&		&		&		&	1.12$\pm$0.03\tablenotemark{j}	&	1.12\tablenotemark{o}	\\
	&		&		&		&		&		&		&		&		&		&	1.2\tablenotemark{o}	\\
	&		&		&		&		&		&		&		&		&		&	0.5\tablenotemark{j}	\\
Final extinction:	&	0.25$\pm$0.02	&	0.8$\pm$0.2	&	0.85$\pm$0.3	&	0.5$\pm$0.2	&	0.2$\pm$0.1	&	0.2$\pm$0.2	&	0.1$\pm$0.1	&	0.73$\pm$0.2	&	1.0$\pm$0.2	&	0.9$\pm$0.3	\\
\enddata
\tablenotetext{a}{Gilmozzi \& Selvelli 2007.}
\tablenotetext{b}{Duerbeck et al. 2002.}
\tablenotetext{c}{Orio et al. 2005.}
\tablenotetext{d}{Kiss et al. 2001.}
\tablenotetext{e}{{\rm L}ynch et al. 2004.}
\tablenotetext{f}{Barlow et al. 1981.}
\tablenotetext{g}{Burstein \& Heiles 1982.}
\tablenotetext{h}{Duerbeck 1988.}
\tablenotetext{i}{Cassatella et al. 1982.}
\tablenotetext{j}{Harrison, Johnson, \& Spyromilio 1993.}
\tablenotetext{k}{Snijders 1987.}
\tablenotetext{l}{Duerbeck 1989.}
\tablenotetext{m}{Sekiguchi et al. 1990.}
\tablenotetext{n}{Gonzalez-Riestra 1992.}
\tablenotetext{o}{Anupama \& Sethi 1994.}

\end{deluxetable}

\clearpage
\begin{deluxetable}{lllllllllll}
\tabletypesize{\scriptsize}
\rotate
\tablecaption{Distances from MMRD Relations and the Companion Star
\label{tbl27}}
\tablewidth{0pt}
\tablehead{
\colhead{}   &
\colhead{T Pyx}   &
\colhead{IM Nor}   &
\colhead{CI Aql}   &
\colhead{V2487 Oph}   &
\colhead{U Sco}   &
\colhead{V394 CrA}   &
\colhead{T CrB}   &
\colhead{RS Oph}   &
\colhead{V745 Sco}   &
\colhead{V3890 Sgr} 
}
\startdata

MMRD distances:	&		&		&		&		&		&		&		&		&		&		\\
~~$V_{peak}$ (mag)	&	6.4	&	8.5	&	9.0	&	9.5	&	7.5	&	7.2	&	2.5	&	4.8	&	9.4	&	8.1	\\
~~$V_{15days}$ (mag)	&	7.7	&	9.2	&	10.6	&	13.7	&	14.0	&	12.7	&	8.4	&	7.9	&	13.2	&	11.2	\\
~~$t_2$ (days)	&	34.6	&	50	&	25.4	&	6.2	&	1.2	&	2.4	&	4.0	&	6.8	&	6.2	&	6.4	\\
~~$t_3$ (days)	&	61.6	&	80.2	&	31.6	&	8.4	&	2.6	&	5.2	&	6.0	&	14.0	&	9.0	&	14.4	\\
~~$E_{B-V}$ (mag)	&	0.25	&	0.8	&	0.85	&	0.5	&	0.2	&	0.2	&	0.1	&	0.73	&	1.0	&	0.9	\\
~~$\mu$(15d)	&	12.9	&	14.4	&	15.8	&	18.9	&	19.2	&	17.9	&	13.6	&	13.1	&	18.4	&	16.4	\\
~~$\mu$($t_2$)	&	13.5	&	15.5	&	16.3	&	19.1	&	18.2	&	17.4	&	12.4	&	14.3	&	19.0	&	17.7	\\
~~$\mu$($t_3$)	&	13.5	&	15.5	&	16.3	&	19.3	&	18.1	&	17.3	&	12.5	&	14.2	&	19.2	&	17.5	\\
~~$\langle\mu\rangle$	&	13.3	&	15.1	&	16.1	&	19.1	&	18.5	&	17.6	&	12.8	&	13.9	&	18.9	&	17.2	\\
~~$d$ (pc)	&	3200	&	3400	&	5000	&	32400	&	37700	&	24400	&	3200	&	2100	&	14100	&	7600	\\
Blackbody distances:	&		&		&		&		&		&		&		&		&		&		\\
~~Method 1	&	\ldots	&	\ldots	&	\ldots	&	\ldots	&	8500	&	\ldots	&	900	&	3200	&	6800	&	7300	\\
~~Method 2	&	\ldots	&	\ldots	&	\ldots	&	\ldots	&	15400	&	\ldots	&	700	&	5400	&	7800	&	4700	\\
~~Method 3	&	\ldots	&	\ldots	&	\ldots	&	\ldots	&	\ldots	&	\ldots	&	1080	&	4000	&	8900	&	9000	\\

\enddata
    
\end{deluxetable}

\clearpage
\begin{deluxetable}{ll}
\tabletypesize{\scriptsize}
\tablecaption{U Sco Distance from Companion Star
\label{tbl28}}
\tablewidth{0pt}
\tablehead{
\colhead{}   &
\colhead{U Sco}  
}
\startdata

$P_{orb}$ (d)	&	1.23	\\
$R_{Roche}$ ($R_{\odot}$)	&	2.3	\\
$E_(B-V)$ (mag)	&	0.2	\\
Method 1:	&		\\
$~~$Spectral Type	&	G5 IV	\\
$~~T_{comp}$ (K)	&	5300	\\
$~~m$ (mag)	&	18.9	\\
$~~A$ (mag)	&	0.62	\\
$~~m_{bol}$ (mag)	&	17.98	\\
$~~M_{bol}$ (mag)	&	3.34		\\
$~~d$ (pc)	&	8500		\\
Method 2:	&		\\
$~~\log [\nu_{max}]$ (Hz)	&	\ldots	\\
$~~f_{\nu 0}(\nu_{max})$ (erg cm$^{-2}$ s$^{-1}$ Hz$^{-1}$)	&	$1.4\times10^{-27}$\\
$~~T_{comp}$ (K)	&	6000	\\
$~~F_{\nu}(\nu_{max})$ (erg cm$^{-2}$ s$^{-1}$ Hz$^{-1}$)	&	$1.3\times10^{-4}$ \\
$~~d$ (pc)	&	15400		\\
Final distance:	&		\\
$~~\langle d \rangle$ (pc)	&	$12000\pm2000$	\\

\enddata
\end{deluxetable}

\clearpage
\begin{deluxetable}{lllllllllll}
\tabletypesize{\scriptsize}
\rotate
\tablecaption{Distances, Extinctions, Galactic Positions, Absolute Magnitudes, and Intrinsic Colors
\label{tbl29}}
\tablewidth{0pt}
\tablehead{
\colhead{}   &
\colhead{T Pyx}   &
\colhead{IM Nor}   &
\colhead{CI Aql}   &
\colhead{V2487 Oph}   &
\colhead{U Sco}   &
\colhead{V394 CrA}   &
\colhead{T CrB}   &
\colhead{RS Oph}   &
\colhead{V745 Sco}   &
\colhead{V3890 Sgr}   
}
\startdata

Distance \& Extinction:	&		&		&		&		&		&		&		&		&		&		\\
$~~$Distance (pc)	&	$3500\pm1000$	&	$3400^{+3400}_{-1700}$	&	$5000^{+5000}_{-2500}$	&	$12000^{+13000}_{-2000}$	&	$12000\pm2000$	&	$10000^{+15000}_{-3000}$	&	$900\pm200$	&	$4200\pm900$	&	$7800\pm1800$	&	$7000\pm1600$	\\
$~~E_{B-V}$ (mag)	&	$0.25\pm0.02$	&	$0.8\pm0.2$	&	$0.85\pm0.3$	&	$0.5\pm0.2$	&	$0.2\pm0.1$	&	$0.2\pm0.2$	&	$0.1\pm0.1$	&	$0.73\pm0.2$	&	$1\pm0.2$	&	$0.9\pm0.3$	\\
Galactic Position:	&		&		&		&		&		&		&		&		&		&		\\
$~~\ell$ (deg)	&	9.5	&	2.5	&	-0.8	&	7.8	&	22.5	&	-7.1	&	48.7	&	11.0	&	-3.4	&	-5.8	\\
$~~b$ (deg)	&	256.8	&	327.1	&	31.7	&	6.6	&	357.3	&	352.5	&	42.4	&	19.5	&	357.0	&	8.9	\\
$~~\theta_{GC}$ (deg)	&	103.0	&	33.0	&	31.7	&	10.2	&	22.7	&	10.3	&	60.8	&	22.3	&	4.5	&	10.6	\\
Height above Plane:	&		&		&		&		&		&		&		&		&		&		\\
\enddata
    
\end{deluxetable}

\clearpage
\begin{deluxetable}{llllllllll}
\tabletypesize{\scriptsize}
\tablecaption{RN colors for the SEDs
\label{tbl30}}
\tablewidth{0pt}
\tablehead{
\colhead{RN}   &
\colhead{JD}   &
\colhead{V (mag)}   &
\colhead{U-V (mag)}   &
\colhead{B-V (mag)}   &
\colhead{V-R (mag)}   &
\colhead{V-I (mag)}   &
\colhead{V-J (mag)}   &
\colhead{V-H (mag)}   &
\colhead{V-K (mag)}   
}
\startdata

T Pyx	&	2453808.61	&	15.56	&	0.90\tablenotemark{a}	&	0.14	&	0.23	&	0.45	&	0.62	&	0.78	&	1.3\tablenotemark{a}	\\
T Pyx	&	2454118.78	&	15.51	&	0.90\tablenotemark{a}	&	0.13	&	0.29	&	0.53	&	0.60	&	0.78	&	1.3\tablenotemark{a}	\\
IM Nor	&	2453590.58	&	18.36	&	\ldots	&	1.02	&	0.56	&	1.17	&	2.15	&	2.45	&	\ldots	\\
IM Nor	&	2453607.52	&	18.33	&	\ldots	&	1.10	&	0.58	&	1.17	&	2.08	&	2.34	&	\ldots	\\
CI Aql	&	2452781.95	&	16.13	&	1.47	&	1.09	&	0.68\tablenotemark{a}	&	1.33	&	2.0\tablenotemark{a}	&	2.3\tablenotemark{a}	&	2.7\tablenotemark{a}	\\
CI Aql	&	2452782.79	&	16.14	&	1.33	&	1.03	&	0.68\tablenotemark{a}	&	1.36	&	2.0\tablenotemark{a}	&	2.3\tablenotemark{a}	&	2.7\tablenotemark{a}	\\
CI Aql	&	2452782.93	&	16.04	&	1.25	&	1.01	&	0.68\tablenotemark{a}	&	1.31	&	2.0\tablenotemark{a}	&	2.3\tablenotemark{a}	&	2.7\tablenotemark{a}	\\
CI Aql	&	2453607.55	&	16.22	&	1.35\tablenotemark{a}	&	1.03	&	0.68	&	1.37	&	2.33	&	2.61	&	2.66	\\
\enddata

\tablenotetext{a}{This color was adopted from the average color on other dates.}
    
\end{deluxetable}

\clearpage
\begin{deluxetable}{llllll}
\tabletypesize{\scriptsize}
\tablecaption{Observing Bands
\label{tbl31}}
\tablewidth{0pt}
\tablehead{
\colhead{Band}   &
\colhead{$\lambda_{center}$ ($\mu$)}   &
\colhead{$\Delta \lambda$ ($\mu$)}   &
\colhead{$\log (\nu_{center)}$}   &
\colhead{$A(\lambda)/A(V)$}   &
\colhead{$f_{\nu}$(0 mag)}   
}
\startdata

U	&	0.36	&	0.07	&	14.92	&	1.56	&	1810	\\
B	&	0.44	&	0.10	&	14.83	&	1.31	&	4260	\\
V	&	0.55	&	0.09	&	14.74	&	1.00	&	3640	\\
R	&	0.64	&	0.22	&	14.67	&	0.749	&	3080	\\
I	&	0.79	&	0.24	&	14.58	&	0.479	&	2550	\\
J	&	1.26	&	0.38	&	14.38	&	0.282	&	1600	\\
H	&	1.60	&	0.37	&	14.27	&	0.176	&	1080	\\
K	&	2.22	&	0.48	&	14.13	&	0.108	&	670	\\

\enddata
    
\end{deluxetable}

\clearpage
\begin{deluxetable}{lllllllll}
\tabletypesize{\scriptsize}
\tablecaption{Extinction-Corrected Magnitudes
\label{tbl32}}
\tablewidth{0pt}
\tablehead{
\colhead{RN}   &
\colhead{$U_0$ (mag)}   &
\colhead{$B_0$ (mag)}   &
\colhead{$V_0$ (mag)}   &
\colhead{$R_0$ (mag)}   &
\colhead{$I_0$ (mag)}   &
\colhead{$J_0$ (mag)}   &
\colhead{$H_0$ (mag)}   &
\colhead{$K_0$ (mag)}   
}
\startdata

T Pyx	&	15.25$\pm$0.11	&	14.68$\pm$0.10	&	14.79$\pm$0.08	&	14.75$\pm$0.07	&	14.74$\pm$0.06	&	14.72$\pm$0.05	&	14.64$\pm$0.05	&	14.18$\pm$0.05	\\
T Pyx	&	15.20$\pm$0.11	&	14.62$\pm$0.10	&	14.74$\pm$0.08	&	14.64$\pm$0.07	&	14.61$\pm$0.06	&	14.69$\pm$0.05	&	14.59$\pm$0.05	&	14.13$\pm$0.05	\\
IM Nor	&	\ldots	&	16.13$\pm$0.81	&	15.88$\pm$0.62	&	15.94$\pm$0.47	&	16.00$\pm$0.30	&	15.51$\pm$0.18	&	15.47$\pm$0.12	&	\ldots	\\
IM Nor	&	\ldots	&	16.18$\pm$0.81	&	15.85$\pm$0.62	&	15.89$\pm$0.47	&	15.97$\pm$0.30	&	15.55$\pm$0.18	&	15.55$\pm$0.12	&	\ldots	\\
CI Aql	&	13.49$\pm$1.45	&	13.77$\pm$1.22	&	13.50$\pm$0.93	&	13.48$\pm$0.70	&	13.54$\pm$0.45	&	13.39$\pm$0.27	&	13.37$\pm$0.17	&	13.15$\pm$0.11	\\
CI Aql	&	13.36$\pm$1.45	&	13.72$\pm$1.22	&	13.51$\pm$0.93	&	13.49$\pm$0.70	&	13.52$\pm$0.45	&	13.40$\pm$0.27	&	13.38$\pm$0.17	&	13.16$\pm$0.11	\\
CI Aql	&	13.18$\pm$1.45	&	13.60$\pm$1.22	&	13.41$\pm$0.93	&	13.39$\pm$0.70	&	13.47$\pm$0.45	&	13.30$\pm$0.27	&	13.28$\pm$0.17	&	13.06$\pm$0.11	\\
CI Aql	&	13.46$\pm$1.45	&	13.80$\pm$1.22	&	13.59$\pm$0.93	&	13.57$\pm$0.70	&	13.59$\pm$0.45	&	13.15$\pm$0.27	&	13.15$\pm$0.17	&	13.28$\pm$0.11	\\
\enddata
    
\end{deluxetable}

\clearpage
\begin{deluxetable}{lllllllll}
\tabletypesize{\scriptsize}
\tablecaption{RN Spectral Energy Distributions
\label{tbl33}}
\tablewidth{0pt}
\tablehead{
\colhead{RN}   &
\colhead{$\log (f_{\nu})$ (U)}   &
\colhead{$\log (f_{\nu})$ (B)}   &
\colhead{$\log (f_{\nu})$ (V)}   &
\colhead{$\log (f_{\nu})$ (R)}   &
\colhead{$\log (f_{\nu})$ (I)}   &
\colhead{$\log (f_{\nu})$ (J)}   &
\colhead{$\log (f_{\nu})$ (H)}   &
\colhead{$\log (f_{\nu})$ (K)}   
}
\startdata

T Pyx	&	-2.84$\pm$0.04	&	-2.24$\pm$0.04	&	-2.35$\pm$0.03	&	-2.41$\pm$0.03	&	-2.49$\pm$0.02	&	-2.68$\pm$0.02	&	-2.82$\pm$0.02	&	-2.84$\pm$0.02	\\
T Pyx	&	-2.82$\pm$0.04	&	-2.22$\pm$0.04	&	-2.33$\pm$0.03	&	-2.37$\pm$0.03	&	-2.44$\pm$0.02	&	-2.67$\pm$0.02	&	-2.80$\pm$0.02	&	-2.82$\pm$0.02	\\
IM Nor	&	\ldots	&	-2.82$\pm$0.33	&	-2.79$\pm$0.25	&	-2.89$\pm$0.19	&	-2.99$\pm$0.12	&	-3.00$\pm$0.07	&	-3.16$\pm$0.05	&	\ldots	\\
IM Nor	&	\ldots	&	-2.84$\pm$0.33	&	-2.78$\pm$0.25	&	-2.87$\pm$0.19	&	-2.98$\pm$0.12	&	-3.02$\pm$0.07	&	-3.19$\pm$0.05	&	\ldots	\\
CI Aql	&	-2.14$\pm$0.58	&	-1.88$\pm$0.49	&	-1.84$\pm$0.37	&	-1.90$\pm$0.28	&	-2.01$\pm$0.18	&	-2.15$\pm$0.11	&	-2.31$\pm$0.07	&	-2.43$\pm$0.04	\\
CI Aql	&	-2.09$\pm$0.58	&	-1.86$\pm$0.49	&	-1.84$\pm$0.37	&	-1.91$\pm$0.28	&	-2.00$\pm$0.18	&	-2.15$\pm$0.11	&	-2.32$\pm$0.07	&	-2.44$\pm$0.04	\\
CI Aql	&	-2.01$\pm$0.58	&	-1.81$\pm$0.49	&	-1.80$\pm$0.37	&	-1.87$\pm$0.28	&	-1.98$\pm$0.18	&	-2.11$\pm$0.11	&	-2.28$\pm$0.07	&	-2.40$\pm$0.04	\\
CI Aql	&	-2.13$\pm$0.58	&	-1.89$\pm$0.49	&	-1.87$\pm$0.37	&	-1.94$\pm$0.28	&	-2.03$\pm$0.18	&	-2.05$\pm$0.11	&	-2.23$\pm$0.07	&	-2.48$\pm$0.04	\\
\enddata

\end{deluxetable}

\clearpage
\begin{deluxetable}{lllllllllll}
\tabletypesize{\scriptsize}
\rotate
\tablecaption{RN Summary
\label{tbl34}}
\tablewidth{0pt}
\tablehead{
\colhead{}   &
\colhead{T Pyx}   &
\colhead{IM Nor}   &
\colhead{CI Aql}   &
\colhead{V2487 Oph}   &
\colhead{U Sco}   &
\colhead{V394 CrA}   &
\colhead{T CrB}   &
\colhead{RS Oph}   &
\colhead{V745 Sco}   &
\colhead{V3890 Sgr}   
}
\startdata

Number of eruptions	&	5	&	2	&	3	&	2	&	9	&	2	&	2	&	8	&	2	&	2	\\
Eruption years	&	1890, 1902,	&	1920, 2002	&	1917, 1941,	&	1900, 1998	&	1863, 1906,	&	1949, 1987	&	1866, 1946	&	1898, 1907,	&	1937, 1989	&	1962, 1990	\\
	&	1920, 1944,	&		&	2000	&		&	1917, 1936,	&		&		&	1933, 1945,	&		&		\\
	&	1966	&		&		&		&	1945, 1969,	&		&		&	1958, 1967,	&		&		\\
	&		&		&		&		&	1979, 1987,	&		&		&	1985, 2006	&		&		\\
	&		&		&		&		&	1999	&		&		&		&		&		\\
$B_{peak}$ (mag)	&	6.7	&	9.2	&	9.4	&	10.1	&	7.8	&	7.0	&	2.5	&	5.3	&	10.5	&	8.9	\\
$V_{peak}$ (mag)	&	6.4	&	8.5	&	9.0	&	9.5	&	7.5	&	7.2	&	2.5	&	4.8	&	9.4	&	8.1	\\
\enddata

\end{deluxetable}

\clearpage
\begin{figure}
\epsscale{1.0}
\plotone{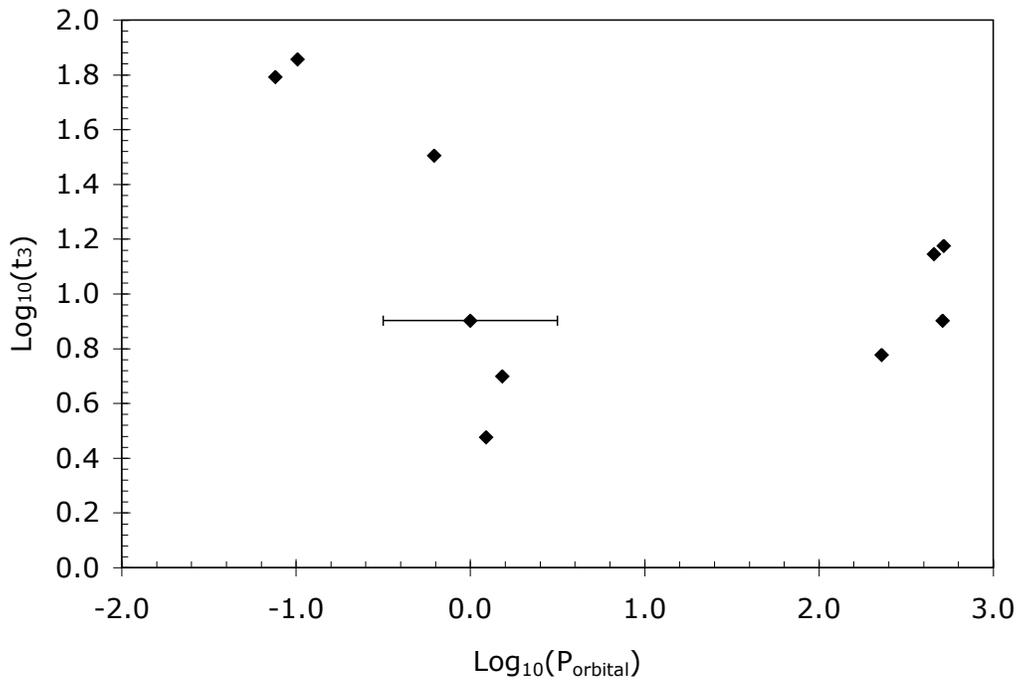}
\caption{
RN groups?  The ten RNe have some distribution of decline rates ($t_3$ in days) and orbital periods ($P_{orbital}$ in days), but there is no apparent clustering.  Yes, we can always draw lines in such diagrams to create divisions (with the best try being to separate out those systems with red giant companions), but to be meaningful the groups must be either well-separated or caused by separate physical processes, while other properties should correlate with the putative division.  None of the prior  divisions amongst the wide variety of properties satisfiy the need for a useful division.}
\end{figure}

\clearpage
\begin{figure}
\epsscale{1.0}
\plotone{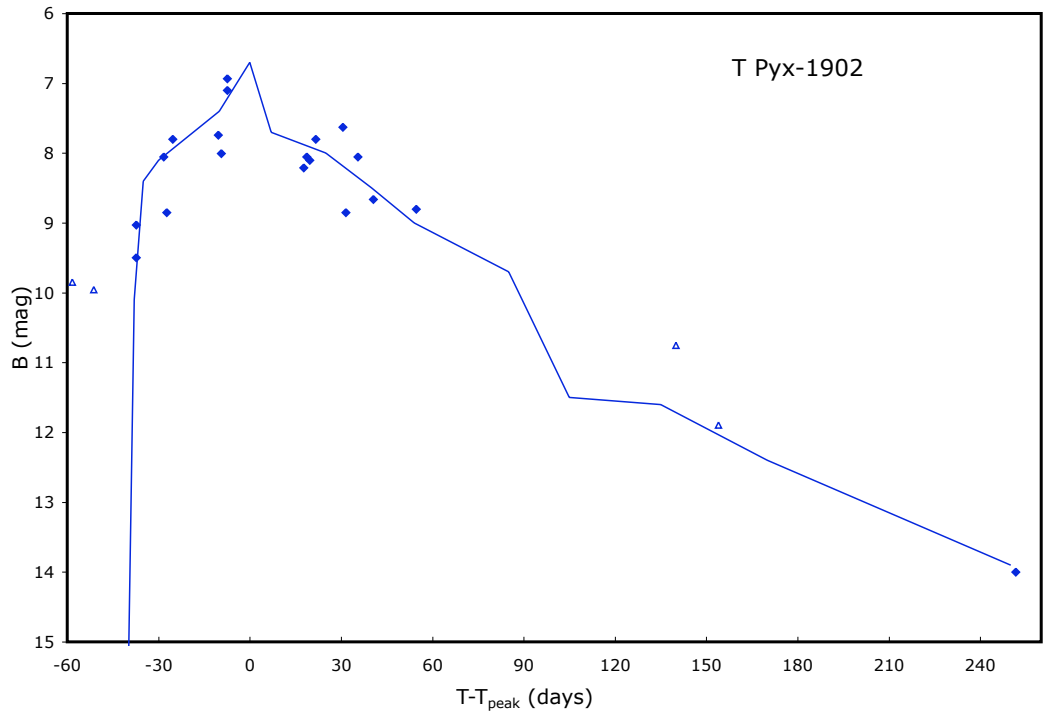}
\caption{
T Pyx in 1902.  The diamonds are for B-band magnitudes, and the triangles indicate upper limits (from Table 4).  The broken-line indicates the template for the B-band (see Table 16).}
\end{figure}

\clearpage
\begin{figure}
\epsscale{1.0}
\plotone{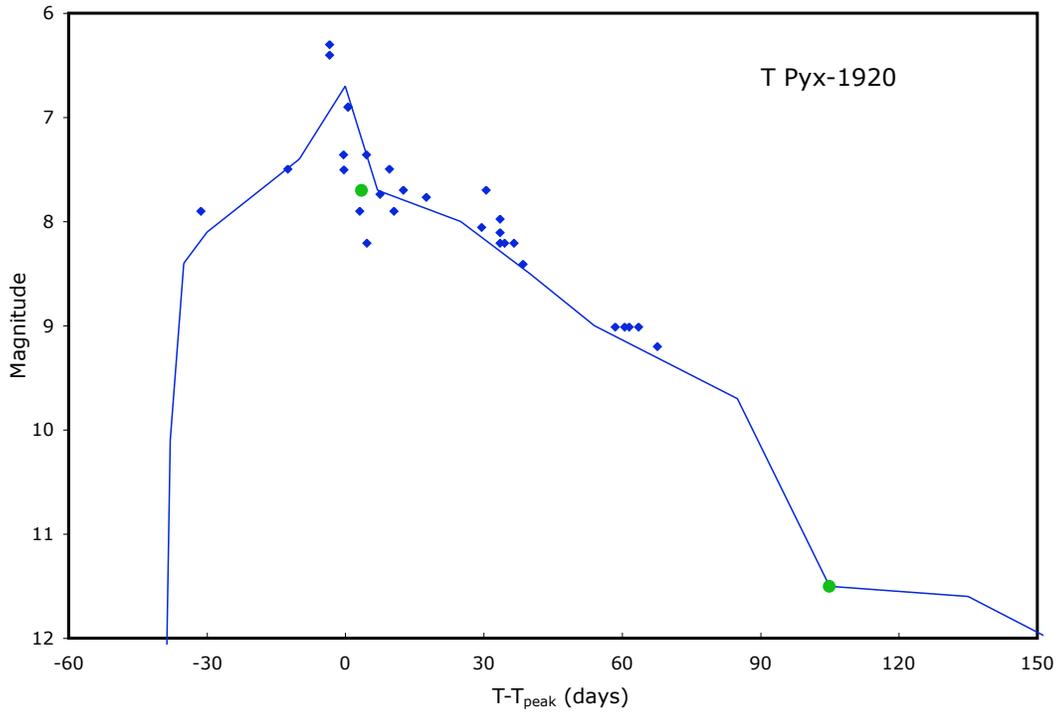}
\caption{
T Pyx in 1920.  The diamonds are for B-band magnitudes, and the broken-line indicates the templates for the B-band.  Two visual observations are indicated with the filled circles.}
\end{figure}

\clearpage
\begin{figure}
\epsscale{1.0}
\plotone{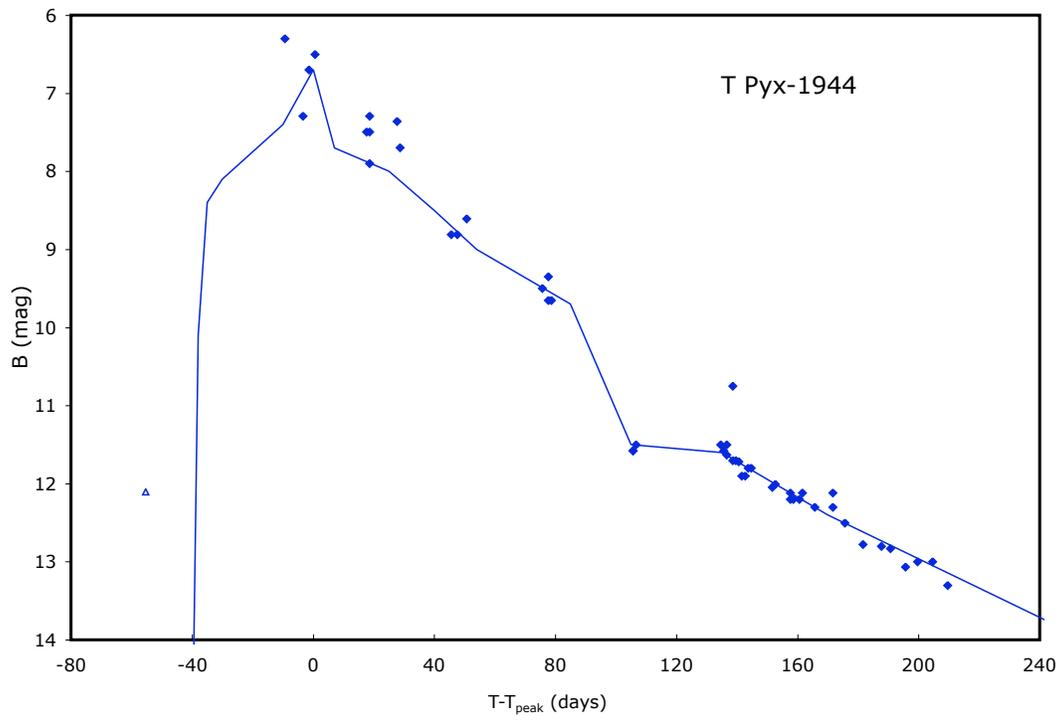}
\caption{
T Pyx in 1944.  The diamonds are for B-band magnitudes, while the empty triangle is an upper limit just before the initial rise.}
\end{figure}

\clearpage
\begin{figure}
\epsscale{1.0}
\plotone{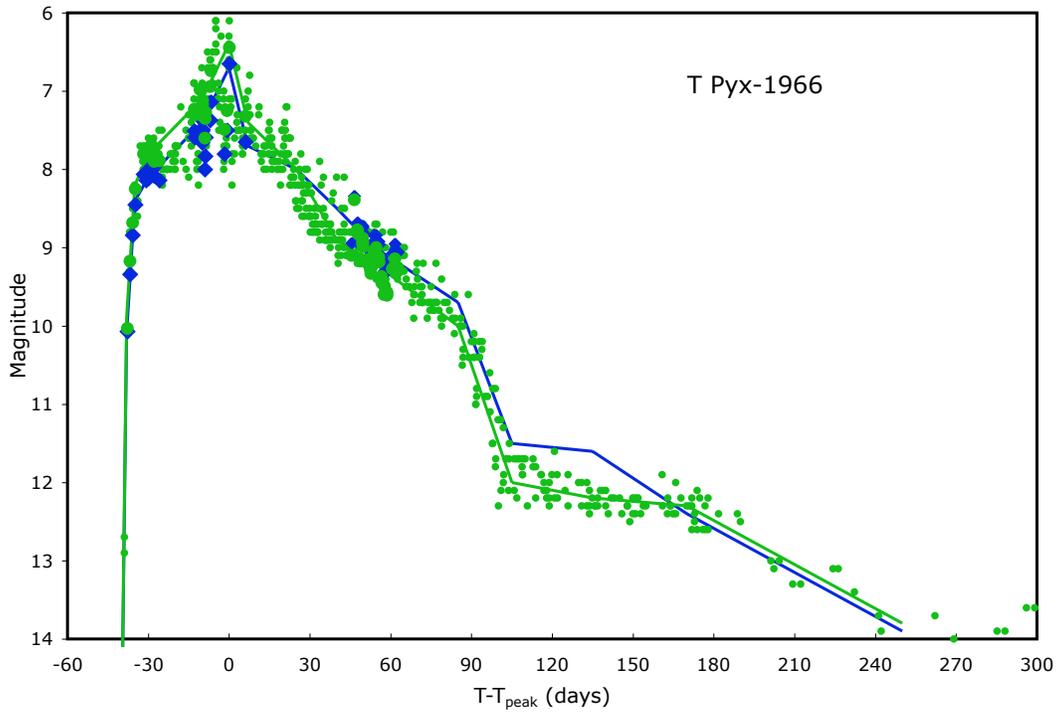}
\caption{
T Pyx in 1966.  The diamonds are for B-band magnitudes, while the circles are for V-band magnitudes.  The small circles are for visual observations, while the large symbols are for photoelectric observations.  The broken-lines indicate the templates for the B-band and the V-band (see Table 16), with the V-band template lying on top around the time of peak.  The T Pyx light curve has observed fast oscillations with an amplitude of $>0.8$ mag on time scales of a day or less, and this causes the large apparent scatter about the templates in the light curves.}
\end{figure}

\clearpage
\begin{figure}
\epsscale{1.0}
\plotone{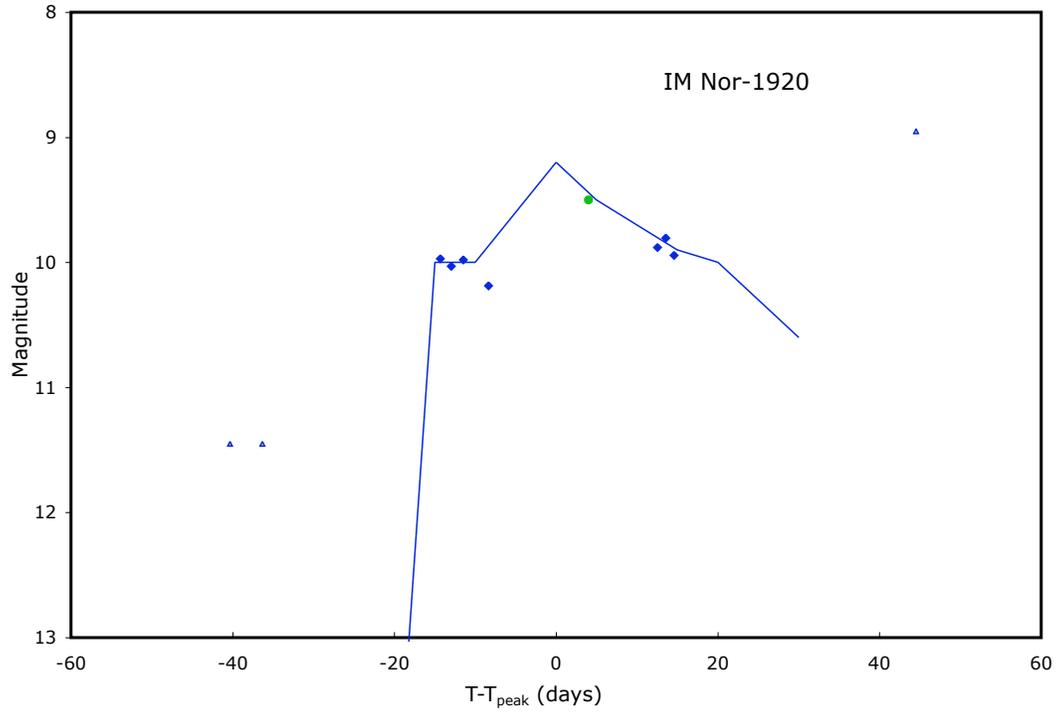}
\caption{
IM Nor in 1920.  The diamonds are for B-band magnitudes, the lone circle is for V-band magnitude, and the empty triangles are for B-band limits.  The broken-line indicates the template for the B-band (see Table 16).}
\end{figure}

\clearpage
\begin{figure}
\epsscale{0.66}
\plotone{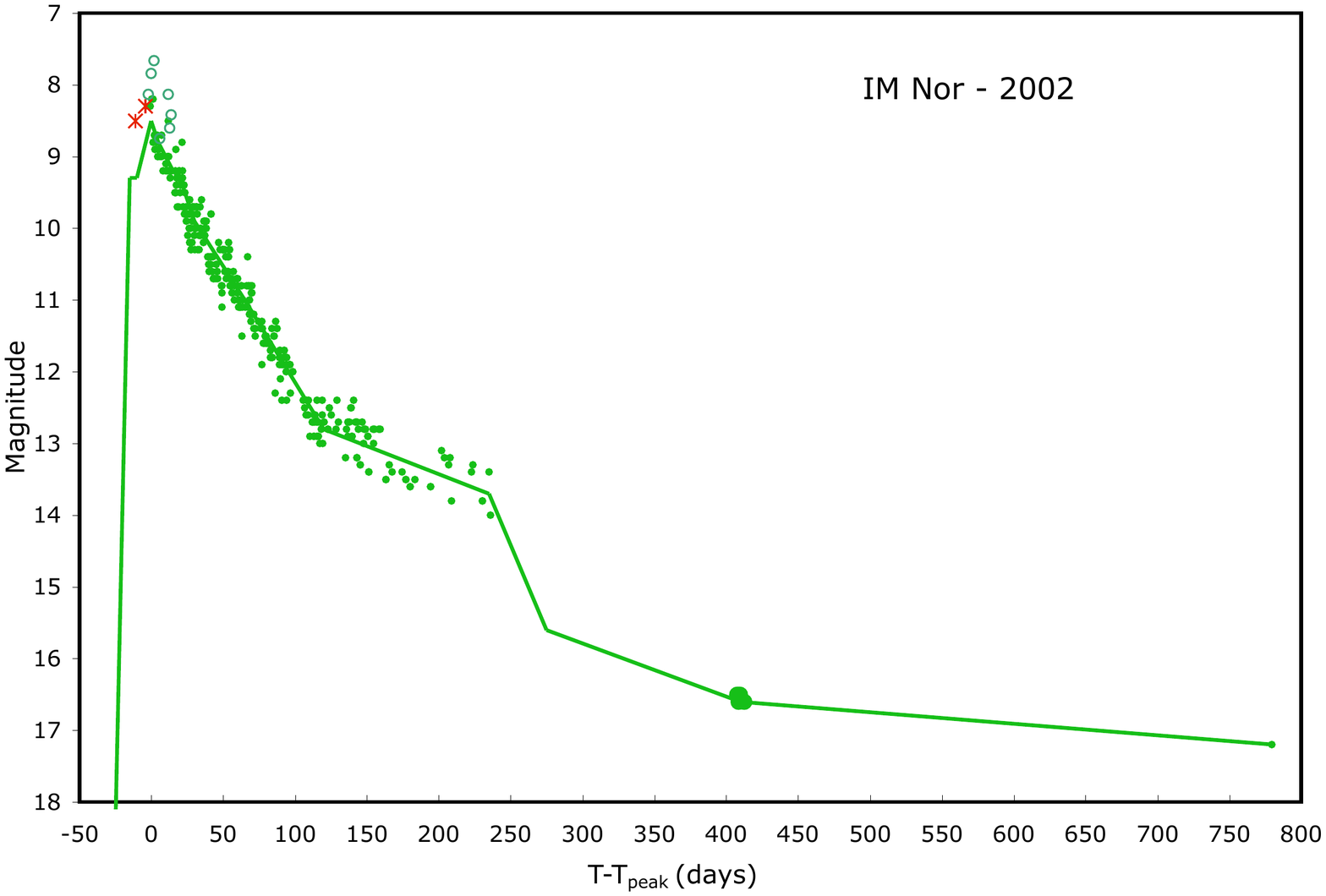}
\plotone{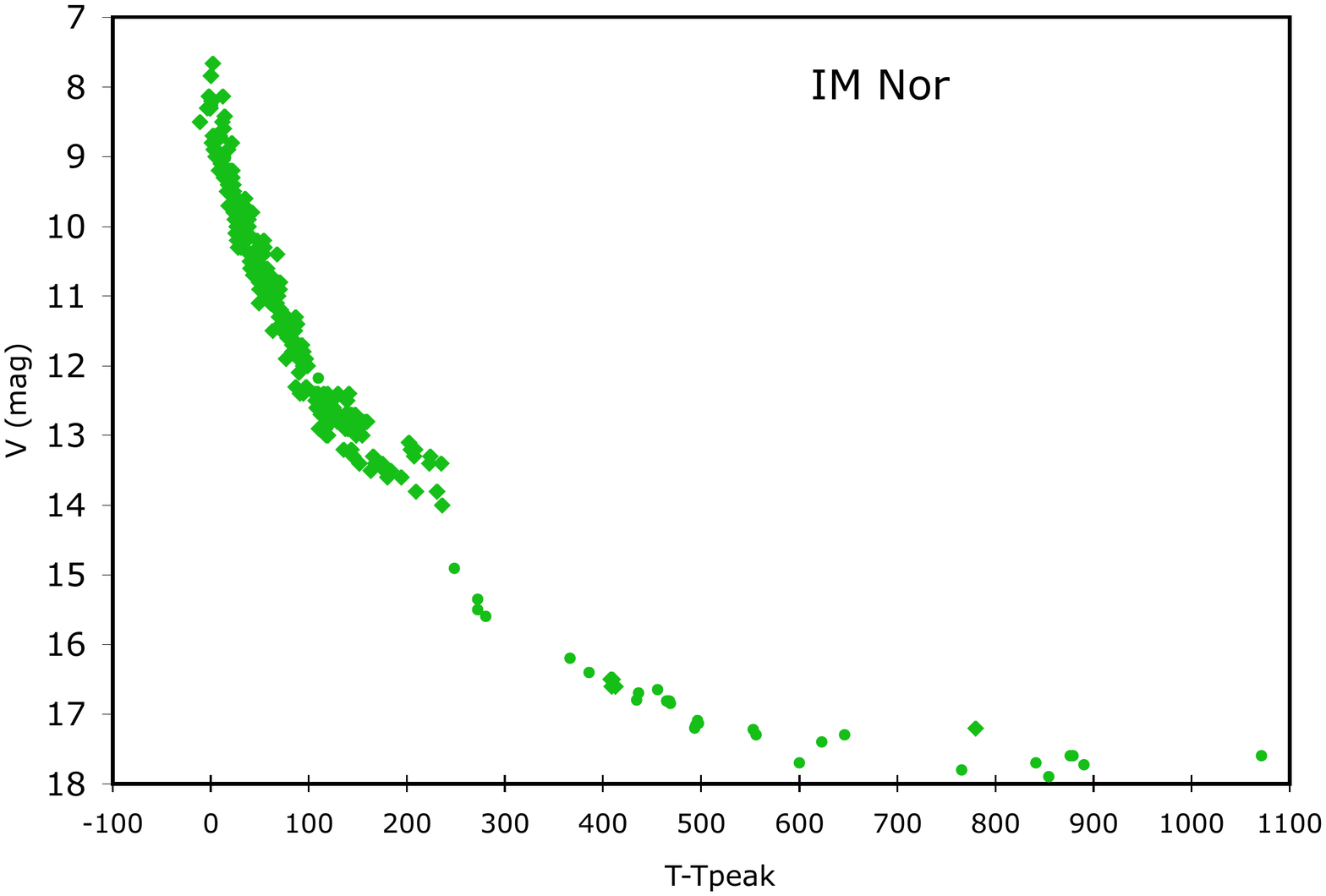}
\caption{
IM Nor in 2002.  In the upper panel, the small filled circles are for visual observations from the AAVSO or the {\it IAU Circulars}, while the large circles are CCD photometry from Woudt \& Warner (2003).  The asterisks at the very start are red band magnitudes from Liller, while the empty circles are other photographic observations of Liller with an unknown color sensitivity.  In the lower panel, the unfiltered-CCD magnitudes of Monard are included with a shift of 0.6 mag so as to match the light curve over the first 400 days.  The inclusion of this data provides a clear picture of the light curve from before the peak all the way past the return to quiescence over 900 days later.  Indeed, with this, we see a sharp drop off starting at 235 days immediately after a flattening from 110-235 days.  With this, it appears that IM Nor has a plateau from 110-235 days after peak.}
\end{figure}

\clearpage
\begin{figure}
\epsscale{1.0}
\plotone{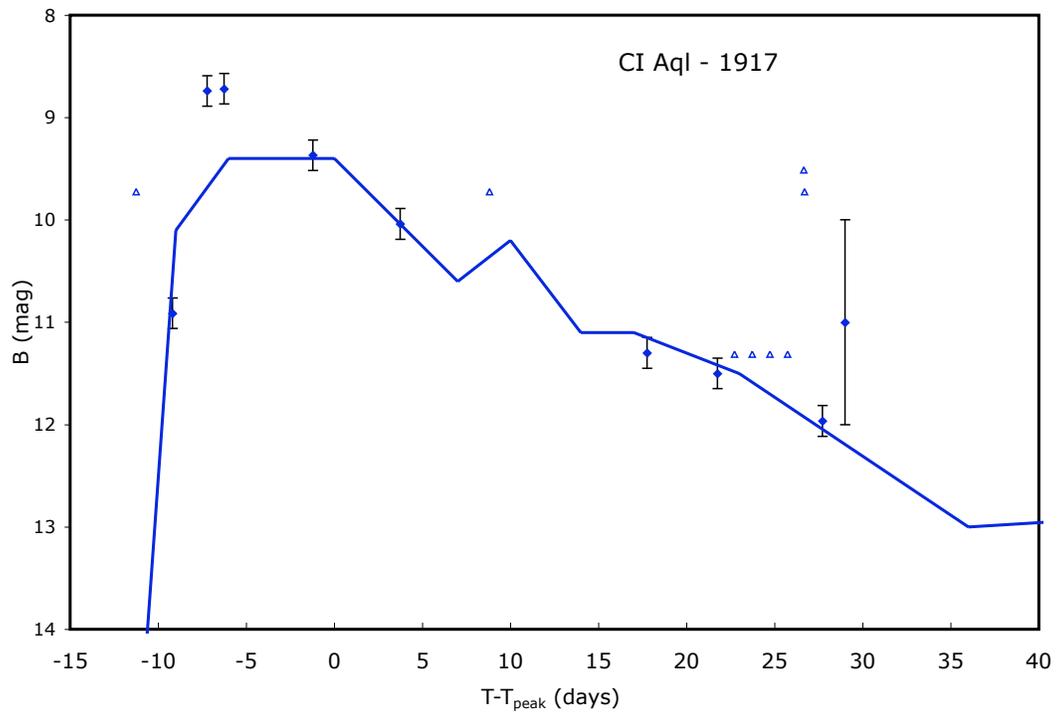}
\caption{
CI Aql in 1917.  The diamonds are B-band magnitudes, the empty triangles are limits, and the broken line the B-band template.}
\end{figure}

\clearpage
\begin{figure}
\epsscale{1.0}
\plotone{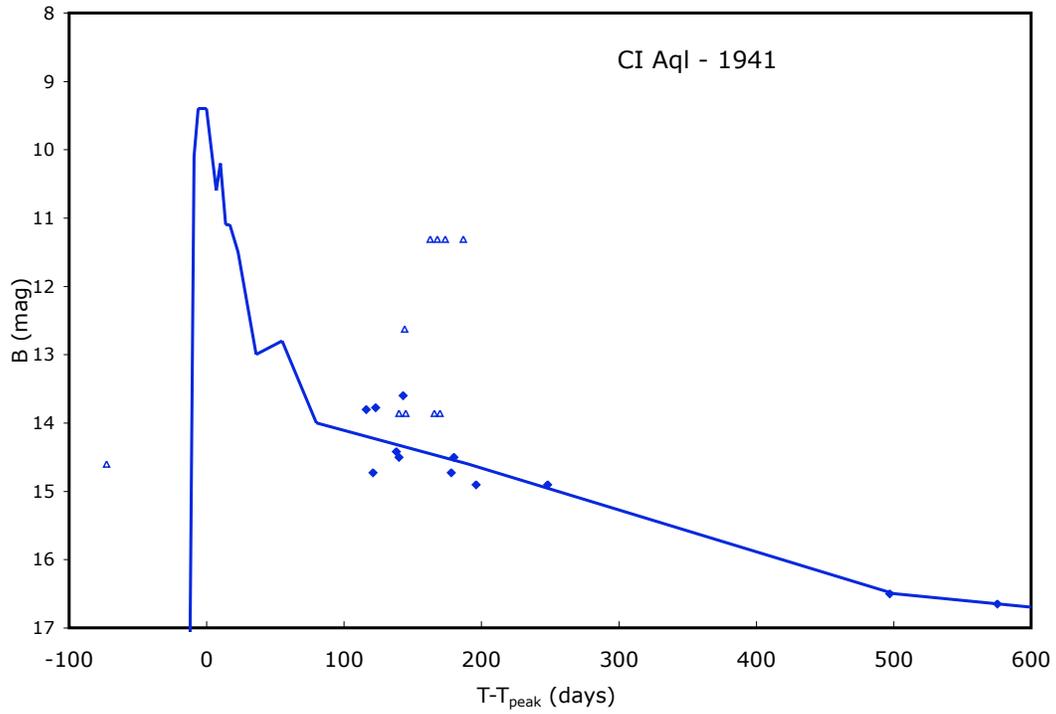}
\caption{
CI Aql in 1941.  The diamonds are B-band magnitudes from Harvard, the empty triangles are limits, and the broken line the B-band template.  The eruption is only seen in its tail, with no plates being taken earlier than roughly 121 days after the peak (at which time CI Aql had an elongation of 120$\degr$ from the Sun).  The seasonal gap was 194 days long, with this not being unusual.  The large size of the seasonal gaps imply that many RN eruptions will be missed.}
\end{figure}

\clearpage
\begin{figure}
\epsscale{1.0}
\plotone{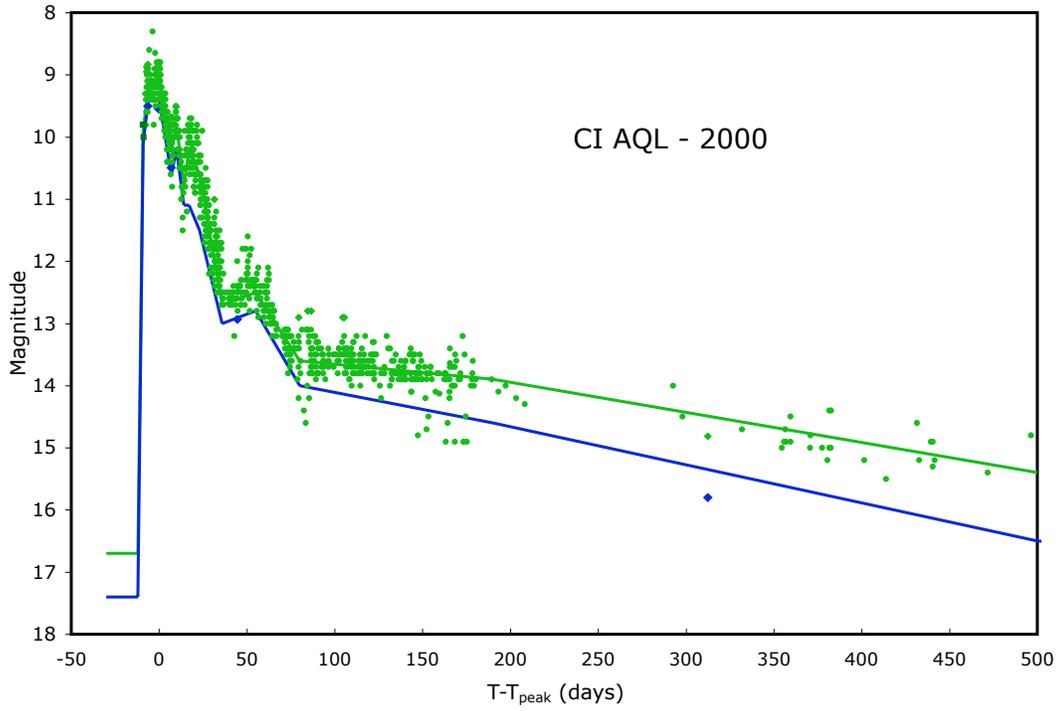}
\caption{
CI Aql in 2000.  The small circles are for V-band magnitudes, while the few diamonds are for B-band magnitudes.  The V-band template is hidden by the densely-sampled light curve (until late in the tail), while the B-band template lies below it.}
\end{figure}

\clearpage
\begin{figure}
\epsscale{1.0}
\plotone{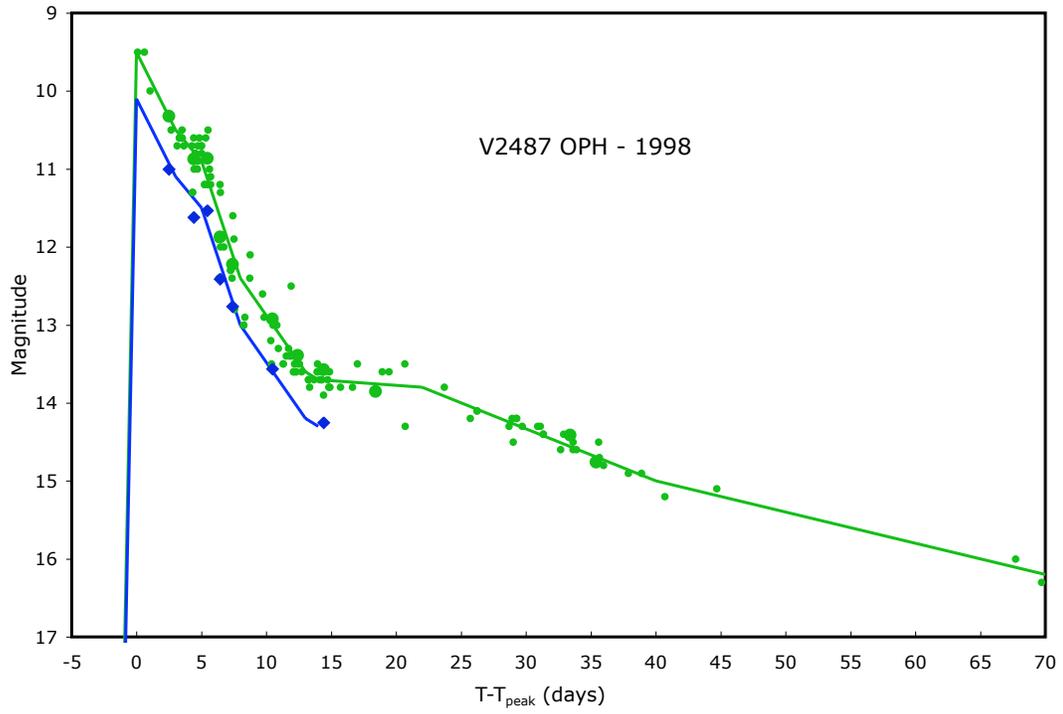}
\caption{
V2487 Oph in 1998.  The diamonds are for B-band, while the circles are for the V-band.  The large symbols are for photometry by D. Hanzl.  The two curves are the light curve templates for the two bands, with the blue being on the bottom.  Note that V2487 Oph is one of the fastest known novae.}
\end{figure}

\clearpage
\begin{figure}
\epsscale{1.0}
\plotone{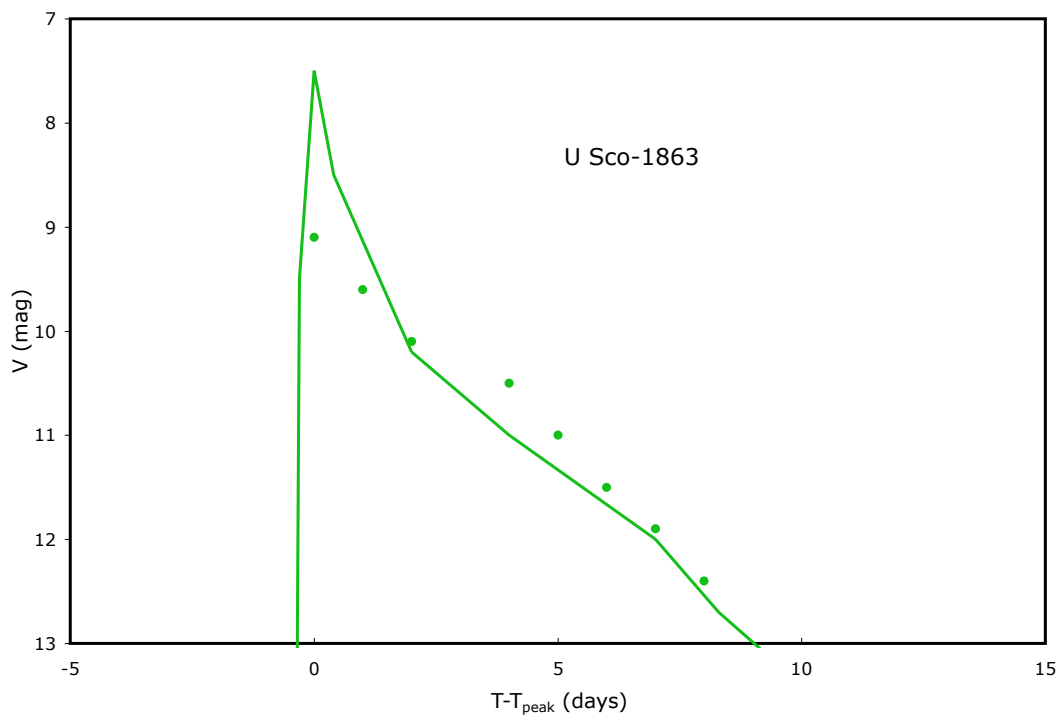}
\caption{
U Sco in 1863.  Pogson's discovery of the 1863 eruption was one of the earliest nova events recorded by `modern' astronomers.  But no one else saw this short lived outburst, and so we only have Pogson's eight visual magnitudes on the light curve.  Pogson states the magnitudes of his comparison stars, and these are on average close to the modern magnitudes for these same stars, so we can take Pogson's magnitudes (the filled circles) for U Sco as being approximately correct.  The V-band light curve template (derived from later eruptions) is sufficiently close to Pogson's data, especially given that Pogson had only just defined the modern magnitude scale.}
\end{figure}

\clearpage
\begin{figure}
\epsscale{1.0}
\plotone{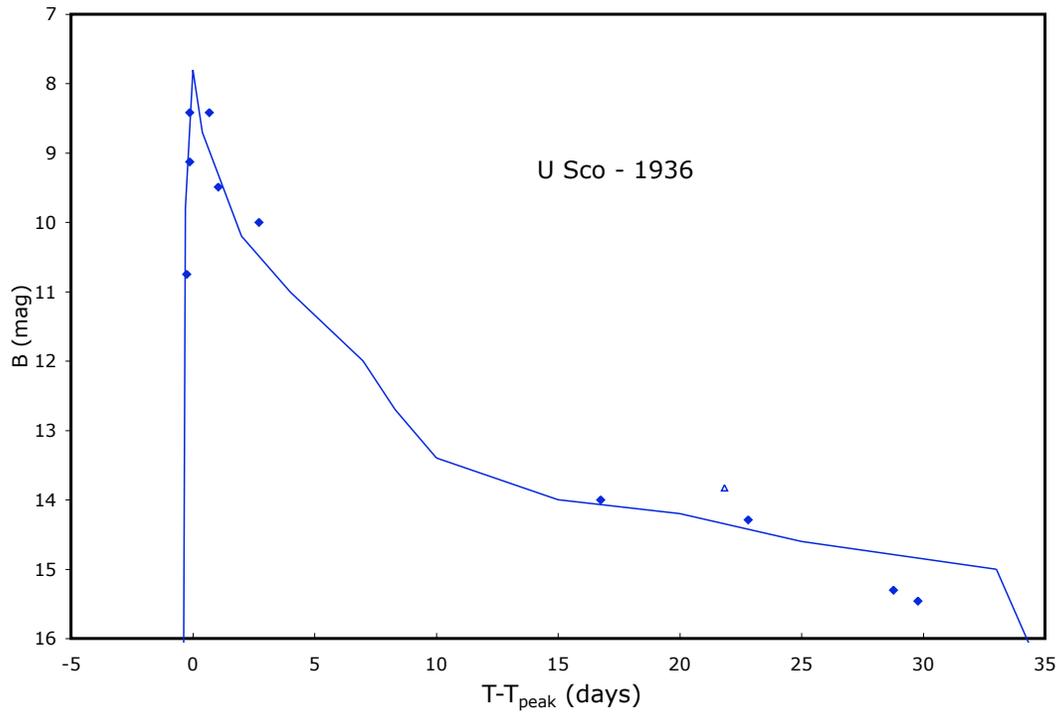}
\caption{
U Sco in 1936.  We have only ten B-band magnitudes (the filled diamonds), and these well define a very narrow peak with a very fast decline and a somewhat slower tail.  The B-band template (the broken line) is placed for the maximum being at JD 2428341.6.}
\end{figure}

\clearpage
\begin{figure}
\epsscale{1.0}
\plotone{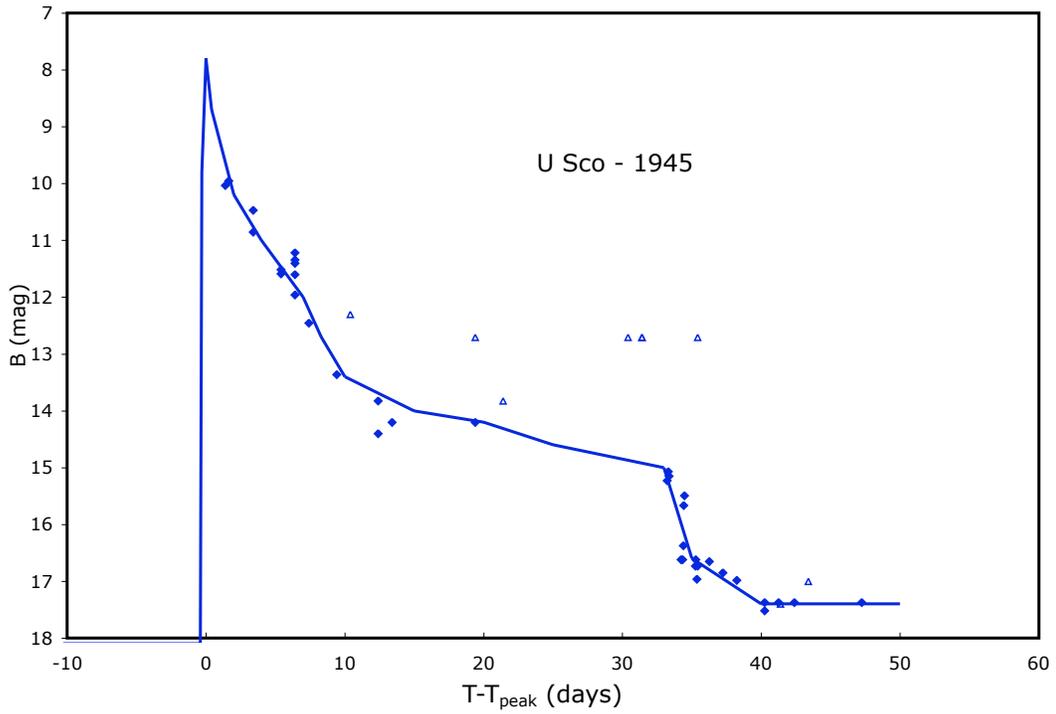}
\caption{
U Sco in 1945.  The B-band magnitudes are indicated with diamonds, the limits with empty triangles, and the template with the broken line.  This light curve is the only one of the nine eruptions that has B-band coverage after 30 days, and we see a well measured sudden drop in brightness at 33 days after peak.}
\end{figure}

\clearpage
\begin{figure}
\epsscale{1.0}
\plotone{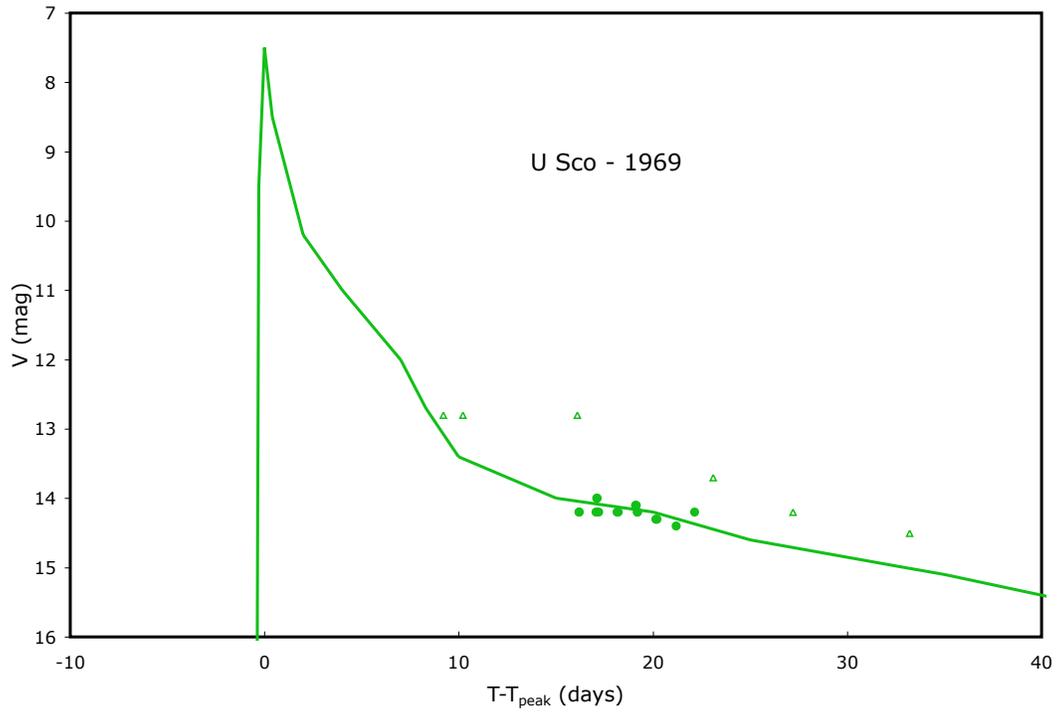}
\caption{
U Sco in 1969.  This eruption was only seen by two legendary observers (A. Jones and F. Bateson) as U Sco was coming out from behind the Sun.  The V-band light curve (with magnitudes indicated with filled circles and limits by empty triangles) is that of the tail of a U Sco eruption whose peak was lost due to the Sun.}
\end{figure}

\clearpage
\begin{figure}
\epsscale{1.0}
\plotone{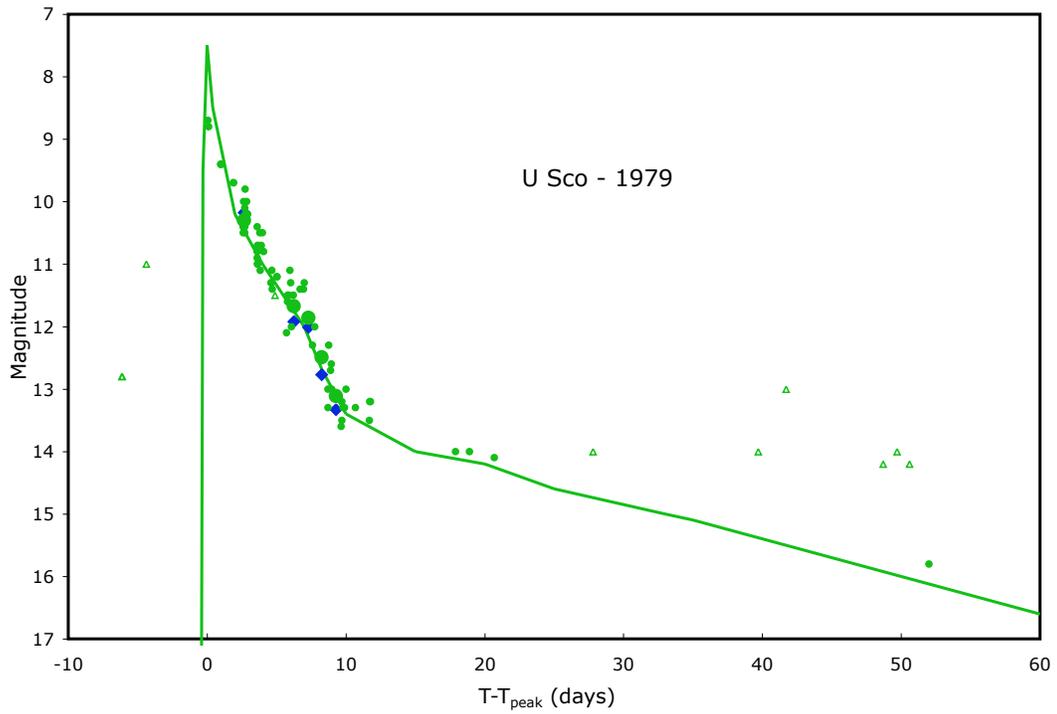}
\caption{
U Sco in 1979.  The RN was only recorded from 0-20 days after peak plus a single magnitude at 52 days.  Small filled circles are for visual data, empty triangles are for limits, large symbols are for the photometry reported in Barlow et al. (1981).}
\end{figure}

\clearpage
\begin{figure}
\epsscale{1.0}
\plotone{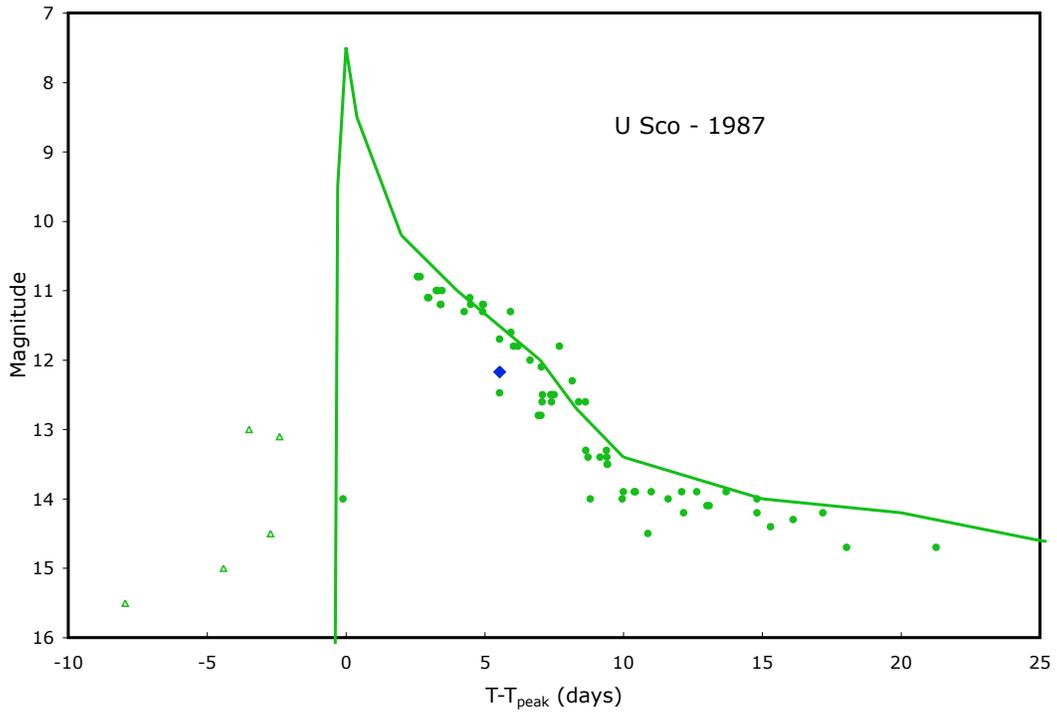}
\caption{
U Sco in 1987.  The peak of this eruption was missed, despite having a positive observation low on the rising branch.  With the exception of one B-band magnitude from Sekiguchi et al. (1988) indicated with a filled diamond, all the observations were made by amateur astronomers.  These V-band magnitudes are indicated with filled circles, while limits are with empty triangles.}
\end{figure}

\clearpage
\begin{figure}
\epsscale{1.0}
\plotone{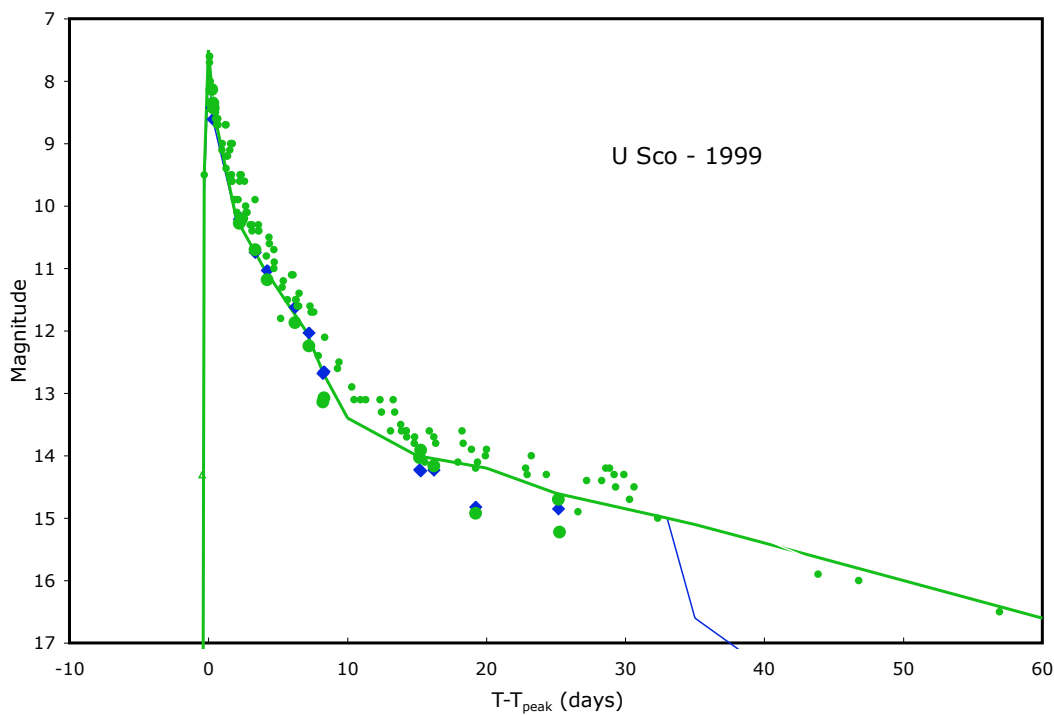}
\caption{
U Sco in 1999.  With this recent eruption, all points on the light curve are from amateur astronomy sources.  The circles are for V-band data, while the diamonds are for B-band data.  Note that the two templates diverge greatly after day 33, with a consequent large shift in color.  In this case (as adopted for the templates displayed in this figure), the sudden dimming would only affect the blue light and not the red light.  The drop in blue brightness is forced only by the 1945 light curve.  Alternatively, the visual brightness does not have any data from day 32 to 44, so it is possible that U Sco underwent a DQ-Her-like dust formation dip of short duration, only to return to a normal decline as indicated by a few V-band magnitudes from the 1979 and 1999 light curves.}
\end{figure}

\clearpage
\begin{figure}
\epsscale{1.0}
\plotone{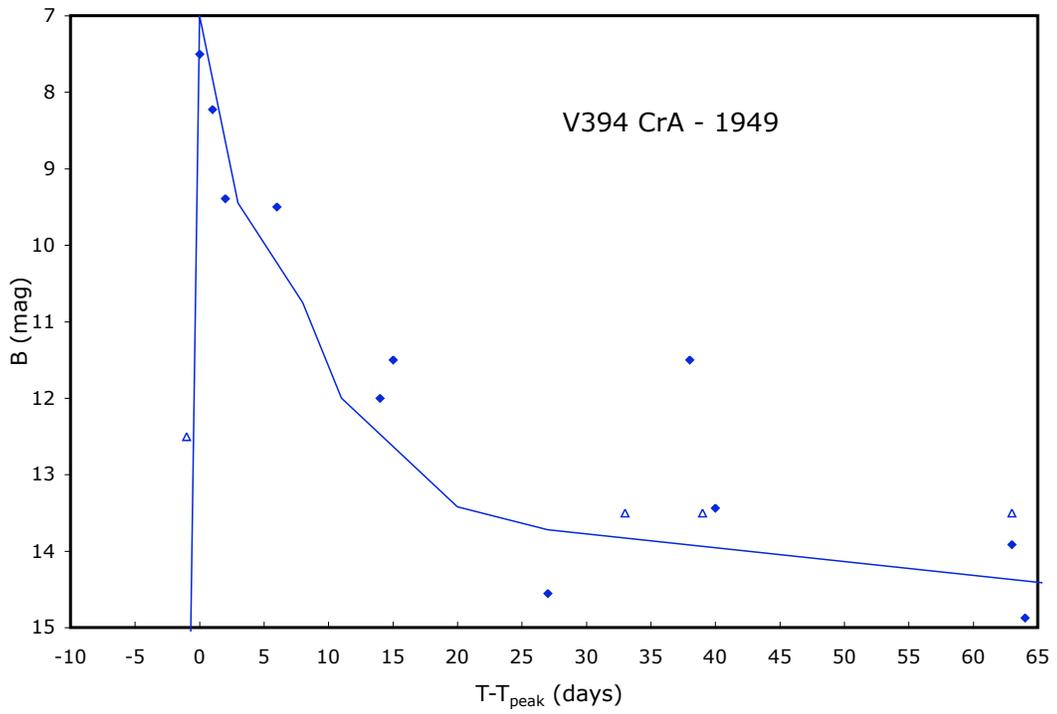}
\caption{
V394 CrA in 1949.  The diamonds are B-band magnitudes, while the triangles are limits (see Table 11).  The curve is the B-band template from Table 16.}
\end{figure}

\clearpage
\begin{figure}
\epsscale{1.0}
\plotone{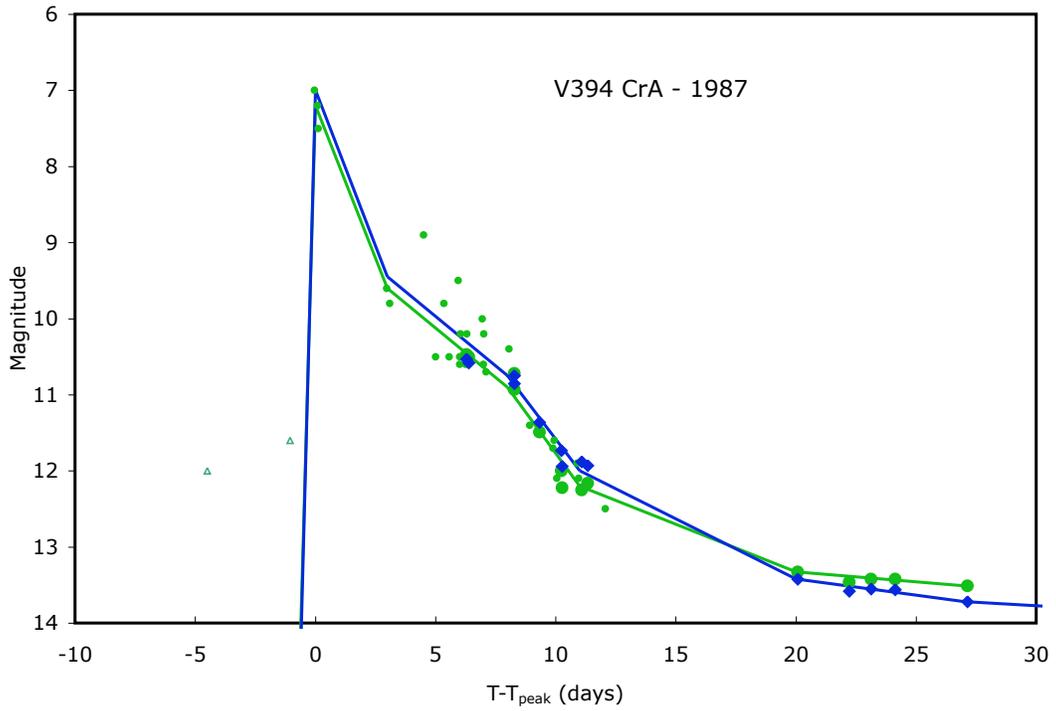}
\caption{
V394 CrA in 1987.  The diamonds are for B-band, while the circles are for the V-band.  The small symbols are for visual observations, while the large symbols are for the photoelectric magnitudes of Sekiguchi et al. (1989).  The two curves are the light curve templates for the two bands.}
\end{figure}

\clearpage
\begin{figure}
\epsscale{1.0}
\plotone{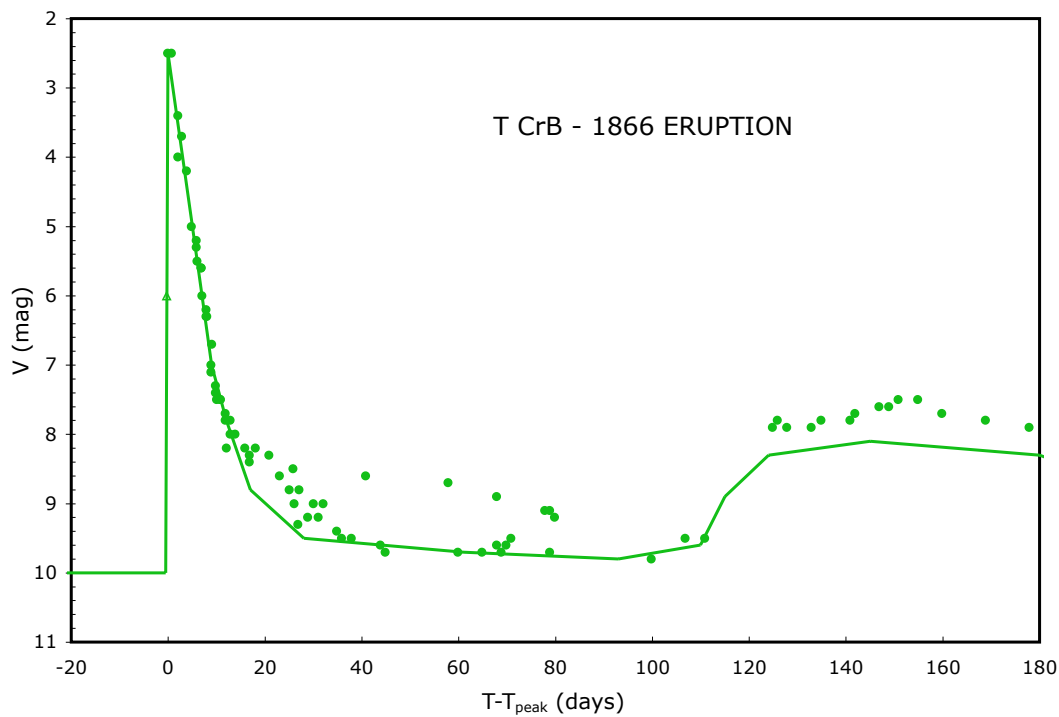}
\caption{
T CrB in 1866.  At this early date, we have only V-band photometry (diamonds).  The one limit (empty triangle) just before the discovery by J. Schmidt sharply constrains the time and magnitude of the peak.  The well-known, poorly-understood, and unique property of the light curve is that the fast nova event is followed by roughly fifty days at the quiescent level followed by a several month rebrightening.  The rebrightening gets above eighth magnitude and has comparable total energy to the initial fast nova event.  There is a series of magnitude estimates from day 21 to 80 by a single observer that lies above everyone else by about one magnitude, and this is certainly some sort of an error.  During the rebrightening, the light curve as observed by various people lies systematically above the V-band template as constructed from the 1946 light curve, and this is likely simply due to the known long-term variations in the red giant companion star.}
\end{figure}

\clearpage
\begin{figure}
\epsscale{1.0}
\plotone{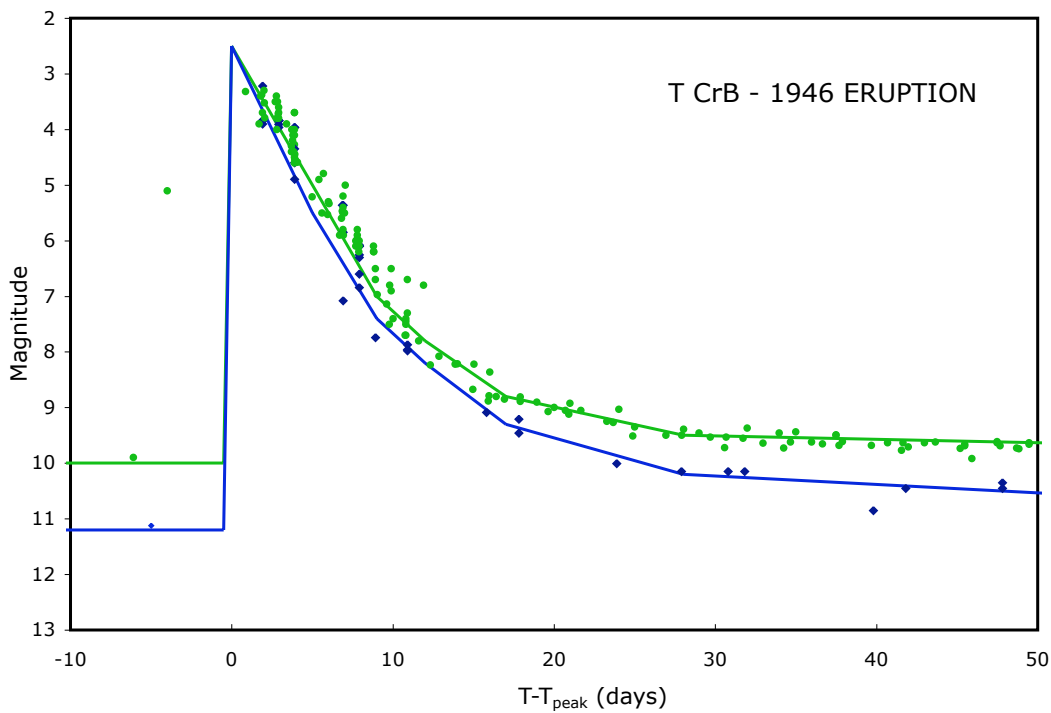}
\caption{
T CrB in 1946.  This light curve shows the B-band (diamonds) and V-band (circles) magnitudes only for the main peak.  Note that the light curve starts out fairly blue, but steadily returns to the much redder color of the underlying red giant companion star.  The peak magnitude is fairly uncertain, as a slightly earlier date would lead to a brighter peak and a slightly later date would lead to a fainter peak.  The peak chosen was simply that which gave a peak of V=2.5 which corresponds to the brightest observed magnitude for the 1866 outburst.  (The reported magnitude of 5.1 on day -4 is likely an error of date.)  The primary peak of T CrB corresponds to a very fast light curve.}
\end{figure}

\clearpage
\begin{figure}
\epsscale{1.0}
\plotone{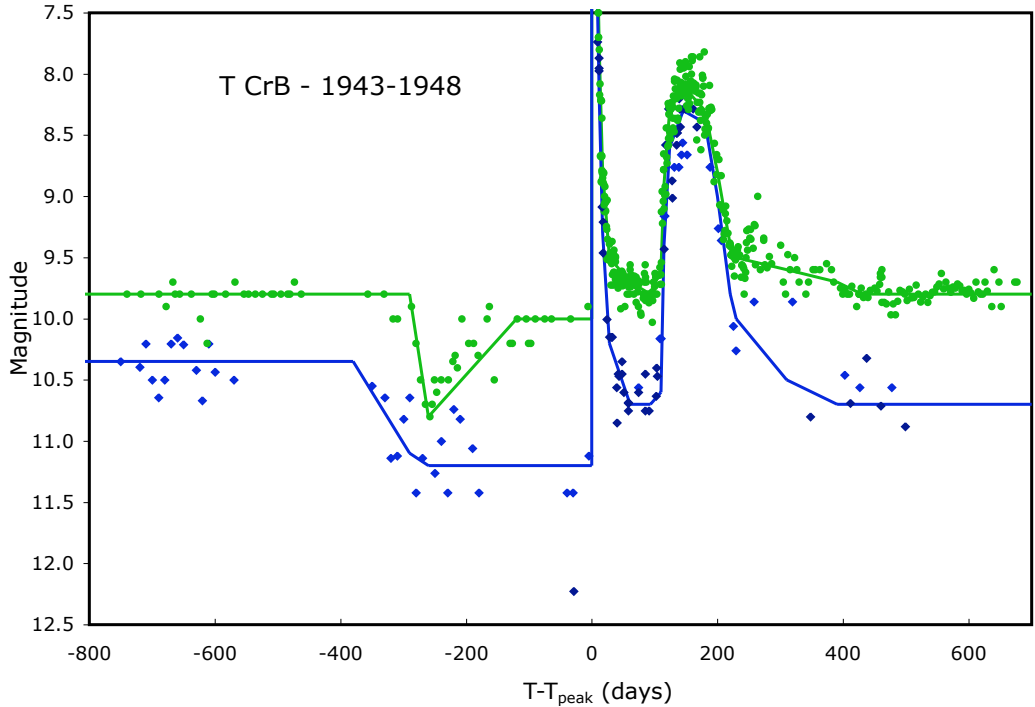}
\caption{
T CrB from 1943 to 1948.  This light curve shows the B-band (diamonds) and V-band (circles) magnitudes for an extended period around the 1946 eruption.  This extended light curve displays two unique and still-yet-unexplained phenomena.  First, the light curve has a significant and long-duration pre-eruption dip, with the colors in the dip changing throughout.  To get such a dip, either the red giant star would have to be dimmed or the accretion luminosity would largely have to turn off.  A further problem is that if the dip is causally connected to the eruption, then we must explain why the triggering of the nova event (related to the accumulation of matter on the surface of the white dwarf) has a dimming (related to either the accretion disk or red giant companion) soon to follow?  Second, after T CrB has returned to quiescence for roughly 50 days, it suddenly rebrightens to eighth magnitude.  Such a rise has never been seen in any other nova event.  Is this rebrightening caused by some accretion disk instability or by a further nova event arising from material falling back on the white dwarf?}
\end{figure}

\clearpage
\begin{figure}
\epsscale{1.0}
\plotone{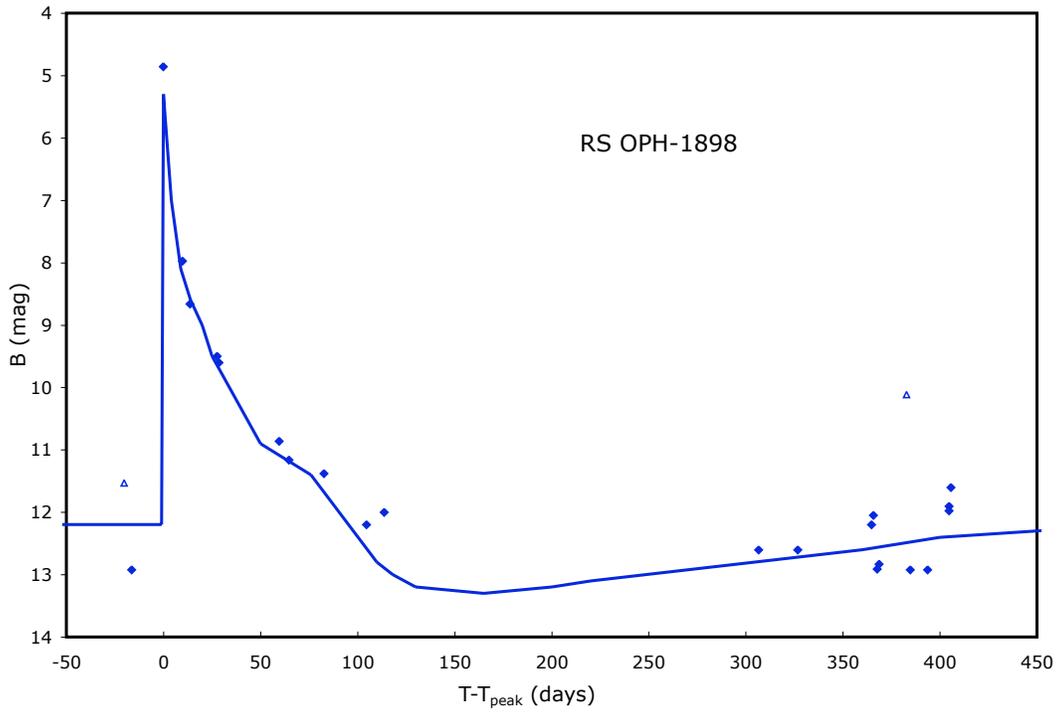}
\caption{
RS Oph in 1898.  As with all the light curves, the diamonds are for the B-band magnitudes, while the empty triangles are for limits.  All the data are from Harvard plates, and the fast-rise fast-decay is seen to match the template well.}
\end{figure}

\clearpage
\begin{figure}
\epsscale{1.0}
\plotone{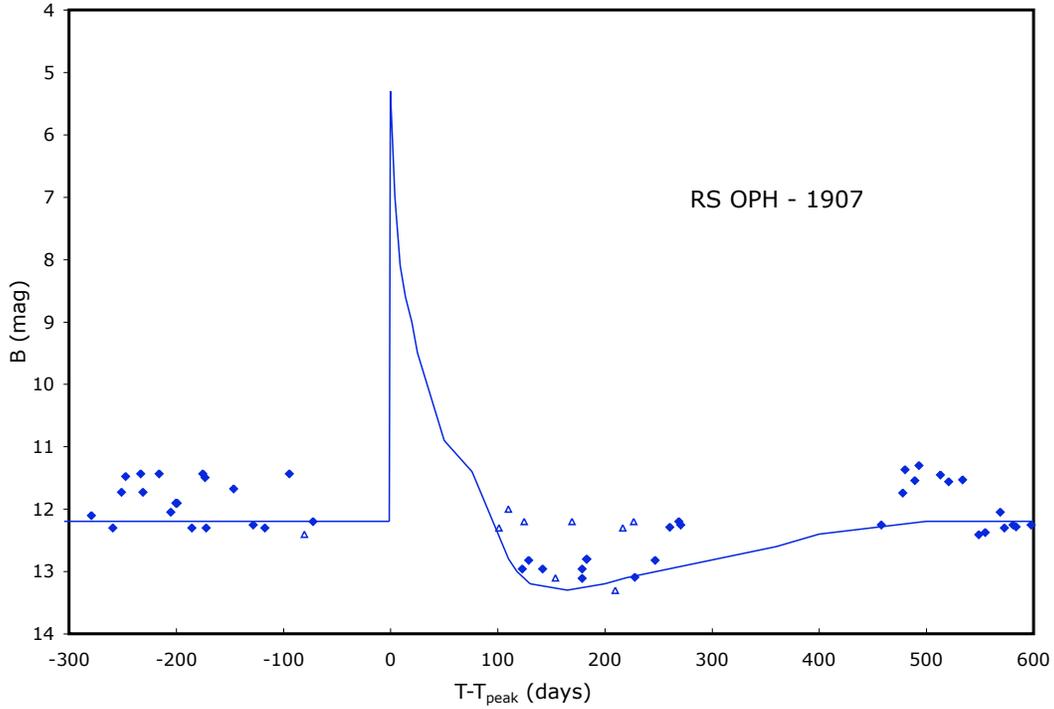}
\caption{
RS Oph in 1907.  The peak of the 1907 eruption was not seen (due to the usual solar gap), but RS Oph was a magnitude fainter than its quiescent brightness for 148 days.  This dip is highly significant, substantially fainter than the normal quiescent level, and completely unprecedented other than immediately after an eruption.  An easy and persuasive conclusion is that RS Oph had an eruption in 1907 with its peak lost in the solar glare, only to have the usual post-eruption dip producing the fading seen in the Harvard plates. }
\end{figure}

\clearpage
\begin{figure}
\epsscale{1.0}
\plotone{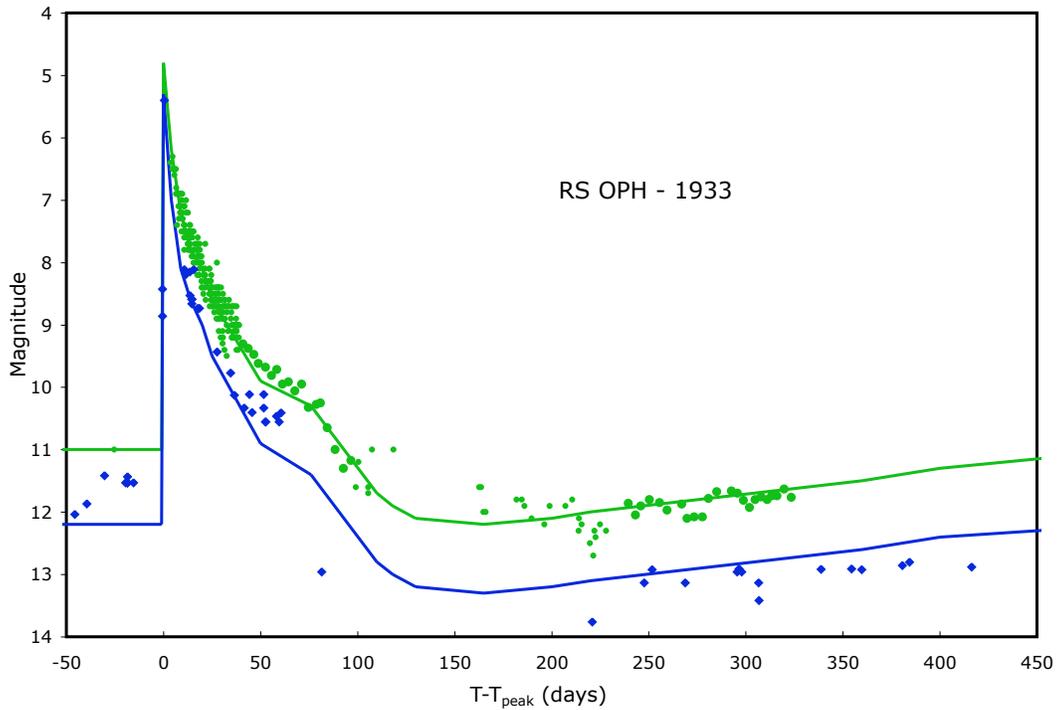}
\caption{
RS Oph in 1933.  This eruption is well observed, both in the blue (diamonds, all from Harvard) and visual (circles, all from AAVSO).  The large circles from days 40-98 and 239-324 are for 3-day averages of AAVSO data.  Two Harvard plates were fortuitously low in the rise and one at the peak, so that we have a good measure of the rise time, which is the last three magnitudes of the rise in 0.9 days.  Note that the post eruption dip has the same B-V color as does the red giant companion star.}
\end{figure}

\clearpage
\begin{figure}
\epsscale{1.0}
\plotone{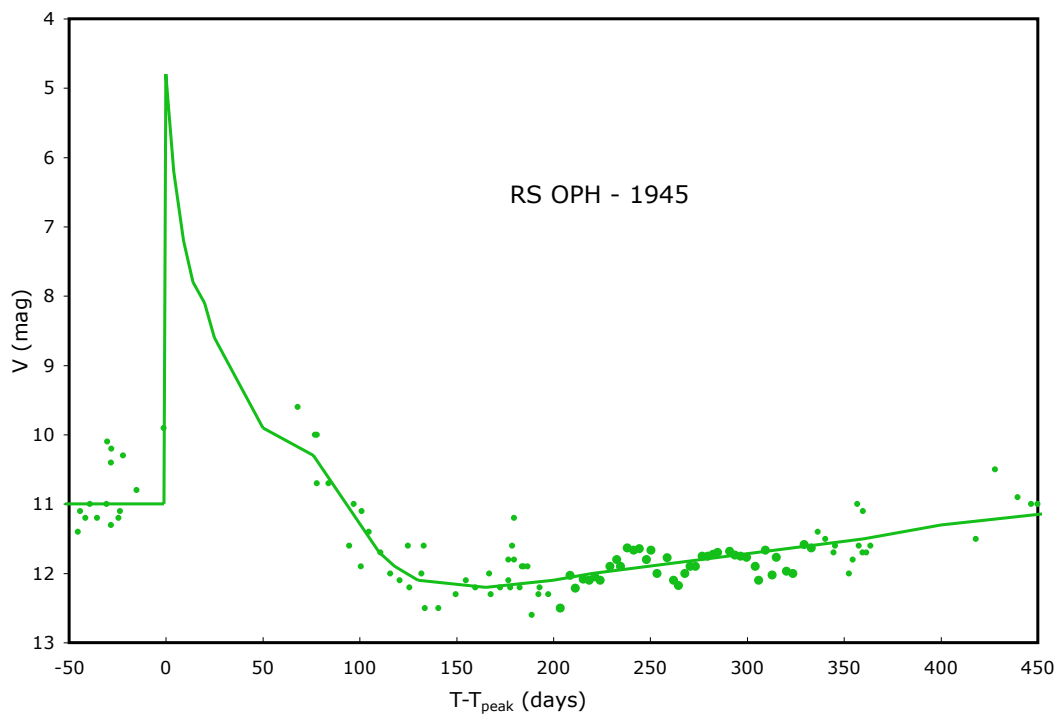}
\caption{
RS Oph in 1945.  Like the 1907 eruption, the peak of the 1945 eruption was missed inside the solar gap when RS Oph was behind the Sun.  Like in 1907, the post-eruption dip is easily seen and diagnostic of the missing eruption.  Unlike the 1907 eruption, the AAVSO V-band light curve (circles) displays a fading tail of the nova eruption.  Indeed, at V=9.6 mag just after the solar gap, RS Oph is brighter than at any other time in its observed history outside of eruption.  The small circles are for individual magnitude estimates and the large circles are for 3-day averages (for days 202-333) with typically 4.4 magnitudes for each point.}
\end{figure}

\clearpage
\begin{figure}
\epsscale{1.0}
\plotone{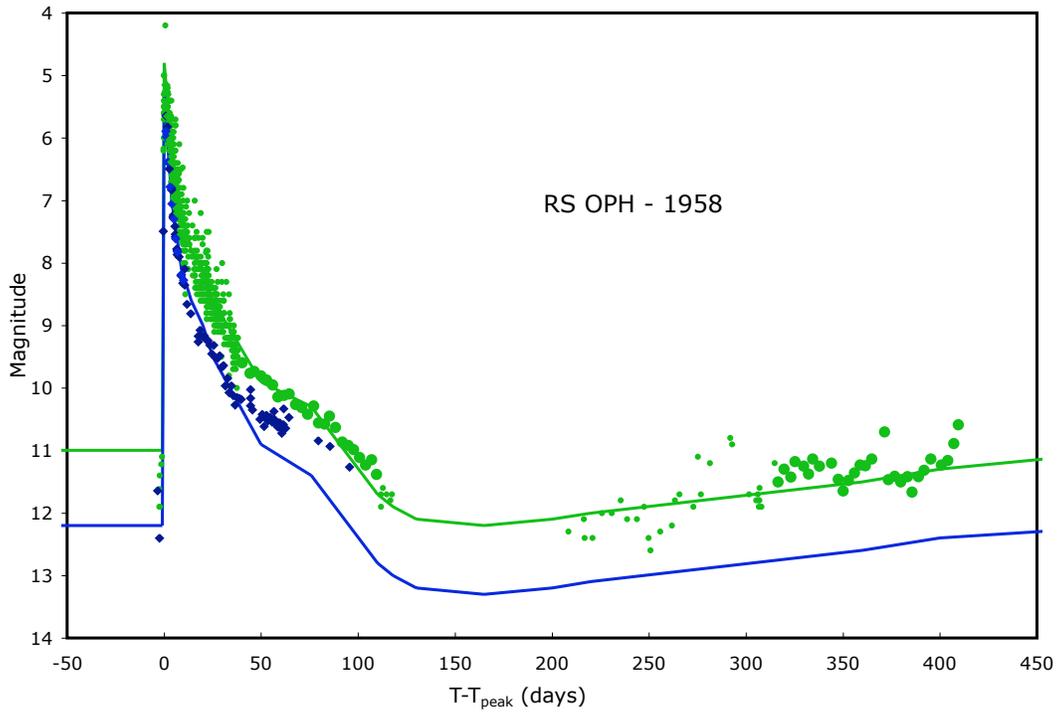}
\caption{
RS Oph in 1958.  The small circles are for individual AAVSO magnitudes, while the large circles are for 3-day averages of AAVSO data.  The time intervals with averages are from 40-110 days and 316-411 days, with averages of 14 and 6 magnitudes per average respectively.  The V-band light curve (circles) rise from quiescence to peak in 0.7 days.  The gap at the bottom of the post-eruption dip is the solar gap.}
\end{figure}

\clearpage
\begin{figure}
\epsscale{1.0}
\plotone{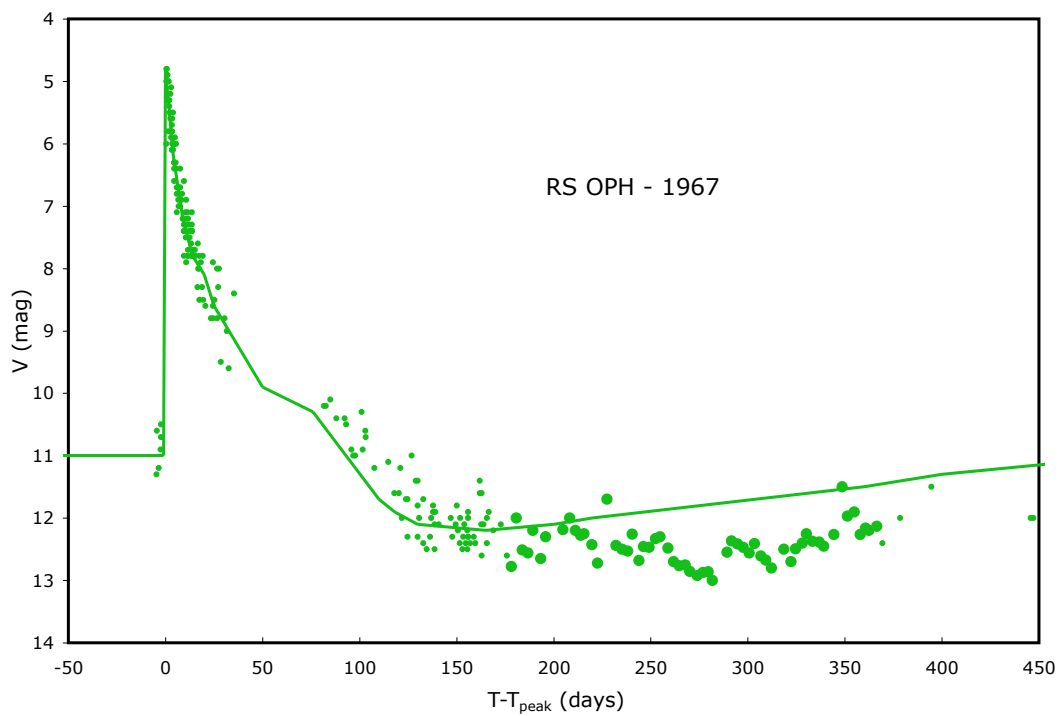}
\caption{
RS Oph in 1967.  All the data is from the AAVSO in the V-band.  The small circles are for individual magnitudes, while the large circles are for 3-day averages for days 177-367.  The V-band light curve well fits the template pretty well.  However, the light curve is somewhat high around day 100, and falls below  the template with a minimum around day 300.  I interpret this as simply being that the red giant companion star is fading and near minimum around day 300.}
\end{figure}

\clearpage
\begin{figure}
\epsscale{1.0}
\plotone{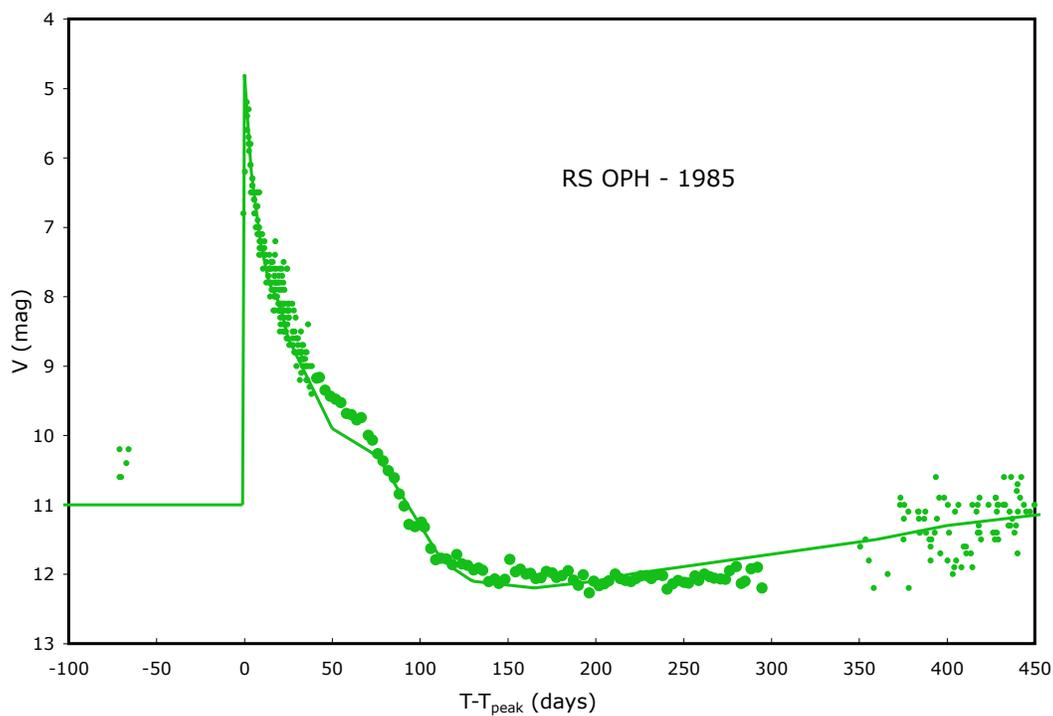}
\caption{
RS Oph in 1985.  The only extant magnitudes from this outburst are visual measures recorded in the AAVSO database.  The small circles are for individual observations, while the large circles are for the 3-day averages typically involving half-a-dozen or so magnitudes.  The amateur observations provide a well-defined V-band light curve, while the professionals left no record.}
\end{figure}

\clearpage
\begin{figure}
\epsscale{1.0}
\plotone{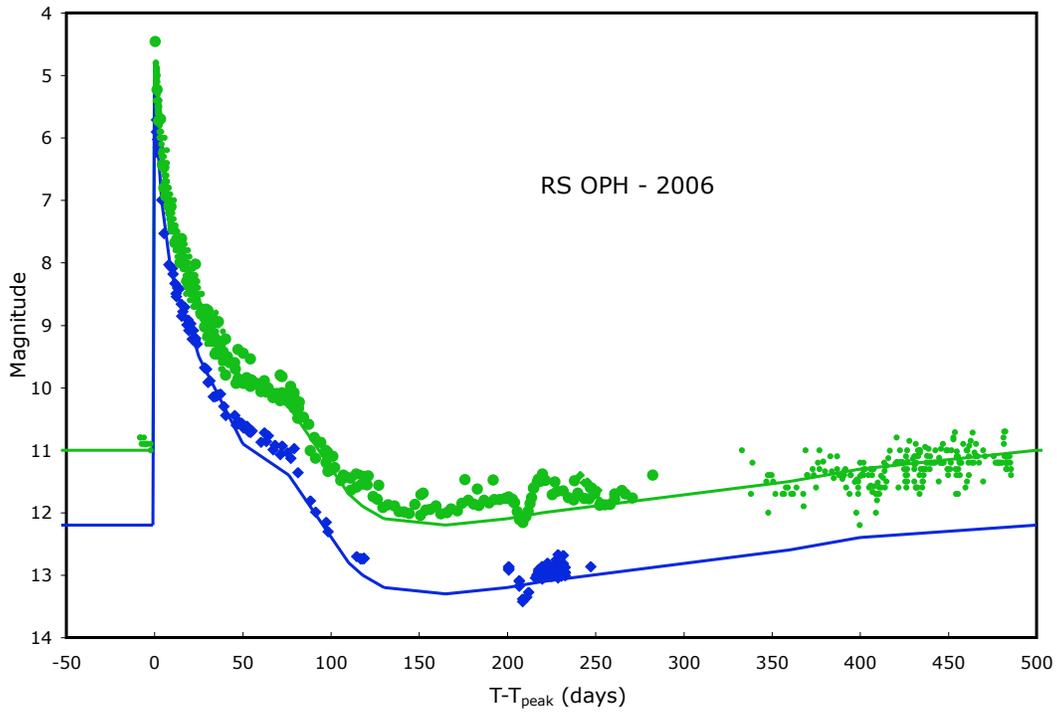}
\caption{
RS Oph in 2006.  The 2006 eruption has a well defined light curve in the blue (diamonds) and visual (circles), with large symbols indicating either CCD measures or averages of 1-day or 3-day.  Interestingly, there is a short-term dip in the light curve around day 210 (near the bottom of the post-eruption dip).  From day 203 to day 220, RS Oph dimmed by about half a magnitude and then recovered its brightness.  This dip has not been previously recognized and I know of no ready explanation.}
\end{figure}

\clearpage
\begin{figure}
\epsscale{1.0}
\plotone{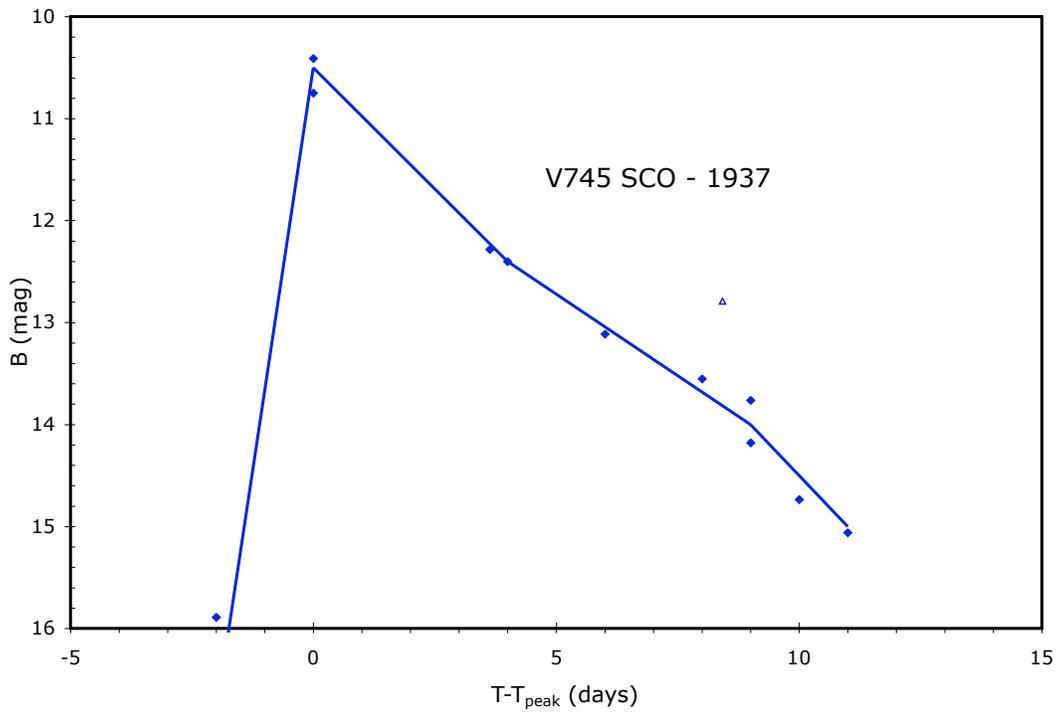}
\caption{
V745 Sco in 1937.  All the B-band magnitudes (diamonds) are either from the Harvard or Leiden plates.  The plates defined a typical fast-raise very-fast-decline light curve.}
\end{figure}

\clearpage
\begin{figure}
\epsscale{1.0}
\plotone{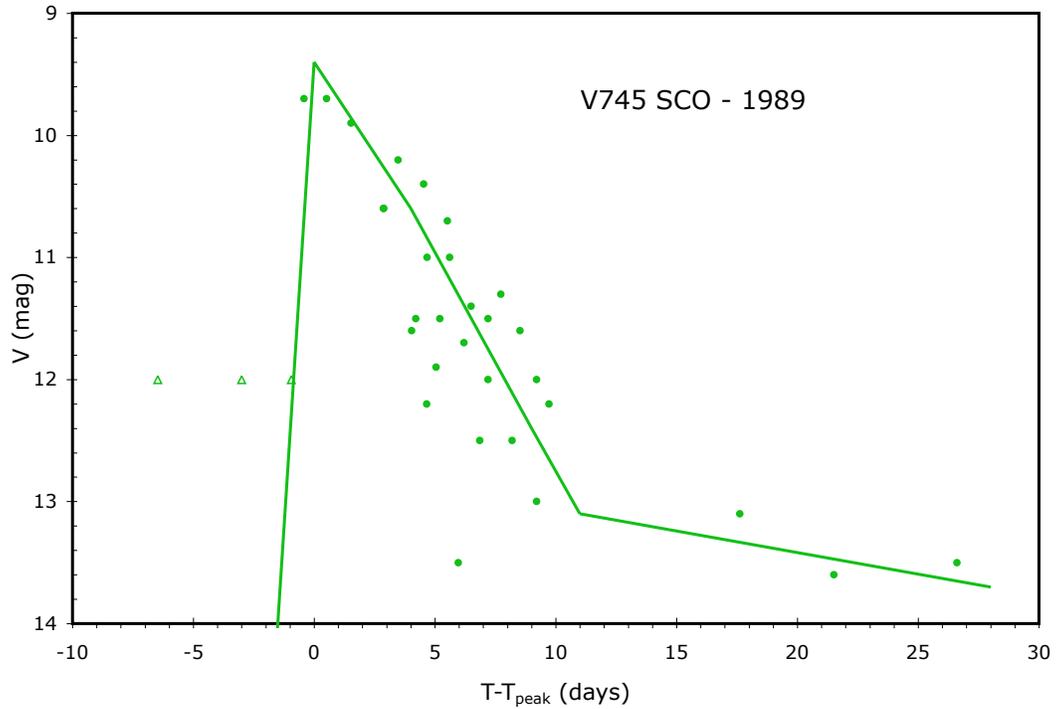}
\caption{
V745 Sco in 1989.  Only V-band magnitudes (filled circles) and some useful limits (empty triangles) are known for this brief and faint event.  The light curve displays a large amount of scatter, where light curves from separate observers have systematic offsets from each other.  No CCD photometry is available, so it is difficult to know where the true light curve lies.  The cause of this scatter is likely inconsistent comparison star sequences and color terms.  The observing details are not known, so I cannot correct the reported magnitudes onto a consistent modern magnitude scale.}
\end{figure}

\clearpage
\begin{figure}
\epsscale{1.0}
\plotone{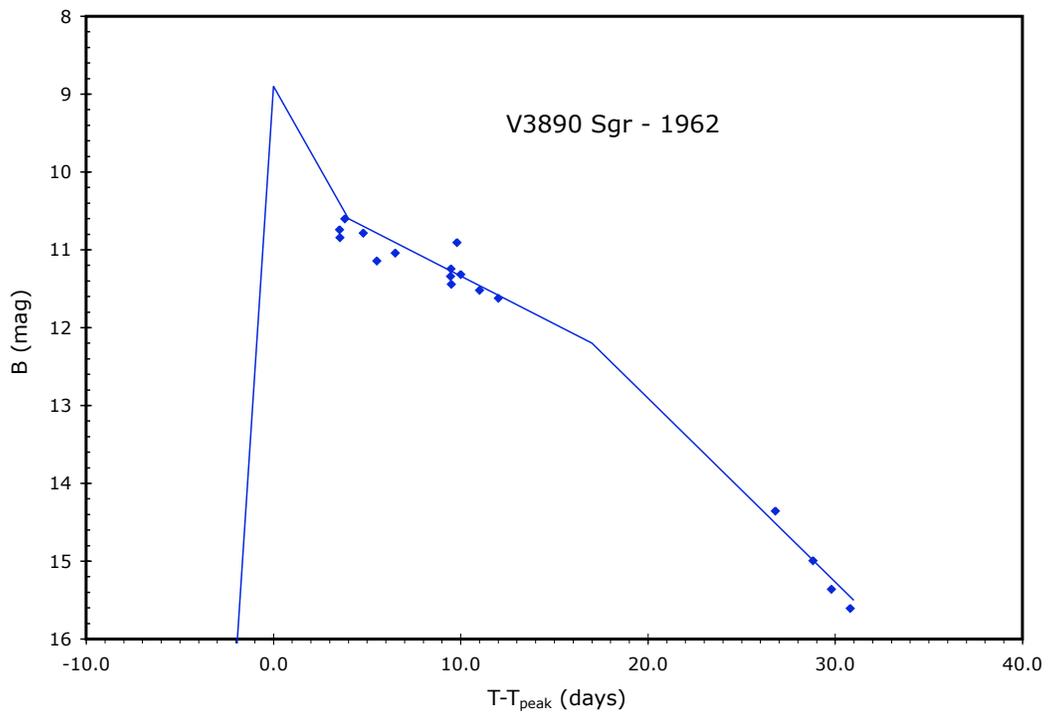}
\caption{
V3890 Sgr in 1962.  All of the B-band light curve (diamonds) are from the Maria Mitchell and Sonneberg plates.  The fit to the template (broken line) is reasonable.}
\end{figure}

\clearpage
\begin{figure}
\epsscale{1.0}
\plotone{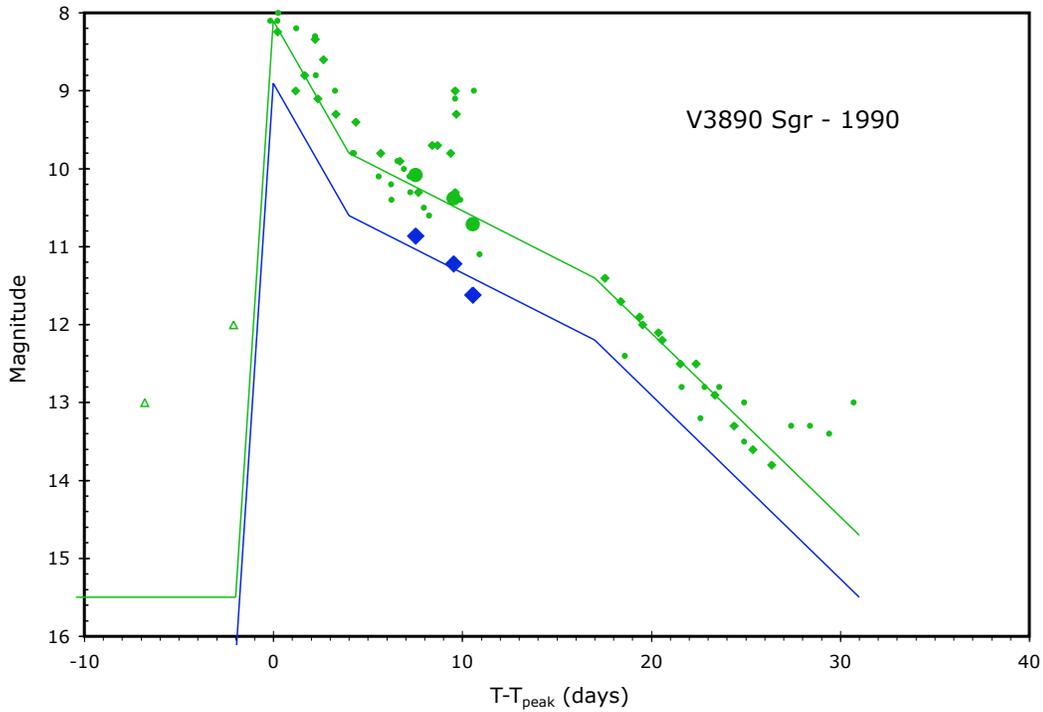}
\caption{
V3890 Sgr in 1990.  The diamonds are for B-band magnitudes, the circles are for V-band magnitudes, and the large symbols are for the CCD magnitudes of Buckley et al. (1990).  The general shape of the B-band template from 4-to-31 days after maximum is determined from the 1990 data, the B-V color around day 10 is determined by the CCD measures, and the peak of the B-band template is taken to be the V-band peak light curve with a constant offset.  As such, any color evolution of the light curve near the peak could result in a peak B magnitude different from that quoted in the template.}
\end{figure}

\clearpage
\begin{figure}
\epsscale{1.0}
\plotone{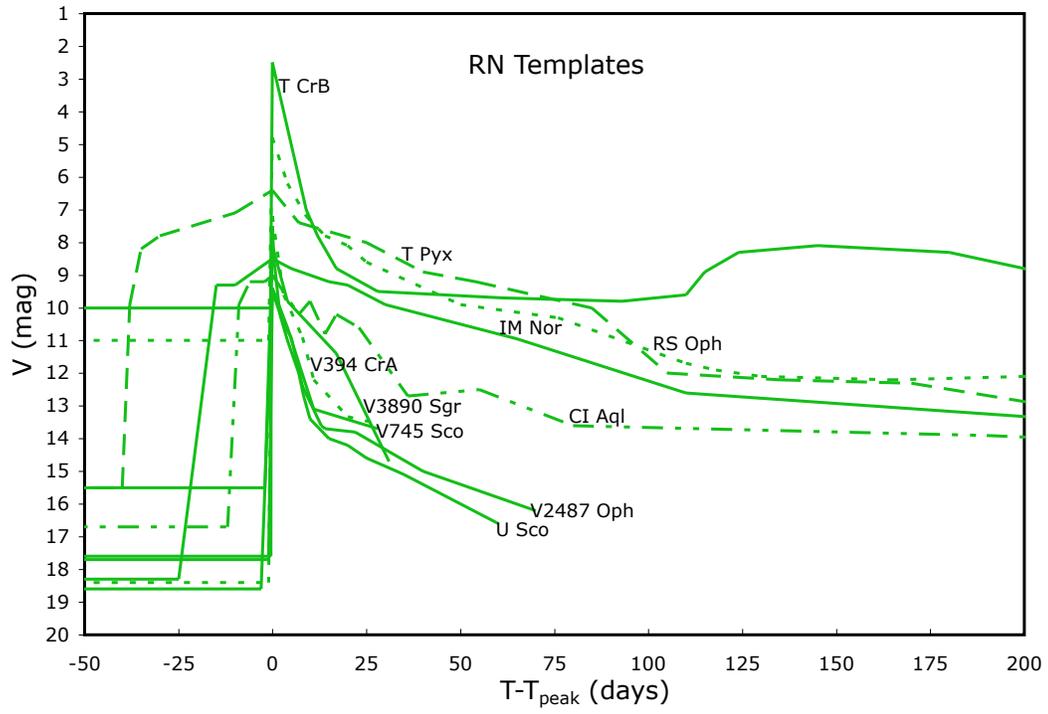}
\caption{
RN templates.  The V-band templates for all ten known galactic RNe are collected here in one plot.}
\end{figure}

\clearpage
\begin{figure}
\epsscale{1.0}
\plotone{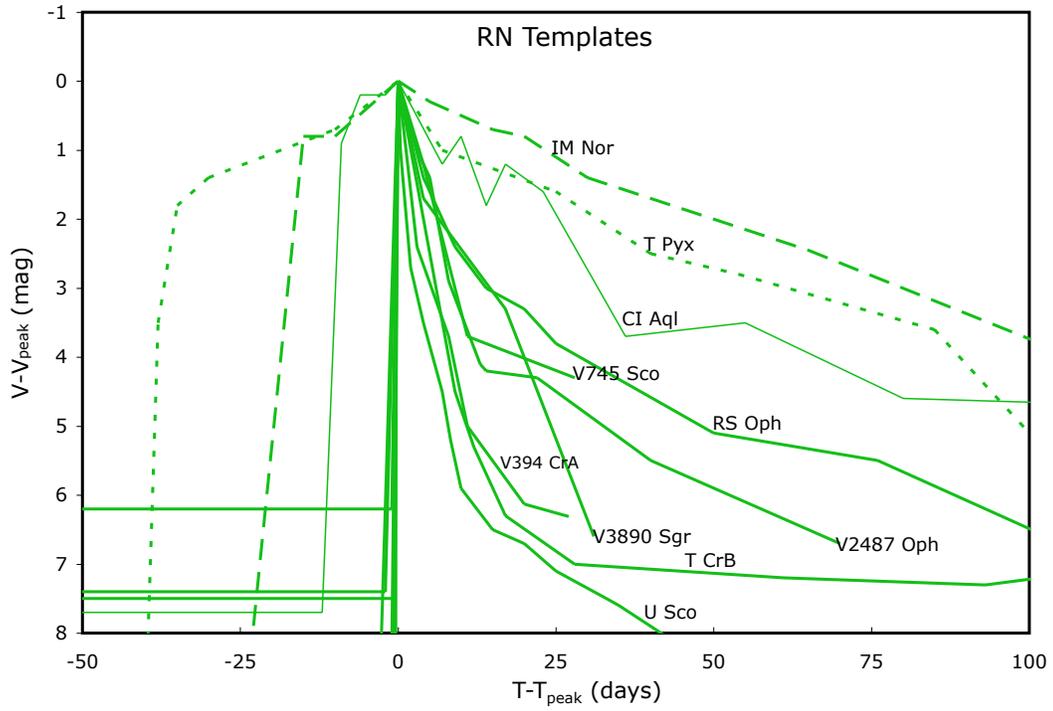}
\caption{
RN templates shifted to a common peak.  The V-band templates from Figure 36 are here all shifted vertically so that they coincide at $V-V_{peak}=0$ and $T-T_{peak}=0$.}
\end{figure}

\clearpage
\begin{figure}
\epsscale{1.0}
\plotone{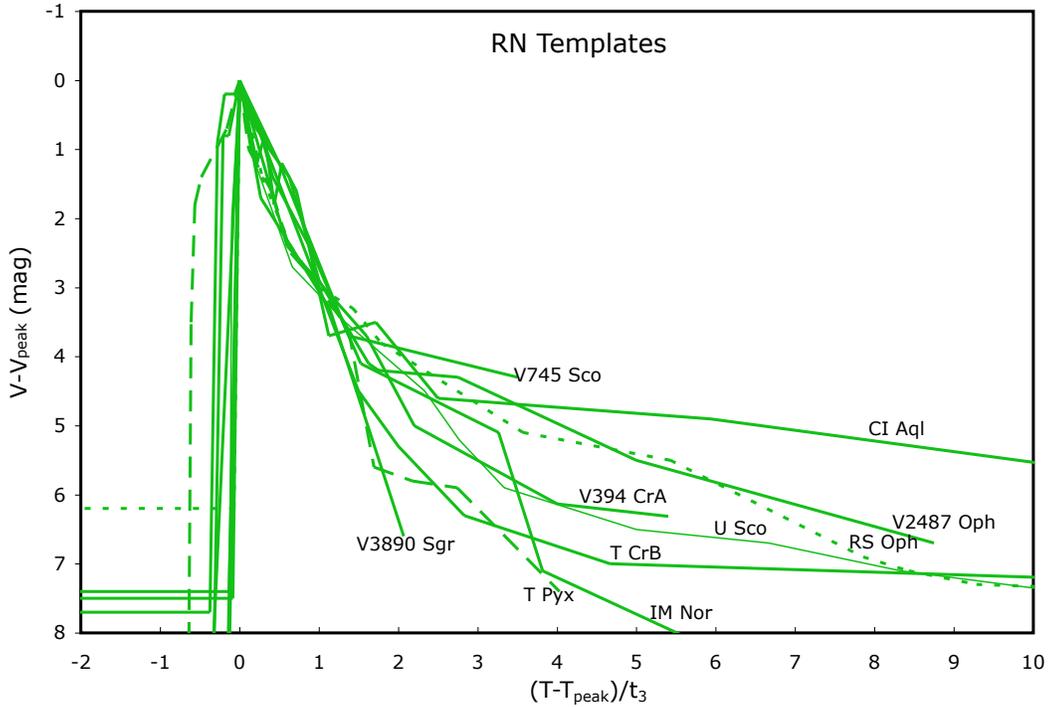}
\caption{
RN templates scaled in time.  The V-band templates are here shifted to a common peak and with the time scaled by the $t_3$ value.  By construction, the shifted templates will all coincide at two points, with 0 and 1 on the horizontal axis.  This plot can be used to compare the {\it shapes} of the light curves.  We see that the RNe have fairly similar shapes, although much of this is by construction.  We do not see any obvious plateaus, nor other features such as dust formation dips.}
\end{figure}

\clearpage
\begin{figure}
\epsscale{1.0}
\plotone{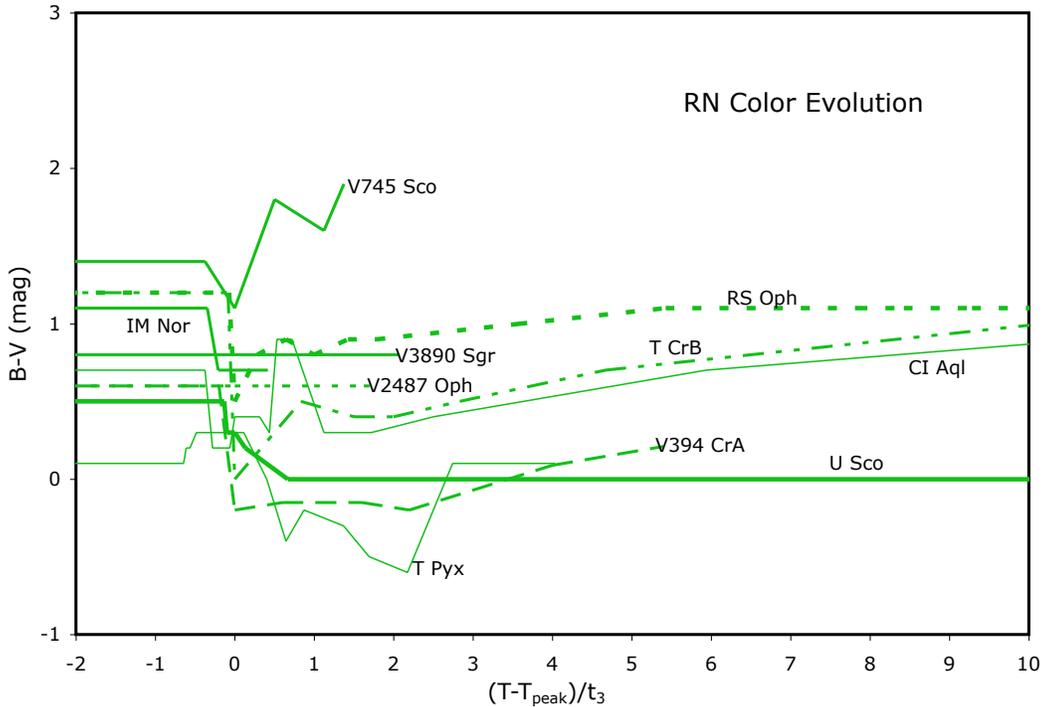}
\caption{
RN color curves.  We see that the RNe generally do not have strong color evolution.  Nevertheless, three of the curves have apparent reddening from the left to the right side of this plot.  Two of these are the case of T CrB and RS Oph for which the blue nova light becomes dominated by the red companion star light as the nova fades.  (V745 Sco also is reddening systematically through the observed portion of its eruption, but the horizontal range on this plot is limited, the change in color is fairly small, and there are large uncertainties in the color evolution.)  But the other case with a likely red giant companion (V3890 Sgr) has little if any color shift, although the resolution might be simply that the fading tails of the eruption were not followed long enough. The other case with an observed reddening trend over much of its tail (CI Aql) does not have a simple explanation that I know of.  Color changes could easily be expected due to varying contributions from emission lines in the B and V bands which could change substantially throughout the eruption.}
\end{figure}

\clearpage
\begin{figure}
\epsscale{1.0}
\plotone{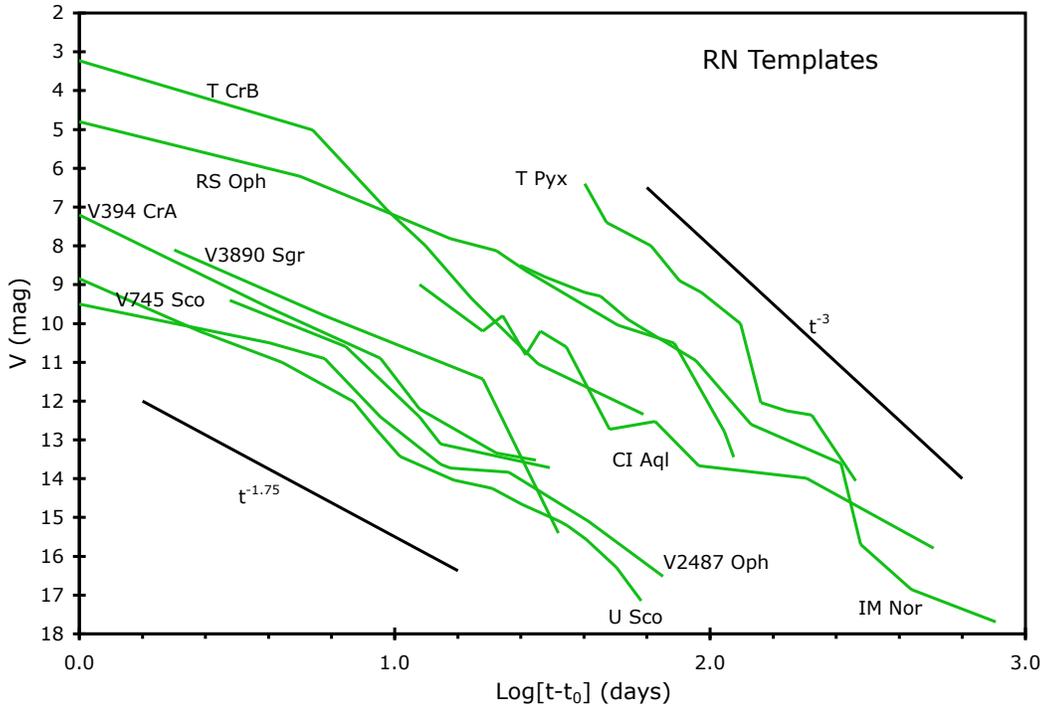}
\caption{
RN light curves as power laws.  The declines for all ten RNe are presented on a plot of V-band magnitude versus $t-t_0$ (the time after the first rise in the eruption).  If the flux varies as a power law, then the light curve would appear as a straight line on this plot.  The power law index ($\alpha$) will equal to the slope in this plot divided by 2.5.  Many of the RNe appear as roughly a broken power law.  However, there is substantial variation about this broken power law, arising from flares and plateaus.  The median $\alpha$ values are -1.4 and -2.9 at early and late times, with break times from 6 to 77 days plus IM Nor at around 235 days.   This is to be compared with the `universal decline law' of Hachisu \& Kato (2007), where the $\alpha$ values should be -1.75 and -3.0 at early and late times, with break times up to 80 days for RNe.}
\end{figure}

\clearpage
\begin{figure}
\epsscale{1.0}
\plotone{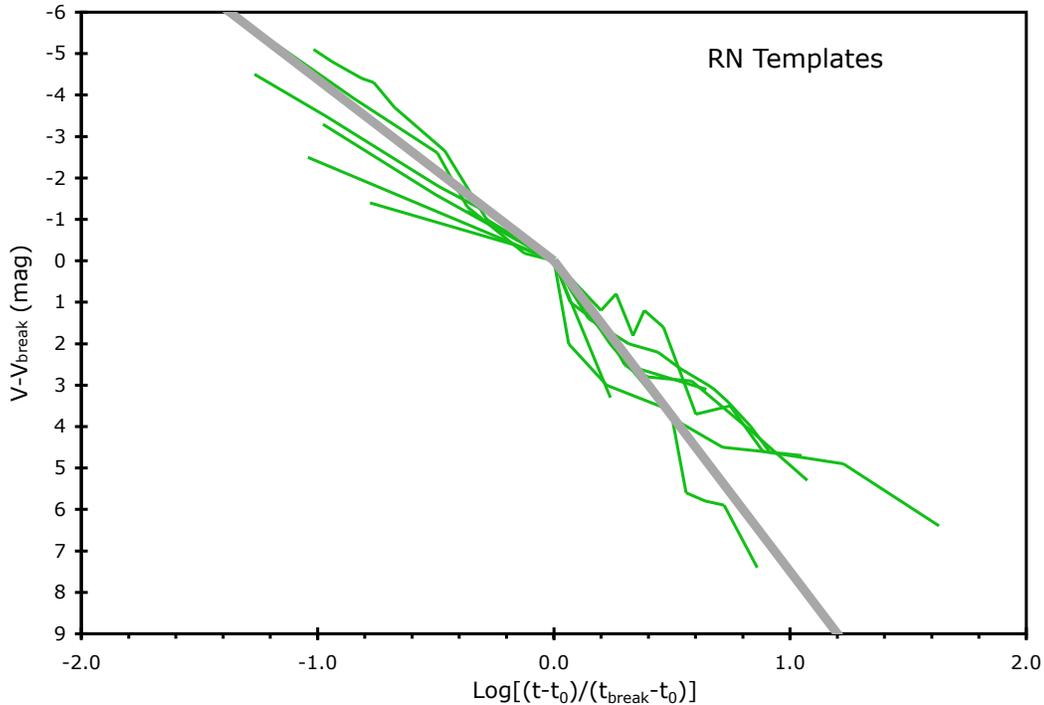}
\caption{
RN light curves superposed.  The light curves from Figure 40 have been shifted vertically and horizontally so that the fitted breaks are placed at the origin.  The theoretical model (with $\alpha$ values of -1.75 and -3.0) are represented by the thick gray line.  We see a general agreement with the `universal decline law' of Hachisu \& Kato (2007), for the predicted early and late time slopes as displayed by the labeled line segments.  This agreement is not perfect, as the RN light curves show variations around the simple predicted broken power law.  Part of the reason is that the observations are using the V-band which includes bright emission line flux (starting during the transition region), whereas the theoretical prediction is for continuum flux alone.}
\end{figure}

\clearpage
\begin{figure}
\epsscale{1.0}
\plotone{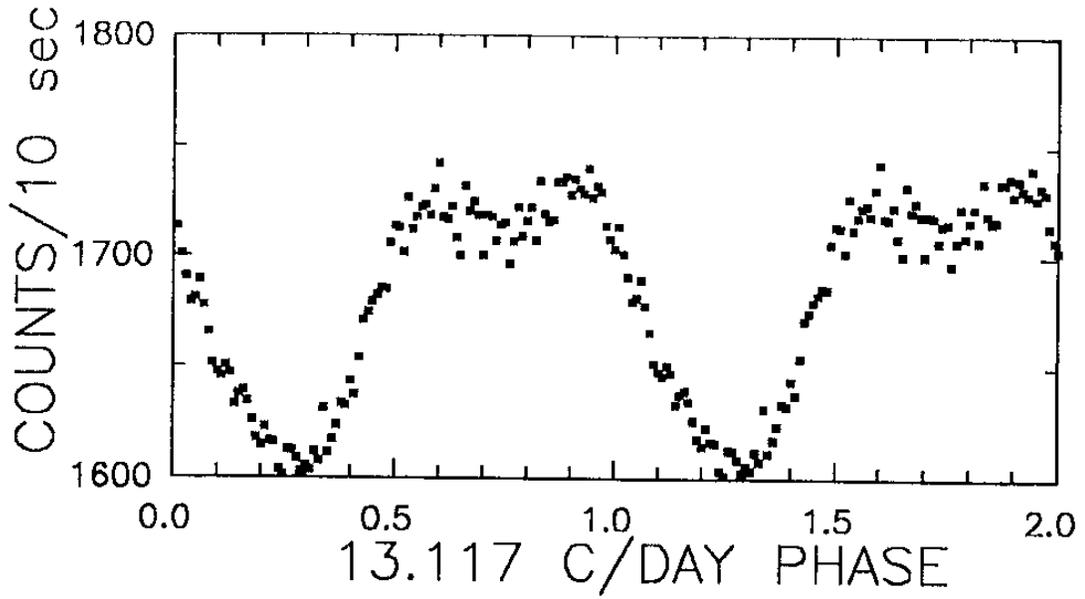}
\caption{
T Pyx light curve.  This figure, taken from Patterson et al. (1998), shows the light curve of T Pyx folded on a period of 13.117 cycles per day (0.076237 days).  The plot shows two full periods, so T Pyx only has one dip per cycle.  We see a broad dip, which is too broad to be from an eclipse alone.  Patterson et al. interpret the dip as resulting from the heating of one face on the companion star.}
\end{figure}

\clearpage
\begin{figure}
\epsscale{0.70}
\plotone{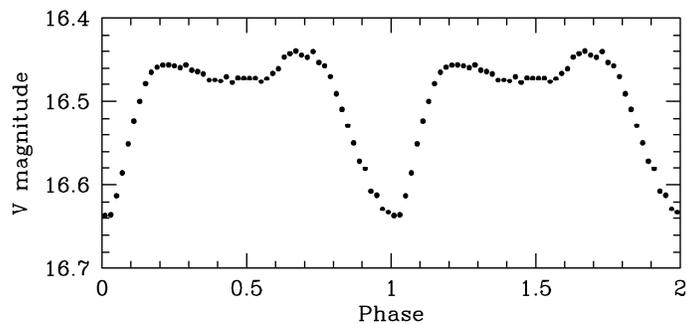}
\epsscale{0.62}
\plotone{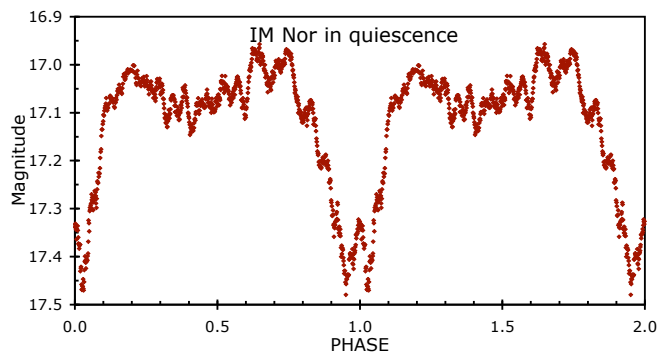}
\caption{
IM Nor light curve.  Both panels are folded on the 0.1026 day period.  The top panel is taken from Woudt \& Warner (2003) and shows the IM Nor on four nights from JD2452696-2452701 in the tail of the eruption.  The bottom panel is a running average of the unfiltered CCD magnitudes of Monard from JD2453179 in quiescence.  We see a broad dip, which is too broad to be from an eclipse alone.  Woudt \& Warner interpret the dip as being primarily a reflection effect from the inside face of the companion star plus a partial eclipse of the accretion disk.  The folded light curve is essentially identical between the two panels, despite the change in irradiation from eruption to quiescence.}
\end{figure}

\clearpage
\begin{figure}
\epsscale{1.0}
\plotone{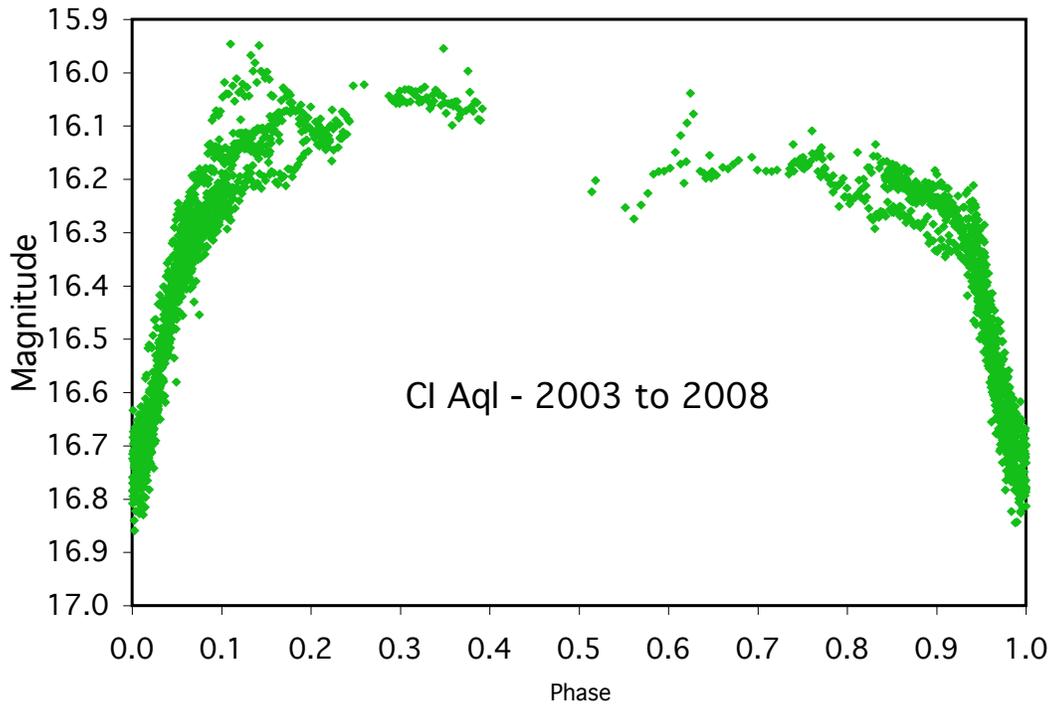}
\caption{
CI Aql light curve.  This figure is constructed with data from 2003-2008 taken at McDonald Observatory and Cerro Tololo.  The light curve is asymmetric, with the extra light centered around phase 0.15.  This extra light is variable, which suggests that it arises in the hot spot.}
\end{figure}

\clearpage
\begin{figure}
\epsscale{1.0}
\plotone{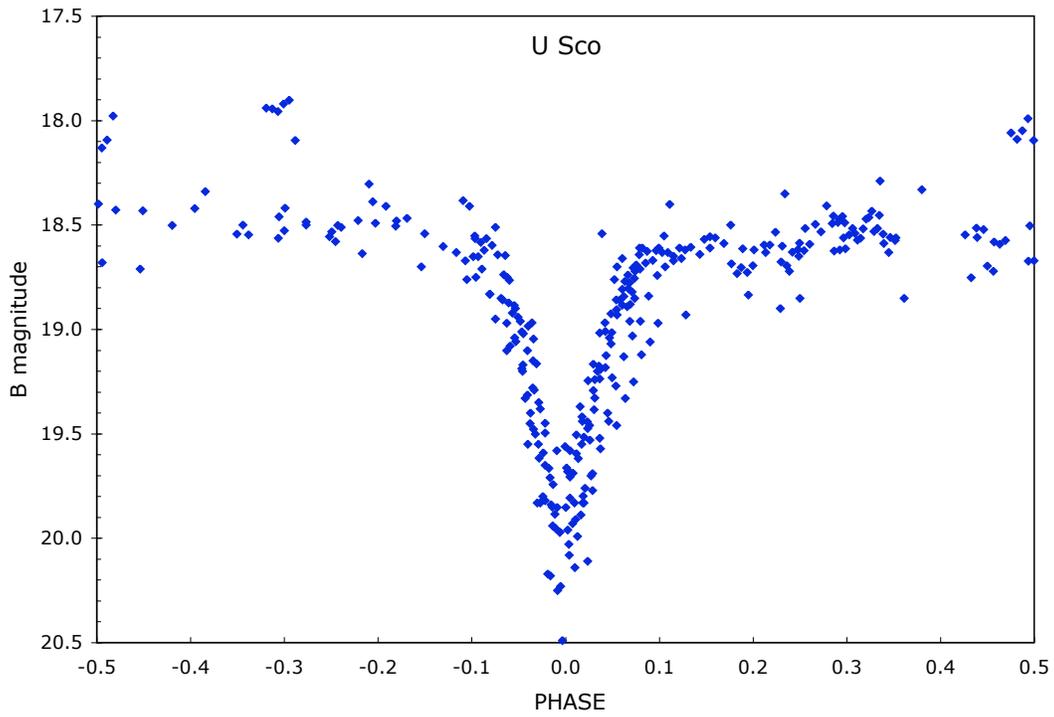}
\caption{
U Sco in the B-band.  U Sco has a prominent and deep eclipse.  Outside of eclipse, the light curve is flat (with no apparent secondary eclipse) with substantial flickering.  The quiescent level appears to change from orbit-to-orbit from B=18.8 to B=18.0, even though the average is close to B=18.5.  This folded light curve is from my Cerro Tololo and Kitt Peak data from 1988-1998.}
\end{figure}

\clearpage
\begin{figure}
\epsscale{1.0}
\plotone{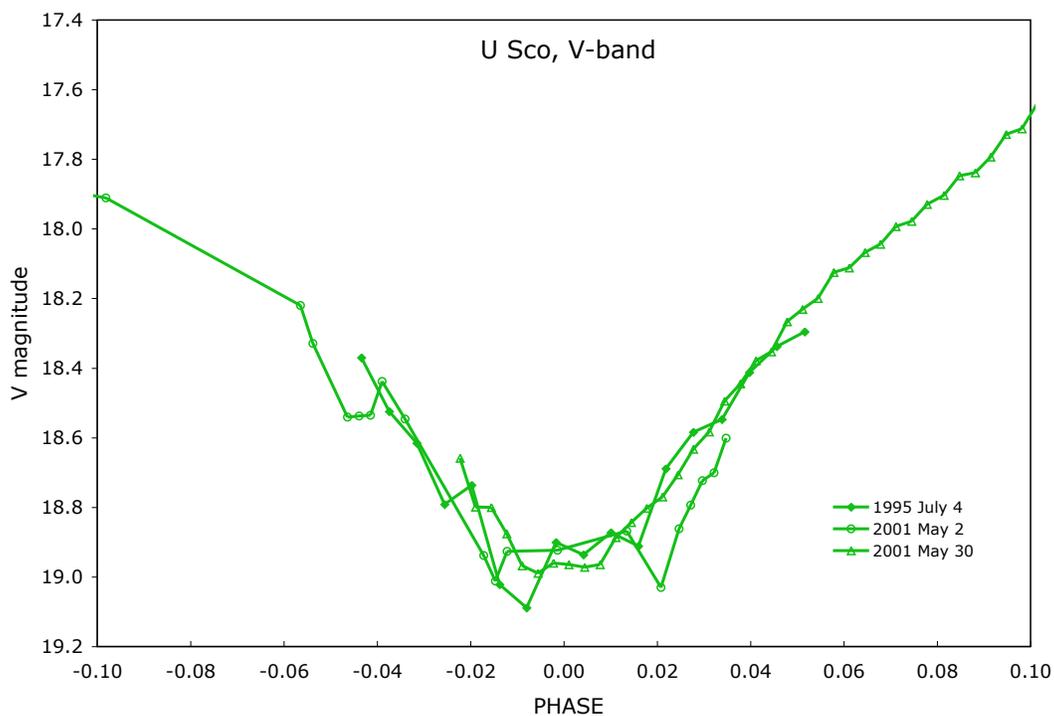}
\caption{
The U Sco eclipse in the V-band.  Three of my many eclipse time series are in the V-band, and these apparently show a flat bottom.  A reasonable conclusion is that the eclipse is total, so that the minimum light is only coming from the non-irradiated side of the companion star.  As such, I conclude that the apparent magnitude of the companion star alone is close to V=18.9 mag, and this will provide the primary input for the best distance determination of U Sco.}
\end{figure}

\clearpage
\begin{figure}
\epsscale{1.0}
\plotone{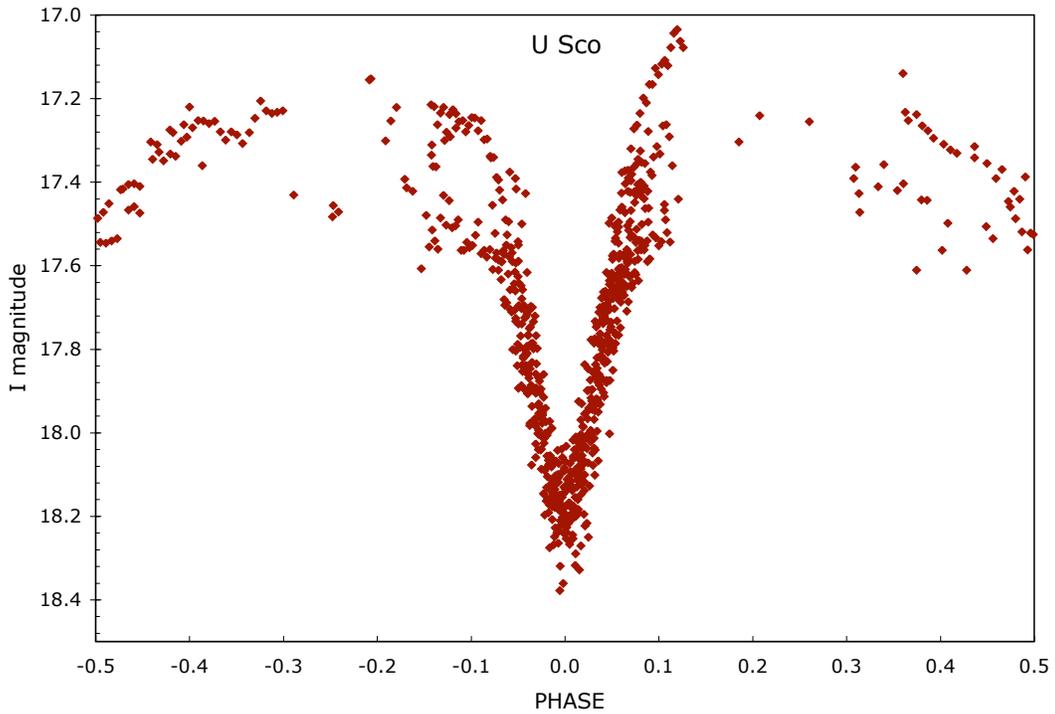}
\caption{
U Sco in the I-band.  This light curve is similar to that in the B-band (see Figure 45), with the most important difference being that the secondary minimum is prominent, with an amplitude of typically around 0.3 mag.  This is not surprising as the companion star is much redder than the accretion disk, so the secondary minimum will become more prominent as the observing band moves to the red.}
\end{figure}

\clearpage
\begin{figure}
\epsscale{1.0}
\plotone{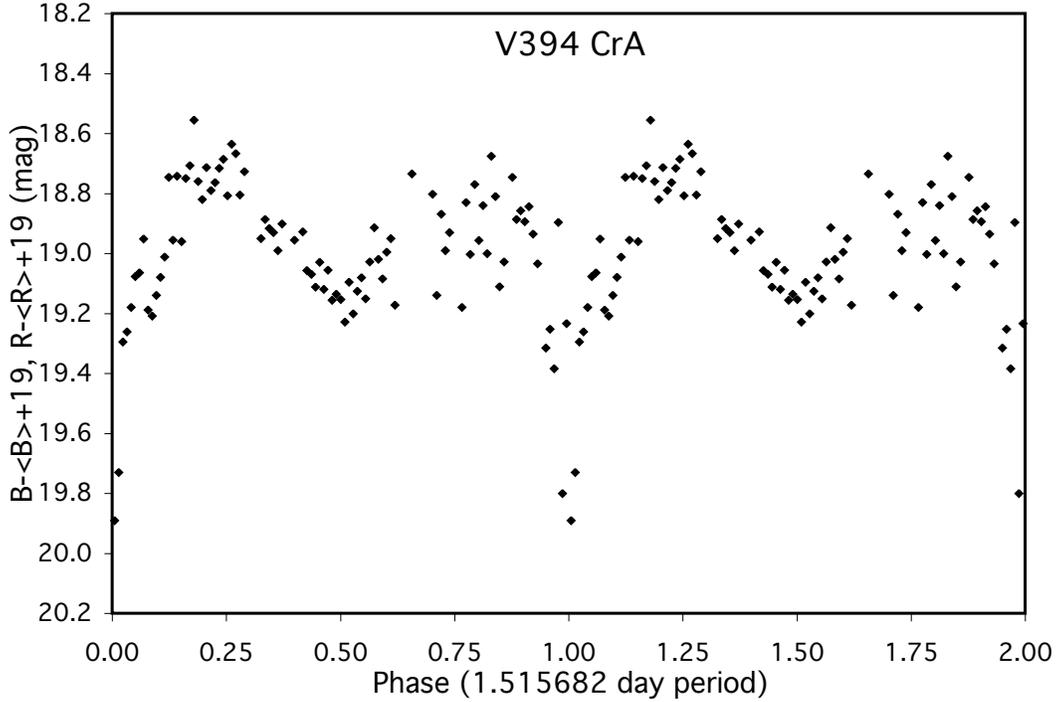}
\caption{
V394 CrA folded and binned light curve for the 1989 and 2005 data only.  V394 CrA has a highly significant photometric modulation (with a period of 0.757841$\pm$0.000004 days) superposed on a light curve that rises and falls on a much longer time scale.  The modulation is most prominent when the system is faintest, as in the years 1989 and 2005.  The stability of this periodicity from 1989-2008 implies that the modulation is associated with the orbital period.  But $P_{orb}$ might be twice this value, because the odd and even maxima and minima are at different levels.  In an effort to subtract out the extra (non-periodic) flux, I have taken the B-band and R-band magnitudes from Table 22 and subtracted out average magnitudes over $40$-day bins, and then added 19.  This figure displays a folded and binned light curve for a period of 1.515682 days for just the data of 1989 and 2005.  The primary minimum (at phase zero) is deeper than the secondary minimum (at phase half), while the maximum after the primary minimum is brighter than the maximum after the secondary minimum.  The light curve displays little variability between phases 0.1-0.6 and substantial variability between phases 0.6-1.1, with this being a clue to the position of the source and directionality of the source of the flickering.}
\end{figure}

\clearpage
\begin{figure}
\epsscale{1.0}
\plotone{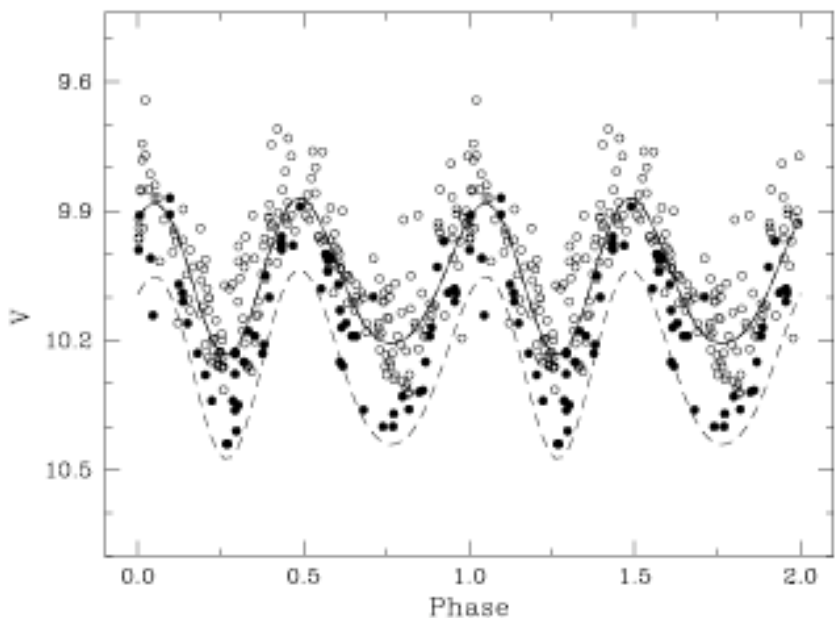}
\caption{
T CrB ellipsoidal oscillations.  This light curve is reproduced from Zamanov et al. (2004).  The figure shows the V-band magnitude as folded with a 227 day orbital period.  These data are from their extensive UBV photometry of T CrB over a three year interval.  The empty circles are for points where T CrB is brighter than U=12 mag, while the filled circles are for points with the system brighter than that limit.  Note that the phase range is doubled, and that T CrB displays two minima per period.  The upper curve is a sine wave fit to all their magnitudes, and thus represents both the red giant plus the accretion disk and the flickering in the hot spot.  The lower curve represents the underlying red giant star contribution.}
\end{figure}

\clearpage
\begin{figure}
\epsscale{1.0}
\plotone{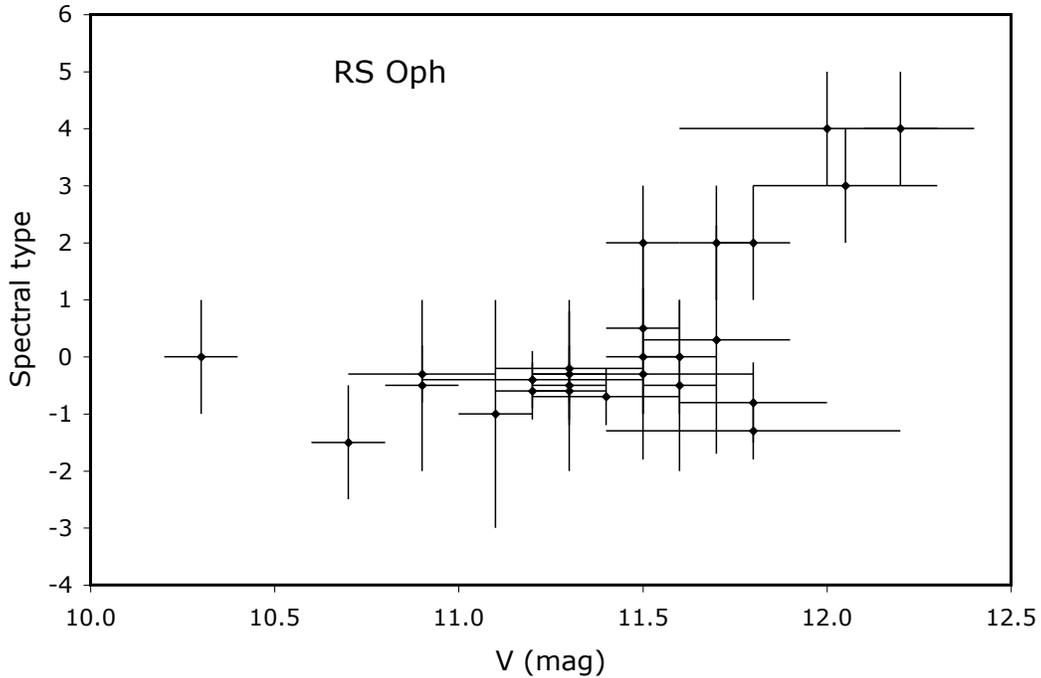}
\caption{
RS Oph spectral type.  The spectral type of RS Oph has been measured many times (see Table 24).  There is no correlation with orbital phase, likely because RS Oph is being viewed not far from pole-on.  But there is a highly significant correlation between the visual magnitude and the spectral type.  When RS Oph is faint the spectral class is M4, while when RS Oph is bright the spectral class in K4, with roughly linear variations between.  This phenomenon is easy to understand, as a bright RS Oph implies a more luminous white dwarf and accretion disk which translates into a greater heating (and earlier spectral class) for the hemisphere facing the white dwarf.  In this plot, I quantified the spectral class as K3 going to -3, K4 going to -2, K5 going to -1, M0 going to 0, M1 going to 1, and so on up to M5 going as 5.}
\end{figure}

\clearpage
\begin{figure}
\epsscale{1.0}
\plotone{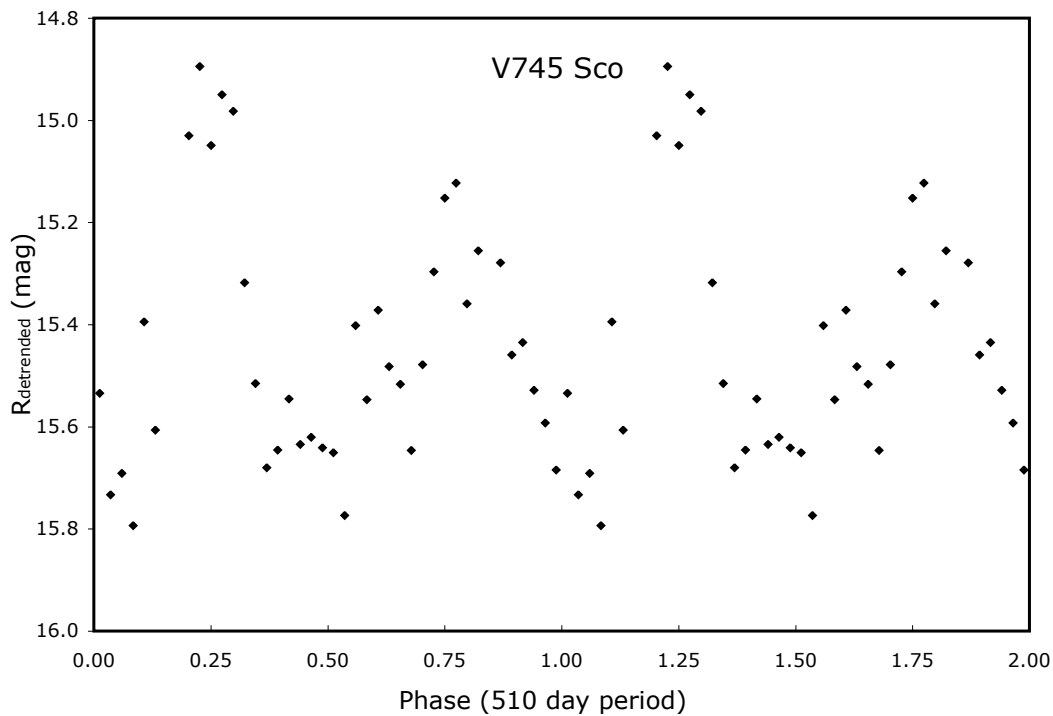}
\caption{
V745 Sco light curve.  The SMARTS telescopes in Chile have given 512 R-band magnitudes over 246 nights (see Table 22 and Schaefer 2009), with these forming nightly averages which were detrended with by normalizing by the yearly average, and then folded on the orbital period and binned in phase.  This light curve shows that the maximum at 0.25 phase is significantly brighter than the maximum at 0.75 phase.  The scatter in the light curve from phase 0.1-0.6 is relatively small, while the scatter for the remainder of the orbit is relatively high.  The minimum at 0.0 phase has all 16 individual magnitudes fainter than 15.9 mag, along with enough not-so-faint points so as to make the binned light curve have comparable depth minima.}
\end{figure}

\clearpage
\begin{figure}
\epsscale{1.0}
\plotone{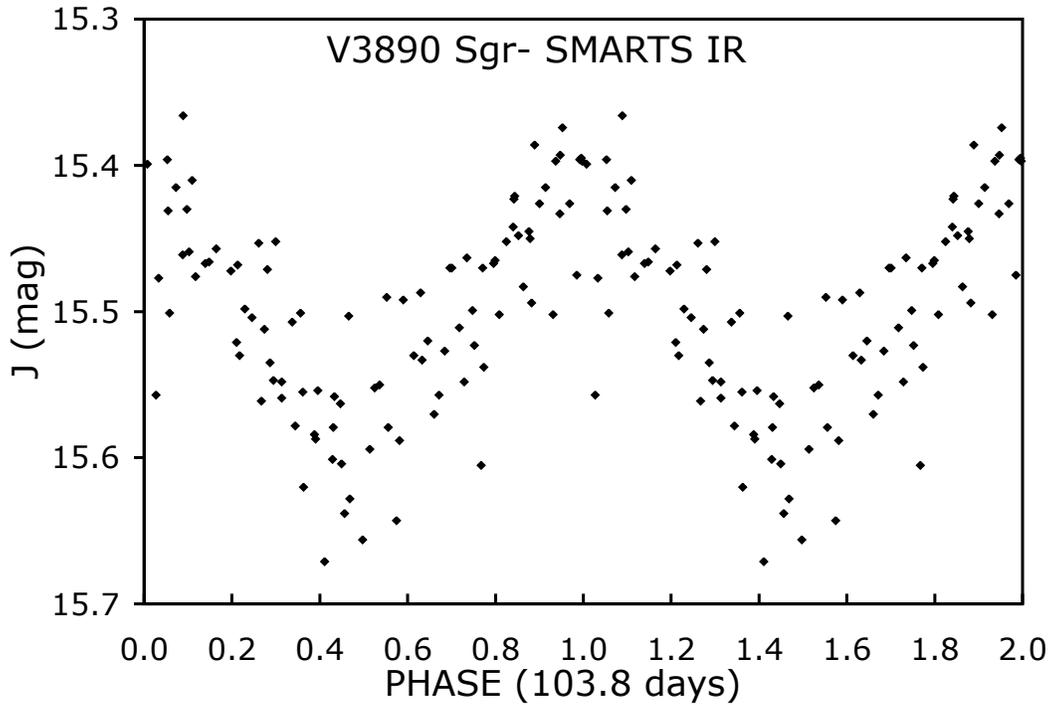}
\caption{
V3890 Sgr light curve folded on the red giant oscillation period.  This folded light curve has 105 J-band magnitudes from the SMARTS 1.3-m telescope in Chile, with each magnitude presented twice for phases 0-1 and 1-2.  The folding period is 103.8 days and the epoch for the zero phase is JD 2454760.  This period and amplitude is typical for oscillations on red giant stars.  Note that the points are partly separated into individual strings, with each string having relatively little scatter.  In this case, the overall scatter in this plot is due to cycle-to-cycle variations and not due to fast variations.}
\end{figure}

\clearpage
\begin{figure}
\epsscale{1.0}
\plotone{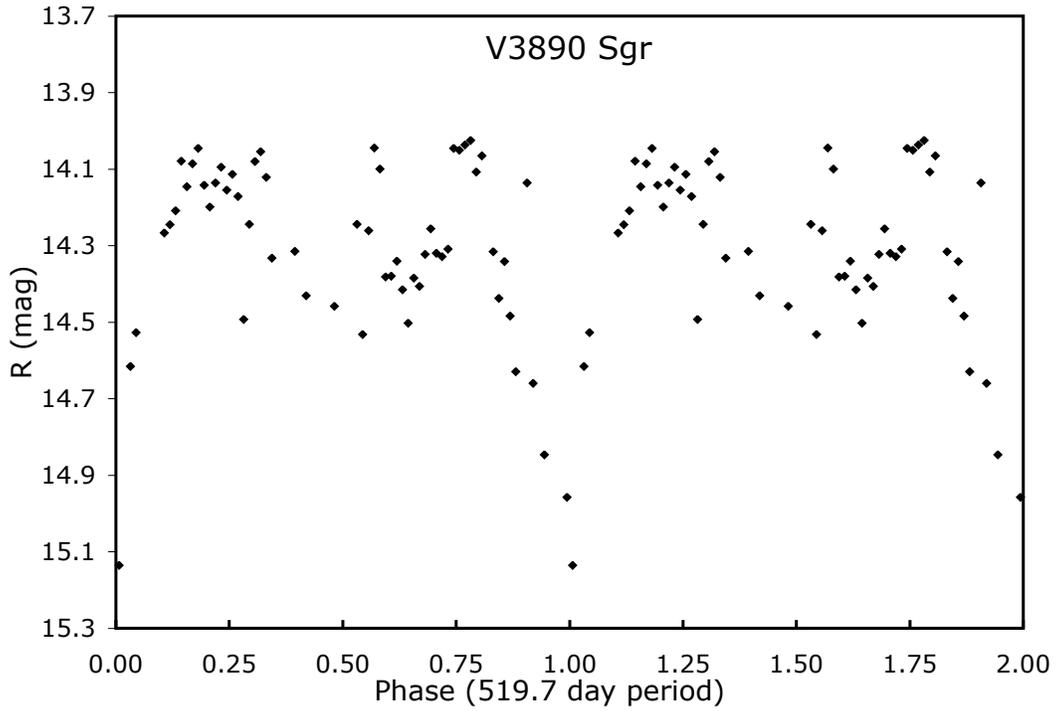}
\caption{
V3890 Sgr light curve folded on the orbital period.  This folded and binned light curve is based on 141 R-band magnitudes obtained with the SMARTS telescopes in Chile.  We see apparent ellipsoidal oscillations where one of the minimum is deeper than the other, likely due to a shallow eclipse of the accretion disk.}
\end{figure}

\clearpage
\begin{figure}
\epsscale{1.0}
\plottwo{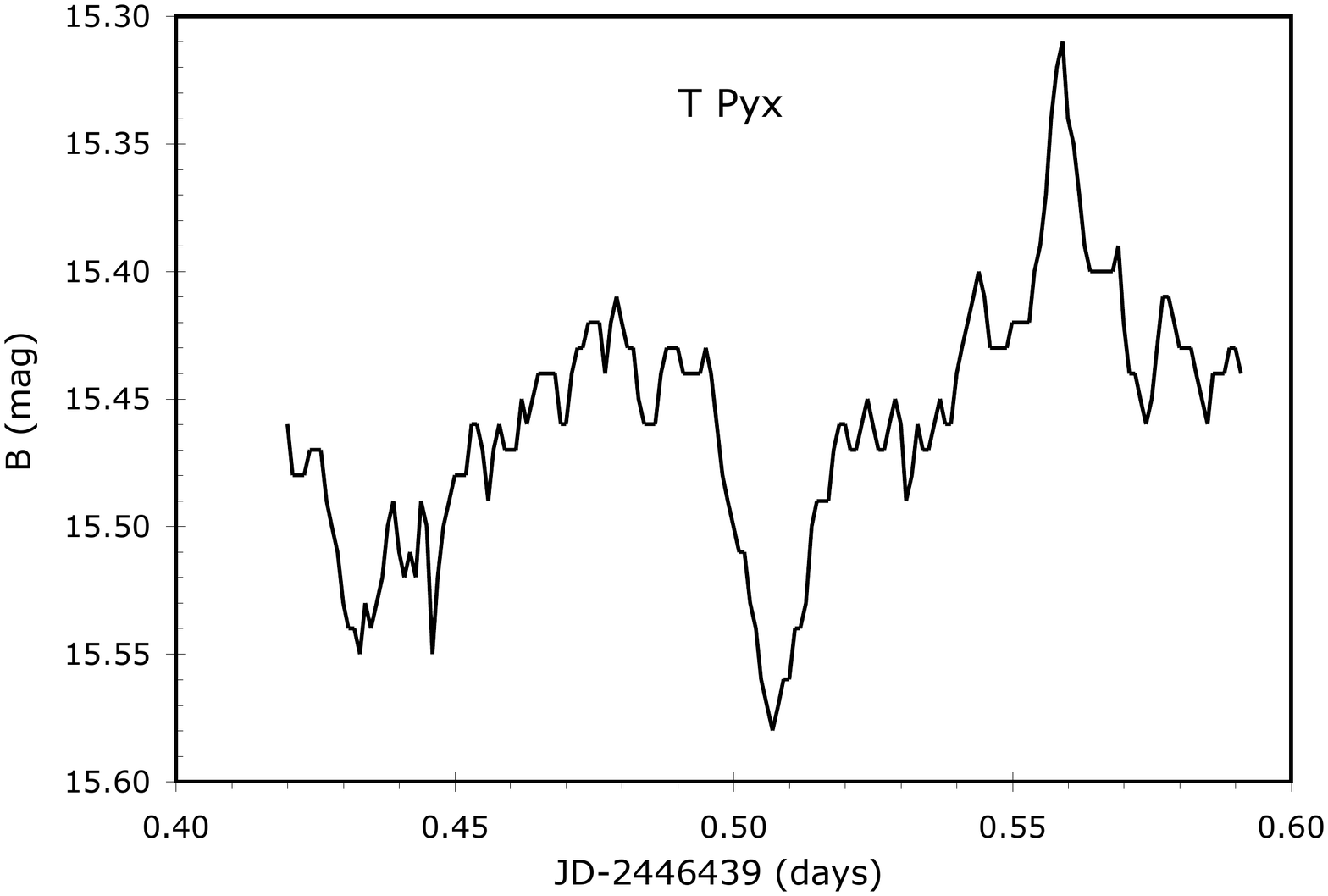}{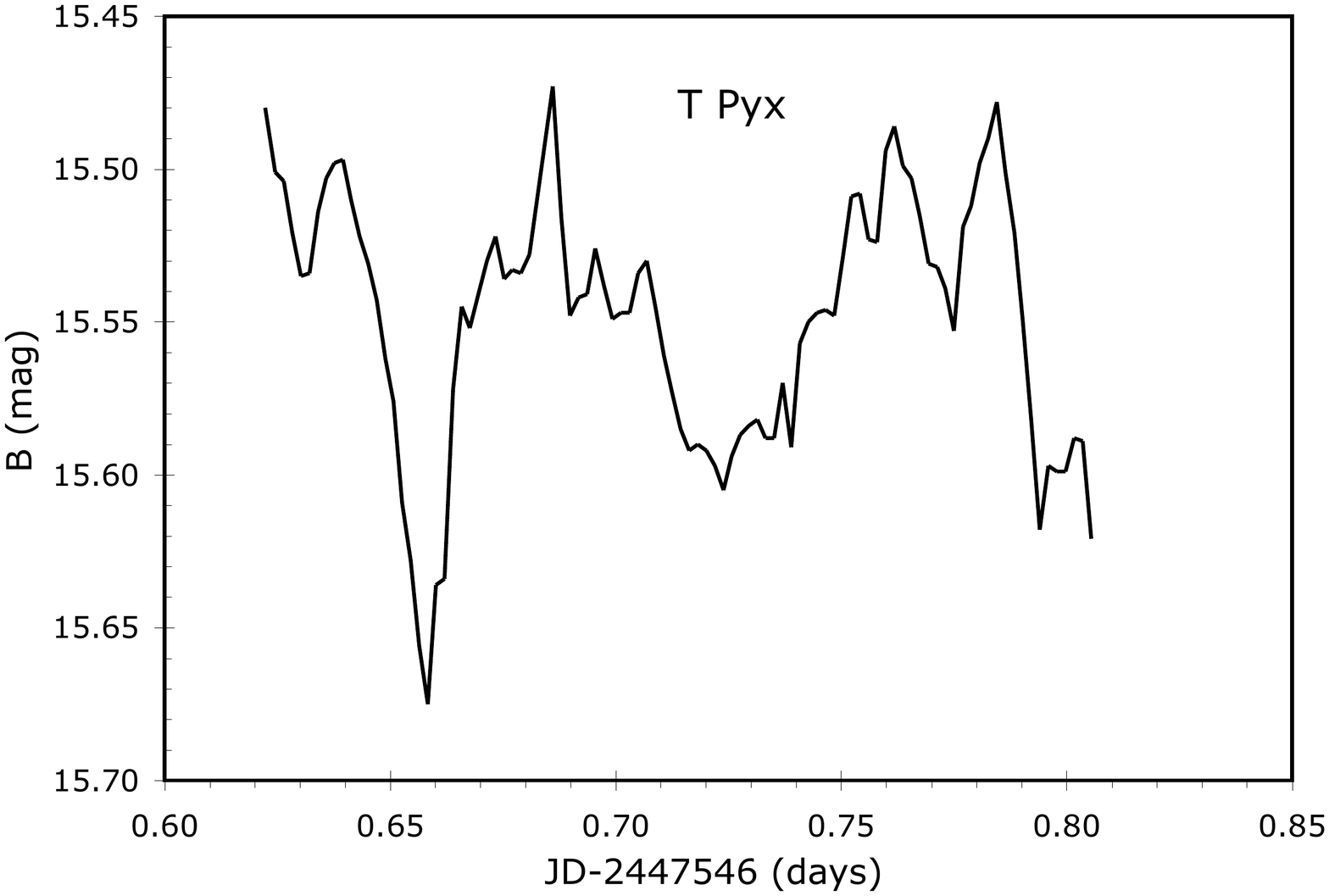}
\caption{
T Pyx flickering.  The two B-band light curves are from data tabulated in Schaefer et al. (1992).  The durations of the two time series are 246 and 264 minutes and covers around 2.3 orbital periods each.  This light curve is directly comparable with the average folded light curve displayed in Figure 42, which covers nearly the same time interval.  The non-averaged light curve displays much normal flickering large enough to generally mask the average shape of the periodic modulation.}
\end{figure}

\clearpage
\begin{figure}
\epsscale{1.0}
\plotone{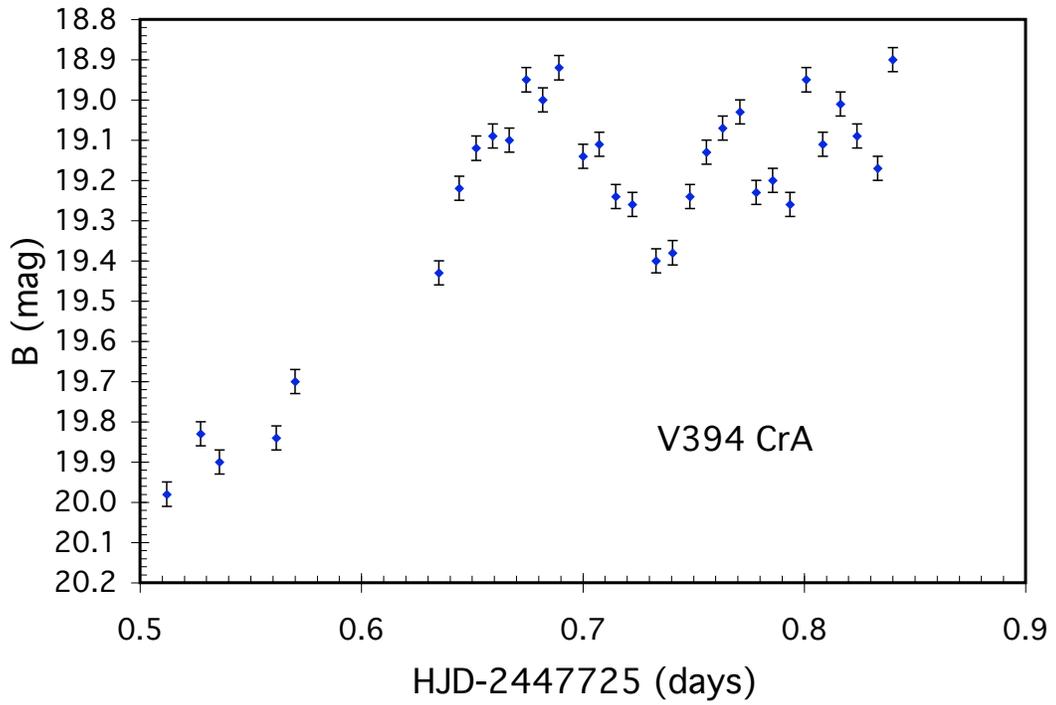}
\caption{
V394 CrA flickering.  This is the best of the existing time series for V394 CrA (taken from Table 22), and we see the usual flickering on time scales from minutes to hours and with amplitudes up to half a magnitude.}
\end{figure}

\clearpage
\begin{figure}
\epsscale{1.0}
\plottwo{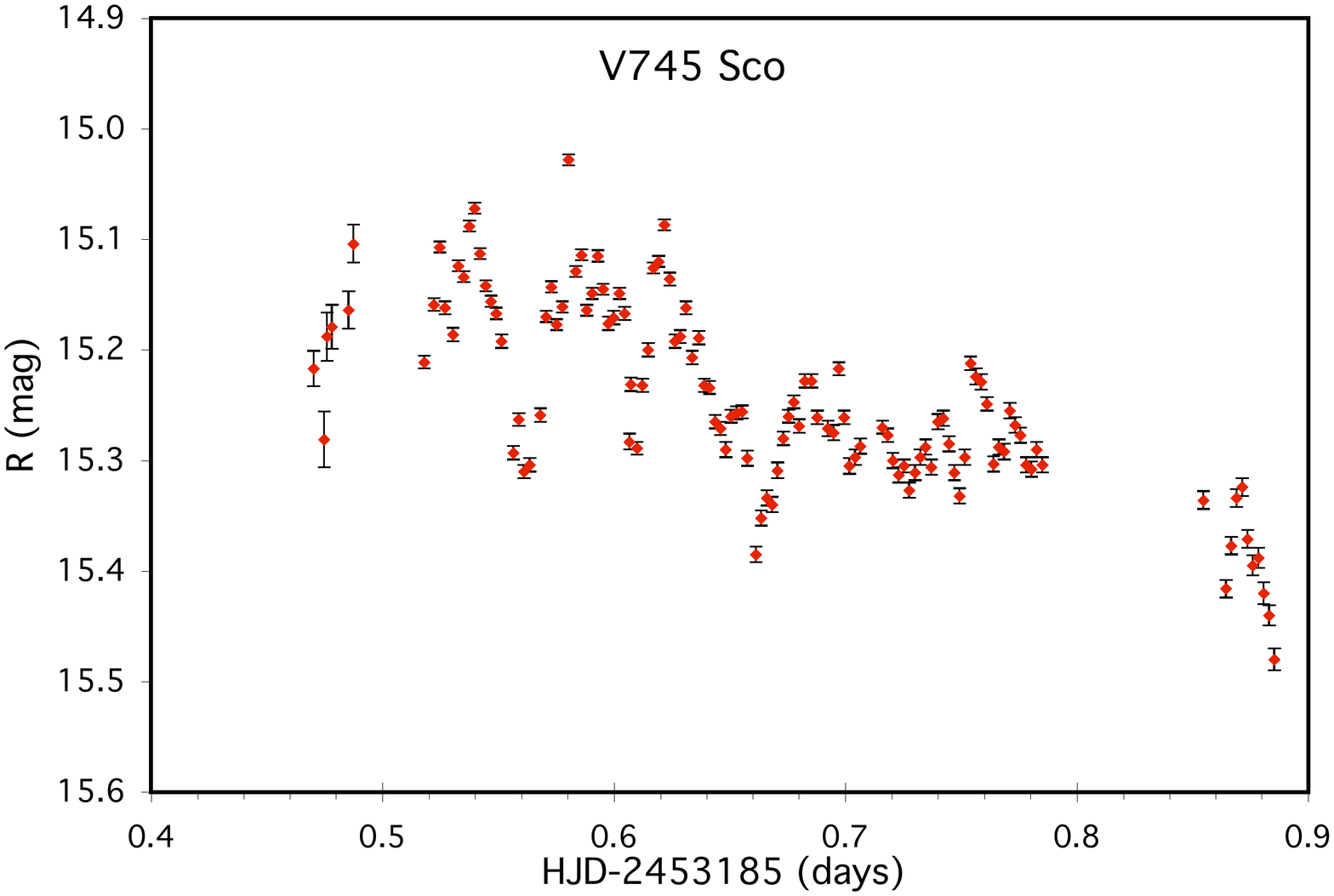}{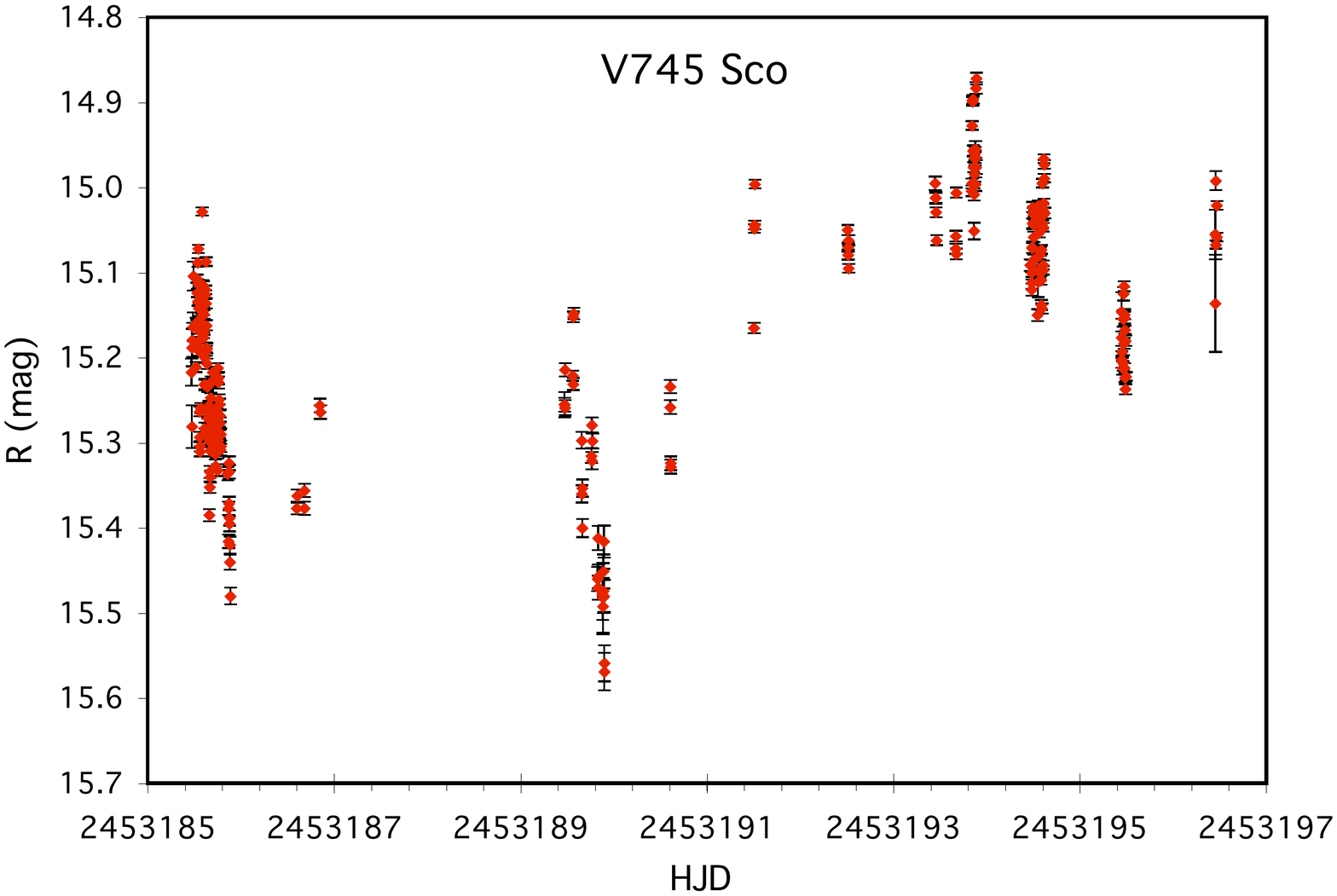}
\caption{
V745 Sco flickering.  This light curve in the R-band was taken on Cerro Tololo with the 1.0-m telescope in July 2004, with the left panel showing a close up of one night and the right panel showing ten nights.  We see the usual flickering, with time scales from tens of minutes up to at least 5 days.}
\end{figure}

\clearpage
\begin{figure}
\epsscale{1.0}
\plottwo{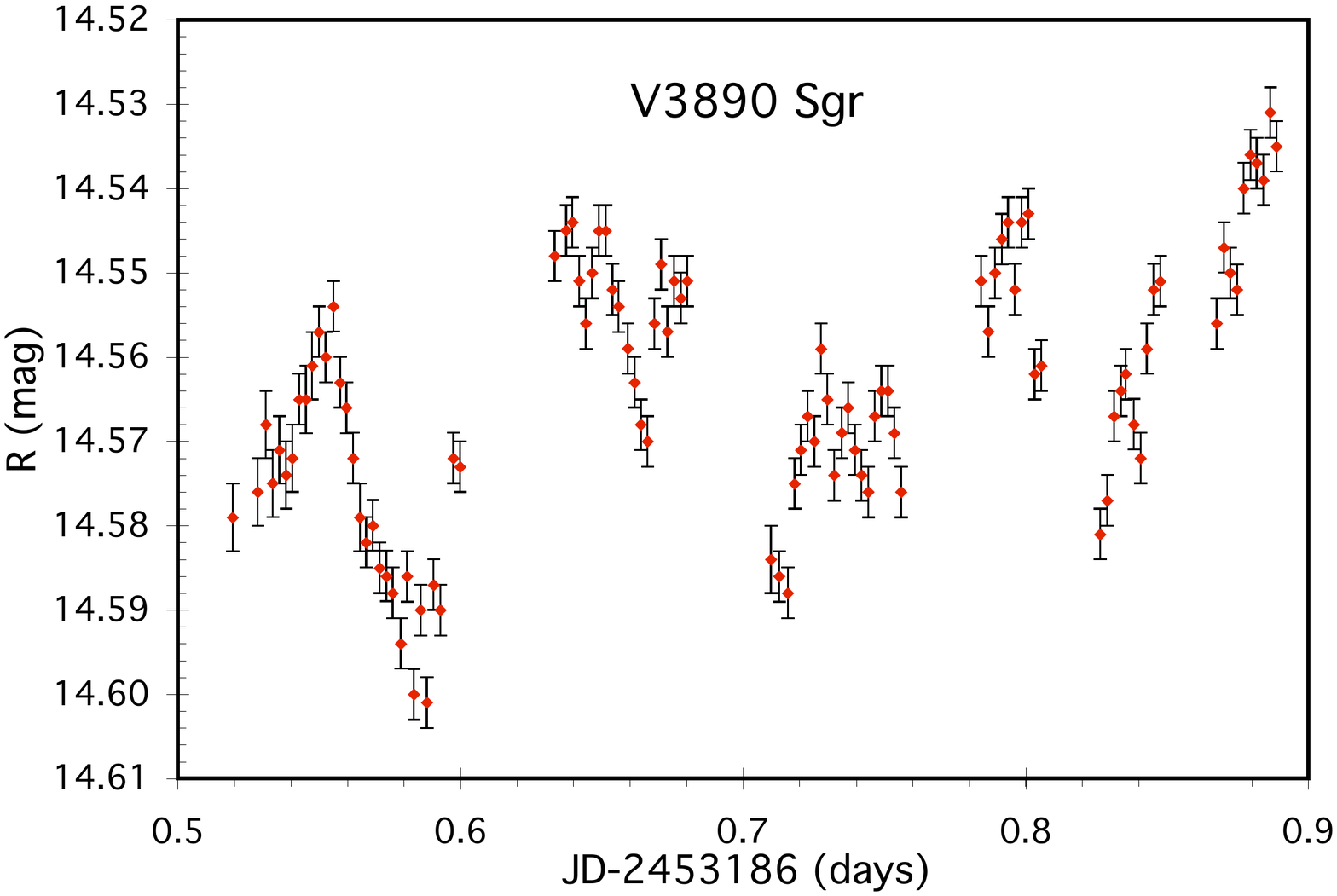}{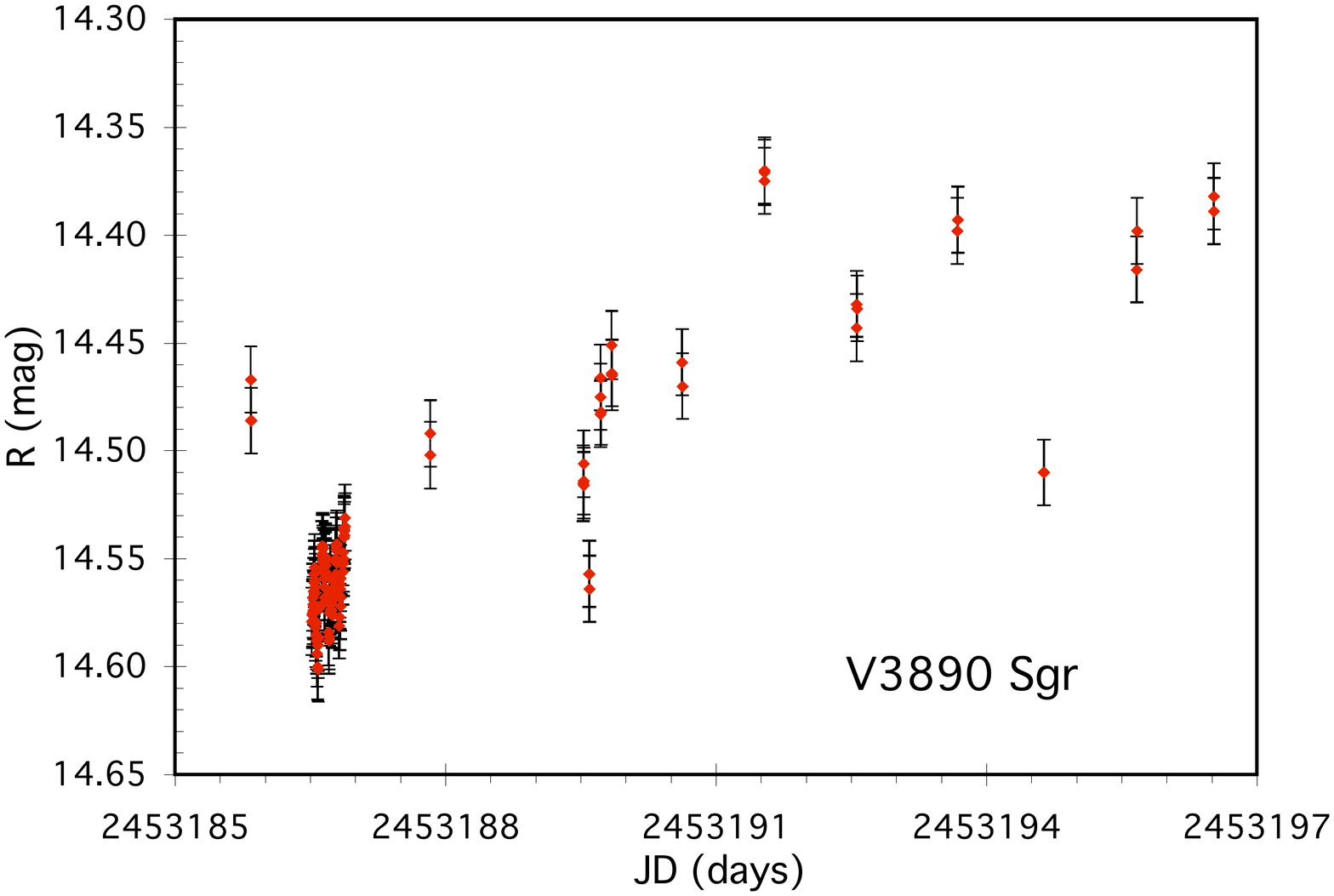}
\caption{
V3890 Sgr flickering.  This light curve in the R-band was taken on Cerro Tololo with the 1.0-m telescope in July 2004, with the left panel showing a close up of one night and the right panel showing eleven nights.  The times series on the one night shows ordinary flickering, with amplitudes of $\sim0.03$ mag and time scales of tens of minutes.  The right panel shows night-to-night changes of order 0.1 mag, superposed on a rising trend of 0.02 mag/day.}
\end{figure}

\clearpage
\begin{figure}
\epsscale{0.8}
\plotone{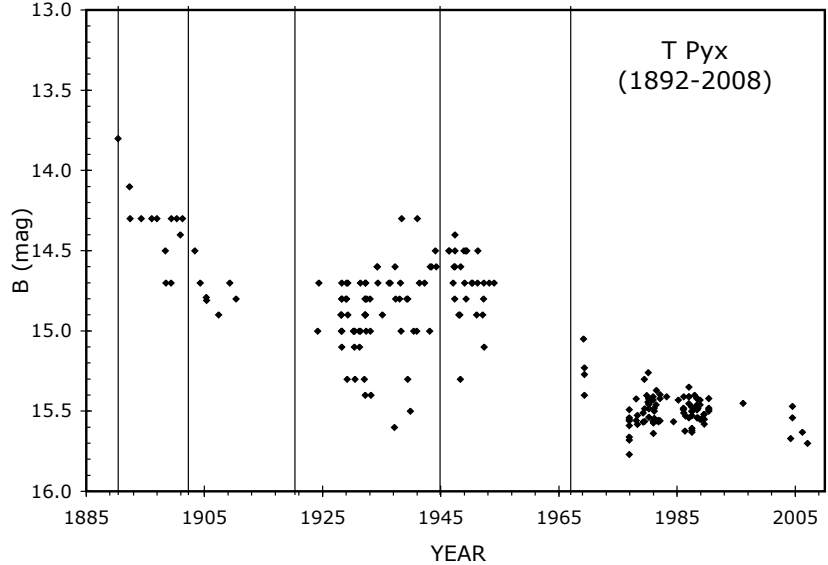}
\caption{
T Pyx in quiescence.  T Pyx has a well observed light curve from 1892 to present (116 years).  Each vertical line is for an eruption of T Pyx, with the eruption magnitudes not being displayed.  We see a light curve with substantial scatter representing the usual flickering.  But this flickering is superposed on highly significant long-term variations.  From 1892 to present, T Pyx has been systematically fading from 13.8 to 15.7 mag.  A variety of lessons can be learned from this graph:  (a) Novae can and do have large-amplitude variability from decades-to-century time scales.  Most novae do not have this quality of long-term light curve, so a reasonable supposition is that many novae also have comparable long-term variations.  With the brightness being dominated by the accretion disk, this light curve is a demonstration of the long-term changes in the rate of material being lost from the companion star.  (b) When T Pyx is bright we see short inter-eruption intervals, while when T Pyx is faint we see long inter-eruption intervals.  This is easily understood by the requirement of nova theory that the trigger occurs when some constant amount of mass has been accumulated, and T Pyx is bright when the mass transfer rate is high.  (c) We see why the long-anticipated eruption expected for $\sim$1988 is long overdue, with the reason being that the accretion rate was greatly lowered soon after the 1967 eruption so that it will take a much longer time to accumulate the critical mass on the surface of the white dwarf.  Schaefer (2005) has made a quantitative analysis to derive the next eruption date as $2052\pm3$ provided that the accretion continues at the same rate as in the last few decades.  In the last few years, T Pyx is apparently slowly fading, with the implication that the next eruption will be long after the year 2052.}
\end{figure}

\clearpage
\begin{figure}
\epsscale{0.8}
\plotone{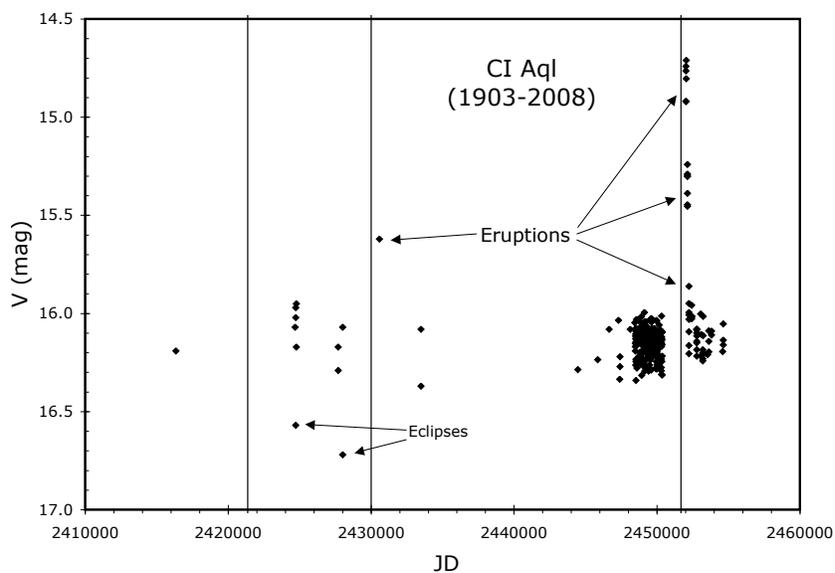}
\caption{
CI Aql in quiescence.  The plot is of V-band magnitudes (with B-band magnitudes converted with $B-V=1.03$ mag) taken from the Harvard plates, Schmidt sky surveys, Robocam, McDonald Observatory, and Cerro Tololo.  Eclipses are excluded except for the two old Harvard plates, while the tails of the 1941 and 2000 eruptions should be ignored.  What we are left with is a flat light curve during quiescence.}
\end{figure}

\clearpage
\begin{figure}
\epsscale{0.8}
\plotone{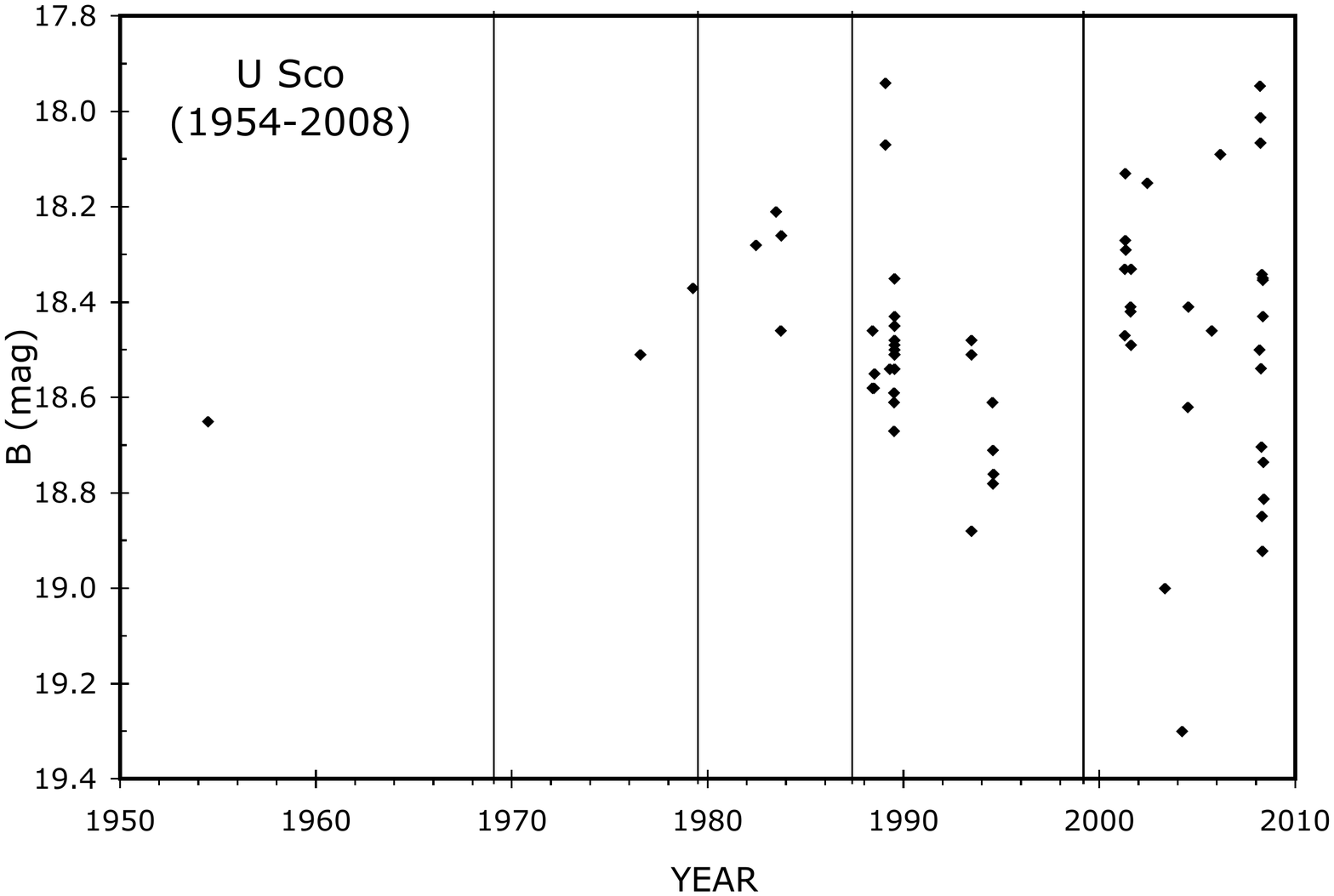}
\caption{
U Sco in quiescence.  We have a reasonable record of U Sco's brightness (outside of eclipse) in the blue band from my own CCD observations since 1988 plus deep Schmidt survey plates from earlier times.  We see that U Sco varies from around 18-19 mag.  This variation has no apparent connection with the times of the eruptions (indicated by the vertical lines).  Schaefer (2005) has used this information to show that the average magnitude is correlated with the inter-eruption interval, and also that the next eruption will be in the year $2009.3\pm1.0$.  This prediction is predicated on the average brightness after 2005 being the same as from 1999-2004.  My new data from 2005-2008 shows that this assumption is accurate to date, thus the predicted date (around spring of 2009) is still the best.}
\end{figure}

\clearpage
\begin{figure}
\epsscale{0.8}
\plotone{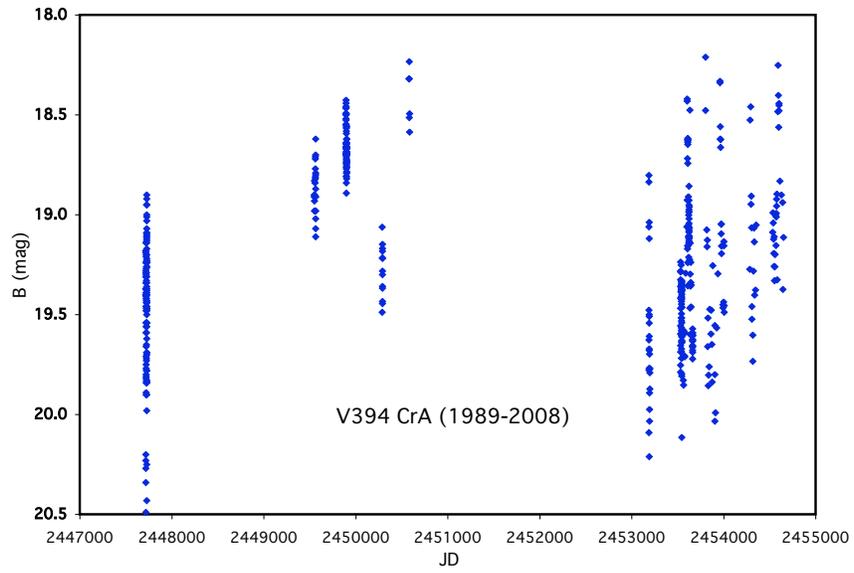}
\caption{
V394 CrA in quiescence.  This recurrent nova has a wide range of variations (from 18.3 to 20.5 mag) on all times scales (from hours to years).}
\end{figure}

\clearpage
\begin{figure}
\epsscale{0.8}
\plotone{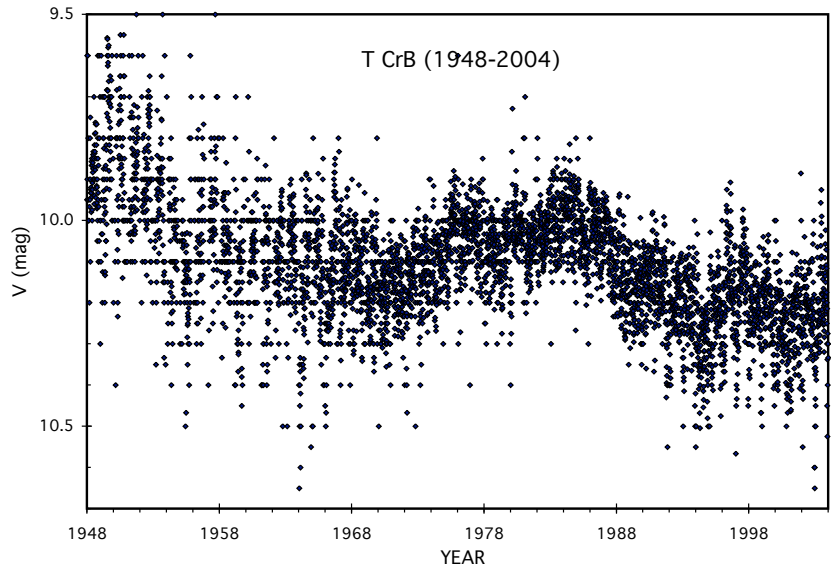}
\caption{
T CrB in quiescence.  This 1948-2004 light curve is from roughly 80,000 magnitudes from the AAVSO archives binned into 0.01 year intervals.  We see a complicated series of variations with decadal time scale and amplitude of roughly a quarter of a magnitude.}
\end{figure}

\clearpage
\begin{figure}
\epsscale{0.8}
\plotone{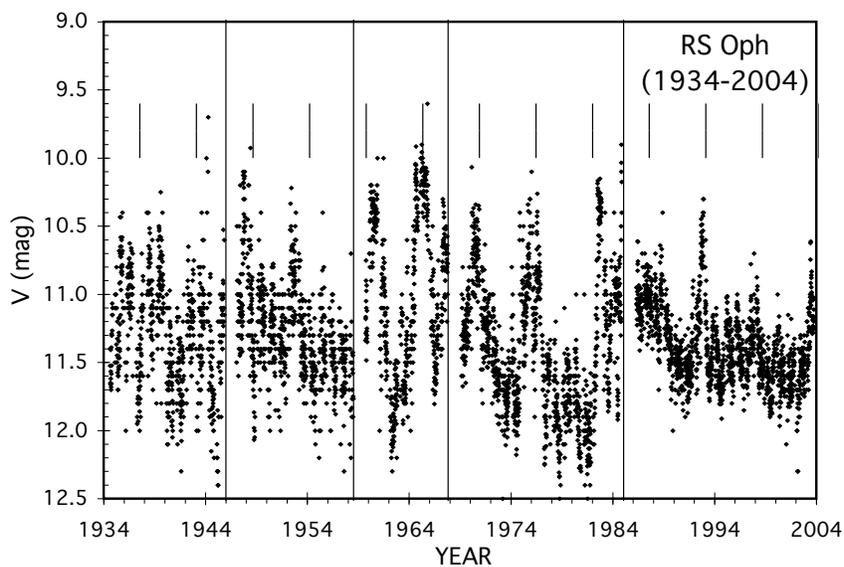}
\caption{
RS Oph in quiescence.  The AAVSO provides a wonderful light curve from 1934-2004 composed of 47,000 magnitudes (in 0.01 year time bins).  The eruptions (indicated in time by the long vertical lines) and the post-eruption dips have been clipped out so we can only see the changes of the system in quiescence.  We see a messy light curve, with flares on all time scales, and decadal episodes of flares and relative calm.  A Fourier transform of this light curve shows a peak at 2030 days, with the putative maxima indicated by the short vertical lines along the top.  This apparent periodicity is only a result of the crude alignment between a handful of flares, and indeed a detailed quantitative analysis shows that for flares as seen for RS Oph with random times we will always view a false periodicity as good as seen in the real light curve.   That is, the periodicity is not significant.}
\end{figure}

\clearpage
\begin{figure}
\epsscale{0.8}
\plotone{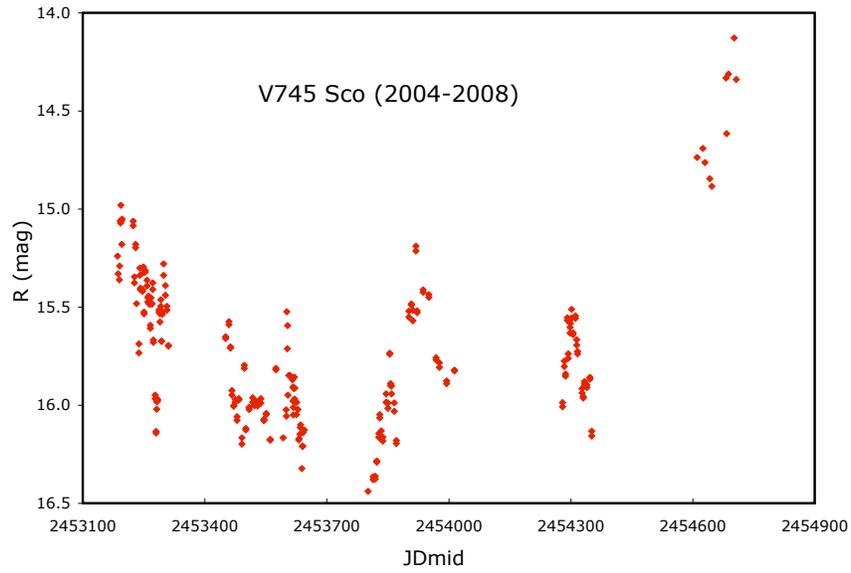}
\caption{
V745 Sco in quiescence.  The only long-term information on V745 Sco comes from my R-band CCD magnitudes from Cerro Tololo, covering 2004-2008 (with observations on-going).  We see that the RN varies on all time scales from weeks to months to years, and recall from Figure 56 that it also varies on time scales of minutes and hours and days.  The range of variations is from 14.1 to 16.5, almost a factor of ten in luminosity.  In comparing this figure with the two panels of Figure 56, I am impressed by the similarity in the light curve shapes with greatly different scales.}
\end{figure}

\clearpage
\begin{figure}
\epsscale{0.8}
\plotone{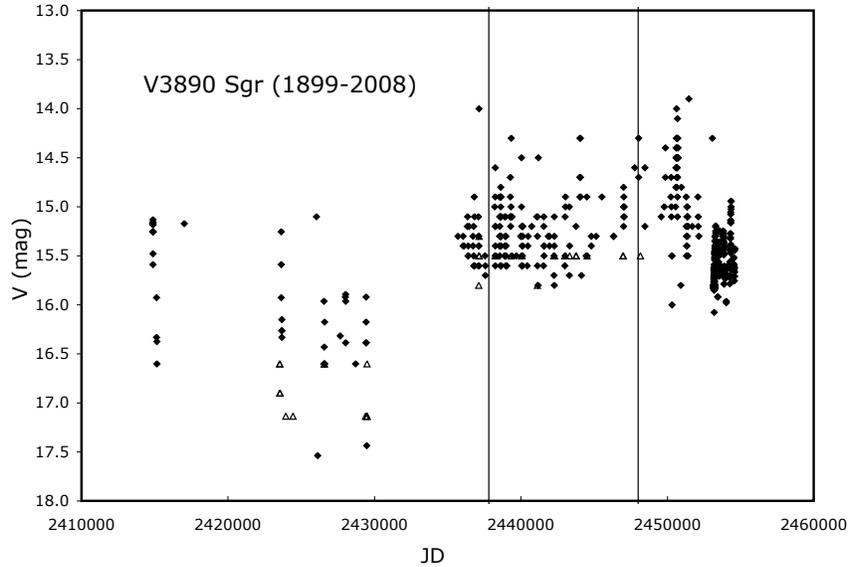}
\caption{
V3890 Sgr in quiescence.  This long-term light curve is constructed from magnitudes corrected to the V-band from the Harvard plates, the Maria Mitchell plates, the AAVSO, and from my SMARTS CCD measures in order from left to right.  The open triangles represent upper limits on the brightness, while the vertical lines represent the times of the two known eruptions.  A substantial problem with seeking secular changes is that the first three data sets have their detection threshold cutting off the distribution of magnitudes.  But the correction for these truncation effects will only increase the amplitude of variations for the first three data sets, and the large range of variations are already inconsistent with the relatively small range seen in the SMARTS data in recent years.  I think that this inconsistency is caused by the amplitude of variations being smaller in the red than in the blue.  I see no confident pre-eruption rises (or dips) nor any secular trends in the brightness following eruptions.}
\end{figure}

\clearpage
\begin{figure}
\epsscale{0.8}
\plotone{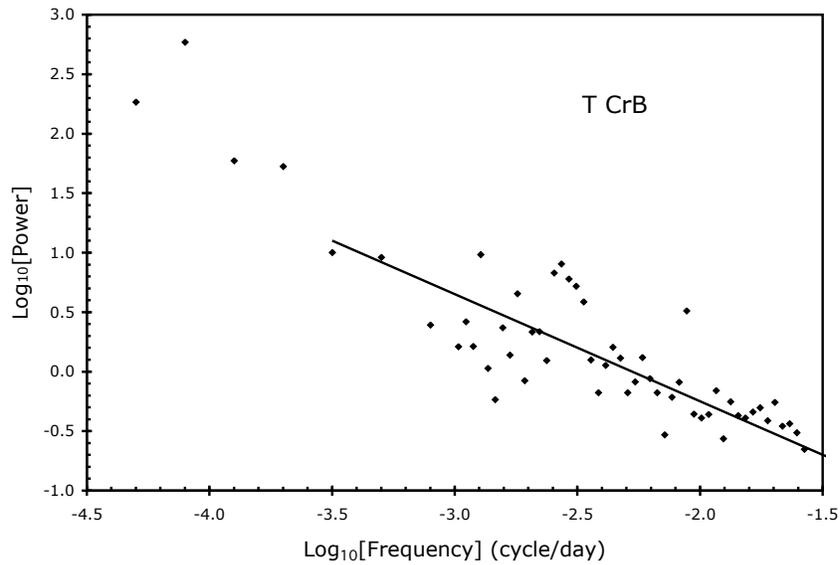}
\caption{
T CrB power density spectrum.  The AAVSO has a large data base with essentially no gaps from 1947 to present, and this can be used create a uniformly-sampled light curve and then a PDS.  This allows for the unique PDS extending to frequencies lower by four orders-of-magnitude than any other non-RN cataclysmic variable.  We see that the PDS has a roughly power law shape extending to low frequencies.  (The lowest frequency slope is likely not significant because there are few cycles in the data for these frequencies and so random variations in the light curve will result in deviations from the true PDS.)  The slope here is -0.9, while the slope for frequencies $>10$ cycles per day is -1.46 on average.  This is saying that T CrB shows significant variability on time scales from seconds to many decades.}
\end{figure}

\clearpage
\begin{figure}
\epsscale{0.8}
\plotone{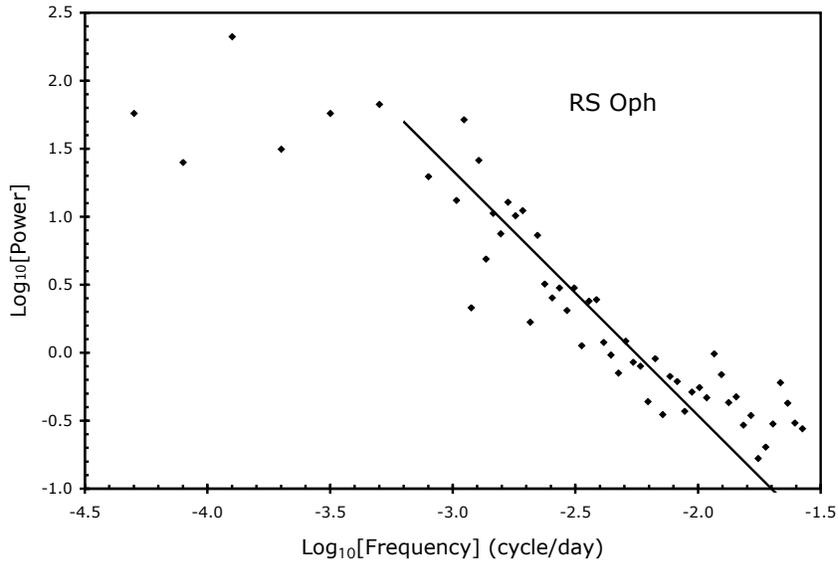}
\caption{
RS Oph power density spectrum.  The AAVSO light curve from 1934 to present had solar and eruption gaps filled so as to produce a uniformly-sampled light curve and then a PDS.  The shape of the PDS is characteristic of other cataclysmic variables ($1/f$ noise perhaps with a break to white noise at the lowest frequencies) except that the power keeps rising to incredibly low frequencies.  The break to white noise might well be insignificant as being simply one realization of the light curve producing deviations from some long-term average PDS shape.  The shape of the PDS is also characteristic of ordinary variations on red giant stars likely caused by convection cells of all sizes.}
\end{figure}

\clearpage
\begin{figure}
\epsscale{0.8}
\plotone{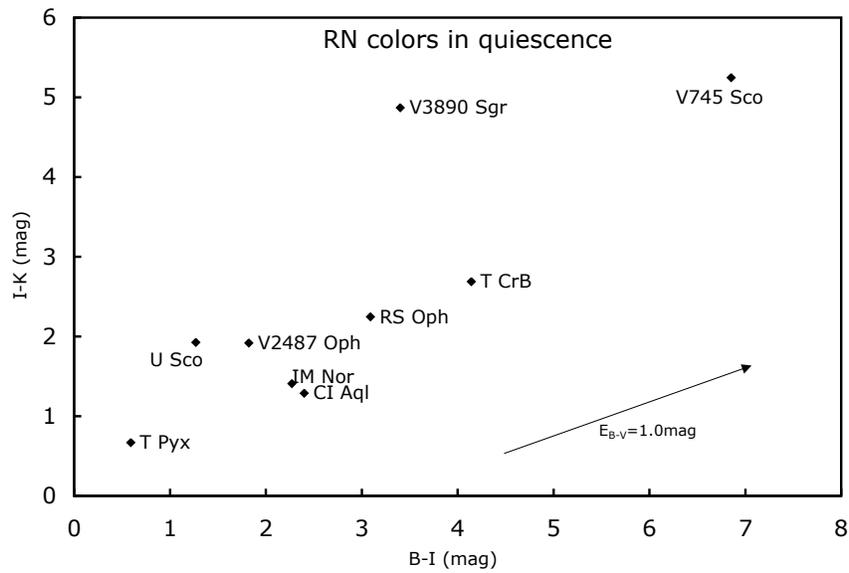}
\caption{
RN colors in quiescence.  The B-I and I-K colors were chosen as they have the broadest wavelength range, and hence the greatest sensitivity.  V394 CrA (with B-I=1.59 mag) is not included as it has never been detected in the JHK bands, so I-K is not known.  V745 Sco is very red (due to its red giant plus a heavy extinction) and has the remarkably large B-K color or 12.1 mag.  The arrow shows the change associated with normal extinction for $E_{B-V}=1.0$ mag.}
\end{figure}

\clearpage
\begin{figure}
\epsscale{0.8}
\plotone{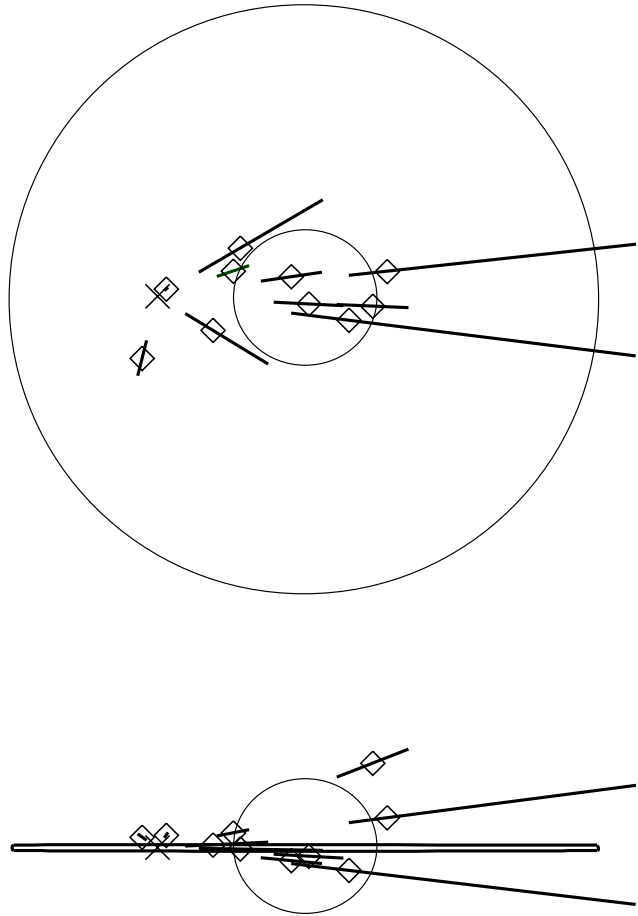}
\caption{
Galactic distribution of RNe.  The two panels show a top and side view of our Milky Way galaxy, represented as a flat disk of radius 15,000 pc and thickness 300 pc to indicate the disk population plus a circle of radius 3,000 pc to indicate the bulge population.  Our Sun's position is represented by the large `X'.  Each RN is represented in position by a large diamond, with the uncertainty in distance indicated by the line segments attached to each diamond.  We see that three of the systems have a large distance from the galactic plane, with U Sco being placed above the galactic center and both V394 CrA and V2487 Oph being on the far side of the galaxy with large uncertainties in distance.  These three RNe are likely part of a bugle population.  The other seven RNe are fairly well concentrated towards the plane like a thick disk population.  With small number statistics, it appears that the RNe are roughly evenly divided between being from thick disk and bulge populations.}
\end{figure}

\clearpage
\begin{figure}
\epsscale{0.8}
\plotone{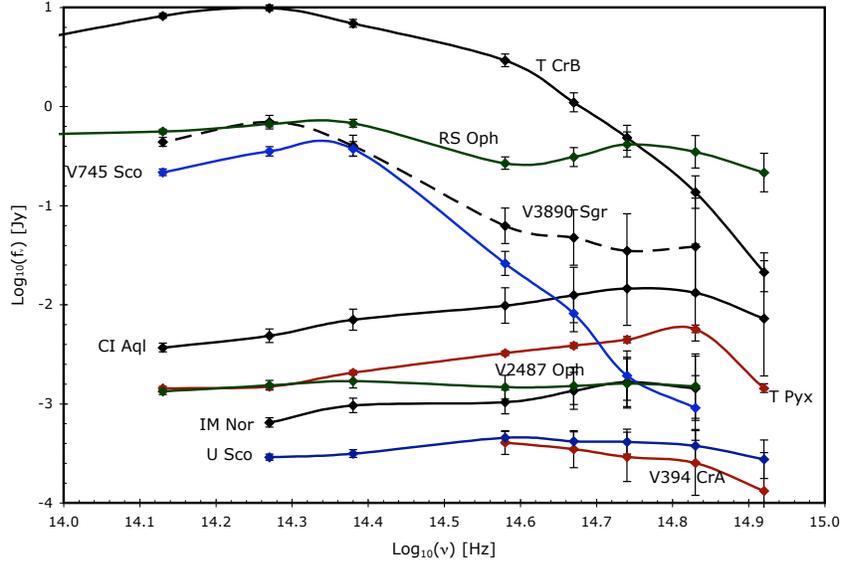}
\caption{
RNe spectral energy distributions.  These are effectively low-resolution spectra of ten RNe, with the flux density (in jansky) as a function of the light frequency (in hertz) on a log-log plot.  The error bars are dominated by the uncertainty in the extinction, and are correlated from point-to-point for a given RN, so that a curve following the upper end of the error bars represents the lowest extinction case and the curve that follows the lower end of the error bars represents the  highest extinction case.  For V3890 Sgr, V745 Sco, T CrB, and RS Oph, we see a broad peak around $\log (\nu)$ from 14.3-14.4 (corresponding to temperatures of $\sim3,800\degr$) produced by the early M giant companion star.  For T Pyx, IM Nor, CI Aql, V2487 Oph, and U Sco, we see the expected $f_{\nu} \propto \nu^{1/3}$ behavior for an accretion disk.}
\end{figure}

\clearpage
\begin{figure}
\epsscale{0.8}
\plotone{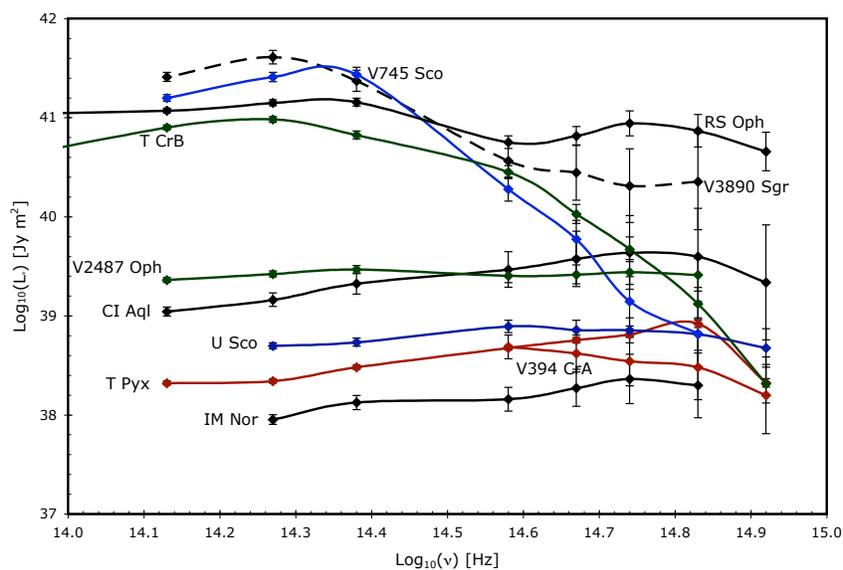}
\caption{
RNe spectral energy distributions corrected for distances.  The observed spectral energy distributions from Figure 70 have been corrected for distances, to get the spectral luminosity distributions.  We see that four systems (T CrB, RS Oph, V745 Sco, and V3890 Sgr) all have similar luminosities for their red giant companion stars, while four of the systems (T Pyx, IM Nor, V394 CrA, and U Sco) all have similar luminosities for their accretion disks.  But we are then left with trying to understand why CI Aql and V2487 Oph are an order of magnitude brighter than the other systems dominated by accretion disks.}
\end{figure}

\end{document}